\documentclass[twocolumn]{aastex7}

\received{November 27, 2024}
\revised{August 19, 2025}
\accepted{August 20, 2025}

\submitjournal{AAS Journals}

\shorttitle{Terrestrial Exoplanet Internal Structure Constraints}
\shortauthors{Ross et al.}

\begin{document}

\title{Terrestrial Exoplanet Internal Structure Constraints Enabled by
Comprehensive Host Star Characterization Reveal that Terrestrial Planets
in Mean-motion Resonances are Water Rich}

\correspondingauthor{Alejandra Ross}

\author[0009-0000-2200-131X]{Alejandra Ross}
\affiliation{William H.\ Miller III Department of Physics \& Astronomy,
Johns Hopkins University, 3400 N Charles St, Baltimore, MD 21218, USA}
\email[show]{alejross12@gmail.com}

\author[0000-0001-6533-6179]{Henrique Reggiani}
\affiliation{Gemini South, Gemini Observatory, NSF's NOIRLab, Casilla 603,
La Serena, Chile}
\email{henrique.reggiani@noirlab.edu}

\author[0000-0001-5761-6779]{Kevin C.\ Schlaufman}
\affiliation{William H.\ Miller III Department of Physics \& Astronomy,
Johns Hopkins University, 3400 N Charles St, Baltimore, MD 21218, USA}
\email{kschlaufman@jhu.edu}

\author[0000-0002-9479-2744]{Mykhaylo Plotnykov}
\affiliation{Department of Physics, University of Toronto, 27 King's
College Cir, Toronto, ON M5S, Canada}
\email{mykhaylo.plotnykov@mail.utoronto.ca}

\author[0000-0003-3993-4030]{Diana Valencia}
\affiliation{Department of Physics, University of Toronto, 27 King's
College Cir, Toronto, ON M5S, Canada}
\email{diana.valencia@utoronto.ca}

\begin{abstract}

\noindent
Exoplanet mass and radius inferences fundamentally rely on host star
mass and radius inferences.  Despite the importance of host star mass,
radius, and elemental abundance inferences for the derivation of exoplanet
internal structure constraints, published constraints have often been
based on inferences that are not self-consistent.  For 24 dwarf stars
hosting terrestrial exoplanets, we use astrometric and photometric data
plus high-resolution spectroscopy to infer accurate, precise, homogeneous,
and physically self-consistent photospheric and fundamental stellar
parameters as well as elemental abundances.  We infer updated planetary
masses and radii using these data plus Doppler and transit observables,
then use the complete data set to derive constraints on the core-mass
fractions of these terrestrial exoplanets.  We find that the population of
resonant or likely formerly resonant terrestrial exoplanets represented
by Kepler-36 b and Kepler-105 c has a significantly lower mean core-mass
fraction than the rest of the terrestrial exoplanets in our sample.
Their resonant configurations suggest that they migrated inwards from
more distant formation locations, and we attribute their low densities
to the incorporation and retention of significant amounts of water during
their formation.  We confirm that the ultra-short-period exoplanets 55 Cnc
e and WASP-47 e have densities inconsistent with pure-rock compositions.
We propose that they are both the stripped cores of mini-Neptunes and
associate their low densities with the presence of significant amounts
of hydrogen, helium, water, and/or other volatiles in their interiors.
We verify that our results are independent of stellar parameter and
elemental abundance inference approach and therefore robust.

\end{abstract}


\section{Introduction}\label{sec:intro}

The planet formation process leaves its mark on the densities and
internal structures of terrestrial planets.  In our own solar system,
Mercury's high density and large core mass fraction have long been used
to argue that it formed differently than Venus, Earth, and Mars, possibly
from one or more giant impacts that led to the stripping of its mantle
\citep[e.g.,][]{Smith1979, Benz1988, Benz2007, Asphaug2014, Ebel2018,
Clement2019, Clement2021a, Clement2021b, Clement2023, Clement2021c}.
The low density and tiny core of Earth's Moon support the idea that
it formed from the ejecta of a collision between the young Earth and
a Mars-mass impactor \citep[e.g.,][]{Canup2001, Canup2004, Canup2008}.
While all of the terrestrial planets in the solar system formed inside
the protosolar nebula's water-ice line, low-density terrestrial exoplanets
formed closer-to or outside their parent protoplanetary disks' water-ice
lines can in principle incorporate significant amounts of water in
their interiors \citep[e.g.,][]{Emsenhuber2021a, Emsenhuber2021b,
Emsenhuber2023, Schlecker2021}.

While terrestrial planet masses, radii, and internal structures can be
directly measured in the solar system, inferences of these quantities for
terrestrial exoplanets first require accurate, precise, homogeneous,
and physically self consistent inferences of the masses, radii,
and compositions of their host stars \citep[e.g.,][]{Plotnykov2024}.
Virtually all Earth-radius exoplanets are discovered with the transit
technique and the observable depth of a transit $\delta \propto
(R_{\text{p}}/R_{\ast})^2$, where $R_{\text{p}}$ and $R_{\ast}$ are
planet and host star radius.  Model-dependent stellar radius inferences
can then be used to infer $R_{\text{p}}$.  Likewise, if $\sin{i}
\approx 1$ as expected for transiting systems the Doppler technique
can be used to measure $M_{\text{p}}/(M_{\text{p}}+M_{\ast}) \approx
M_{\text{p}}/M_{\ast}$ where $M_{\text{p}}$ and $M_{\ast}$ are planet and
host star mass.  Model-dependent stellar mass inferences can then be used
to infer $M_{\text{p}}$.  Because (1) the bulk Earth has roughly the same
refractory abundances observed in the solar photosphere and measured in
meteorites \citep[e.g.,][]{Asplund2021} and (2) the elemental abundances
of terrestrial exoplanets cannot be directly measured, it is often
assumed that the abundances of the refractory elements in a terrestrial
exoplanet roughly correspond to the refractory abundances of its host star
\citep[e.g.,][]{Thiabaud2015, Dorn2017}.\footnote{\citet{Plotnykov2020}
have questioned this assumption for presumed terrestrial exoplanets in
a population sense.}

Interior structure constraints for the population of terrestrial
exoplanets orbiting solar-type dwarf stars have been based on planetary
masses and radii \citep[e.g.,][]{Unterborn2016, Ligi2019, Adibekyan2021,
Unterborn2023}, host star photospheric abundances as a proxy for
planetary composition \citep[e.g.,][]{Santos2015}, or all of these
\citep[e.g.,][]{Dorn2015, Dorn2017, Dorn2019, Plotnykov2020, Liu2020,
Wang2022, Zhao2023, Zhao2024, Haldemann2024, Brinkman2024, Weeks2025}.
\citet{Adibekyan2021} reported a significant correlation between
mass--radius relation-based core-mass fraction and host star metallicity
as well as evidence for a distinct population of super-Mercury terrestrial
exoplanets.  This existence of a significant population of super-Mercury
terrestrial exoplanets comprising approximately 20\% of the terrestrial
exoplanet population was also reported by \citet{Unterborn2023}.
Alternatively, \citet{Wang2022} found no evidence for super-Mercury type
terrestrial exoplanets and instead concluded that with the exception of
the strongly oxidized and therefore low-density exoplanets Kepler-10 b
and Kepler-37 b/c most terrestrial exoplanets have Earth-like core-mass
fractions.

Some of these previous studies had limitations that prevented them from
achieving the best possible constraints on the diversity of terrestrial
exoplanet internal structures.  Many inferred host star abundances using
photospheric stellar parameters that differed from those used to infer the
host star masses and radii needed to turn Doppler and transit observables
into planetary masses and radii \citep[e.g.,][]{Dorn2017, Dorn2019,
Adibekyan2021, Unterborn2023, Zhao2023, Zhao2024}.  This systematic can
increase the apparent dispersion in the core-mass fraction distribution
and/or bias core-mass fraction inferences.  Either of these effects can
distort our understanding of the true diversity of terrestrial exoplanet
internal structures.  Another limitation of previous studies has been
their focus on internal planetary structures in isolation, divorced from
the larger context of the exoplanet system in which those exoplanets
are found.  This disconnect can make emerging subtle relationships
between terrestrial exoplanet internal structure and system architecture
impossible to discern.

In this article, we study the distribution of terrestrial exoplanet
internal structure using our own accurate, precise, homogeneous, and
physically self-consistent stellar parameters to infer updated planetary
parameters.  We find in some cases evidence of a correlation or linear
relationship between host star metallicity and core-mass fraction, though
that evidence only appears in data from one of the four stellar parameter
sources we investigate.  We find that the average core-mass fraction
in our sample of terrestrial exoplanets with precise mass and radius
inferences is consistent with the Earth's core-mass fraction.  We identify
no evidence for a significant population of high core-mass fraction
super-Mercury exoplanets in our sample, but do discern two outlying low
core-mass fraction populations: (1) terrestrial exoplanets in or likely
formerly in mean-motion resonances and (2) massive ultra-short-period
(USP) terrestrial exoplanets orbiting very metal-rich stars with orbital
periods $P < 1$ day in systems with giant exoplanet companions within
about 0.1 AU.  We describe the construction of our samples in Section
\ref{sec:sample_selection} and our analyses in Section \ref{sec:analysis}.
We discuss the implications of our work in Section \ref{sec:discussion}
and summarize our conclusions in Section \ref{sec:conclusion}.

\section{Sample Selection}\label{sec:sample_selection}

We initialize our sample selection process with the \citet{Plotnykov2020}
list of candidate terrestrial exoplanets with mass and radius
uncertainties smaller than 25\%.  Because the most accurate and
precise exoplanet host star parameters can only be obtained for dwarf
stars with photospheric temperatures in the range $4500~\text{K}
\lesssim T \lesssim 6200~\text{K}$, we limit our sample to those
exoplanets orbiting stars in the color range $0.72 < (G_{\text{BP}}
- G_{\text{RP}})_{0} < 1.43$.  For this selection, we use Gaia Data
Release (DR2) $G_{\text{BP}} - G_{\text{RP}}$ colors, the STructuring by
Inversion the Local Interstellar Medium \citep[Stilism;][]{Stilism2014,
Stilism2018, Stilism2017} three-dimensional reddening maps, and
\citet{Casagrande2018} reddening coefficients.  The color range $0.72
< (G_{\text{BP}} - G_{\text{RP}})_{0} < 1.43$ corresponds to the
effective temperature range $4440~\text{K} \lesssim T_{\text{eff}}
\lesssim 6050~\text{K}$ or to spectral types between F9V and K5V
\citep{Pecaut2013}.\footnote{\url{https://www.pas.rochester.edu/~emamajek/EEM\_dwarf\_UBVIJHK\_colors\_Teff.txt}}
It is very challenging to infer photospheric elemental abundances for
stars outside this range.  For example, warmer stars generally rotate
quickly causing line profiles to overlap, while cooler stars evince strong
molecular absorption that makes setting the continuum level impossible.
In either case, the equivalent width measurements that provide the
observational data for photospheric elemental abundance inferences become
quite challenging.

For the stars that meet our dereddened color cut, we then search
for publicly available archival high-resolution optical spectra with
signal-to-noise ratio $\text{S/N} \gtrsim 50$ per pixel at $\lambda
\approx 6000$ \AA.  From the European Southern Observatory (ESO)
Science Archive Facility, we select spectra collected with the Echelle
SPectrograph for Rocky Exoplanets and Stable Spectroscopic Observations
\citep[ESPRESSO;][]{ESPRESSO2021}, the Fiber-fed Extended Range
Optical Spectrograph \citep[FEROS;][]{FEROS1999}, and the Ultraviolet
and Visual Echelle Spectrograph \citep[UVES;][]{UVES2000}.  From the
Canadian Astronomical Data Centre \citep[CADC;][]{CADC1994}, we select
spectra collected with the Echelle SpectroPolarimetric Device for
the Observation of Stars \citep[ESPaDOnS;][]{ESPADONS2003}.  From the
Telescopio Nazionale Galileo archive, we select spectra collected with
the High Accuracy Radial velocity Planet Searcher for the Northern
hemisphere \citep[HARPS-N;][]{Cosentino2012}.  From the Keck Observatory
Archive (KOA), we select spectra collected with HIgh Resolution
Echelle Spectrometer \citep[HIRES;][]{HIRES1994}.  We supplement these
archival spectra with our own Astrophysical Research Consortium Echelle
Spectrograph \citep[ARCES;][]{ARCES2003} spectra for HD 80653, K2-131,
and K2-291 that enable us to infer oxygen abundances for those stars
using the oxygen triplet at 7770 \AA.

We were unable to identify reducible spectra for Kepler-30, Kepler-80,
Kepler-99, Kepler-105, Kepler-406, KOI-1831, or TOI-1444.  The data
available for K2-216 had insufficient S/N for our purposes, and
no spectra were available in the archives we searched for KOI-1599.
As we will describe below, high-quality stellar parameters for many of
these stars are available from \citet{Brewer2016}, \citet{Brewer2018},
or the third and fourth phases of the Sloan Digital Sky Survey
\citep[SDSS;][]{Eisenstein2011,Blanton2017} as part of its Apache Point
Observatory Galactic Evolution Experiment \cite[APOGEE;][]{Majewski2017}.

\begin{figure*}
\plotone{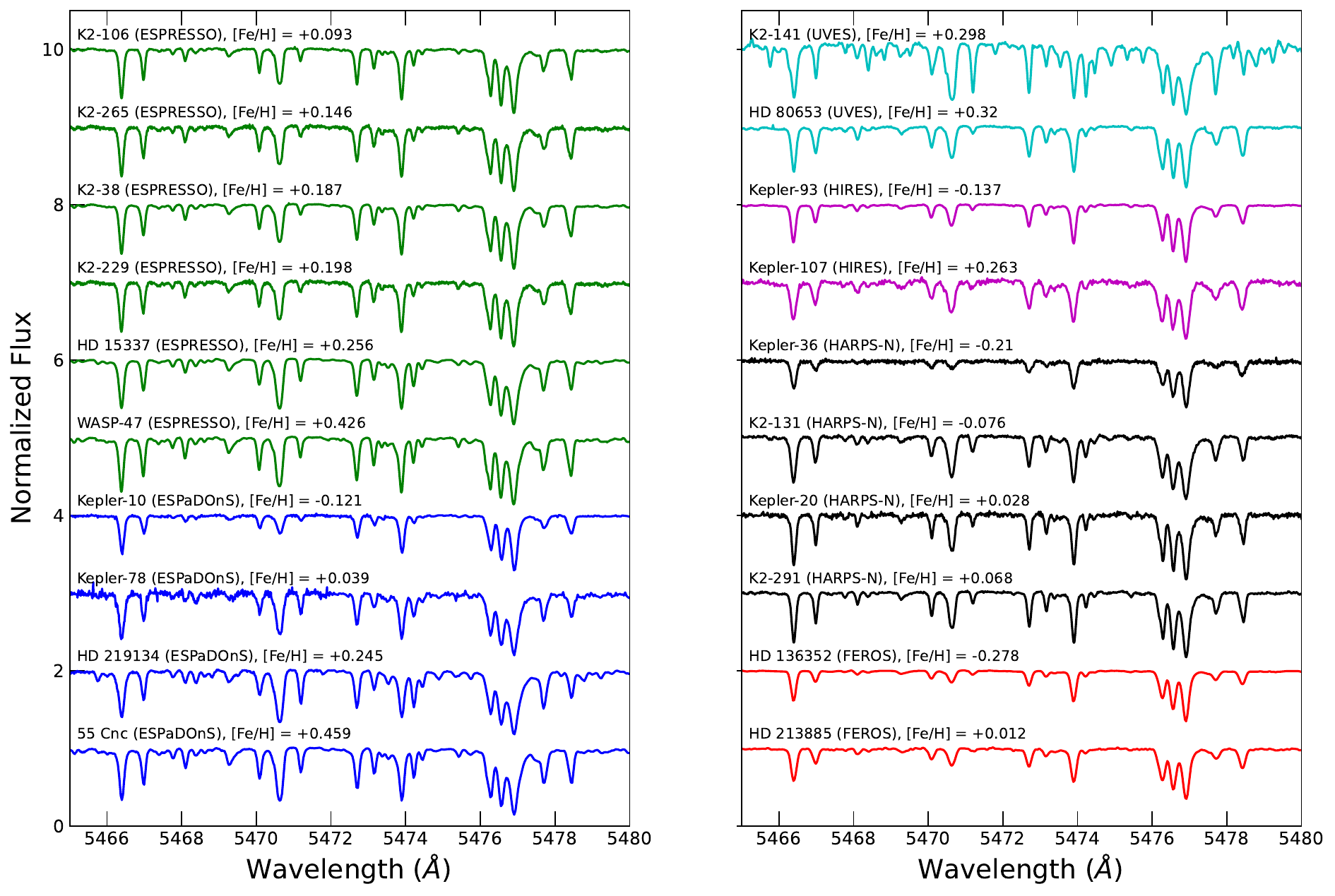}
\caption{Normalized spectrum for each star in our archival sample,
grouped by instrument and sorted by [Fe/H].}
\label{fig:spectra_compare}
\end{figure*}

We list in Table \ref{tab:sample_list} all of the stars in our
analysis sample with data from our archival search, SDSS DR17,
or \citet{Brewer2016}/\citet{Brewer2018}.  We plot in Figure
\ref{fig:spectra_compare} spectra in the range $5465~\text{\AA} \leq
\lambda \leq 5480~\text{\AA}$ for the stars with optical data from our
archival search.  While the spectra we use for our analyses come from
at least six different instruments with a range of wavelength coverage,
spectral resolution, and S/N, all of the spectra are suitable for stellar
parameter and elemental abundance inferences.  As we will describe in
detail in Section \ref{sec:analysis}, we use an updated version of the
\citet{Yana2019} linelist that was constructed using only lines that
are unblended in the spectra of solar-type dwarf stars for all spectral
resolutions in excess of 20,000.  All of our input spectra exceed that
spectral resolution threshold.  Likewise, though our input spectra have
a range of wavelength coverage, more than 90\% of the transitions in
our linelist are in the intersection of wavelength coverage across all
of our spectra.  Consequently, any residual systematic uncertainties
resulting from these heterogeneous data will be very small.

\startlongtable
\begin{deluxetable*}{lcccCCCc}
\tablecaption{Sample Spectra\label{tab:sample_list}}
\tablewidth{0pt}
\tabletypesize{\scriptsize}
\tablehead{
\colhead{Designation} & \colhead{R.A.} & \colhead{Dec.} & \colhead{Instrument} & \colhead{S/N} & \colhead{S/N} & \colhead{S/N} & \colhead{Program IDs}\\
\colhead{} & \colhead{(hh:mm:ss.ss)} & \colhead{(dd:mm:ss.s)} & \colhead{} & \colhead{(4500 \AA)} & \colhead{(6000 \AA)} & \colhead{(Median)$^a$} & \colhead{}
}
\startdata
\multicolumn{8}{l}{\textbf{This Study}} \\
K2-106	& 00:52:19.14	& +10:47:40.9	& ESPRESSO	& 27	& 53	& \cdots	& 0103.C-0289; 0104.C-0044\\
HD 15337	& 02:27:28.38	& --27:38:06.7	& ESPRESSO	& 1134	& 2133	& \cdots	& 104.20U8\\
K2-291	& 05:05:46.99	& +21:32:55.0	& HARPS-N$^b$	& $106$	& $ 170$	& \cdots	& OPT17B\_59; A36TAC\_12\\
55 Cnc	& 08:52:35.81	& +28:19:51.0	& ESPaDOnS	& 347	& 572	& \cdots	& 15AB01\\
HD 80653	& 09:21:21.42	& +14:22:04.5	& UVES$^b$	& \cdots$^c$	& 156	& \cdots	& 106.20ZM\\
K2-131	& 12:11:00.38	& --09:45:54.8	& HARPS-N$^b$	& $85$	& $135$	& \cdots	& A34TAC\_44; A34TAC\_10\\
K2-229	& 12:27:29.58	& --06:43:18.8	& ESPRESSO	& 61	& 108	& \cdots	& 105.202T\\
HD 136352	& 15:21:48.15	& --48:19:03.5	& FEROS	& \cdots	& \cdots	& 288	& 095.A-9029\\
K2-38	& 16:00:08.06	& --23:11:21.3	& ESPRESSO	& 43	& 76	& \cdots	& 1102.C-0744; 1102.C-0958; 1104.C-0350\\
Kepler-10	& 19:02:43.06	& +50:14:28.7	& ESPaDOnS	& 61	& 87	& \cdots	& 13AB06; 15AB01\\
Kepler-20	& 19:10:47.52	& +42:20:19.3	& HARPS-N	& $55$	& $80$	& \cdots	& GTO\\
Kepler-36	& 19:25:00.04	& +49:13:54.6	& HARPS-N	& $60$	& $ 64$	& \cdots	& GTO\\
Kepler-93	& 19:25:40.39	& +38:40:20.4	& HIRES	& 220	& 257	& \cdots	& N169; Z148Hr\\
Kepler-78	& 19:34:58.01	& +44:26:54.0	& ESPaDOnS	& 38	& 65	& \cdots	& 13AD94\\
Kepler-107	& 19:48:06.77	& +48:12:31.0	& HIRES	& 41	& 62	& \cdots	& Y198Hr\\
WASP-47	& 22:04:48.73	& --12:01:08.0	& ESPRESSO	& 26	& 72	& \cdots	& 0103.C-0422\\
HD 213885	& 22:35:56.32	& --59:51:52.1	& FEROS	& \cdots	& \cdots	& 110	& 0101.A-9008\\
K2-265	& 22:48:07.56	& --14:29:40.8	& ESPRESSO	& 61	& 108	& \cdots	& 105.202T\\
HD 219134	& 23:13:16.98	& +57:10:06.1	& ESPaDOnS	& 128	& 296	& \cdots	& 07bo03\\
K2-141	& 23:23:39.97	& --01:11:21.4	& UVES	& \cdots$^c$	& 109	& \cdots	& 0102.C-0226\\
\hline
\multicolumn{8}{l}{\textbf{APOGEE DR17 \citep{APOGEEDR17} sample}}\\
K2-216	& 00:45:55.26	& +06:20:49.1	& APOGEE	& \cdots	& \cdots	& 215	& $\cdots$\\
K2-106	& 00:52:19.14	& +10:47:40.9	& APOGEE	& \cdots	& \cdots	& 149	& $\cdots$\\
K2-38	& 16:00:08.06	& --23:11:21.3	& APOGEE	& \cdots	& \cdots	& 382	& $\cdots$\\
Kepler-10	& 19:02:43.06	& +50:14:28.7	& APOGEE	& \cdots	& \cdots	& 197	& $\cdots$\\
Kepler-20	& 19:10:47.52	& +42:20:19.3	& APOGEE	& \cdots	& \cdots	& 75	& $\cdots$\\
Kepler-36	& 19:25:00.04	& +49:13:54.6	& APOGEE	& \cdots	& \cdots	& 118	& $\cdots$\\
Kepler-93	& 19:25:40.39	& +38:40:20.4	& APOGEE	& \cdots	& \cdots	& 1908	& $\cdots$\\
Kepler-99	& 19:49:24.96	& +41:18:00.2	& APOGEE	& \cdots	& \cdots	& 221	& $\cdots$\\
KOI-1599	& 19:53:29.72	& +40:37:06.4	& APOGEE	& \cdots	& \cdots	& 113	& $\cdots$\\
WASP-47	& 22:04:48.73	& --12:01:08.0	& APOGEE	& \cdots	& \cdots	& 178	& $\cdots$\\
K2-265	& 22:48:07.56	& --14:29:40.8	& APOGEE	& \cdots	& \cdots	& 239	& $\cdots$\\
K2-141	& 23:23:39.97	& --01:11:21.5	& APOGEE	& \cdots	& \cdots	& 61	& $\cdots$\\
\hline
\multicolumn{8}{l}{\textbf{\citet{Brewer2016}/\citet{Brewer2018} sample}}\\
55 Cnc	& 08:52:35.81	& +28:19:51.0	& HIRES	& \cdots	& 330	& \cdots	& $\cdots$\\
K2-38	& 16:00:08.06	& --23:11:21.3	& HIRES	& \cdots	& 90	& \cdots	& $\cdots$\\
Kepler-10	& 19:02:43.06	& +50:14:28.7	& HIRES	& \cdots	& 47	& \cdots	& $\cdots$\\
Kepler-20	& 19:10:47.52	& +42:20:19.3	& HIRES	& \cdots	& 156	& \cdots	& $\cdots$\\
Kepler-105	& 19:11:32.95	& +46:16:34.4	& HIRES	& \cdots	& 75	& \cdots	& $\cdots$\\
Kepler-36	& 19:25:00.04	& +49:13:54.6	& HIRES	& \cdots	& 242	& \cdots	& $\cdots$\\
Kepler-93	& 19:25:40.39	& +38:40:20.4	& HIRES	& \cdots	& 223	& \cdots	& $\cdots$\\
Kepler-406	& 19:27:23.54	& +44:58:05.7	& HIRES	& \cdots	& 166	& \cdots	& $\cdots$\\
Kepler-78	& 19:34:58.01	& +44:26:54.0	& HIRES	& \cdots	& 228	& \cdots	& $\cdots$\\
Kepler-107	& 19:48:06.77	& +48:12:31.0	& HIRES	& \cdots	& 49	& \cdots	& $\cdots$\\
Kepler-99	& 19:49:24.96	& +41:18:00.2	& HIRES	& \cdots	& 174	& \cdots	& $\cdots$\\
K2-265	& 22:48:07.56	& --14:29:40.8	& HIRES	& \cdots	& 88	& \cdots	& $\cdots$\\
\enddata
\tablenotetext{a}{Median S/N over an entire spectrum}
\tablenotetext{b}{ARCES spectrum used for oxygen abundance inference}
\tablenotetext{c}{The lower limit of the wavelength coverage of these
UVES spectra is greater than 4500 \AA.}
\end{deluxetable*}

\section{Analysis}\label{sec:analysis}

\subsection{Host Star Parameters}\label{sec:stellarparams}

We infer accurate, precise, homogeneous, and physically self-consistent
photospheric and fundamental stellar parameters using both spectroscopy
and isochrones with the approach described in \citet{Reggiani2022,
Reggiani2024}.  Briefly, our methodology first uses the classical
spectroscopy-only approach to photospheric stellar parameter inference
by simultaneously minimizing for individual line-based iron abundance
inferences, the difference between \ion{Fe}{1} \& \ion{Fe}{2}-based
abundances, as well as their dependencies on transition excitation
potential and measured reduced equivalent width.  We then fit a grid of
isochrones to the resulting photospheric stellar parameters effective
temperature $T_{\text{eff}}$, surface gravity $\log{g}$, and metallicity
$[\text{Fe/H}]$ as well as astrometric, photometric, and reddening data.
We next use ($T_{\text{eff}}$, $\log{g}$) ordered pairs from the posterior
produced in the previous step as constraints on another round of reduced
equivalent width balance.  If the $[\text{Fe/H}]$ inferences from both
this step and the previous step agree, then we stop our iterations.
If they do not yet agree, then we continue iterating between the
comprehensive Bayesian fit and reduced equivalent width balance until
all photospheric stellar parameters have converged (typically less than
three iterations are required).  The resulting photospheric stellar
parameters are thereby informed by all available data and are guaranteed
to be physically self consistent in that are produced by a star that
satisfies the equations of stellar structure and evolution.

We use \texttt{iSpec} \citep{iSpec1,iSpec2} for our spectroscopic
analyses.  We first normalize spectra and zero their radial velocities
using the semi-automated routines provided by \texttt{iSpec} for those
purposes.  \texttt{iSpec} cross correlates a normalized observed spectrum
against a comprehensive sample of atomic absorption lines present in
the Kurucz linelist.  Although this linelist we use for radial velocity
measurement is not well curated and unsuited for precise radial velocity
inferences, it is suitable for the purpose of placing an observed spectrum
in its restframe for an abundance analysis.

We next measure the equivalent widths of many \ion{Fe}{1} and
\ion{Fe}{2} atomic absorption lines based on an updated version of the
\citet{Yana2019} linelist comprised of transitions found to be insensitive
to stellar activity in solar-analog stars.  We use \texttt{iSpec} to
fit Gaussian profiles to atomic absorption lines in small wavelength
intervals referred to as ``line masks'' in \texttt{iSpec}.  We visually
verified each line mask and the resulting equivalent width measurement in
each line mask.  We exclude from our analyses absorption lines for which
the fit failed due to blending, line saturation, or some other reason.
We ultimately exclude from our analyses saturated absorption lines
with successfully measured equivalent widths in excess of 120 m\AA.
We report our measured \ion{Fe}{1} and \ion{Fe}{2} equivalent widths and
the atomic data we used for their interpretation in Table \ref{tab:ews}.

\begin{deluxetable*}{cCcCCCC}
\tablecaption{Atomic Data and Equivalent Widths\label{tab:ews}}
\tabletypesize{\small}
\tablewidth{0pt}
\tablehead{
\colhead{Designation} & \colhead{Wavelength} & \colhead{Species} &
\colhead{Excitation Potential} & \colhead{log($gf$)} &
\colhead{EW} & \colhead{$\log_\epsilon(X)$} \\ \colhead{}
 & \colhead{(\text{\AA})} & \colhead{} & \colhead{(eV)} & \colhead{} & \colhead{(\text{m\AA})} & \colhead{}
}
\startdata
K2-106  & 7771.944      & \ion{O}{1}    & 9.146 & 0.370 & 52.7  & 9.016 \\
K2-106  & 7774.166      & \ion{O}{1}    & 9.146 & 0.220 & 45.3  & 9.018 \\
K2-106  & 7775.388      & \ion{O}{1}    & 9.146 & 0.000 & 35.7  & 9.017 \\
K2-106  & 5711.088      & \ion{Mg}{1}   & 4.345 & -1.729        & 119.7 & 7.632 \\
K2-106  & 6318.717      & \ion{Mg}{1}   & 5.108 & -1.945        & 57.7  & 7.671 \\
K2-106  & 6319.236      & \ion{Mg}{1}   & 5.108 & -2.165        & 40.4  & 7.628 \\
K2-106  & 5488.983      & \ion{Si}{1}   & 5.614 & -1.690        & 25.1  & 7.578 \\
K2-106  & 5517.540      & \ion{Si}{1}   & 5.080 & -2.496        & 13.1  & 7.564 \\
K2-106  & 5645.611      & \ion{Si}{1}   & 4.929 & -2.040        & 40.2  & 7.649 \\
K2-106  & 5665.554      & \ion{Si}{1}   & 4.920 & -1.940        & 44.7  & 7.620 \\
\enddata
\tablecomments{This table is published in its entirety in the
machine-readable format.  A portion is shown here for guidance regarding
its form and content.}
\end{deluxetable*}

For each star in our samples we collect Gaia astrometry and high-quality
photometry.  In particular, we collect
\begin{enumerate}
\item
Gaia DR3 parallaxes \citep{Gaia2016, Gaia2021, Lindegren2021a,
Lindegren2021b, Fabricius2021, Torra2021};
\item
Galaxy Evolution Explorer \citep[GALEX;][]{Galex2005} far-ultraviolet
(FUV) and near-ultraviolet (NUV) photometry from the GUVcat\_AIS
\citep{Bianchi2017}, or, if the system is in the Kepler field, the
GALEX-CAUSE Kepler (GCK) survey \citep{Olmedo2015};
\item
SkyMapper DR2 $u$, $v$, $g$, $r$, $i$, and $z$ data \citep{Onken2019};
\item
SDSS DR17 $u$, $g$, $r$, $i$, and $z$ photometry \citep{Fukugita1996,
Gunn1998, York2000, Gunn2006, Doi2010, APOGEEDR17};
\item
Tycho-2 $B_{T}$ and $V_{T}$ photometry \citep{Tycho2000};
\item
Kepler Input Catalog (KIC) $g$, $r$, $i$, and $z$ photometry
\citep{Brown2011};
\item
Gaia DR2 $G$ photometry \citep{Gaia2018, Arenou2018, Evans2018,
Hambly2018, Riello2018};
\item
Two-micron All-sky Survey (2MASS) $J$, $H$, and $K_{s}$ photometry
\citep{2MASS2006}; and
\item
Wide-field Infrared Survey Explorer (WISE) All-sky \citep{Wise2010}
or AllWISE \citep{Mainzer2011} $W1$, $W2$, $W3$, and $W4$ photometry.
Because the procedures used in the construction of the WISE All-sky
catalog were better for saturated sources than the procedures used in
the construction of the AllWISE catalog, we use WISE All-sky data when $2
\leq W1 < 8$ and $1.5 \leq W2 < 7$.  We use AllWISE data when $W1 \geq 8$
and $W2 \geq 7$.
\end{enumerate}
We use the data quality flags listed in \citet{Hamer2022} to ensure we
include only the highest quality photometry in our analyses.  We list
the data we use in our analyses in Table \ref{tab:photometry}.

We use the methodology described in detail in \citet{Reggiani2022,
Reggiani2024} to analyze these data.  We assume \citet{Asplund2021} solar
abundances, \citet{Castelli2003} 1D plane-parallel solar-composition
ATLAS9 model atmospheres, and local thermodynamic equilibrium (LTE).
We then use the \texttt{q2} \citep{q2} wrapper to the 2019 version of
the \texttt{MOOG} radiative transfer code \citep{Sneden1973} to infer an
initial guess for the photospheric stellar parameters for each star in
our sample.  We collect estimates of the line-of-sight extinction to each
star from the three-dimensional Bayestar19 \citep{Green2014,Green2019}
or Stilism extinction maps.  Using the \texttt{isochrones}
\citep{isochrones} package, we execute with \texttt{MultiNest}
\citep{Feroz2008,Feroz2009,Feroz2019} a simultaneous Bayesian fit of the
Modules for Experiments in Stellar Evolution \citep[MESA;][]{MESA1,
MESA2, MESA3, MESA4, MESA5, MESA6} Isochrones \& Stellar Tracks
\cite[MIST;][]{MIST0, MIST1} isochrone grid to the data described above
for each star.  We use a log uniform age prior between 1 Gyr and 13.721
Gyr (i.e., the age of the Universe), a uniform reddening prior between
the estimated reddening value minus/plus three times its uncertainty,
and a distance prior proportional to volume over the 3-$\sigma$ distance
range implied by the Gaia DR3 parallax measurement.  Because the model
atmospheres we use to interpret our equivalent width measurements do
not extend beyond $[\text{Fe/H}] = +0.5$, we impose an upper bound on
our metallicity prior $[\text{Fe/H}] = +0.5$ in our joint analyses.
We iterate between the spectroscopic and isochrones analyses until the
metallicities from both approaches agree within their uncertainties.

\begin{figure*}
\digitalasset
\plotone{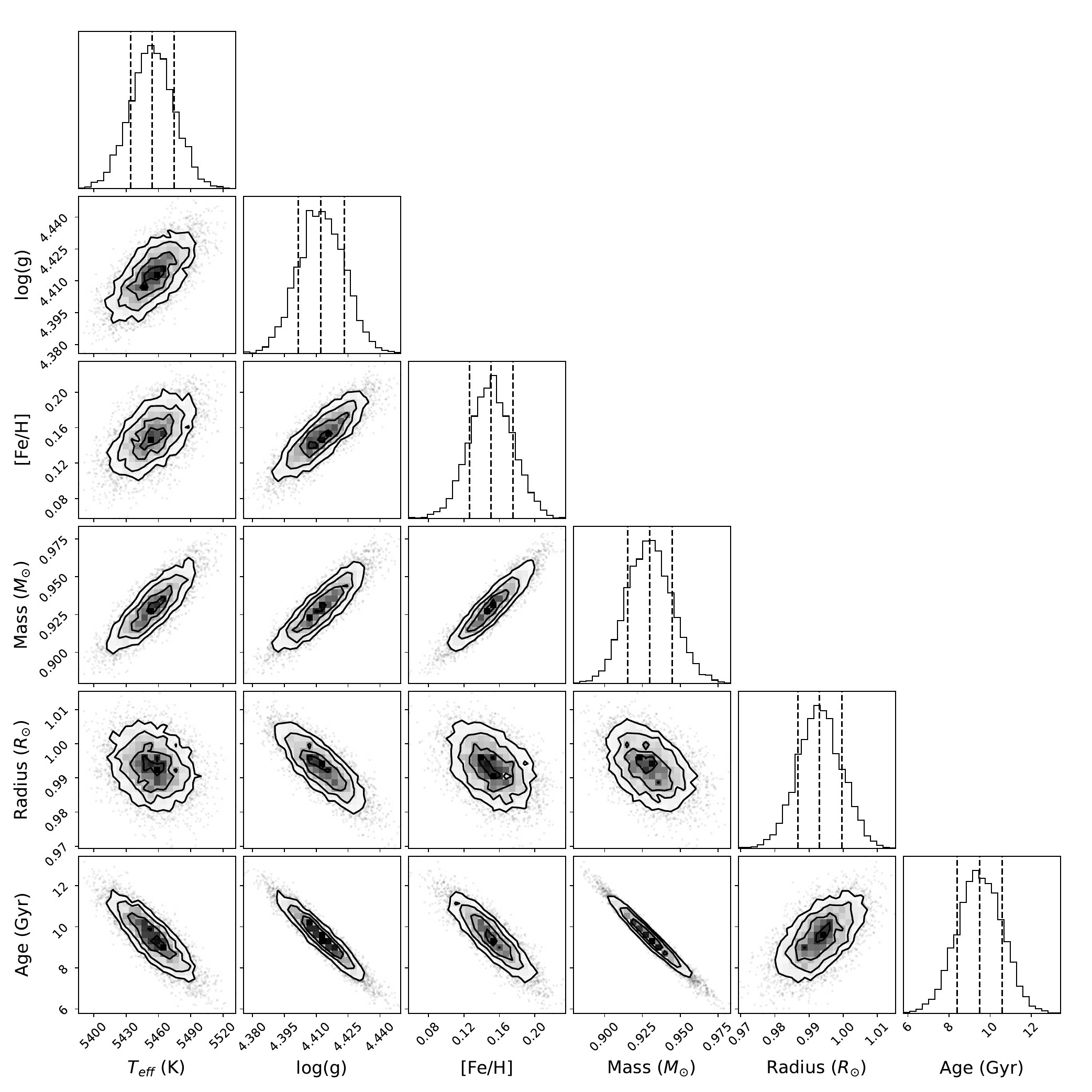}
\caption{Samples from the photospheric and fundamental stellar
parameter posterior resulting from our Bayesian analysis of astrometric,
photometric, and spectroscopic data for K2-106.  The complete figure set
(44 images) is available in the online journal.}
\label{fig:K2-106_corner}
\end{figure*}

We report our adopted photospheric and fundamental stellar parameters
in Table \ref{tab:stellar_params}.  The uncertainties we report in Table
\ref{tab:stellar_params} are only random uncertainties and do not include
any contribution from possible systematic uncertainties.  Nevertheless,
for the subsample of stars in the intersection of our analysis, APOGEE
DR17, and \citet{Brewer2016}/\citet{Brewer2018} our photospheric and
fundamental stellar parameters are in accord with those independent
analyses.  The implication is that any systematic uncertainties present
must be small.  We plot in Figure \ref{fig:K2-106_corner} an example
corner plot for the photospheric and fundamental stellar parameter
posterior that results from our analysis of the star K2-106.

\begin{longrotatetable}
\movetabledown=2cm
\begin{deluxetable*}{lRRRCRRRcRRRcc}
\tablecaption{System Parameters\label{tab:stellar_params}}
\tabletypesize{\tiny}
\tablewidth{0pt}
\tablehead{
\colhead{Designation} & \colhead{$T_{\text{eff}}$} & \colhead{$\log{g}$} & \colhead{[Fe/H]} & \colhead{$\xi$} & \colhead{Age} & \begin{tabular}{c} Stellar \\ Mass \end{tabular} & \begin{tabular}{c} Stellar \\ Radius \end{tabular} & \colhead{Planet} & \begin{tabular}{c} Planet \\ Mass \end{tabular} & \begin{tabular}{c} Planet \\ Radius \end{tabular} & \begin{tabular}{c} Planet \\ Density \end{tabular} & \colhead{Transit Observables} & \colhead{Mass Observables}\\
\colhead{} & \colhead{(K)} & \colhead{(cm s$^{-2}$)} & \colhead{(dex)} & \colhead{(km s$^{-1}$)} & \colhead{(Gyr)} & \colhead{($M_{\odot}$)} & \colhead{($R_{\odot}$)} & \colhead{} & \colhead{($M_{\oplus}$)} & \colhead{($R_{\oplus}$)} & \colhead{(g cm$^{-3}$)} & \colhead{} & \colhead{}
}
\startdata
\multicolumn{6}{l}{\textbf{This Study}} \\
K2-106	& 5455_{-20}^{+19}	& 4.41_{-0.01}^{+0.01}	& +0.09_{-0.01}^{+0.01}	& 0.74_{-0.03}^{+0.02}	& 9.48_{-1.08}^{+1.11}	& 0.93_{-0.02}^{+0.01}	& 0.99_{-0.01}^{+0.01}	& b	& 8.31_{-0.70}^{+0.84}	& 1.83_{-0.08}^{+0.14}	& 7.38_{-1.52}^{+1.47}	& \citet{Mayo2018}	& \citet{Bonomo2023}\\
HD 15337	& 5178_{-21}^{+22}	& 4.53_{-0.01}^{+0.01}	& +0.26_{-0.01}^{+0.01}	& 0.73_{-0.03}^{+0.02}	& 5.10_{-1.92}^{+1.66}	& 0.90_{-0.02}^{+0.02}	& 0.85_{-0.01}^{+0.01}	& b	& 7.63_{-0.87}^{+0.85}	& 1.72_{-0.04}^{+0.04}	& 8.16_{-1.10}^{+1.17}	& \citet{Dumusque2019}	& \citet{Dumusque2019}\\
K2-291	& 5550_{-14}^{+18}	& 4.50_{-0.01}^{+0.01}	& +0.07_{-0.01}^{+0.01}	& 1.03_{-0.01}^{+0.01}	& 4.39_{-1.39}^{+1.36}	& 0.92_{-0.02}^{+0.02}	& 0.89_{-0.01}^{+0.01}	& b	& 6.45_{-1.14}^{+1.13}	& 1.57_{-0.03}^{+0.06}	& 9.00_{-1.77}^{+1.80}	& \citet{Kosiarek2019}	& \citet{Kosiarek2019}\\
55 Cnc	& 5232_{-15}^{+15}	& 4.40_{-0.01}^{+0.01}	& +0.46_{-0.01}^{+0.01}	& 0.81_{-0.02}^{+0.01}	& 12.30_{-1.34}^{+0.82}	& 0.92_{-0.01}^{+0.01}	& 1.00_{-0.01}^{+0.01}	& e	& 8.10_{-0.32}^{+0.33}	& 1.99_{-0.02}^{+0.02}	& 5.66_{-0.31}^{+0.34}	& \citet{Bourrier2018}	& \citet{Bourrier2018}\\
HD 80653	& 5901_{-35}^{+32}	& 4.34_{-0.01}^{+0.01}	& +0.32_{-0.02}^{+0.01}	& 1.07_{-0.01}^{+0.01}	& 2.61_{-0.47}^{+0.71}	& 1.19_{-0.02}^{+0.01}	& 1.22_{-0.01}^{+0.01}	& b	& 5.75_{-0.33}^{+0.34}	& 1.62_{-0.06}^{+0.05}	& 7.50_{-0.83}^{+0.93}	& \citet{Frustagli2020}	& \citet{Bonomo2023}\\
K2-131	& 5054_{-14}^{+12}	& 4.59_{-0.01}^{+0.01}	& -0.08_{-0.01}^{+0.01}	& 1.15_{-0.02}^{+0.01}	& 2.26_{-0.89}^{+1.60}	& 0.83_{-0.01}^{+0.01}	& 0.76_{-0.01}^{+0.01}	& b	& 8.38_{-1.43}^{+1.63}	& 1.75_{-0.11}^{+0.21}	& 8.36_{-2.49}^{+2.93}	& \citet{Mayo2018}	& \citet{Bonomo2023}\\
K2-229	& 5294_{-10}^{+10}	& 4.58_{-0.01}^{+0.01}	& +0.2_{-0.01}^{+0.01}	& 0.52_{-0.01}^{+0.01}	& 1.04_{-0.03}^{+0.09}	& 0.89_{-0.01}^{+0.01}	& 0.80_{-0.01}^{+0.01}	& b	& 2.73_{-0.47}^{+0.45}	& 1.25_{-0.06}^{+0.12}	& 7.37_{-1.95}^{+2.14}	& \citet{Mayo2018}	& \citet{Dai2019}\\
HD 136352	& 5764_{-23}^{+20}	& 4.35_{-0.02}^{+0.02}	& -0.28_{-0.01}^{+0.01}	& 1.02_{-0.02}^{+0.01}	& 10.71_{-1.37}^{+1.23}	& 0.88_{-0.02}^{+0.03}	& 1.03_{-0.01}^{+0.01}	& b	& 4.74_{-0.38}^{+0.39}	& 1.63_{-0.03}^{+0.03}	& 6.04_{-0.61}^{+0.66}	& \citet{Delrez2021}	& \citet{Delrez2021}\\
K2-38	& 5599_{-15}^{+15}	& 4.33_{-0.01}^{+0.01}	& +0.19_{-0.01}^{+0.01}	& 0.95_{-0.01}^{+0.01}	& 8.23_{-0.34}^{+0.35}	& 1.02_{-0.01}^{+0.01}	& 1.15_{-0.01}^{+0.01}	& b	& 7.61_{-1.07}^{+1.06}	& 1.60_{-0.08}^{+0.13}	& 9.87_{-2.22}^{+2.48}	& \citet{Sinukoff2016}	& \citet{Bonomo2023}\\
Kepler-10	& 5714_{-20}^{+20}	& 4.33_{-0.01}^{+0.01}	& -0.12_{-0.01}^{+0.01}	& 0.88_{-0.01}^{+0.03}	& 10.19_{-0.95}^{+0.95}	& 0.92_{-0.02}^{+0.02}	& 1.08_{-0.01}^{+0.01}	& b	& 3.27_{-0.30}^{+0.30}	& 1.48_{-0.02}^{+0.02}	& 5.55_{-0.55}^{+0.55}	& \citet{Fogtmann2014}	& \citet{Bonomo2023}\\
Kepler-20	& 5486_{-21}^{+17}	& 4.47_{-0.01}^{+0.01}	& +0.03_{-0.01}^{+0.01}	& 0.87_{-0.03}^{+0.01}	& 7.93_{-1.34}^{+1.48}	& 0.89_{-0.02}^{+0.02}	& 0.91_{-0.01}^{+0.01}	& b	& 9.38_{-1.46}^{+1.25}	& 1.77_{-0.01}^{+0.04}	& 9.16_{-1.51}^{+1.32}	& \citet{Buchhave2016}	& \citet{Bonomo2023}\\
Kepler-36	& 5989_{-32}^{+19}	& 4.02_{-0.01}^{+0.01}	& -0.21_{-0.02}^{+0.01}	& 1.32_{-0.03}^{+0.01}	& 6.60_{-0.28}^{+0.34}	& 1.09_{-0.02}^{+0.02}	& 1.70_{-0.01}^{+0.01}	& b	& 3.65_{-0.08}^{+0.09}	& 1.55_{-0.04}^{+0.05}	& 5.33_{-0.48}^{+0.42}	& \citet{Thompson2018}	& \citet{Vissapragada2020}\\
Kepler-93	& 5656_{-25}^{+23}	& 4.44_{-0.01}^{+0.01}	& -0.14_{-0.01}^{+0.01}	& 0.94_{-0.03}^{+0.02}	& 8.48_{-1.18}^{+1.21}	& 0.88_{-0.02}^{+0.02}	& 0.94_{-0.01}^{+0.01}	& b	& 4.55_{-0.50}^{+0.51}	& 1.51_{-0.01}^{+0.01}	& 7.31_{-0.81}^{+0.83}	& \citet{Ballard2014}	& \citet{Bonomo2023}\\
Kepler-78	& 4980_{-14}^{+16}	& 4.60_{-0.01}^{+0.01}	& +0.04_{-0.01}^{+0.01}	& 0.85_{-0.02}^{+0.03}	& 1.50_{-0.37}^{+0.65}	& 0.82_{-0.01}^{+0.01}	& 0.75_{-0.01}^{+0.01}	& b	& 1.78_{-0.30}^{+0.31}	& 1.17_{-0.06}^{+0.15}	& 5.86_{-1.83}^{+1.84}	& \citet{Sanchis2013}	& \citet{Bonomo2023}\\
Kepler-107	& 5928_{-37}^{+23}	& 4.22_{-0.01}^{+0.01}	& +0.26_{-0.01}^{+0.01}	& 1.23_{-0.02}^{+0.01}	& 3.72_{-0.32}^{+0.36}	& 1.25_{-0.01}^{+0.01}	& 1.44_{-0.01}^{+0.01}	& c	& 10.09_{-2.01}^{+2.03}	& 1.59_{-0.02}^{+0.02}	& 13.86_{-2.81}^{+2.87}	& \citet{Bonomo2019}	& \citet{Bonomo2023}\\
WASP-47	& 5452_{-16}^{+13}	& 4.32_{-0.01}^{+0.01}	& +0.43_{-0.01}^{+0.01}	& 0.98_{-0.01}^{+0.01}	& 8.58_{-0.41}^{+0.45}	& 1.03_{-0.01}^{+0.01}	& 1.16_{-0.01}^{+0.01}	& e	& 6.75_{-0.54}^{+0.56}	& 1.85_{-0.02}^{+0.02}	& 5.87_{-0.50}^{+0.52}	& \citet{Vanderburg2017}	& \citet{Bryant2022}\\
HD 213885	& 5889_{-20}^{+20}	& 4.33_{-0.02}^{+0.01}	& +0.01_{-0.01}^{+0.01}	& 1.08_{-0.01}^{+0.01}	& 7.65_{-0.83}^{+0.89}	& 0.96_{-0.02}^{+0.02}	& 1.10_{-0.01}^{+0.01}	& b	& 8.18_{-0.60}^{+0.62}	& 1.75_{-0.05}^{+0.05}	& 8.41_{-0.93}^{+1.02}	& \citet{Espinoza2020}	& \citet{Espinoza2020}\\
K2-265	& 5405_{-11}^{+10}	& 4.48_{-0.01}^{+0.01}	& +0.15_{-0.01}^{+0.01}	& 0.67_{-0.02}^{+0.01}	& 5.67_{-0.68}^{+0.80}	& 0.95_{-0.01}^{+0.01}	& 0.93_{-0.01}^{+0.01}	& b	& 6.74_{-0.89}^{+0.85}	& 1.62_{-0.04}^{+0.04}	& 8.61_{-1.22}^{+1.32}	& \citet{Lam2018}	& \citet{Lam2018}\\
HD 219134	& 4907_{-26}^{+26}	& 4.57_{-0.01}^{+0.01}	& +0.24_{-0.01}^{+0.01}	& 0.59_{-0.04}^{+0.05}	& 6.00_{-2.82}^{+2.97}	& 0.80_{-0.02}^{+0.02}	& 0.76_{-0.01}^{+0.01}	& b	& 4.70_{-0.17}^{+0.17}	& 1.58_{-0.05}^{+0.05}	& 6.61_{-0.64}^{+0.72}	& \citet{Gillon2017}	& \citet{Gillon2017}\\
HD 219134	& 4907_{-26}^{+26}	& 4.57_{-0.01}^{+0.01}	& +0.24_{-0.01}^{+0.01}	& 0.59_{-0.04}^{+0.05}	& 6.00_{-2.82}^{+2.97}	& 0.80_{-0.02}^{+0.02}	& 0.76_{-0.01}^{+0.01}	& c	& 4.34_{-0.21}^{+0.21}	& 1.48_{-0.04}^{+0.04}	& 7.31_{-0.67}^{+0.75}	& \citet{Gillon2017}	& \citet{Gillon2017}\\
K2-141	& 4521_{-11}^{+12}	& 4.62_{-0.01}^{+0.01}	& +0.3_{-0.01}^{+0.01}	& 0.45_{-0.03}^{+0.04}	& 1.36_{-0.28}^{+0.63}	& 0.79_{-0.01}^{+0.01}	& 0.72_{-0.01}^{+0.01}	& b	& 5.34_{-0.34}^{+0.34}	& 1.59_{-0.04}^{+0.04}	& 7.24_{-0.64}^{+0.70}	& \citet{Malavolta2018}	& \citet{Bonomo2023}\\
\hline
\multicolumn{6}{l}{\textbf{APOGEE DR17 \citep{APOGEEDR17} sample}} \\
K2-216	& 4610_{-10}^{+10}	& 4.63_{-0.01}^{+0.01}	& -0.05_{-0.01}^{+0.01}	& 0.55_{-0.15}^{+0.15}	& 4.27_{-1.36}^{+1.43}	& 0.71_{-0.01}^{+0.01}	& 0.68_{-0.01}^{+0.01}	& b	& 8.15_{-1.64}^{+1.55}	& 1.63_{-0.05}^{+0.16}	& 9.71_{-2.65}^{+2.75}	& \citet{Persson2018}	& \citet{Persson2018}\\
K2-106	& 5471_{-10}^{+10}	& 4.40_{-0.01}^{+0.01}	& +0.09_{-0.01}^{+0.01}	& 0.47_{-0.15}^{+0.15}	& 10.58_{-0.44}^{+0.41}	& 0.91_{-0.01}^{+0.01}	& 1.00_{-0.01}^{+0.01}	& b	& 8.23_{-0.73}^{+0.82}	& 1.84_{-0.08}^{+0.15}	& 7.14_{-1.49}^{+1.47}	& \citet{Mayo2018}	& \citet{Bonomo2023}\\
K2-38	& 5546_{-14}^{+14}	& 4.32_{-0.01}^{+0.01}	& +0.26_{-0.01}^{+0.01}	& 0.34_{-0.15}^{+0.15}	& 9.32_{-0.35}^{+0.36}	& 1.00_{-0.01}^{+0.01}	& 1.15_{-0.01}^{+0.01}	& b	& 7.52_{-1.06}^{+1.07}	& 1.61_{-0.08}^{+0.13}	& 9.72_{-2.23}^{+2.49}	& \citet{Sinukoff2016}	& \citet{Bonomo2023}\\
Kepler-10	& 5722_{-21}^{+23}	& 4.34_{-0.01}^{+0.01}	& -0.11_{-0.01}^{+0.01}	& 0.36_{-0.15}^{+0.15}	& 10.04_{-0.62}^{+0.64}	& 0.92_{-0.01}^{+0.01}	& 1.08_{-0.01}^{+0.01}	& b	& 3.28_{-0.30}^{+0.28}	& 1.47_{-0.02}^{+0.02}	& 5.63_{-0.54}^{+0.54}	& \citet{Fogtmann2014}	& \citet{Bonomo2023}\\
Kepler-20	& 5421_{-13}^{+18}	& 4.46_{-0.01}^{+0.01}	& +0.02_{-0.01}^{+0.01}	& 0.32_{-0.15}^{+0.15}	& 8.77_{-0.62}^{+0.59}	& 0.89_{-0.01}^{+0.01}	& 0.91_{-0.01}^{+0.01}	& b	& 9.36_{-1.43}^{+1.21}	& 1.77_{-0.02}^{+0.05}	& 9.01_{-1.43}^{+1.32}	& \citet{Buchhave2016}	& \citet{Bonomo2023}\\
Kepler-36	& 5929_{-22}^{+24}	& 4.00_{-0.01}^{+0.01}	& -0.23_{-0.01}^{+0.01}	& 0.74_{-0.15}^{+0.15}	& 7.32_{-0.25}^{+0.23}	& 1.05_{-0.01}^{+0.01}	& 1.71_{-0.01}^{+0.01}	& b	& 3.77_{-0.08}^{+0.08}	& 1.56_{-0.04}^{+0.05}	& 5.40_{-0.48}^{+0.43}	& \citet{Thompson2018}	& \citet{Vissapragada2020}\\
Kepler-93	& 5608_{-14}^{+20}	& 4.43_{-0.01}^{+0.01}	& -0.16_{-0.01}^{+0.01}	& 0.34_{-0.15}^{+0.15}	& 9.46_{-0.58}^{+0.47}	& 0.87_{-0.01}^{+0.01}	& 0.94_{-0.01}^{+0.01}	& b	& 4.52_{-0.50}^{+0.50}	& 1.51_{-0.01}^{+0.01}	& 7.21_{-0.81}^{+0.80}	& \citet{Ballard2014}	& \citet{Bonomo2023}\\
Kepler-99	& 4719_{-10}^{+10}	& 4.58_{-0.01}^{+0.01}	& +0.18_{-0.01}^{+0.01}	& 0.62_{-0.15}^{+0.15}	& 8.18_{-1.22}^{+1.22}	& 0.78_{-0.01}^{+0.01}	& 0.75_{-0.01}^{+0.01}	& b	& 6.44_{-1.32}^{+1.30}	& 1.54_{-0.03}^{+0.04}	& 9.72_{-2.02}^{+2.16}	& \citet{Thompson2018}	& \citet{Marcy2014}\\
KOI-1599	& 5697_{-24}^{+23}	& 4.21_{-0.02}^{+0.01}	& -0.15_{-0.01}^{+0.01}	& 0.30_{-0.15}^{+0.15}	& 12.32_{-0.52}^{+0.53}	& 0.91_{-0.01}^{+0.01}	& 1.25_{-0.02}^{+0.02}	& .01	& 4.12_{-0.26}^{+0.26}	& 2.45_{-0.07}^{+0.07}	& 1.54_{-0.16}^{+0.17}	& \citet{Panichi2019}	& \citet{Panichi2019}\\
WASP-47	& 5389_{-11}^{+12}	& 4.29_{-0.01}^{+0.01}	& +0.38_{-0.01}^{+0.01}	& 0.33_{-0.15}^{+0.15}	& 11.80_{-0.35}^{+0.32}	& 0.98_{-0.01}^{+0.01}	& 1.17_{-0.01}^{+0.01}	& e	& 6.50_{-0.51}^{+0.53}	& 1.87_{-0.02}^{+0.02}	& 5.45_{-0.46}^{+0.48}	& \citet{Vanderburg2017}	& \citet{Bryant2022}\\
K2-265	& 5328_{-10}^{+10}	& 4.42_{-0.01}^{+0.01}	& +0.02_{-0.01}^{+0.01}	& 0.33_{-0.15}^{+0.15}	& 13.44_{-0.08}^{+0.04}	& 0.85_{-0.01}^{+0.01}	& 0.94_{-0.01}^{+0.01}	& b	& 6.26_{-0.78}^{+0.83}	& 1.65_{-0.04}^{+0.04}	& 7.64_{-1.11}^{+1.19}	& \citet{Lam2018}	& \citet{Lam2018}\\
K2-141	& 4518_{-10}^{+10}	& 4.59_{-0.01}^{+0.01}	& +0.09_{-0.01}^{+0.01}	& 0.33_{-0.15}^{+0.15}	& 11.52_{-1.62}^{+1.26}	& 0.71_{-0.01}^{+0.01}	& 0.70_{-0.01}^{+0.01}	& b	& 5.01_{-0.32}^{+0.32}	& 1.57_{-0.04}^{+0.03}	& 7.16_{-0.65}^{+0.70}	& \citet{Malavolta2018}	& \citet{Bonomo2023}\\
\hline
\multicolumn{6}{l}{\textbf{\citet{Brewer2016} and \citet{Brewer2018} sample}} \\
55 Cnc	& 5173_{-10}^{+10}	& 4.40_{-0.01}^{+0.01}	& +0.4_{-0.01}^{+0.01}	& \cdots	& 13.43_{-0.09}^{+0.04}	& 0.90_{-0.01}^{+0.01}	& 0.99_{-0.01}^{+0.01}	& e	& 8.00_{-0.30}^{+0.32}	& 1.97_{-0.02}^{+0.02}	& 5.75_{-0.28}^{+0.30}	& \citet{Bourrier2018}	& \citet{Bourrier2018}\\
K2-38	& 5592_{-17}^{+16}	& 4.33_{-0.01}^{+0.01}	& +0.24_{-0.01}^{+0.01}	& \cdots	& 8.43_{-0.42}^{+0.47}	& 1.01_{-0.01}^{+0.01}	& 1.15_{-0.01}^{+0.01}	& b	& 7.61_{-1.10}^{+1.08}	& 1.61_{-0.08}^{+0.13}	& 9.85_{-2.29}^{+2.49}	& \citet{Sinukoff2016}	& \citet{Bonomo2023}\\
Kepler-10	& 5684_{-15}^{+17}	& 4.31_{-0.01}^{+0.01}	& -0.18_{-0.02}^{+0.02}	& \cdots	& 12.37_{-0.56}^{+0.54}	& 0.88_{-0.01}^{+0.01}	& 1.09_{-0.01}^{+0.01}	& b	& 3.18_{-0.29}^{+0.28}	& 1.49_{-0.02}^{+0.02}	& 5.29_{-0.50}^{+0.52}	& \citet{Fogtmann2014}	& \citet{Bonomo2023}\\
Kepler-20	& 5491_{-18}^{+15}	& 4.48_{-0.01}^{+0.01}	& +0.03_{-0.01}^{+0.01}	& \cdots	& 6.46_{-0.76}^{+0.68}	& 0.91_{-0.01}^{+0.01}	& 0.91_{-0.01}^{+0.01}	& b	& 9.57_{-1.51}^{+1.20}	& 1.76_{-0.01}^{+0.05}	& 9.39_{-1.51}^{+1.36}	& \citet{Buchhave2016}	& \citet{Bonomo2023}\\
Kepler-105	& 5947_{-17}^{+19}	& 4.40_{-0.01}^{+0.01}	& -0.11_{-0.01}^{+0.01}	& \cdots	& 4.68_{-0.43}^{+0.43}	& 1.00_{-0.01}^{+0.01}	& 1.05_{-0.01}^{+0.01}	& c	& 3.99_{-1.46}^{+1.56}	& 1.74_{-0.05}^{+0.24}	& 3.71_{-1.55}^{+1.90}	& \citet{Thompson2018}	& \citet{Hadden2017}\\
Kepler-36	& 6006_{-15}^{+13}	& 4.03_{-0.01}^{+0.01}	& -0.16_{-0.01}^{+0.01}	& \cdots	& 6.34_{-0.14}^{+0.16}	& 1.10_{-0.01}^{+0.01}	& 1.69_{-0.01}^{+0.01}	& b	& 3.60_{-0.07}^{+0.07}	& 1.54_{-0.03}^{+0.05}	& 5.35_{-0.47}^{+0.41}	& \citet{Thompson2018}	& \citet{Vissapragada2020}\\
Kepler-93	& 5676_{-23}^{+19}	& 4.45_{-0.01}^{+0.01}	& -0.17_{-0.01}^{+0.01}	& \cdots	& 7.83_{-0.65}^{+0.72}	& 0.89_{-0.01}^{+0.01}	& 0.93_{-0.01}^{+0.01}	& b	& 4.59_{-0.52}^{+0.50}	& 1.50_{-0.01}^{+0.01}	& 7.44_{-0.84}^{+0.82}	& \citet{Ballard2014}	& \citet{Bonomo2023}\\
Kepler-406	& 5593_{-20}^{+17}	& 4.37_{-0.01}^{+0.01}	& +0.2_{-0.01}^{+0.01}	& \cdots	& 7.96_{-0.52}^{+0.56}	& 0.99_{-0.01}^{+0.01}	& 1.08_{-0.01}^{+0.01}	& b	& 6.10_{-1.25}^{+1.26}	& 1.57_{-0.10}^{+0.14}	& 8.50_{-2.40}^{+2.89}	& \citet{Thompson2018}	& \citet{Marcy2014}\\
Kepler-78	& 5012_{-18}^{+15}	& 4.60_{-0.01}^{+0.01}	& +0.0_{-0.01}^{+0.01}	& \cdots	& 2.31_{-0.82}^{+0.95}	& 0.81_{-0.01}^{+0.01}	& 0.75_{-0.01}^{+0.01}	& b	& 1.77_{-0.29}^{+0.30}	& 1.16_{-0.06}^{+0.15}	& 5.90_{-1.79}^{+1.84}	& \citet{Sanchis2013}	& \citet{Bonomo2023}\\
Kepler-107	& 5813_{-15}^{+14}	& 4.19_{-0.01}^{+0.01}	& +0.28_{-0.02}^{+0.02}	& \cdots	& 5.02_{-0.20}^{+0.20}	& 1.20_{-0.01}^{+0.01}	& 1.46_{-0.01}^{+0.01}	& c	& 9.79_{-1.92}^{+2.00}	& 1.61_{-0.02}^{+0.02}	& 12.95_{-2.59}^{+2.67}	& \citet{Bonomo2019}	& \citet{Bonomo2023}\\
Kepler-99	& 4703_{-10}^{+10}	& 4.58_{-0.01}^{+0.01}	& +0.27_{-0.01}^{+0.01}	& \cdots	& 6.36_{-1.54}^{+1.70}	& 0.80_{-0.01}^{+0.01}	& 0.76_{-0.01}^{+0.01}	& b	& 6.56_{-1.42}^{+1.32}	& 1.55_{-0.04}^{+0.04}	& 9.66_{-2.12}^{+2.13}	& \citet{Thompson2018}	& \citet{Marcy2014}\\
K2-265	& 5328_{-12}^{+13}	& 4.42_{-0.01}^{+0.01}	& +0.07_{-0.01}^{+0.01}	& \cdots	& 12.65_{-0.60}^{+0.50}	& 0.87_{-0.01}^{+0.01}	& 0.95_{-0.01}^{+0.01}	& b	& 6.35_{-0.80}^{+0.79}	& 1.66_{-0.04}^{+0.04}	& 7.62_{-1.11}^{+1.15}	& \citet{Lam2018}	& \citet{Lam2018}
\enddata
\tablenotetext{a}{Microturbulent velocities from SDSS DR17 with
uncertainties from \citet{Jonsson2020}}
\tablecomments{The photospheric stellar parameter uncertainties in this
table indicate random uncertainties alone, so systematic uncertainties
of 50 K, 0.03 dex, and 0.04 dex in $T_{\text{eff}}$, $\log{g}$, and
$[\text{Fe/H}]$ could be added in quadrature to the random uncertainties
in the table to account for the possible impact of systematic
uncertainties on our photospheric stellar parameter inferences.}
\end{deluxetable*}
\end{longrotatetable}

\figsetstart
\figsetnum{2}
\figsettitle{Stellar Parameter Posteriors}

\figsetgrpstart
\figsetgrpnum{2.1}
\figsetgrptitle{A.\ Ross et al.\ K2-106}
\figsetplot{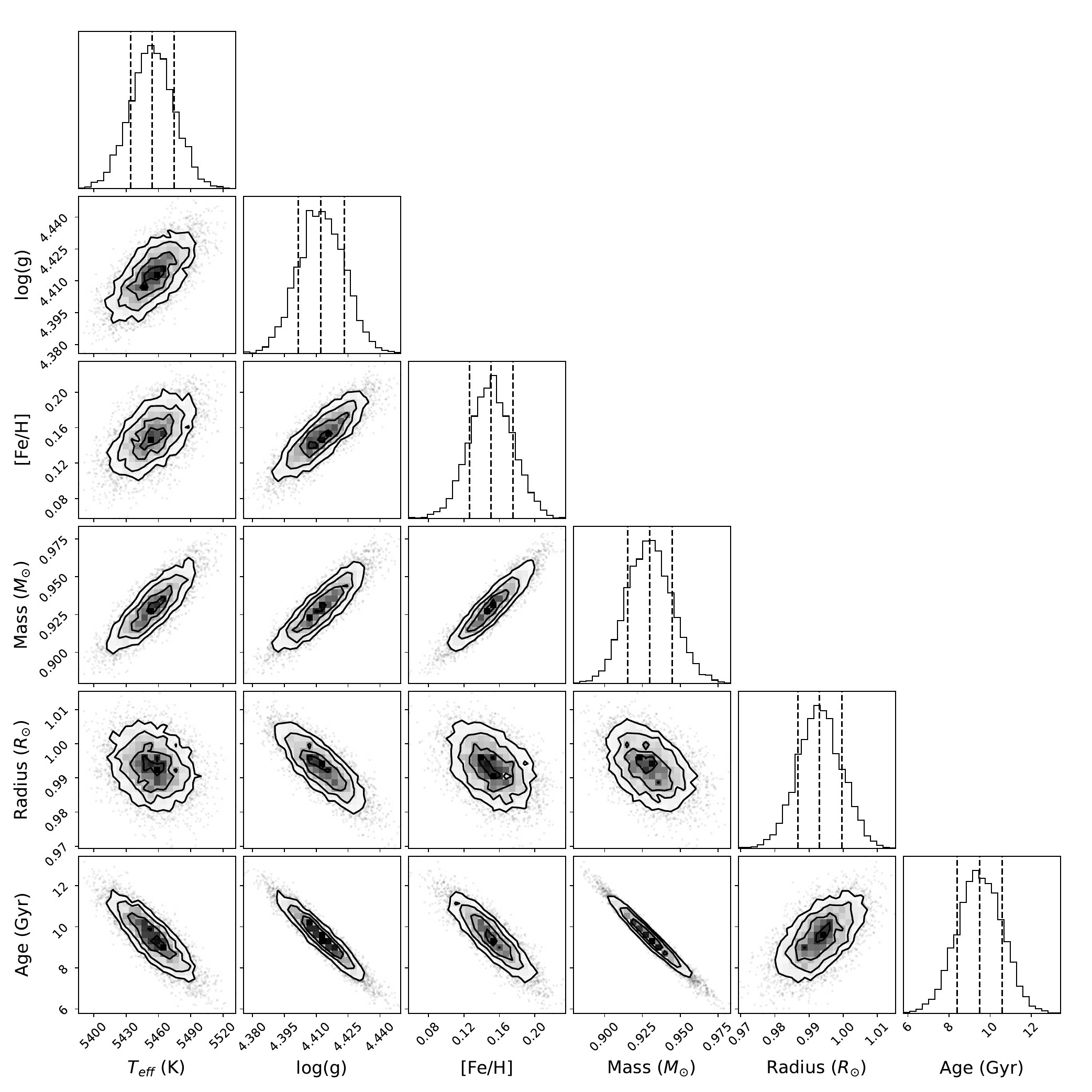}
\figsetgrpnote{Samples from the photospheric and fundamental stellar
parameter posterior resulting from our Bayesian analysis of astrometric,
photometric, and spectroscopic data.}
\figsetgrpend

\figsetgrpstart
\figsetgrpnum{2.2}
\figsetgrptitle{A.\ Ross et al.\ HD 15337}
\figsetplot{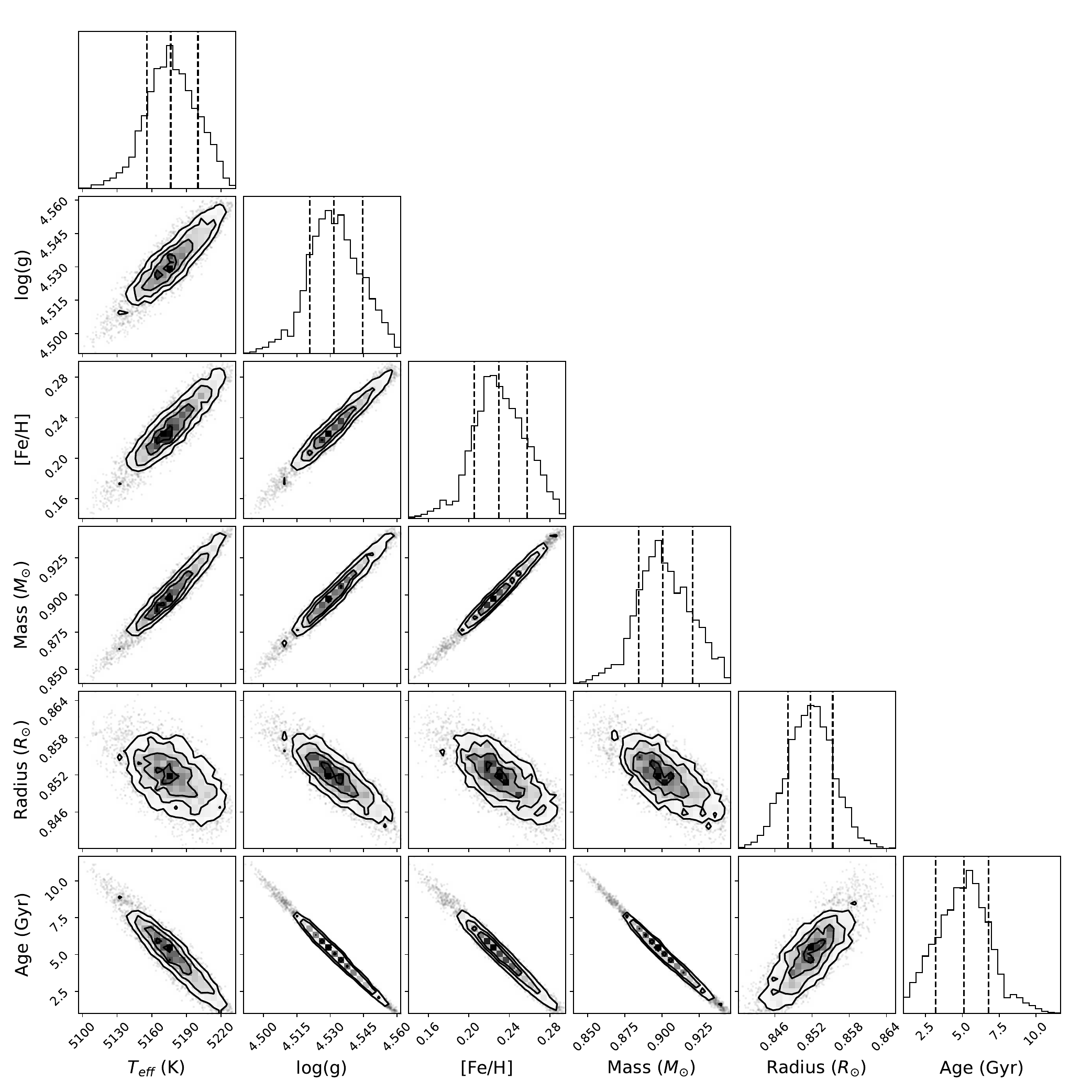}
\figsetgrpnote{Samples from the photospheric and fundamental stellar
parameter posterior resulting from our Bayesian analysis of astrometric,
photometric, and spectroscopic data.}
\figsetgrpend

\figsetgrpstart
\figsetgrpnum{2.3}
\figsetgrptitle{A.\ Ross et al.\ K2-291}
\figsetplot{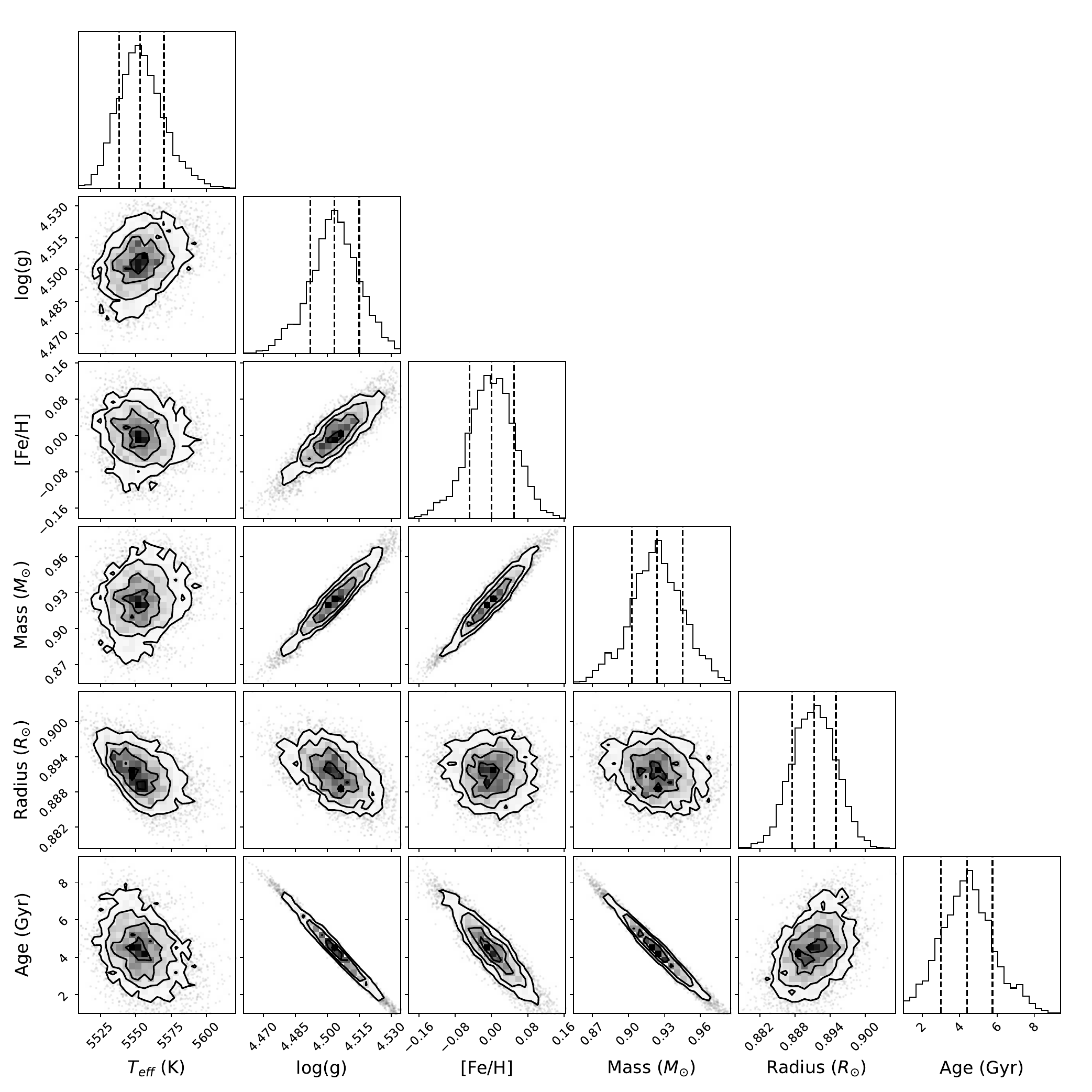}
\figsetgrpnote{Samples from the photospheric and fundamental stellar
parameter posterior resulting from our Bayesian analysis of astrometric,
photometric, and spectroscopic data.}
\figsetgrpend

\figsetgrpstart
\figsetgrpnum{2.4}
\figsetgrptitle{A.\ Ross et al.\ 55 Cnc}
\figsetplot{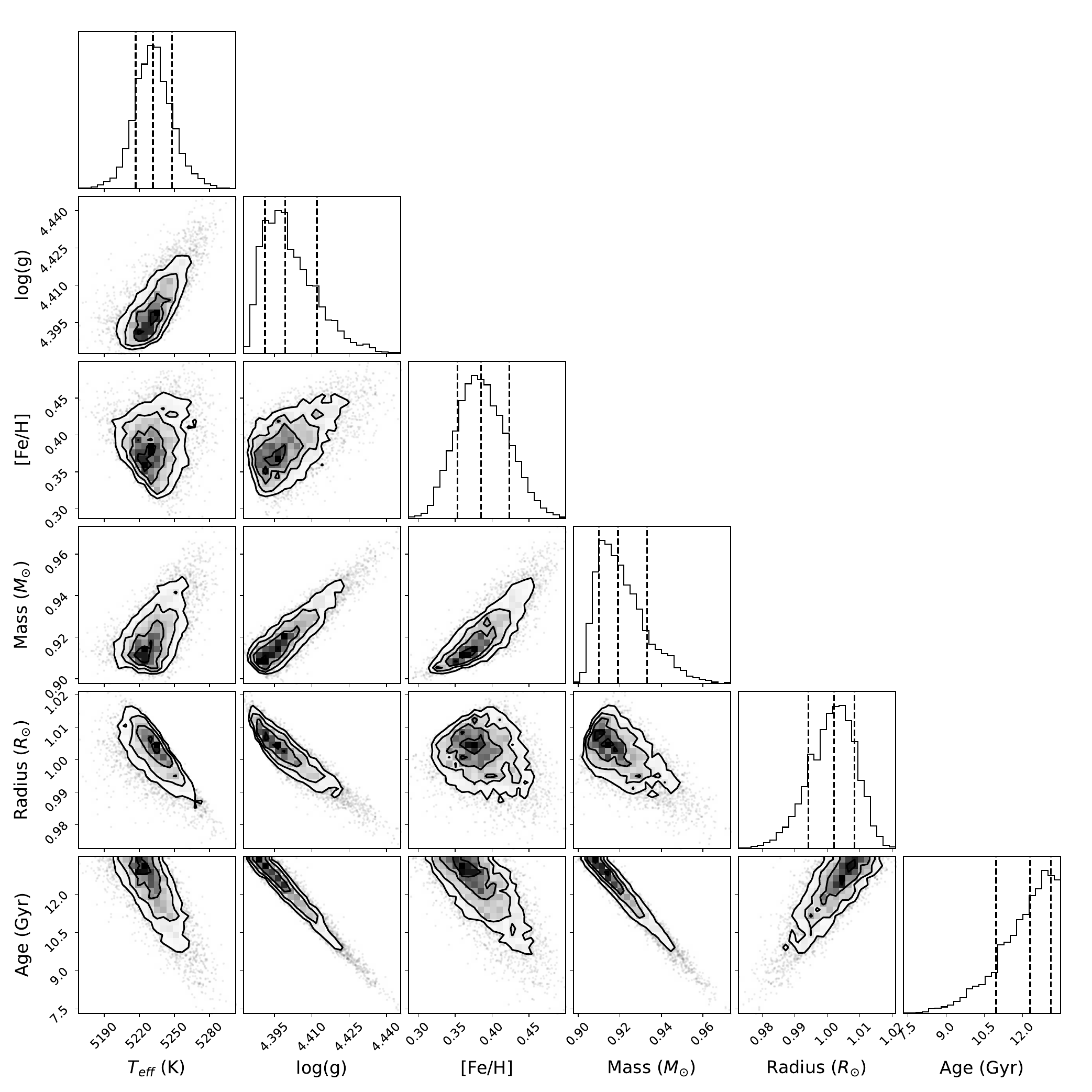}
\figsetgrpnote{Samples from the photospheric and fundamental stellar
parameter posterior resulting from our Bayesian analysis of astrometric,
photometric, and spectroscopic data.}
\figsetgrpend

\figsetgrpstart
\figsetgrpnum{2.5}
\figsetgrptitle{A.\ Ross et al.\ HD 80653}
\figsetplot{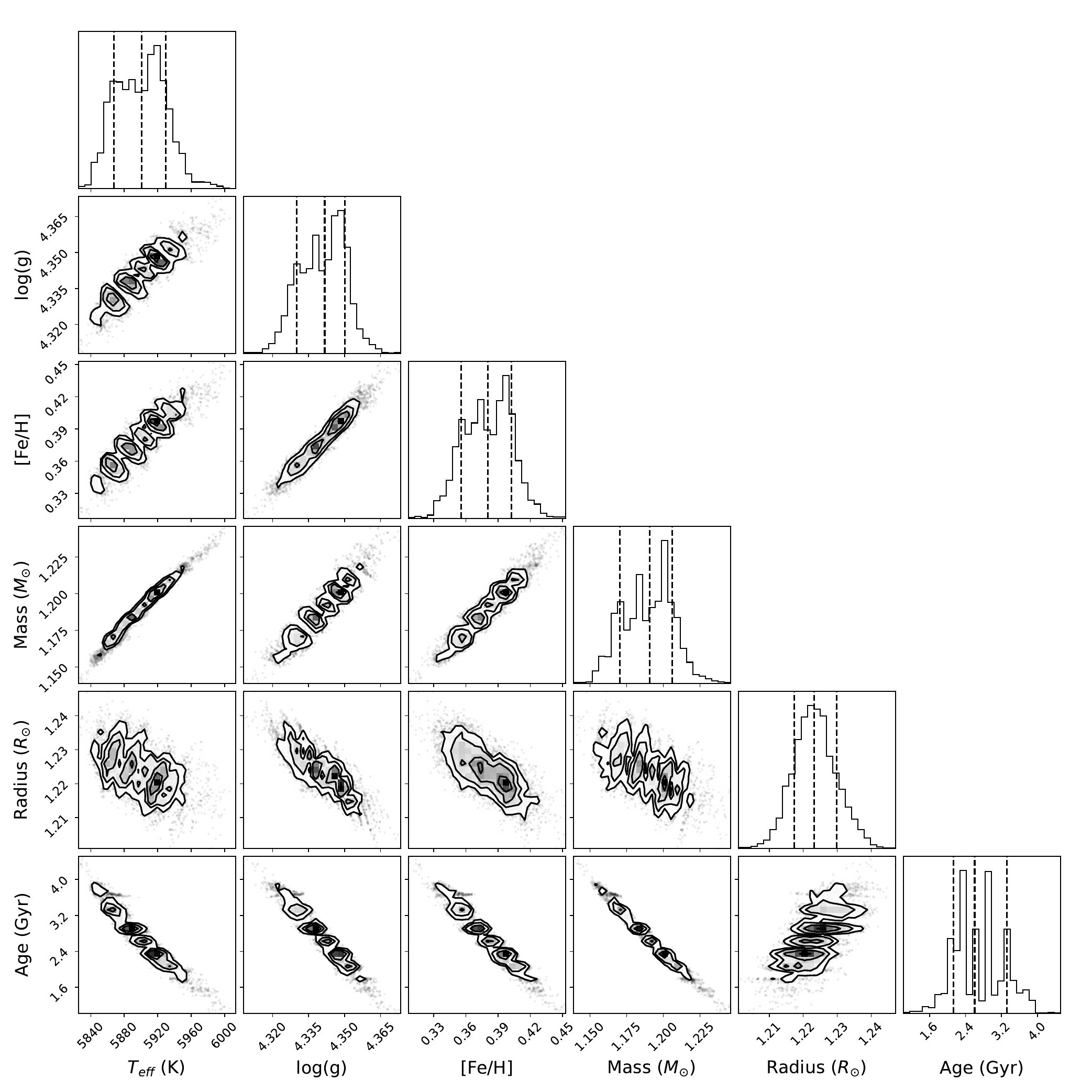}
\figsetgrpnote{Samples from the photospheric and fundamental stellar
parameter posterior resulting from our Bayesian analysis of astrometric,
photometric, and spectroscopic data.}
\figsetgrpend

\figsetgrpstart
\figsetgrpnum{2.6}
\figsetgrptitle{A.\ Ross et al.\ K2-131}
\figsetplot{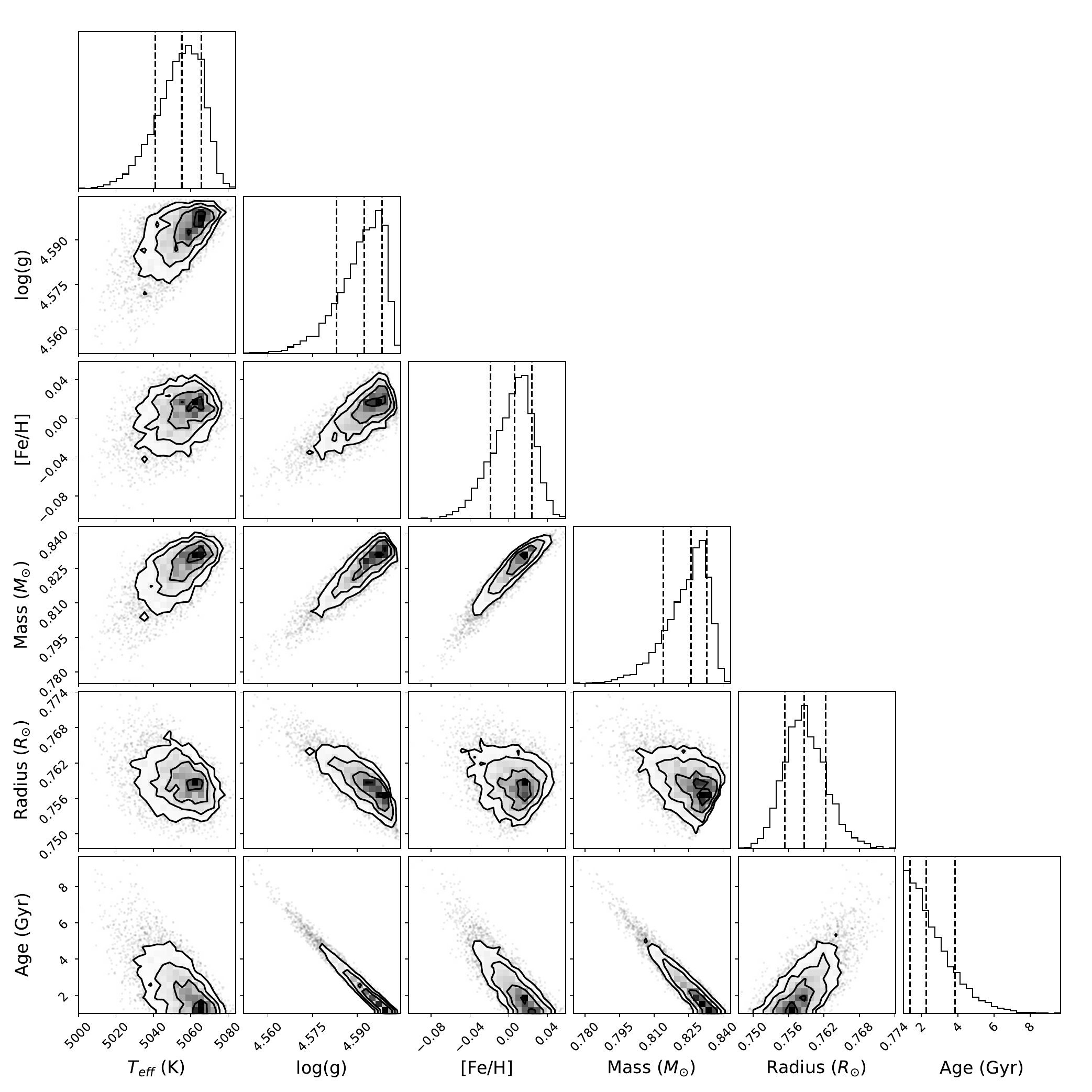}
\figsetgrpnote{Samples from the photospheric and fundamental stellar
parameter posterior resulting from our Bayesian analysis of astrometric,
photometric, and spectroscopic data.}
\figsetgrpend

\figsetgrpstart
\figsetgrpnum{2.7}
\figsetgrptitle{A.\ Ross et al.\ K2-229}
\figsetplot{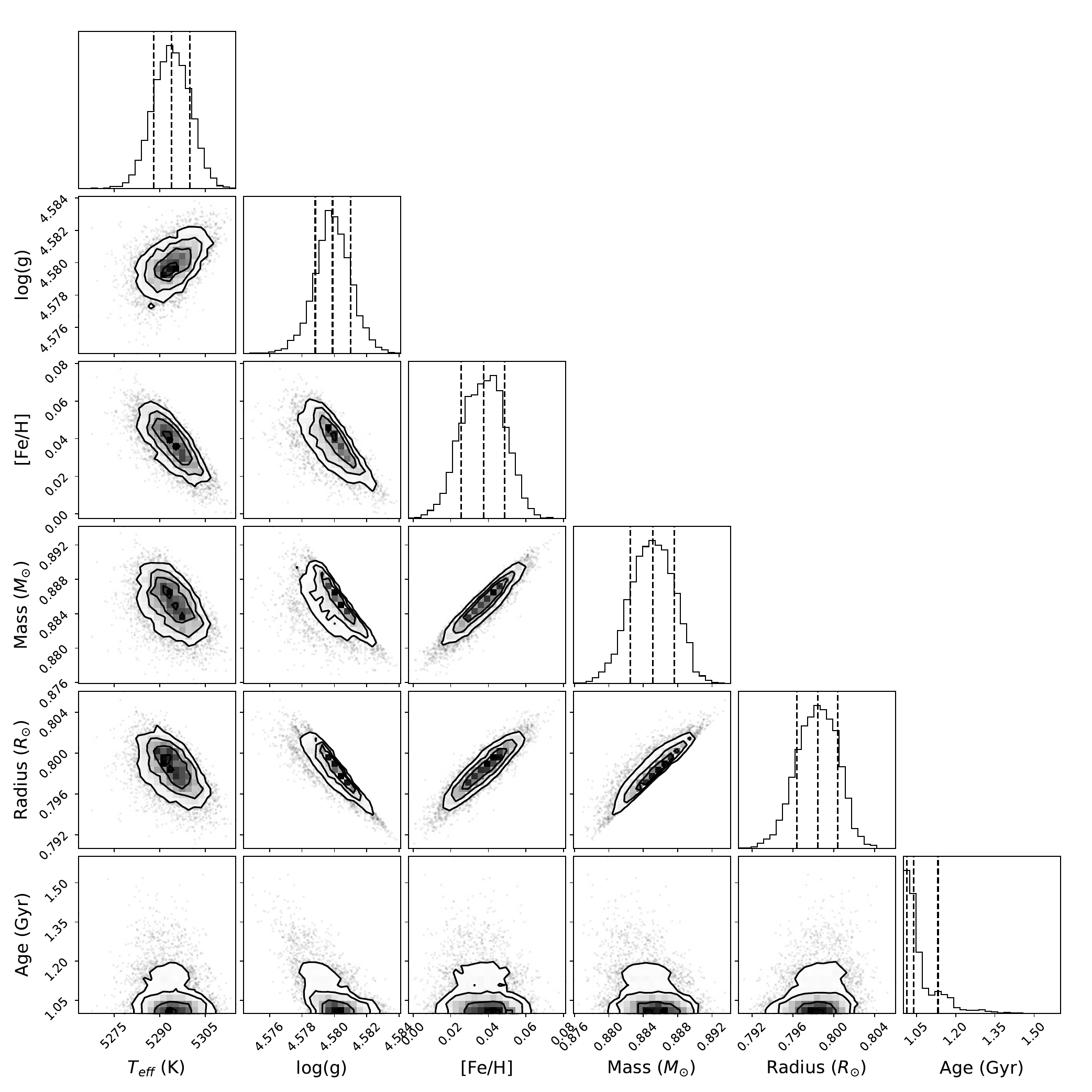}
\figsetgrpnote{Samples from the photospheric and fundamental stellar
parameter posterior resulting from our Bayesian analysis of astrometric,
photometric, and spectroscopic data.}
\figsetgrpend

\figsetgrpstart
\figsetgrpnum{2.8}
\figsetgrptitle{A.\ Ross et al.\ HD 136352}
\figsetplot{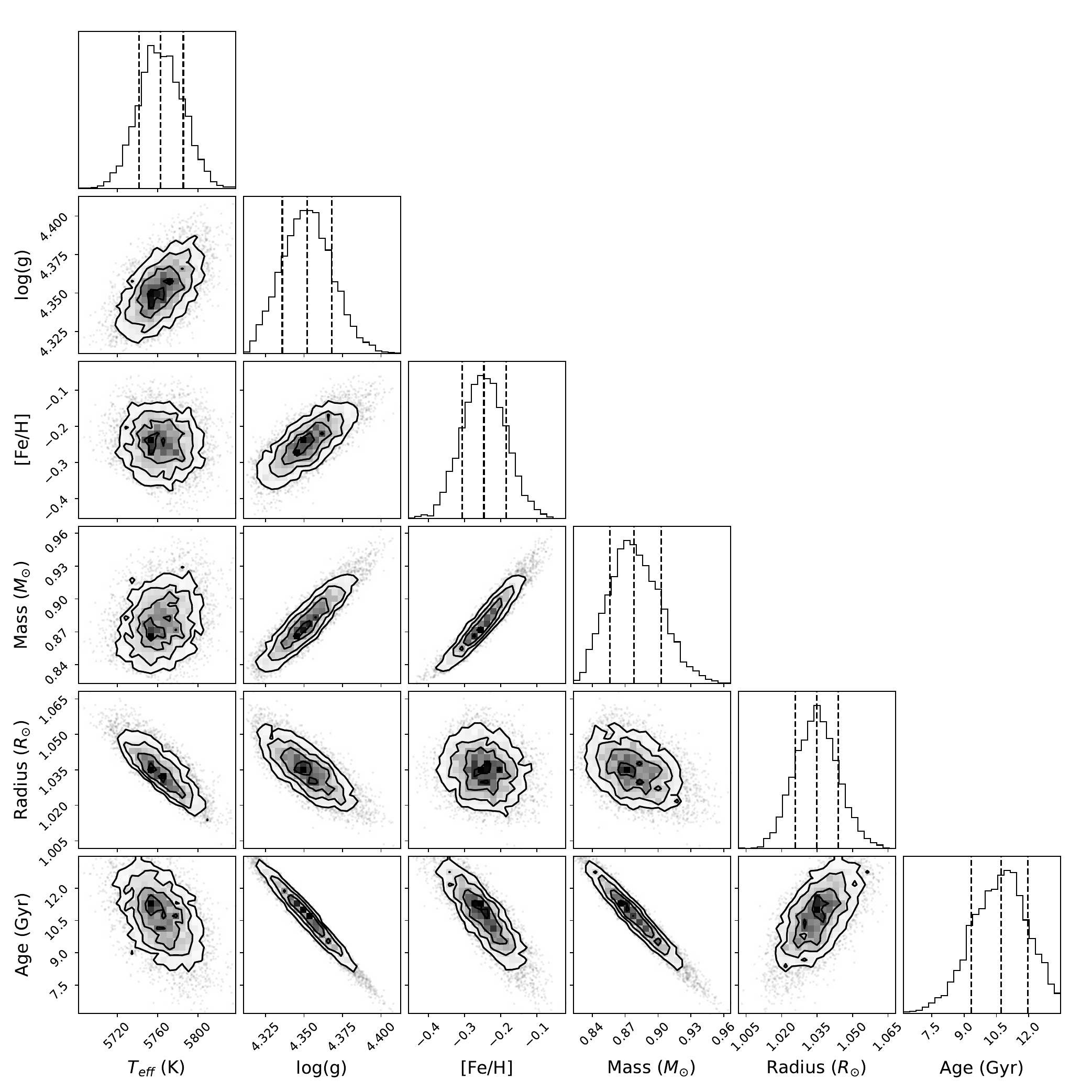}
\figsetgrpnote{Samples from the photospheric and fundamental stellar
parameter posterior resulting from our Bayesian analysis of astrometric,
photometric, and spectroscopic data.}
\figsetgrpend

\figsetgrpstart
\figsetgrpnum{2.9}
\figsetgrptitle{A.\ Ross et al.\ K2-38}
\figsetplot{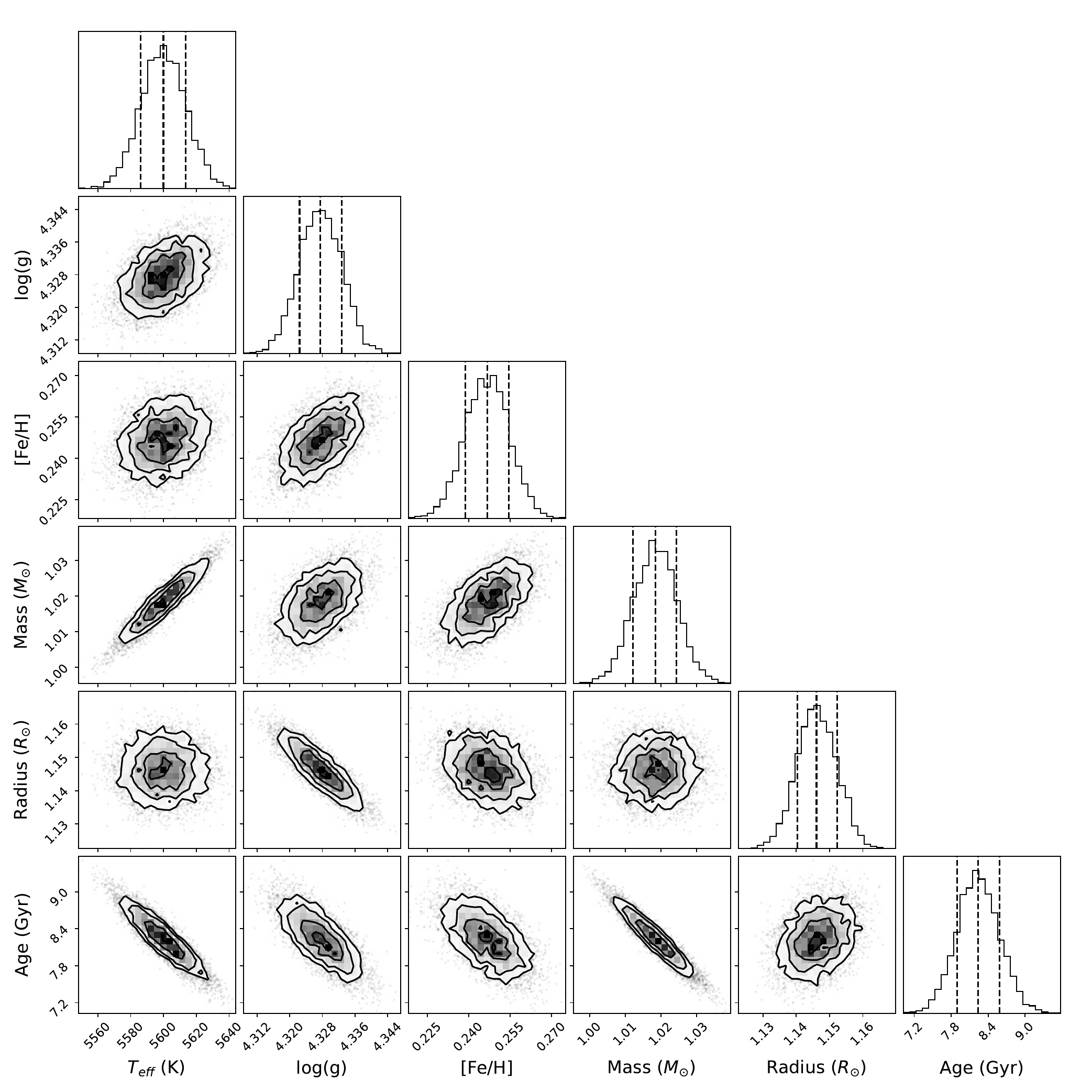}
\figsetgrpnote{Samples from the photospheric and fundamental stellar
parameter posterior resulting from our Bayesian analysis of astrometric,
photometric, and spectroscopic data.}
\figsetgrpend

\figsetgrpstart
\figsetgrpnum{2.10}
\figsetgrptitle{A.\ Ross et al.\ Kepler-10}
\figsetplot{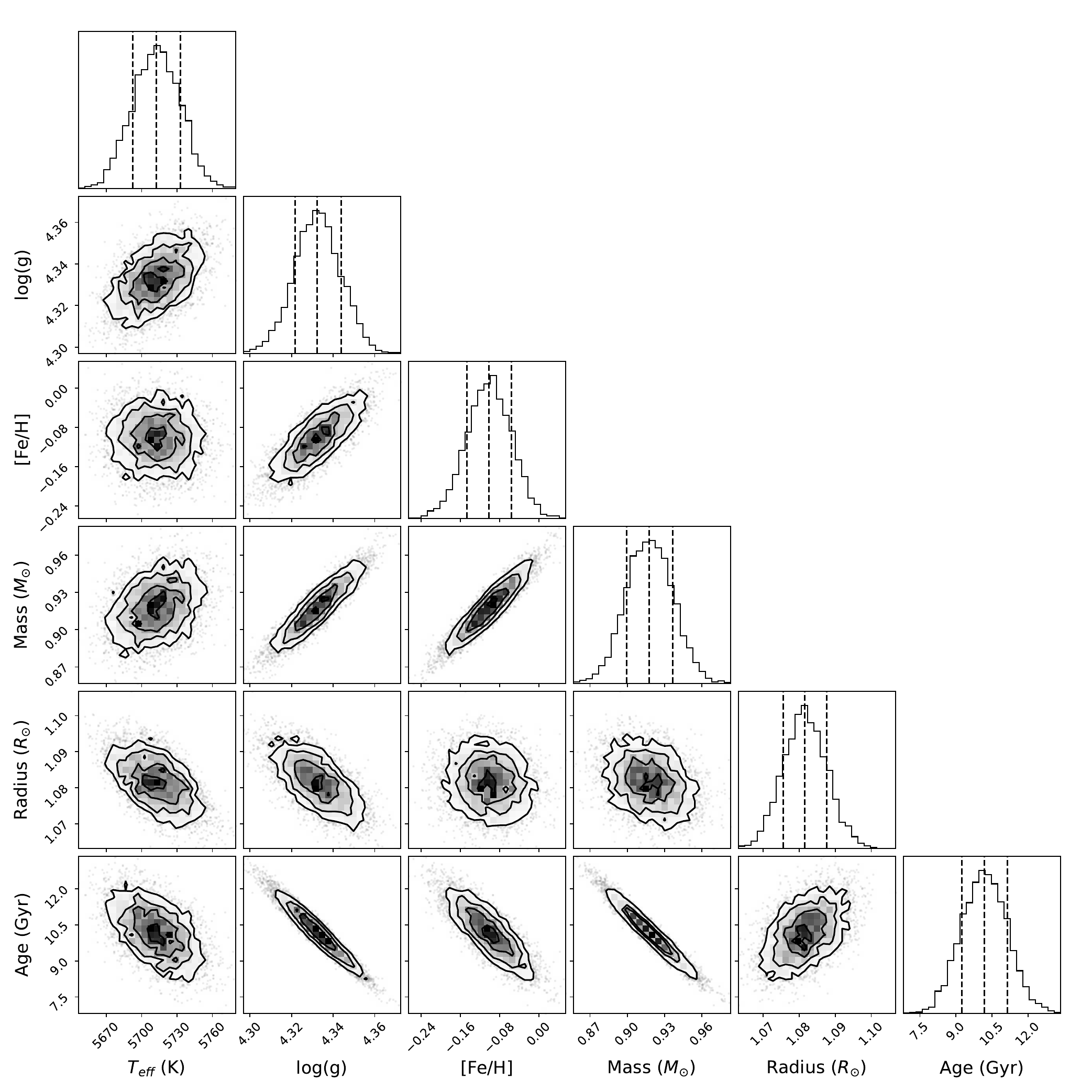}
\figsetgrpnote{Samples from the photospheric and fundamental stellar
parameter posterior resulting from our Bayesian analysis of astrometric,
photometric, and spectroscopic data.}
\figsetgrpend

\figsetgrpstart
\figsetgrpnum{2.11}
\figsetgrptitle{A.\ Ross et al.\ Kepler-20}
\figsetplot{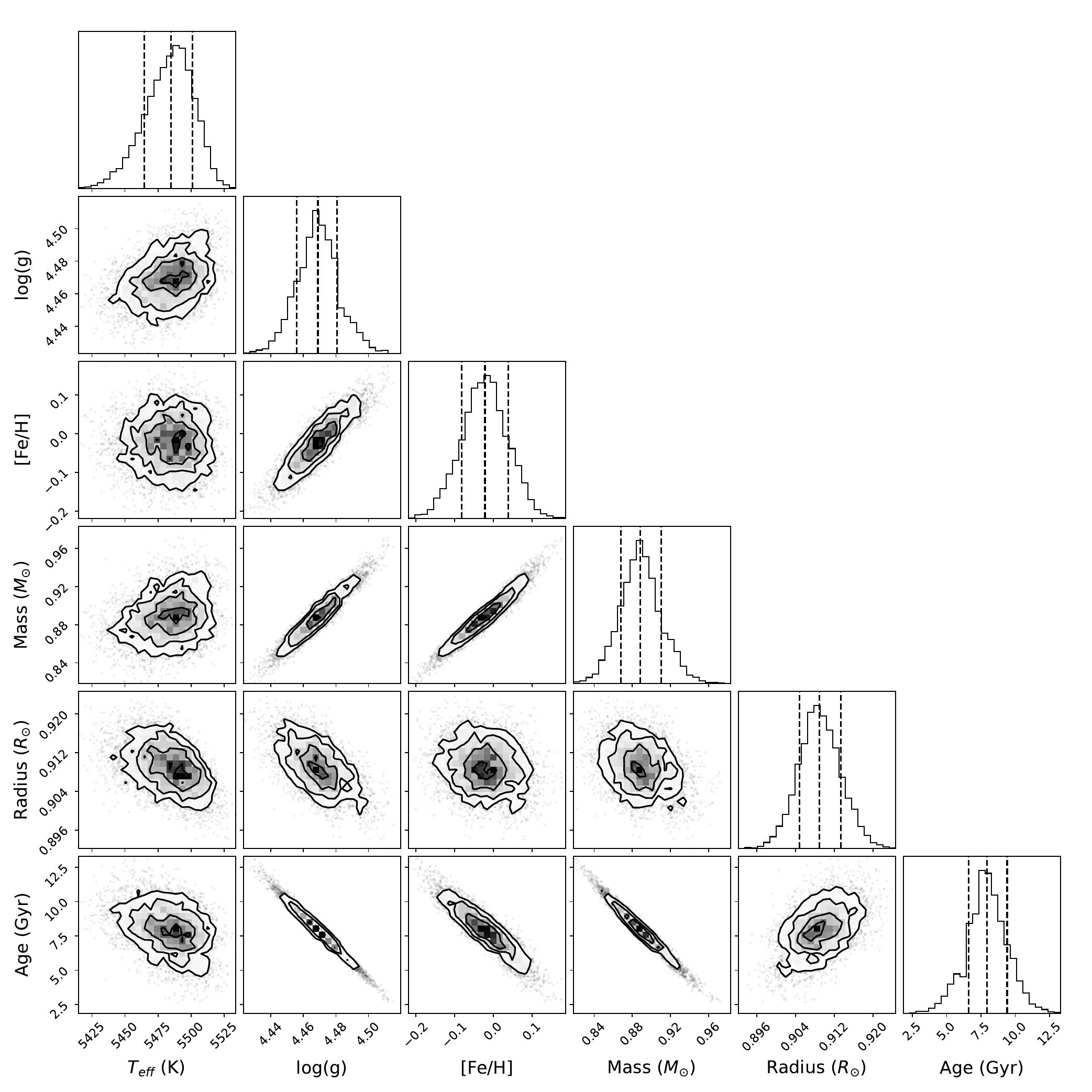}
\figsetgrpnote{Samples from the photospheric and fundamental stellar
parameter posterior resulting from our Bayesian analysis of astrometric,
photometric, and spectroscopic data.}
\figsetgrpend

\figsetgrpstart
\figsetgrpnum{2.12}
\figsetgrptitle{A.\ Ross et al.\ Kepler-36}
\figsetplot{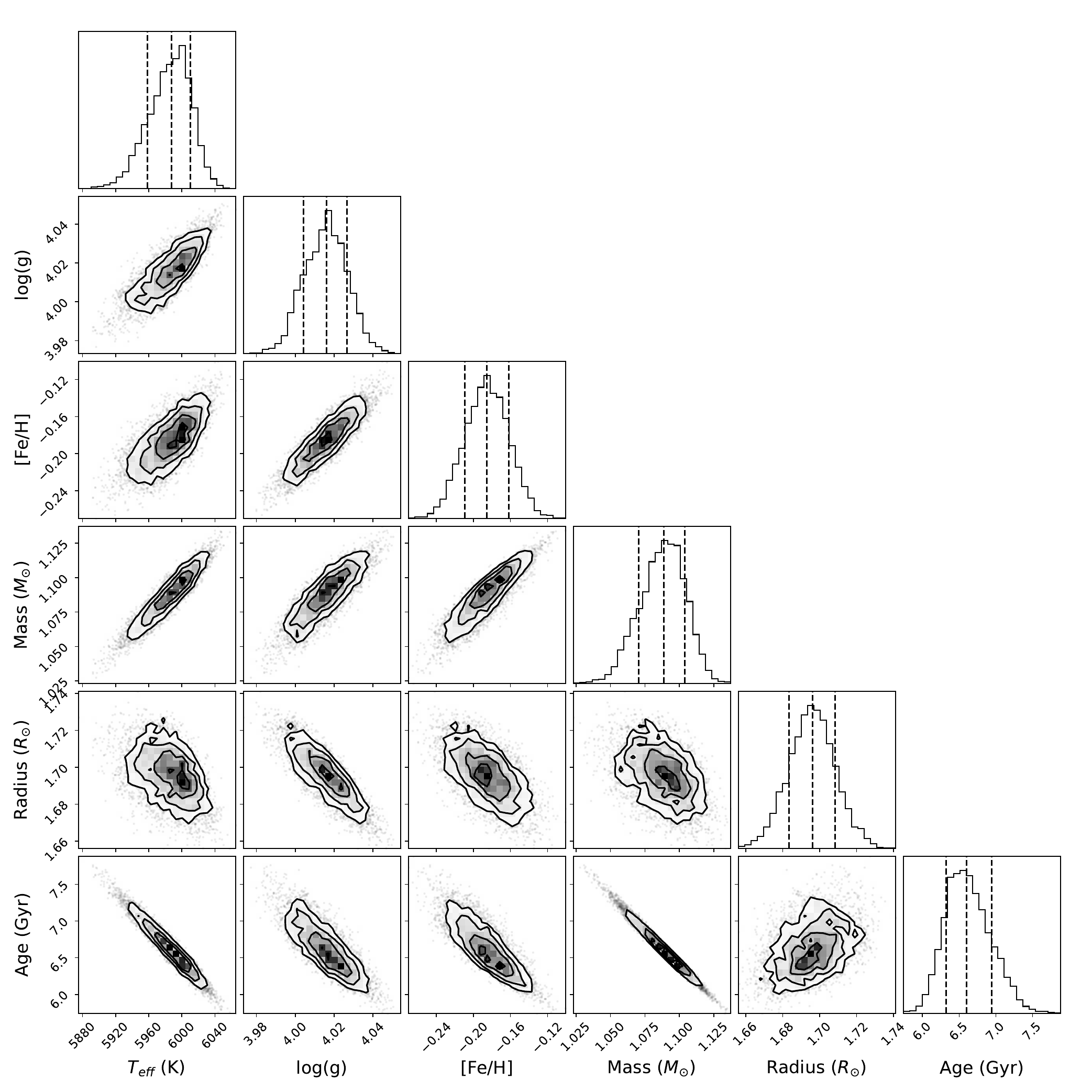}
\figsetgrpnote{Samples from the photospheric and fundamental stellar
parameter posterior resulting from our Bayesian analysis of astrometric,
photometric, and spectroscopic data.}
\figsetgrpend

\figsetgrpstart
\figsetgrpnum{2.13}
\figsetgrptitle{A.\ Ross et al.\ Kepler-93}
\figsetplot{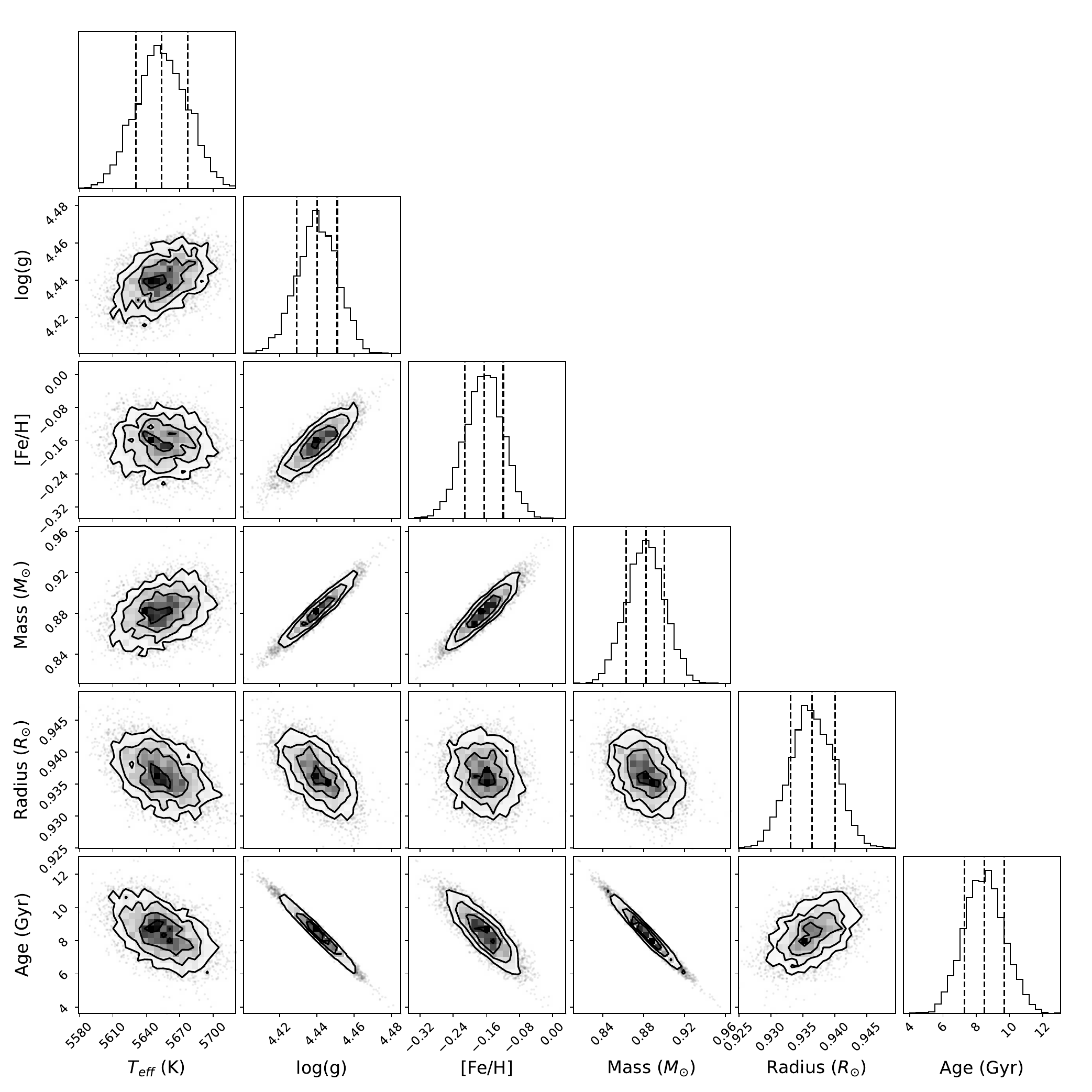}
\figsetgrpnote{Samples from the photospheric and fundamental stellar
parameter posterior resulting from our Bayesian analysis of astrometric,
photometric, and spectroscopic data.}
\figsetgrpend

\figsetgrpstart
\figsetgrpnum{2.14}
\figsetgrptitle{A.\ Ross et al.\ Kepler-78}
\figsetplot{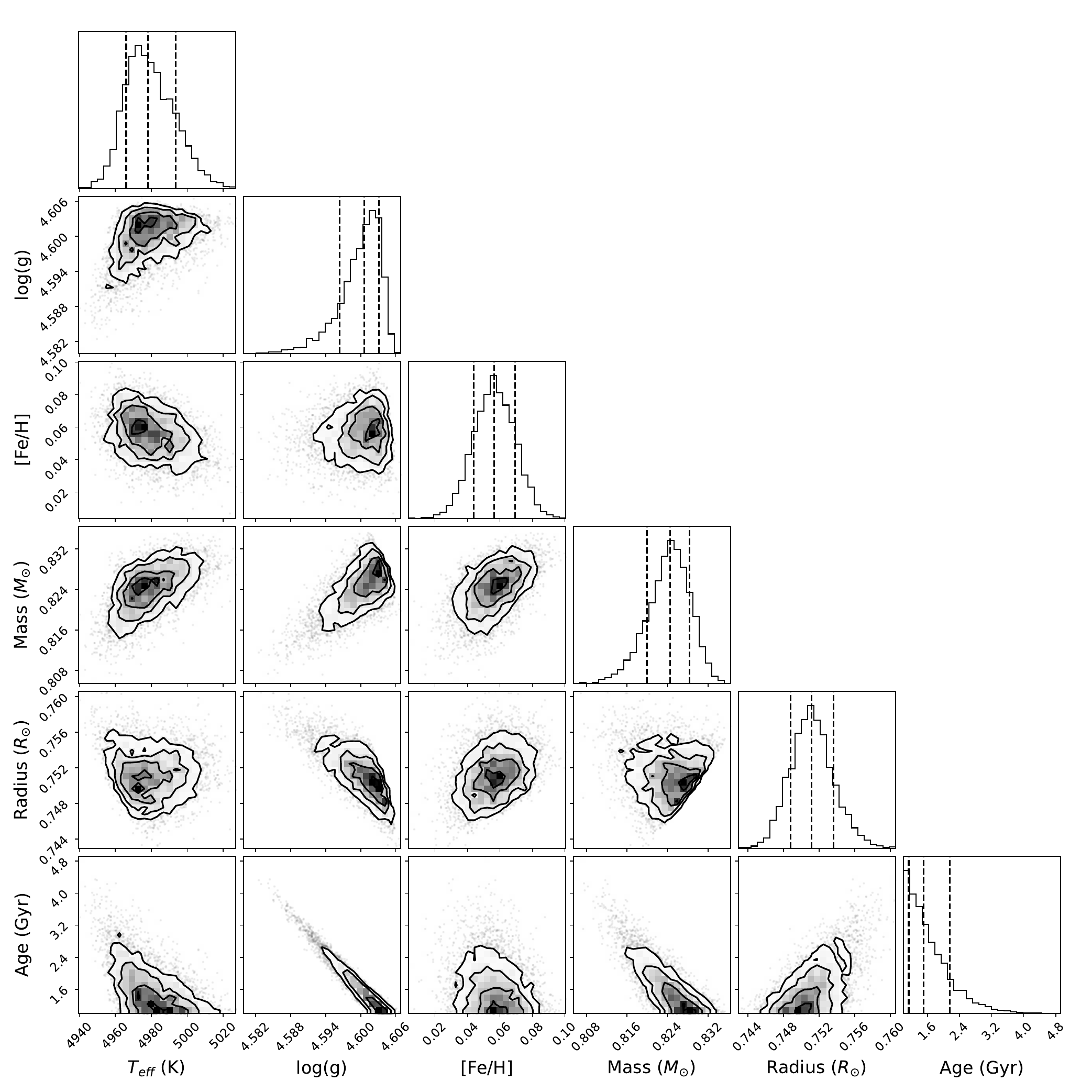}
\figsetgrpnote{Samples from the photospheric and fundamental stellar 
parameter posterior resulting from our Bayesian analysis of astrometric,
photometric, and spectroscopic data.}
\figsetgrpend

\figsetgrpstart
\figsetgrpnum{2.15}
\figsetgrptitle{A.\ Ross et al.\ Kepler-107}
\figsetplot{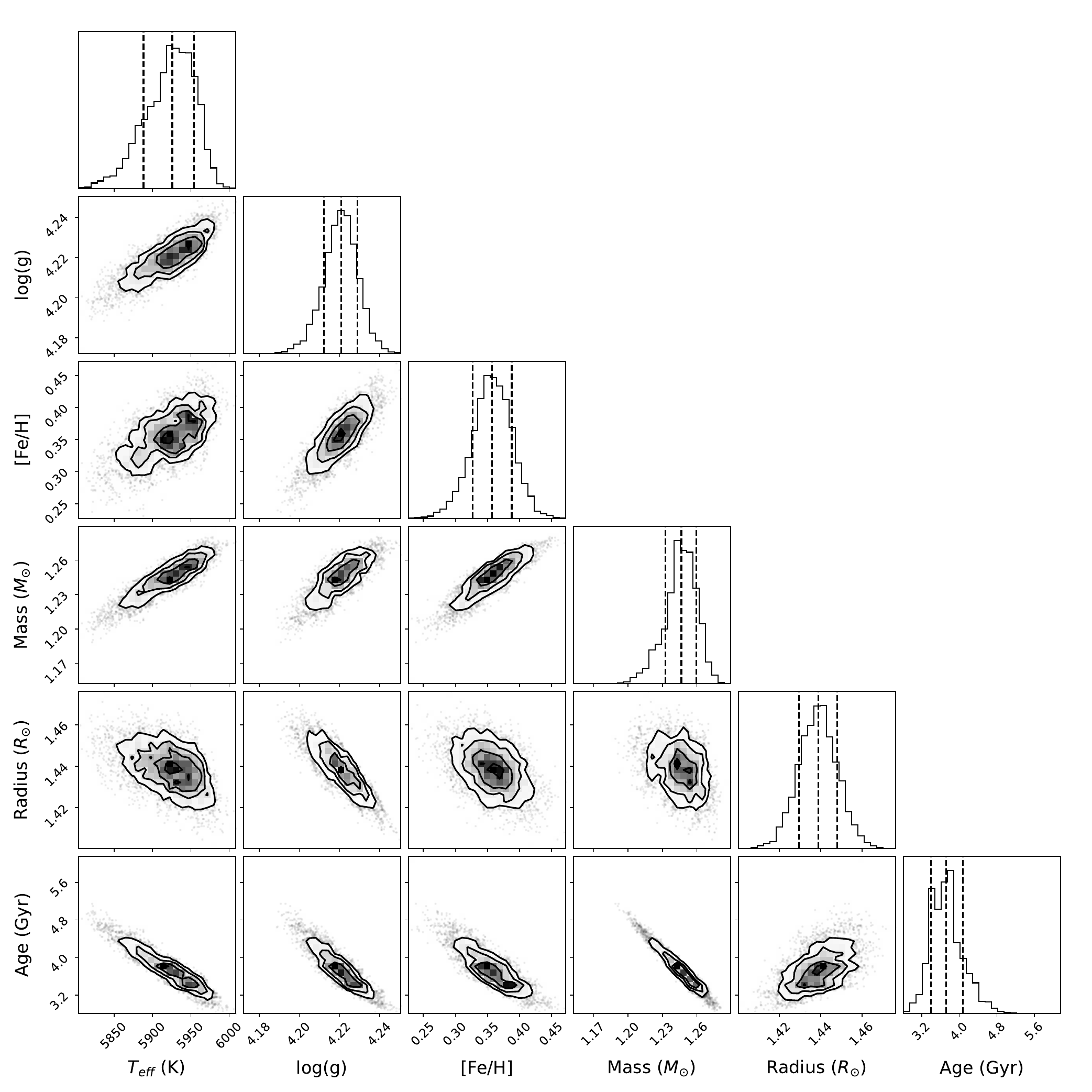}
\figsetgrpnote{Samples from the photospheric and fundamental stellar 
parameter posterior resulting from our Bayesian analysis of astrometric,
photometric, and spectroscopic data.}
\figsetgrpend

\figsetgrpstart
\figsetgrpnum{2.16}
\figsetgrptitle{A.\ Ross et al.\ WASP-47}
\figsetplot{WASP-47_rossetal_cornerplot.pdf}
\figsetgrpnote{Samples from the photospheric and fundamental stellar 
parameter posterior resulting from our Bayesian analysis of astrometric,
photometric, and spectroscopic data.}
\figsetgrpend

\figsetgrpstart
\figsetgrpnum{2.17}
\figsetgrptitle{A.\ Ross et al.\ HD 213885}
\figsetplot{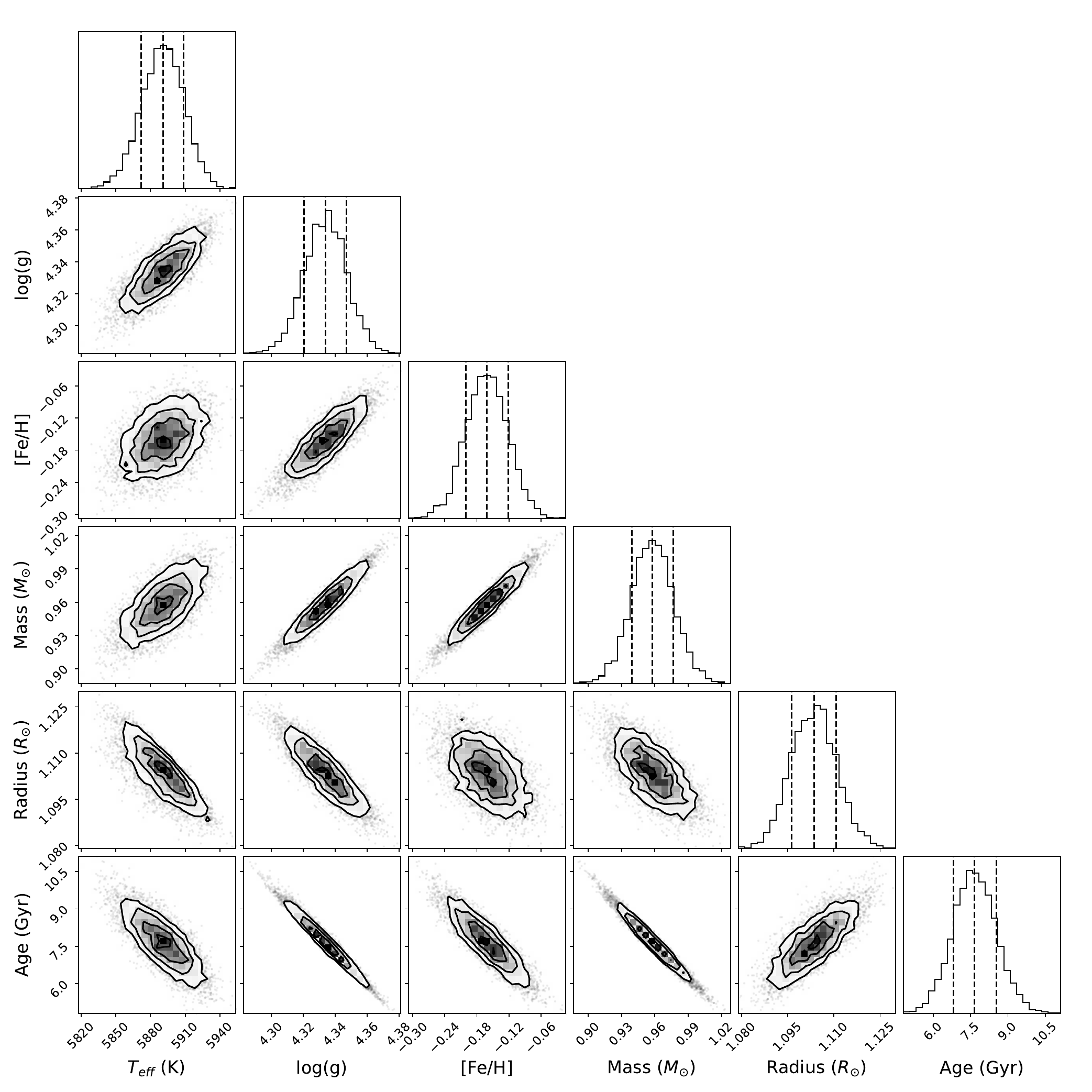}
\figsetgrpnote{Samples from the photospheric and fundamental stellar
parameter posterior resulting from our Bayesian analysis of astrometric,
photometric, and spectroscopic data.}
\figsetgrpend

\figsetgrpstart
\figsetgrpnum{2.18}
\figsetgrptitle{A.\ Ross et al.\ K2-265}
\figsetplot{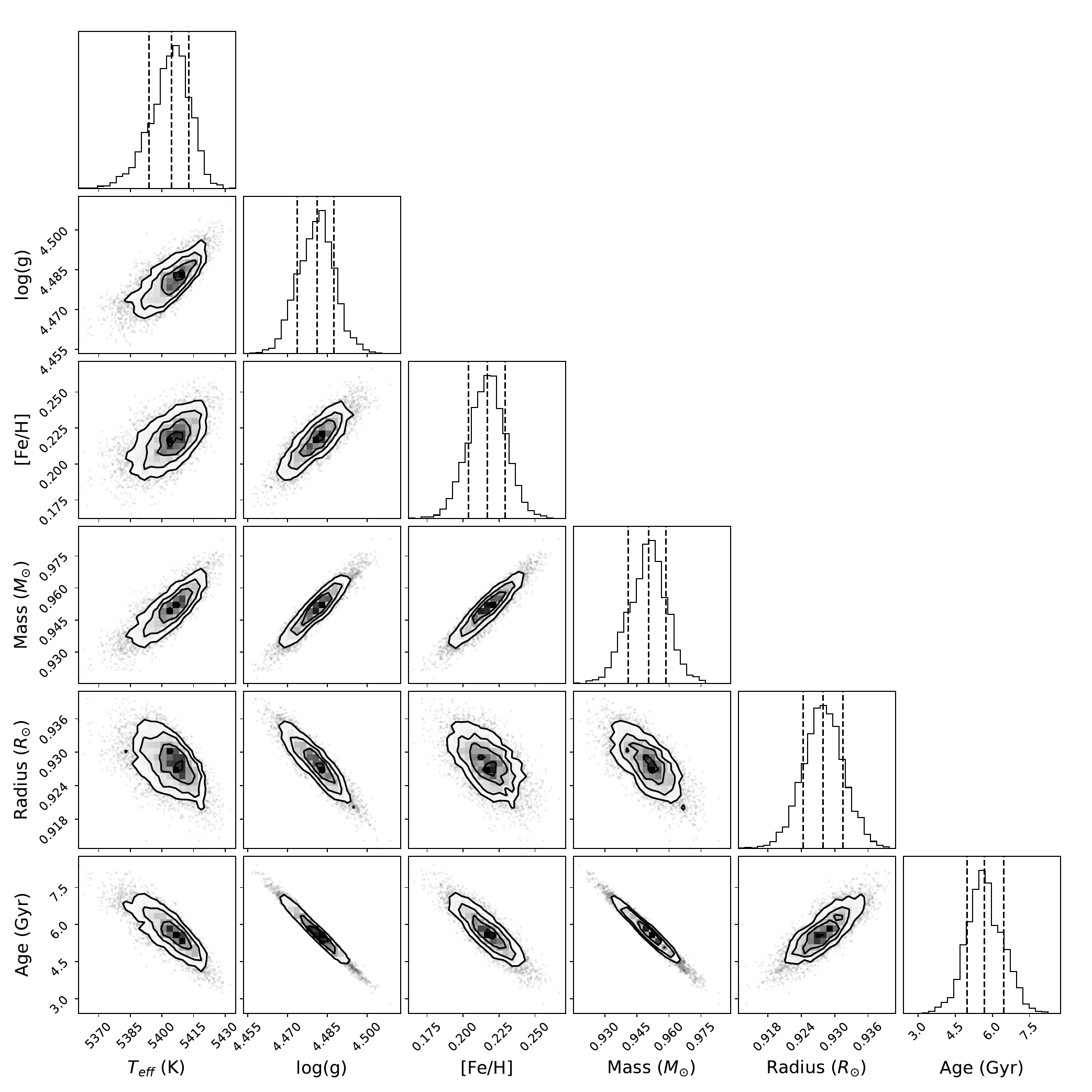}
\figsetgrpnote{Samples from the photospheric and fundamental stellar
parameter posterior resulting from our Bayesian analysis of astrometric,
photometric, and spectroscopic data.}
\figsetgrpend

\figsetgrpstart
\figsetgrpnum{2.19}
\figsetgrptitle{A.\ Ross et al.\ HD 219134}
\figsetplot{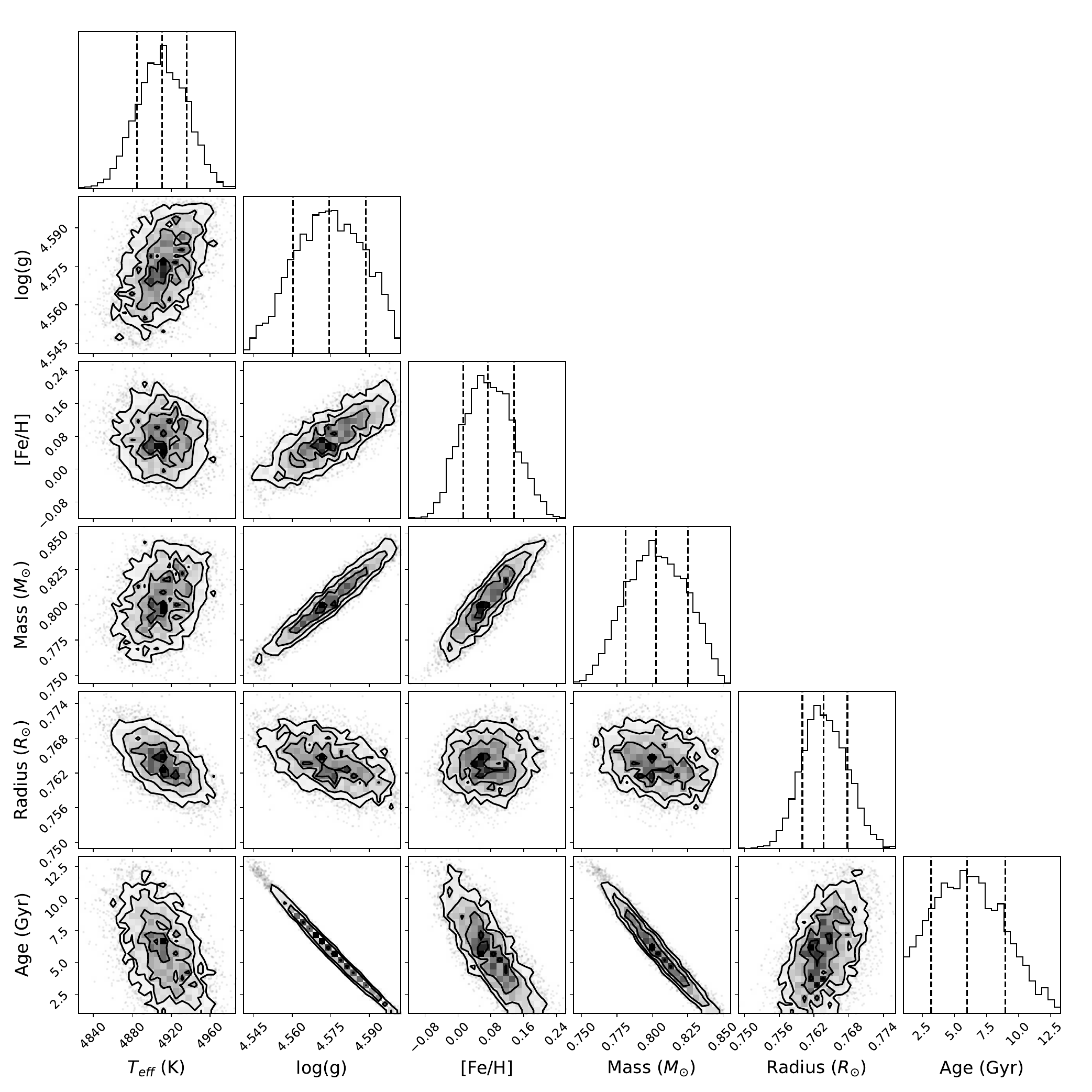}
\figsetgrpnote{Samples from the photospheric and fundamental stellar
parameter posterior resulting from our Bayesian analysis of astrometric,
photometric, and spectroscopic data.}
\figsetgrpend

\figsetgrpstart
\figsetgrpnum{2.20}
\figsetgrptitle{A.\ Ross et al.\ K2-141}
\figsetplot{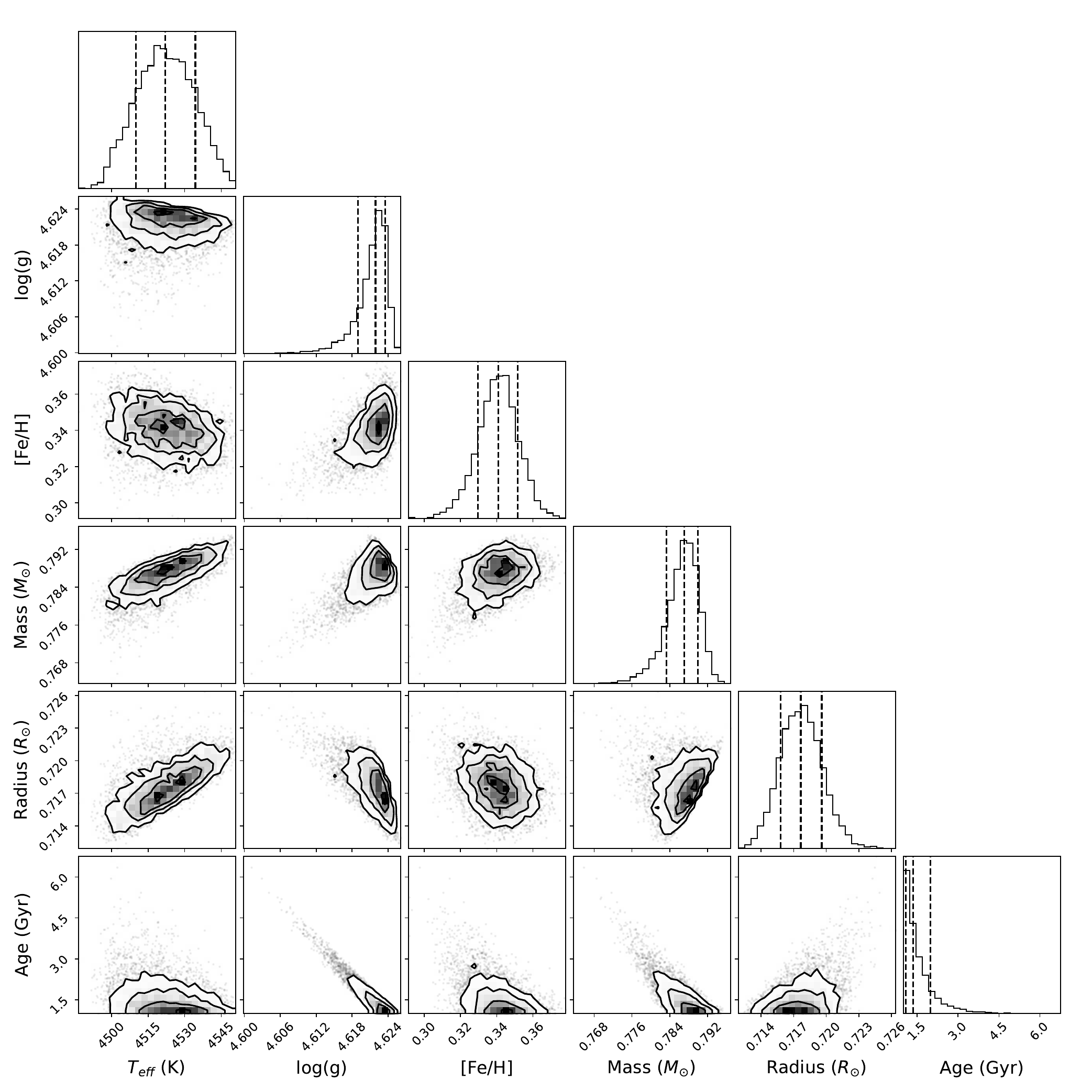}
\figsetgrpnote{Samples from the photospheric and fundamental stellar
parameter posterior resulting from our Bayesian analysis of astrometric,
photometric, and spectroscopic data.}
\figsetgrpend

\figsetgrpstart
\figsetgrpnum{2.21}
\figsetgrptitle{APOGEE DR17 K2-141}
\figsetplot{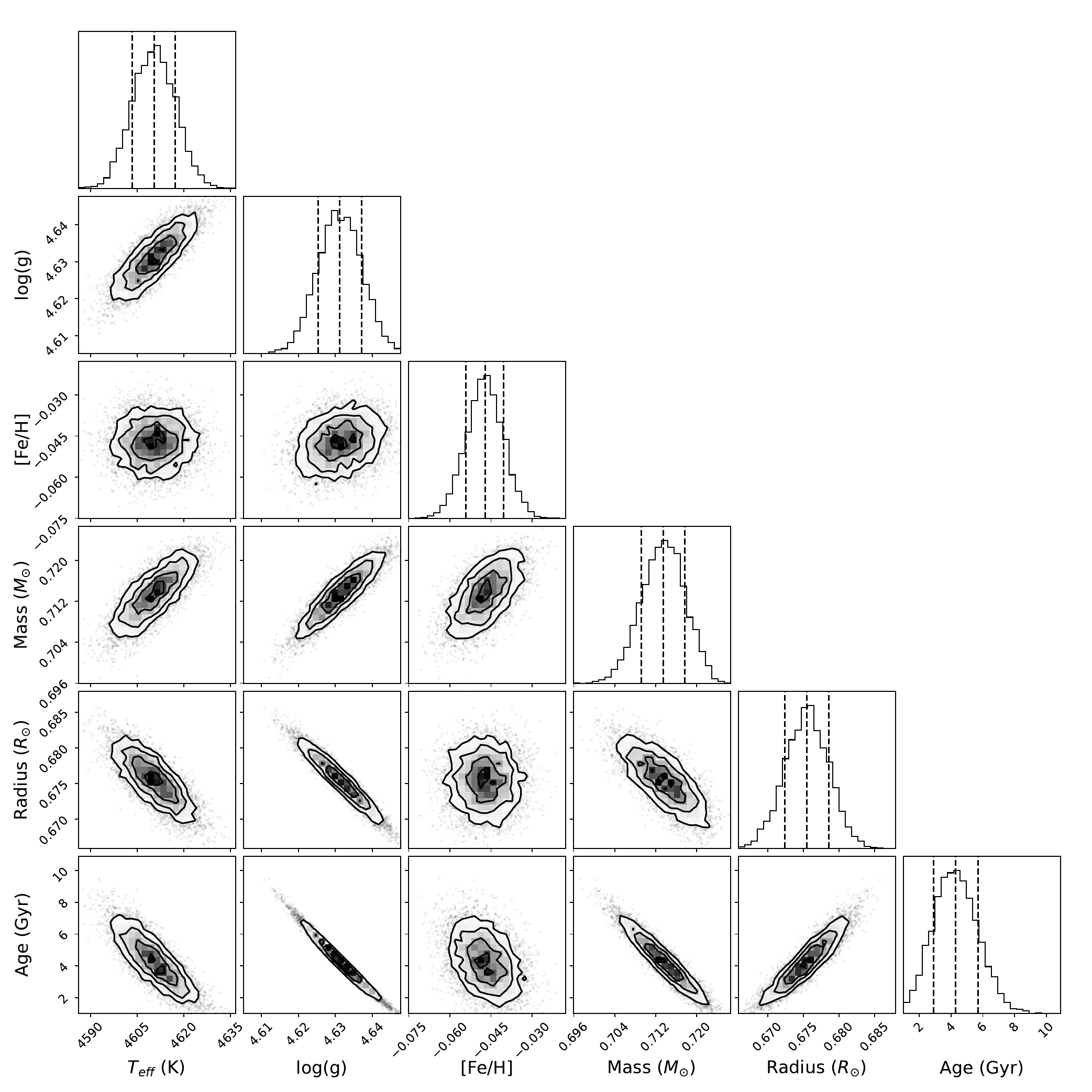}
\figsetgrpnote{Samples from the photospheric and fundamental stellar
parameter posterior resulting from our Bayesian analysis of astrometric,
photometric, and spectroscopic data.}
\figsetgrpend

\figsetgrpstart
\figsetgrpnum{2.22}
\figsetgrptitle{APOGEE DR17 K2-141}
\figsetplot{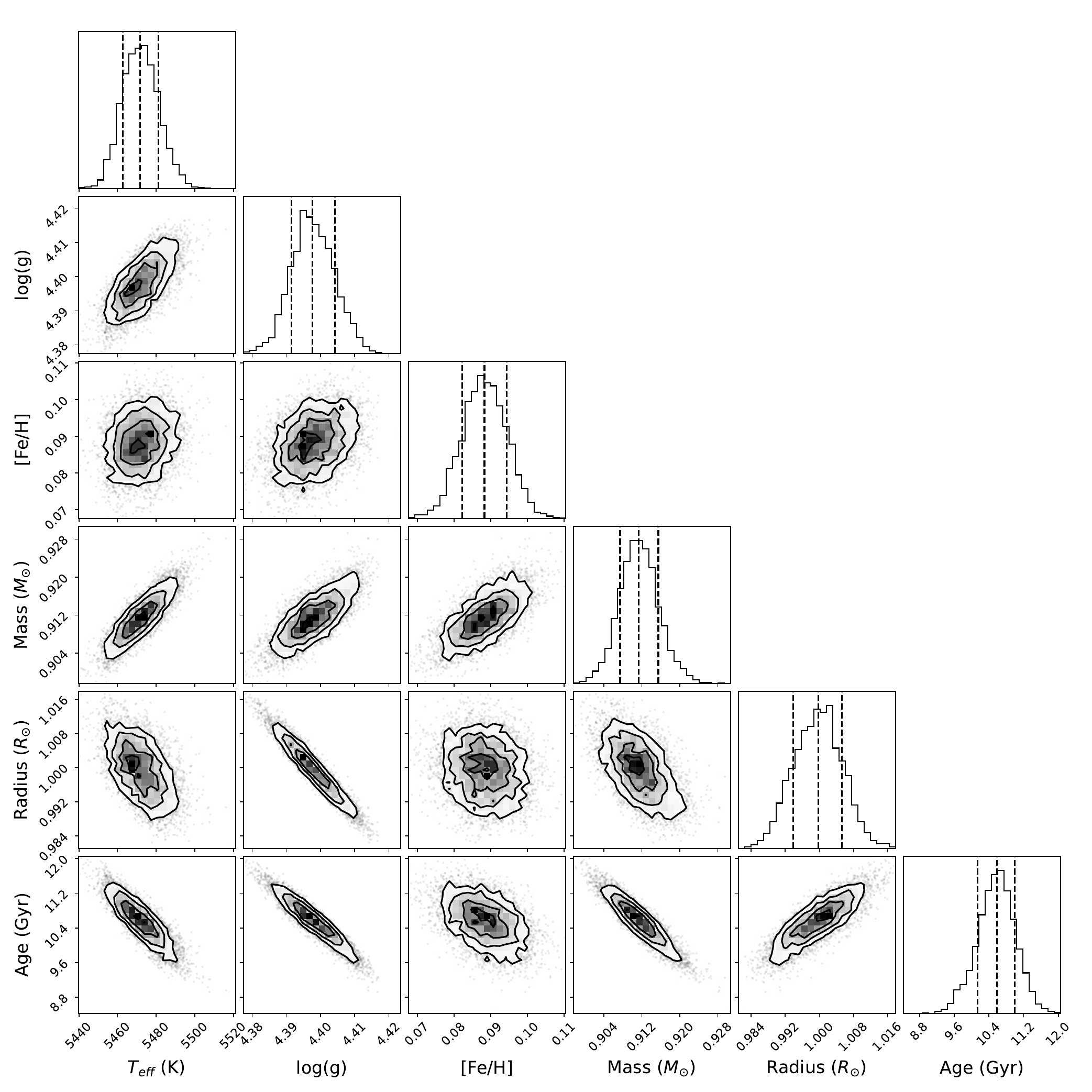}
\figsetgrpnote{Samples from the photospheric and fundamental stellar
parameter posterior resulting from our Bayesian analysis of astrometric,
photometric, and spectroscopic data.}
\figsetgrpend

\figsetgrpstart
\figsetgrpnum{2.23}
\figsetgrptitle{APOGEE DR17 K2-141}
\figsetplot{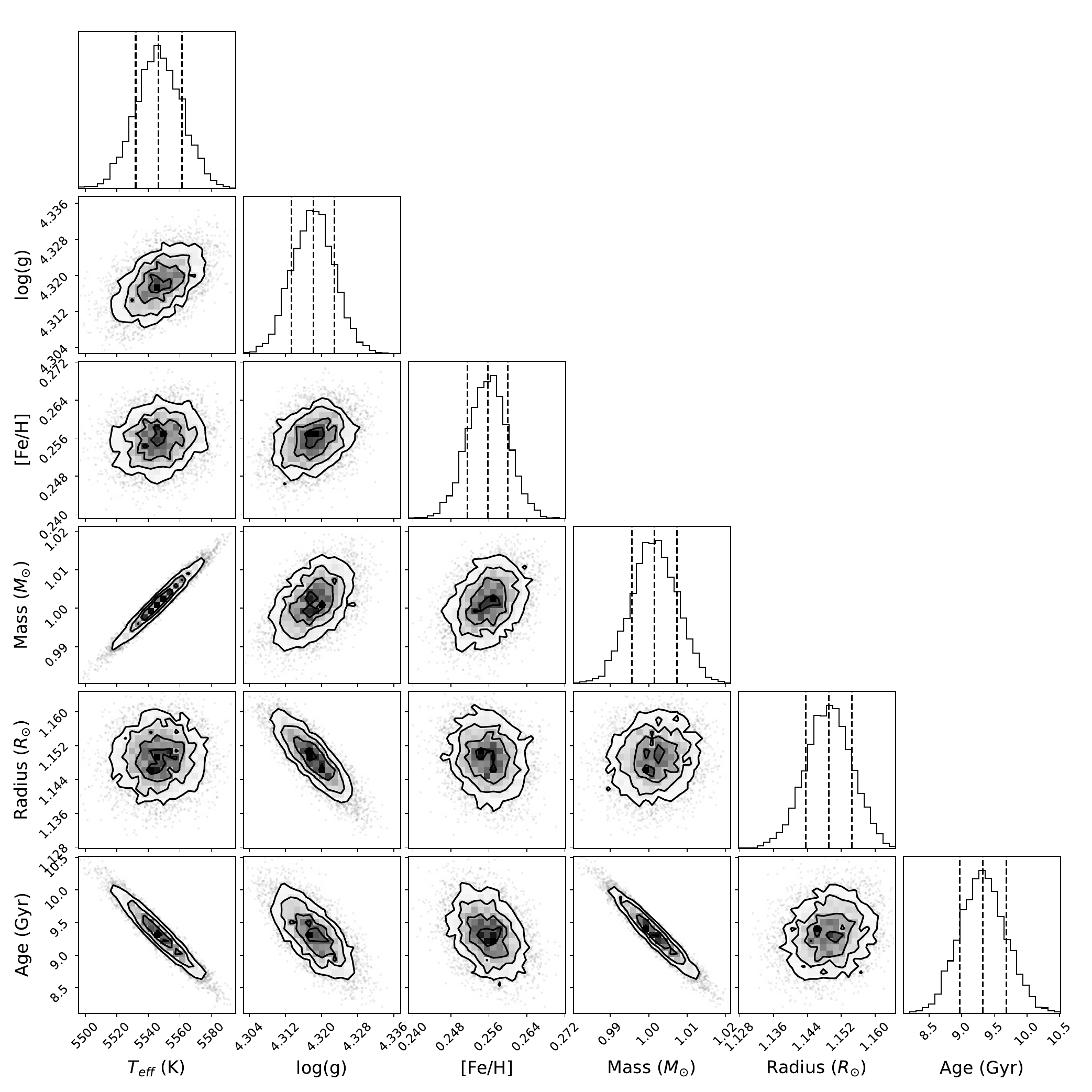}
\figsetgrpnote{Samples from the photospheric and fundamental stellar
parameter posterior resulting from our Bayesian analysis of astrometric,
photometric, and spectroscopic data.}
\figsetgrpend

\figsetgrpstart
\figsetgrpnum{2.24}
\figsetgrptitle{APOGEE DR17 Kepler-10}
\figsetplot{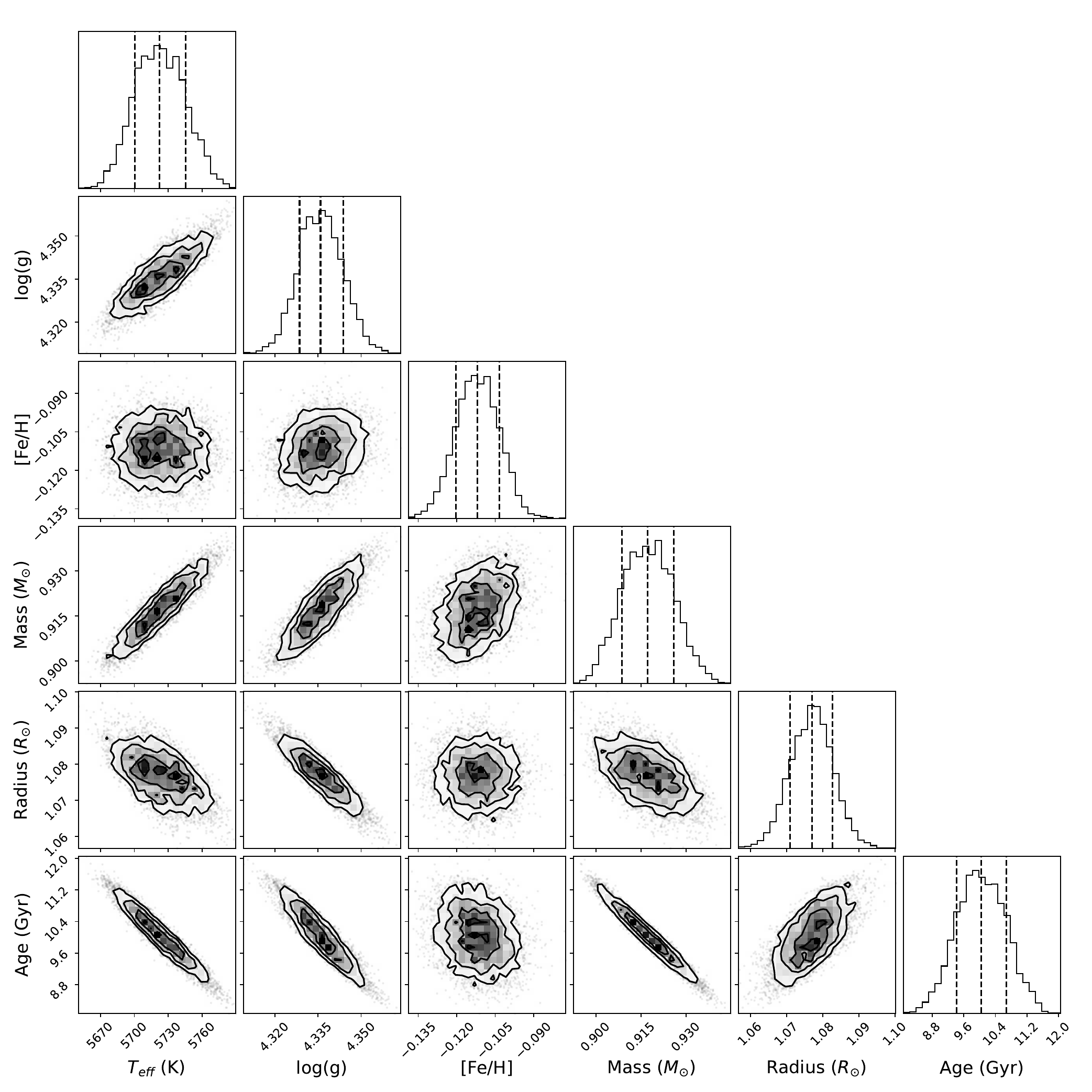}
\figsetgrpnote{Samples from the photospheric and fundamental stellar
parameter posterior resulting from our Bayesian analysis of astrometric,
photometric, and spectroscopic data.}
\figsetgrpend

\figsetgrpstart
\figsetgrpnum{2.25}
\figsetgrptitle{APOGEE DR17 Kepler-20}
\figsetplot{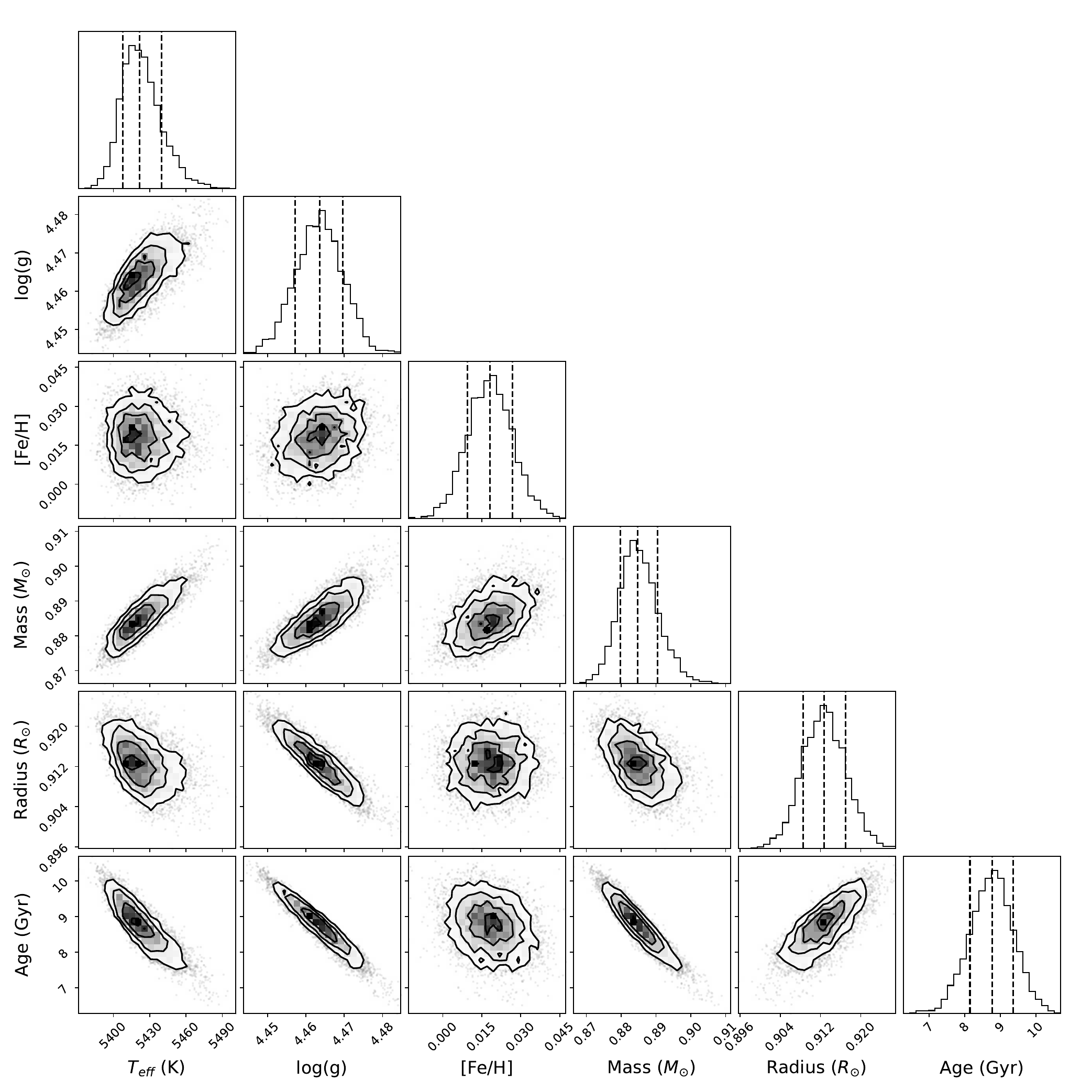}
\figsetgrpnote{Samples from the photospheric and fundamental stellar
parameter posterior resulting from our Bayesian analysis of astrometric,
photometric, and spectroscopic data.}
\figsetgrpend

\figsetgrpstart
\figsetgrpnum{2.26}
\figsetgrptitle{APOGEE DR17 Kepler-36}
\figsetplot{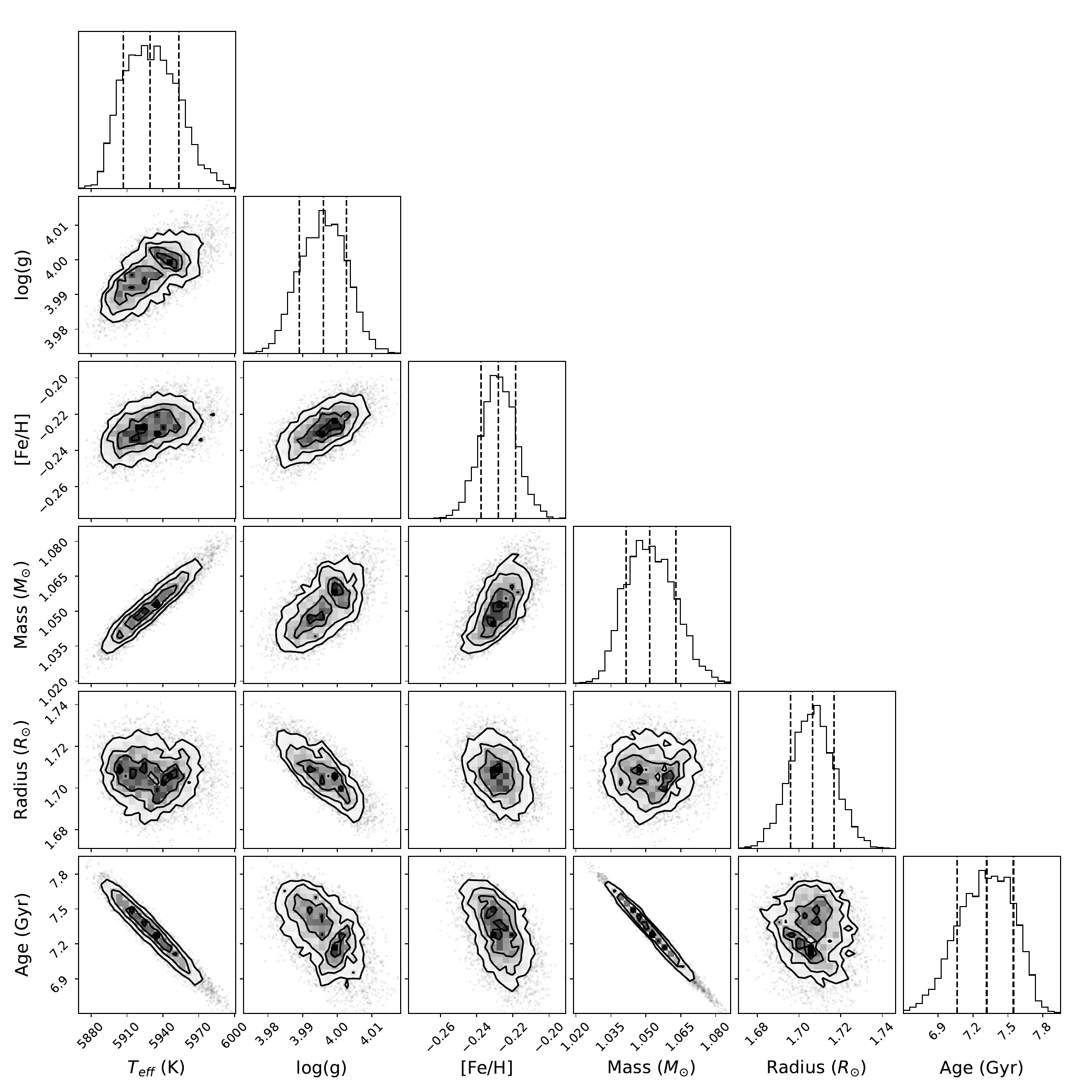}
\figsetgrpnote{Samples from the photospheric and fundamental stellar
parameter posterior resulting from our Bayesian analysis of astrometric,
photometric, and spectroscopic data.}
\figsetgrpend

\figsetgrpstart
\figsetgrpnum{2.27}
\figsetgrptitle{APOGEE DR17 Kepler-93}
\figsetplot{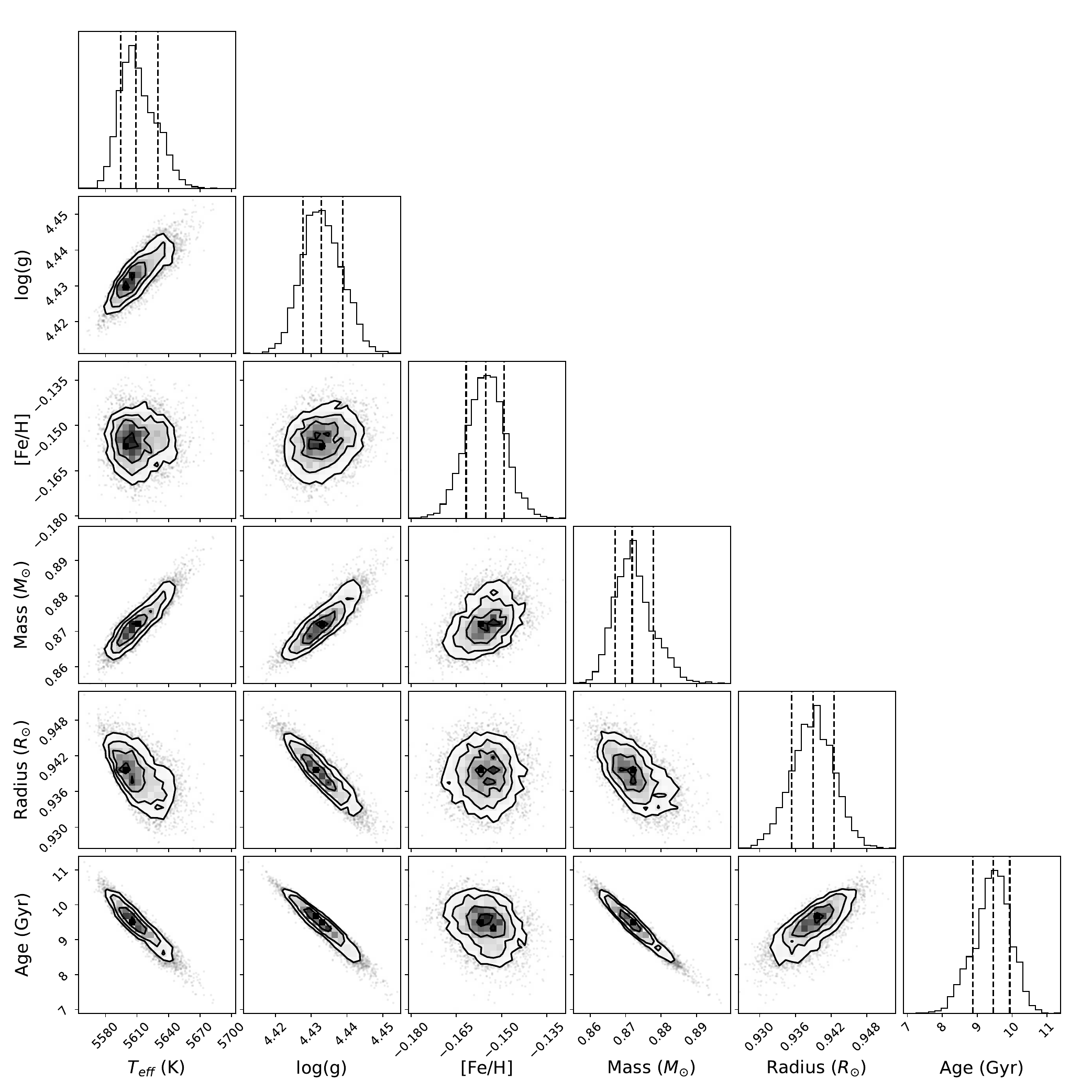}
\figsetgrpnote{Samples from the photospheric and fundamental stellar
parameter posterior resulting from our Bayesian analysis of astrometric,
photometric, and spectroscopic data.}
\figsetgrpend

\figsetgrpstart
\figsetgrpnum{2.28}
\figsetgrptitle{APOGEE DR17 Kepler-99}
\figsetplot{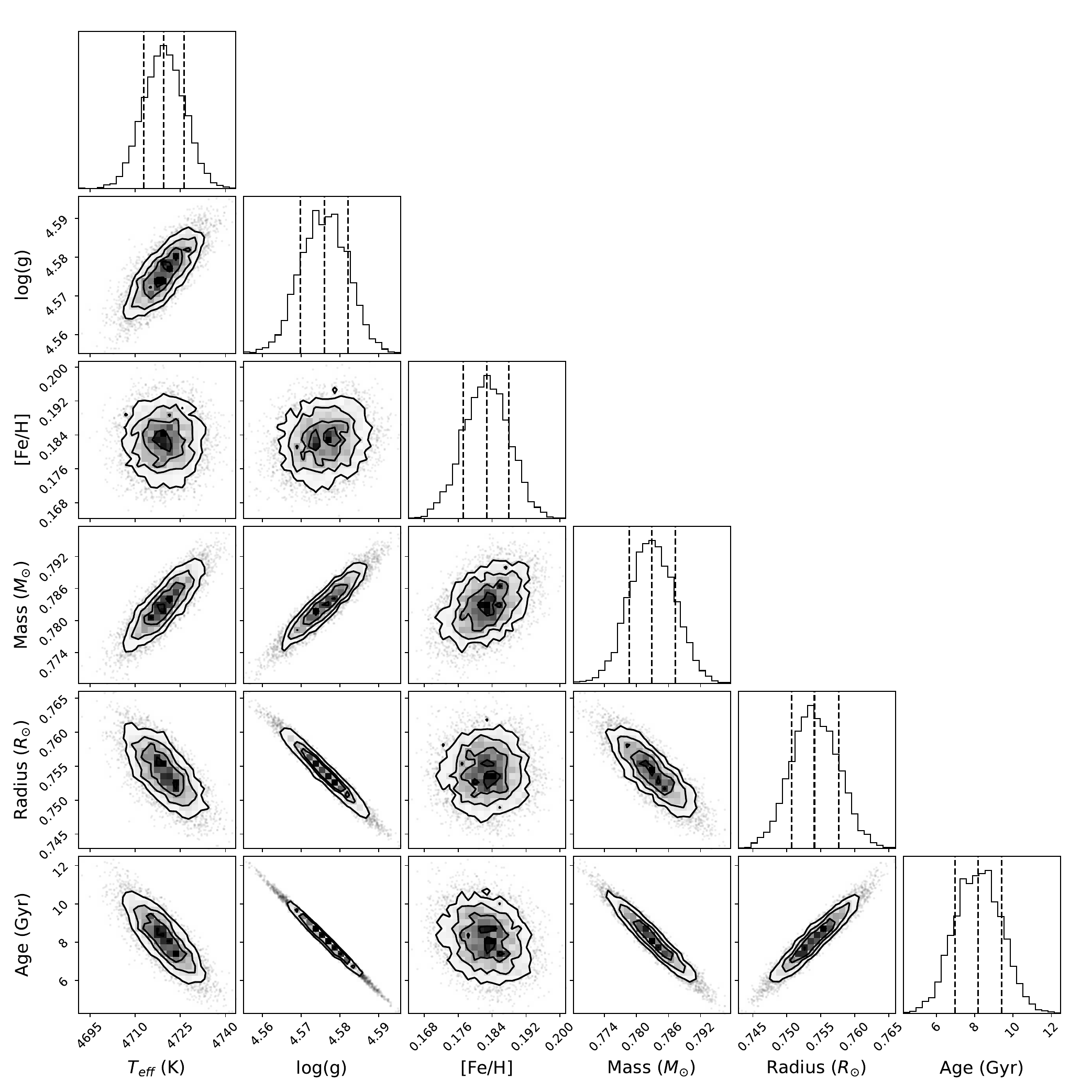}
\figsetgrpnote{Samples from the photospheric and fundamental stellar
parameter posterior resulting from our Bayesian analysis of astrometric,
photometric, and spectroscopic data.}
\figsetgrpend

\figsetgrpstart
\figsetgrpnum{2.29}
\figsetgrptitle{APOGEE DR17 KOI-1599}
\figsetplot{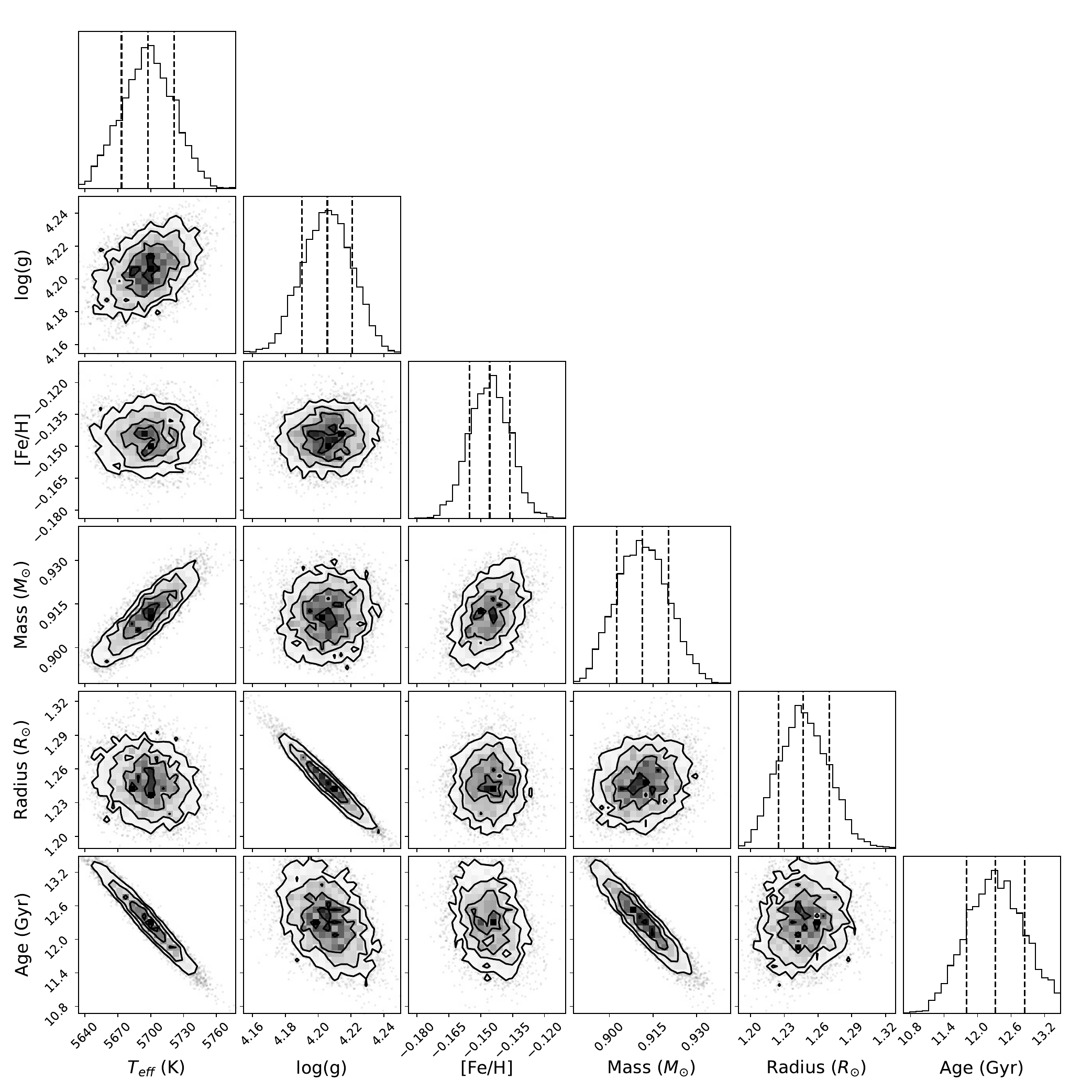}
\figsetgrpnote{Samples from the photospheric and fundamental stellar
parameter posterior resulting from our Bayesian analysis of astrometric,
photometric, and spectroscopic data.}
\figsetgrpend

\figsetgrpstart
\figsetgrpnum{2.30}
\figsetgrptitle{APOGEE DR17 WASP-47}
\figsetplot{WASP-47_apogee_cornerplot.pdf}
\figsetgrpnote{Samples from the photospheric and fundamental stellar
parameter posterior resulting from our Bayesian analysis of astrometric,
photometric, and spectroscopic data.}
\figsetgrpend

\figsetgrpstart
\figsetgrpnum{2.31}
\figsetgrptitle{APOGEE DR17 K2-265}
\figsetplot{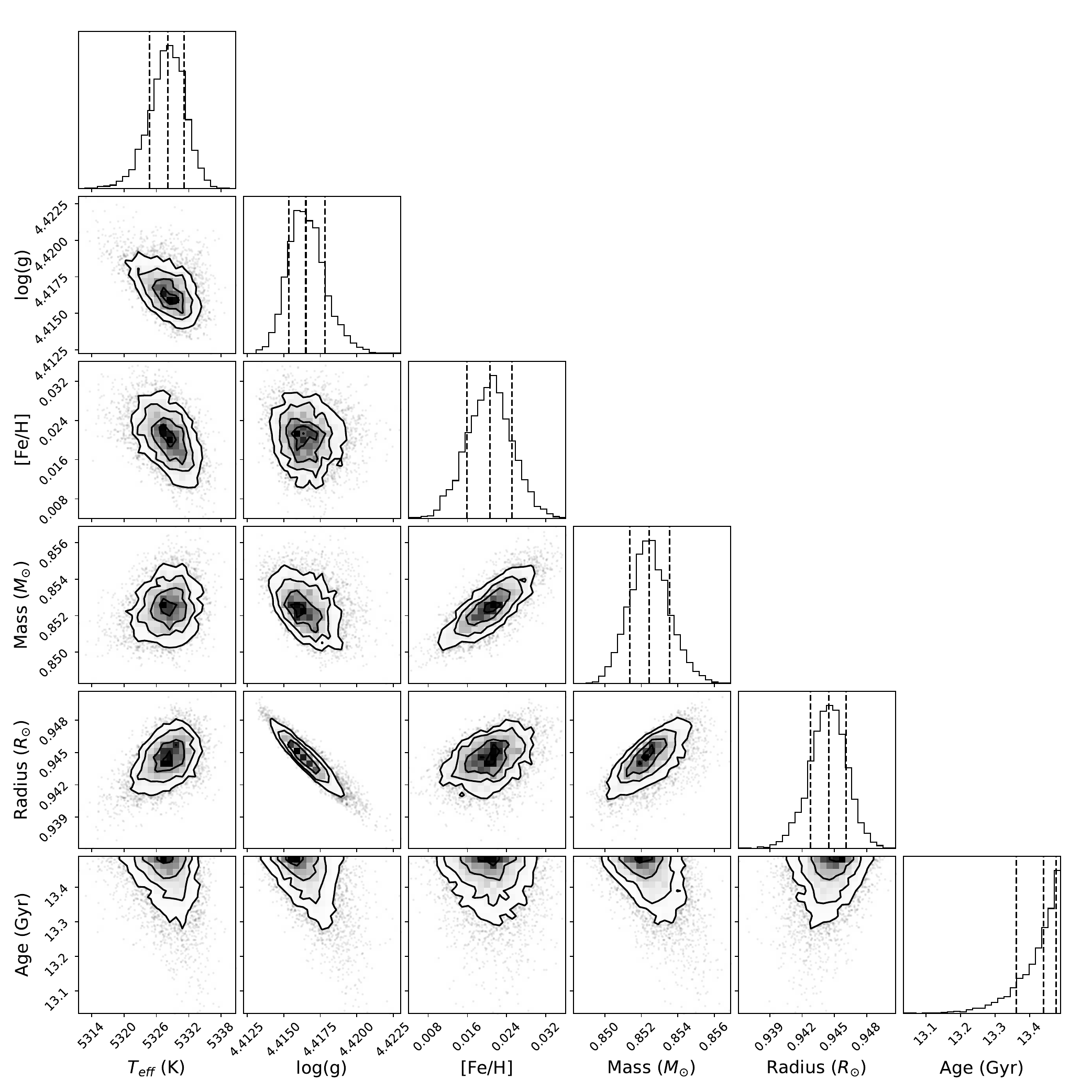}
\figsetgrpnote{Samples from the photospheric and fundamental stellar
parameter posterior resulting from our Bayesian analysis of astrometric,
photometric, and spectroscopic data.}
\figsetgrpend

\figsetgrpstart
\figsetgrpnum{2.32}
\figsetgrptitle{APOGEE DR17 K2-141}
\figsetplot{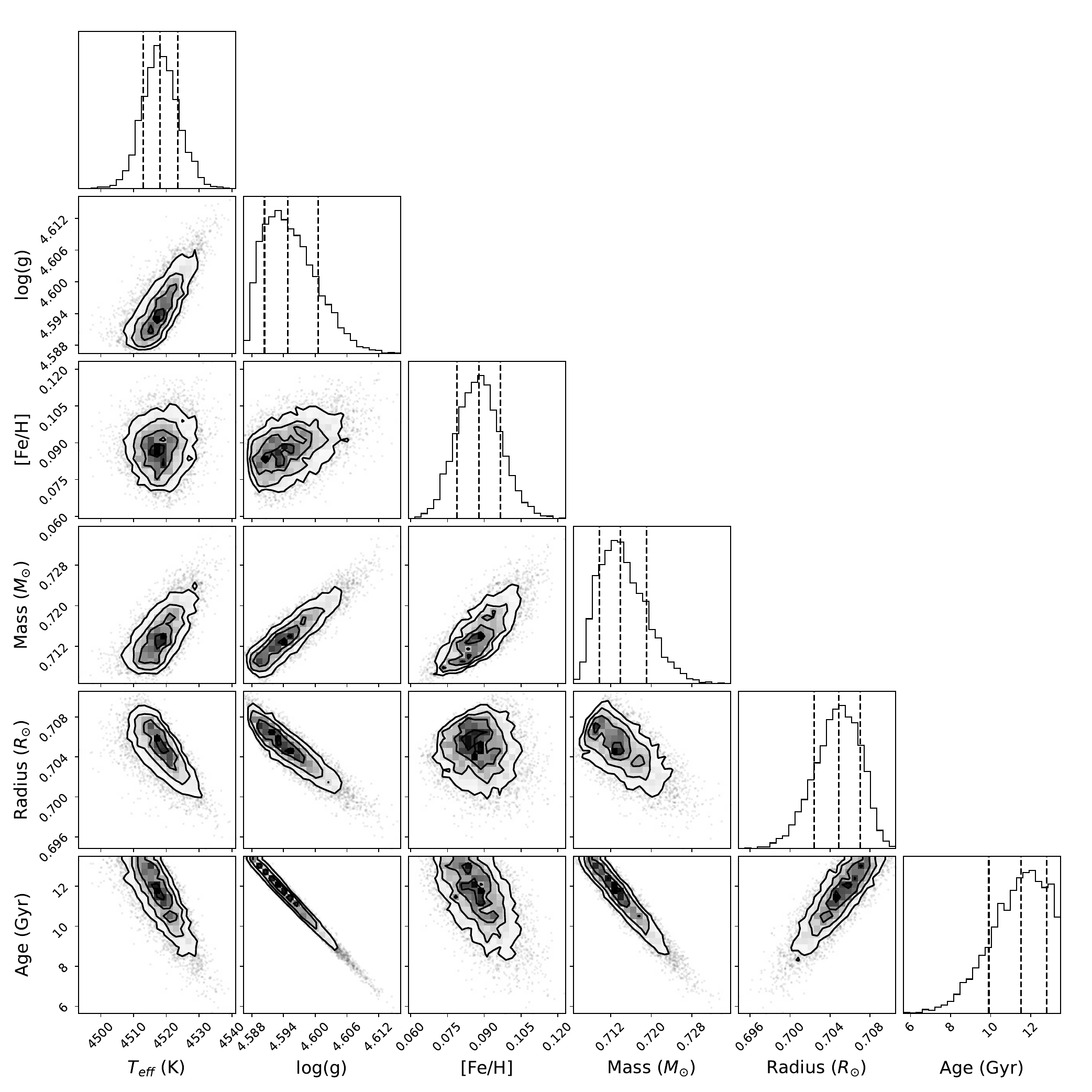}
\figsetgrpnote{Samples from the photospheric and fundamental stellar
parameter posterior resulting from our Bayesian analysis of astrometric,
photometric, and spectroscopic data.}
\figsetgrpend

\figsetgrpstart
\figsetgrpnum{2.33}
\figsetgrptitle{J.\ M.\ Brewer et al.\ 55 Cnc}
\figsetplot{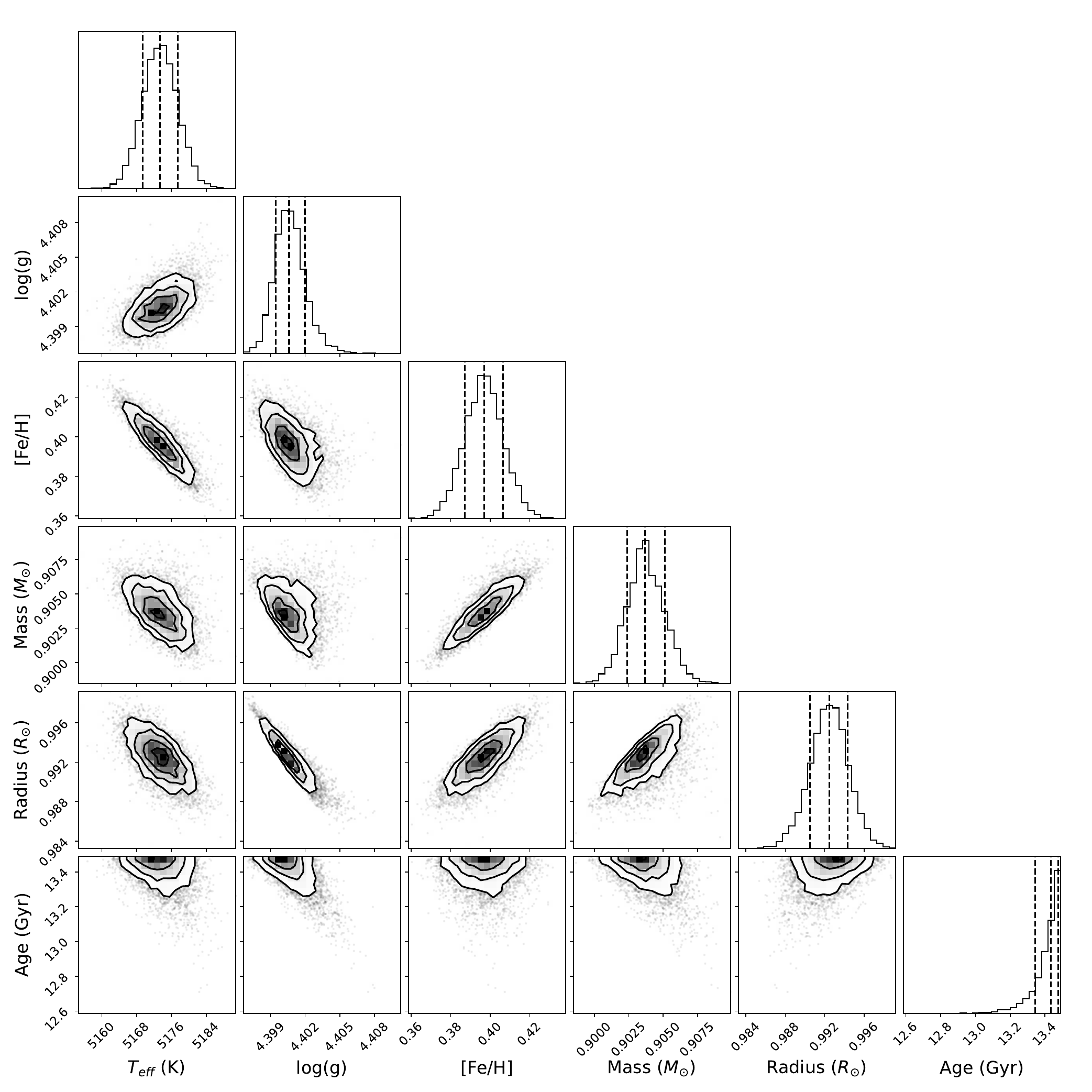}
\figsetgrpnote{Samples from the photospheric and fundamental stellar
parameter posterior resulting from our Bayesian analysis of astrometric,
photometric, and spectroscopic data.}
\figsetgrpend

\figsetgrpstart
\figsetgrpnum{2.34}
\figsetgrptitle{J.\ M.\ Brewer et al.\ K2-38}
\figsetplot{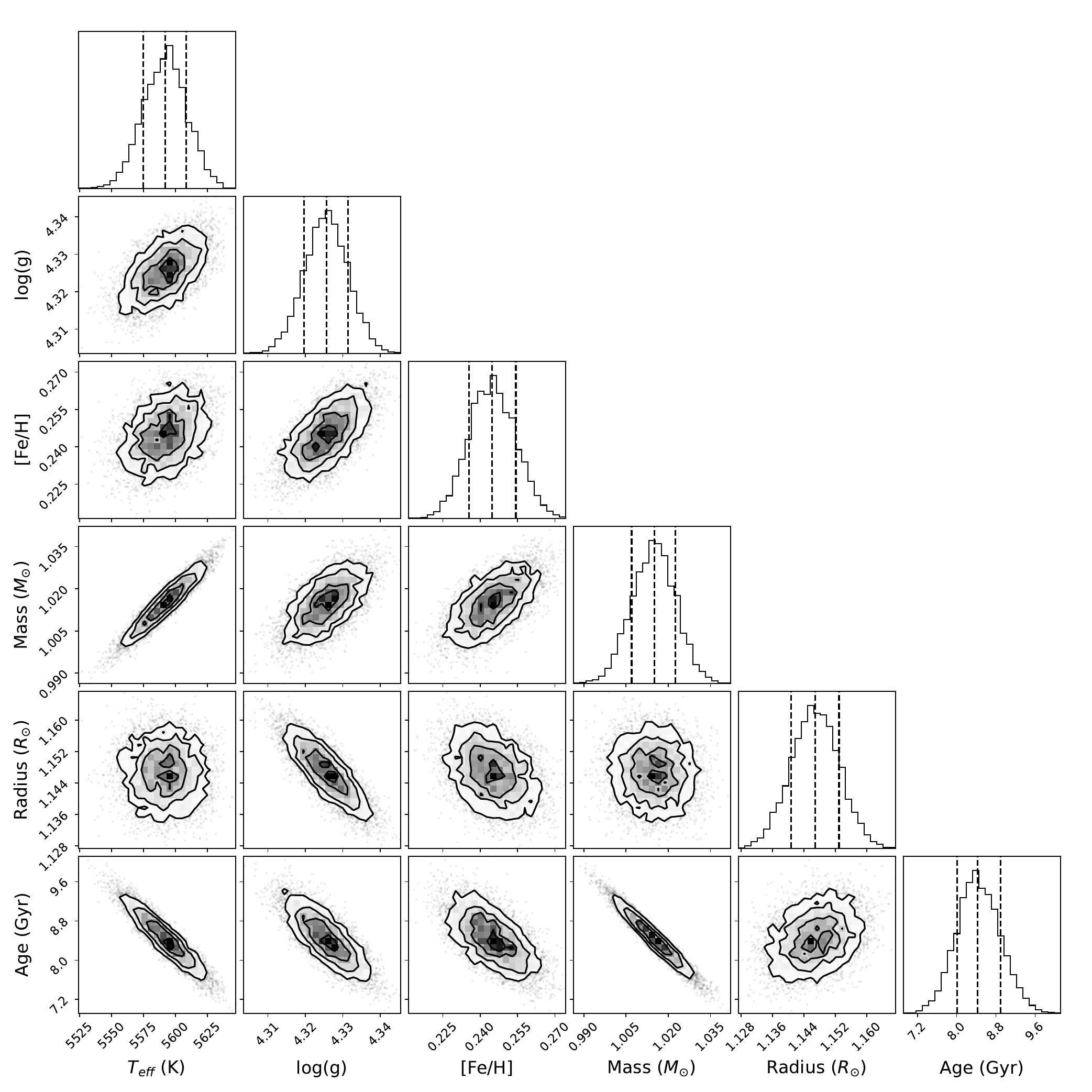}
\figsetgrpnote{Samples from the photospheric and fundamental stellar
parameter posterior resulting from our Bayesian analysis of astrometric,
photometric, and spectroscopic data.}
\figsetgrpend

\figsetgrpstart
\figsetgrpnum{2.35}
\figsetgrptitle{J.\ M.\ Brewer et al.\ Kepler-10}
\figsetplot{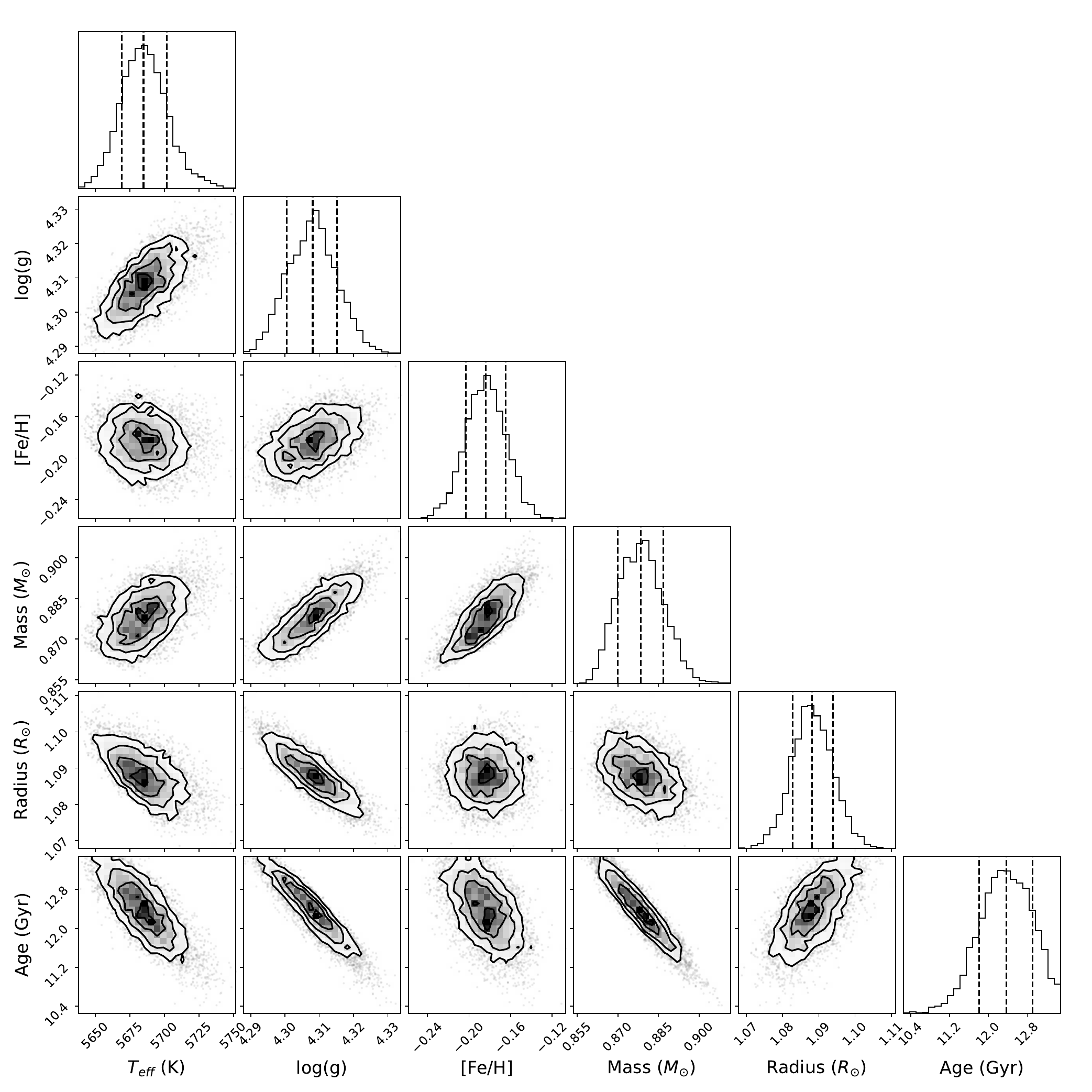}
\figsetgrpnote{Samples from the photospheric and fundamental stellar
parameter posterior resulting from our Bayesian analysis of astrometric,
photometric, and spectroscopic data.}
\figsetgrpend

\figsetgrpstart
\figsetgrpnum{2.36}
\figsetgrptitle{J.\ M.\ Brewer et al.\ Kepler-20}
\figsetplot{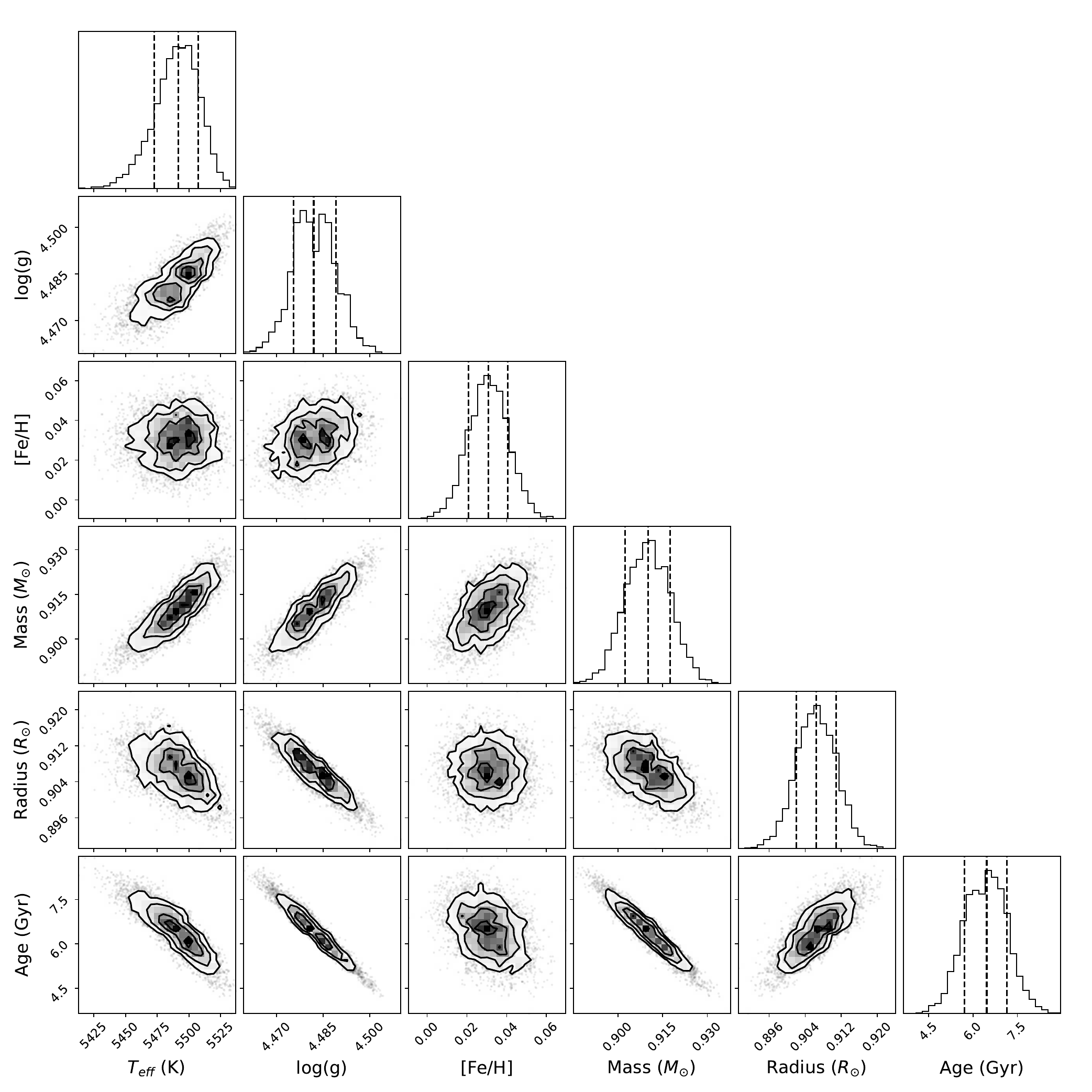}
\figsetgrpnote{Samples from the photospheric and fundamental stellar
parameter posterior resulting from our Bayesian analysis of astrometric,
photometric, and spectroscopic data.}
\figsetgrpend

\figsetgrpstart
\figsetgrpnum{2.37}
\figsetgrptitle{J.\ M.\ Brewer et al.\ Kepler-105}
\figsetplot{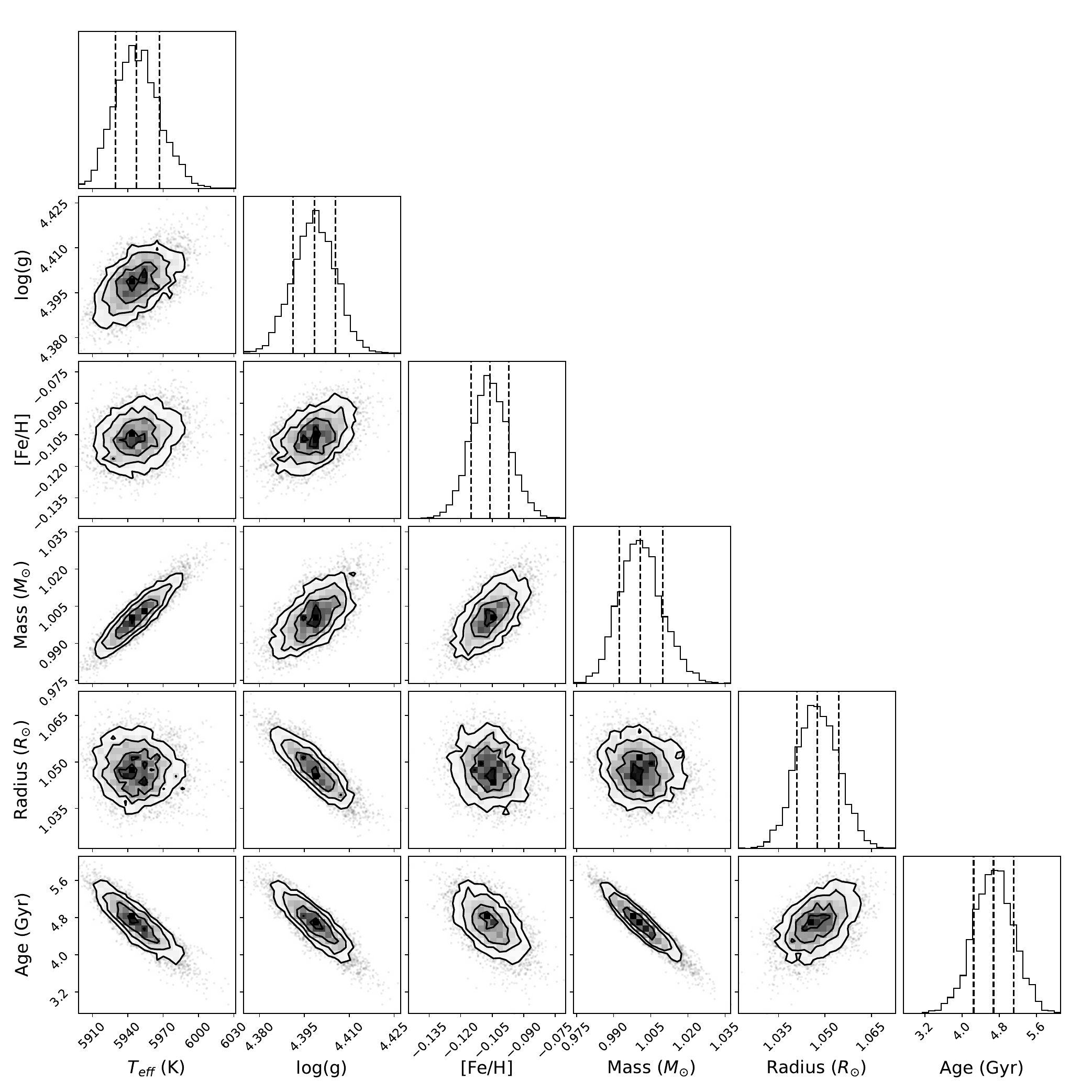}
\figsetgrpnote{Samples from the photospheric and fundamental stellar
parameter posterior resulting from our Bayesian analysis of astrometric,
photometric, and spectroscopic data.}
\figsetgrpend

\figsetgrpstart
\figsetgrpnum{2.38}
\figsetgrptitle{J.\ M.\ Brewer et al.\ Kepler-36}
\figsetplot{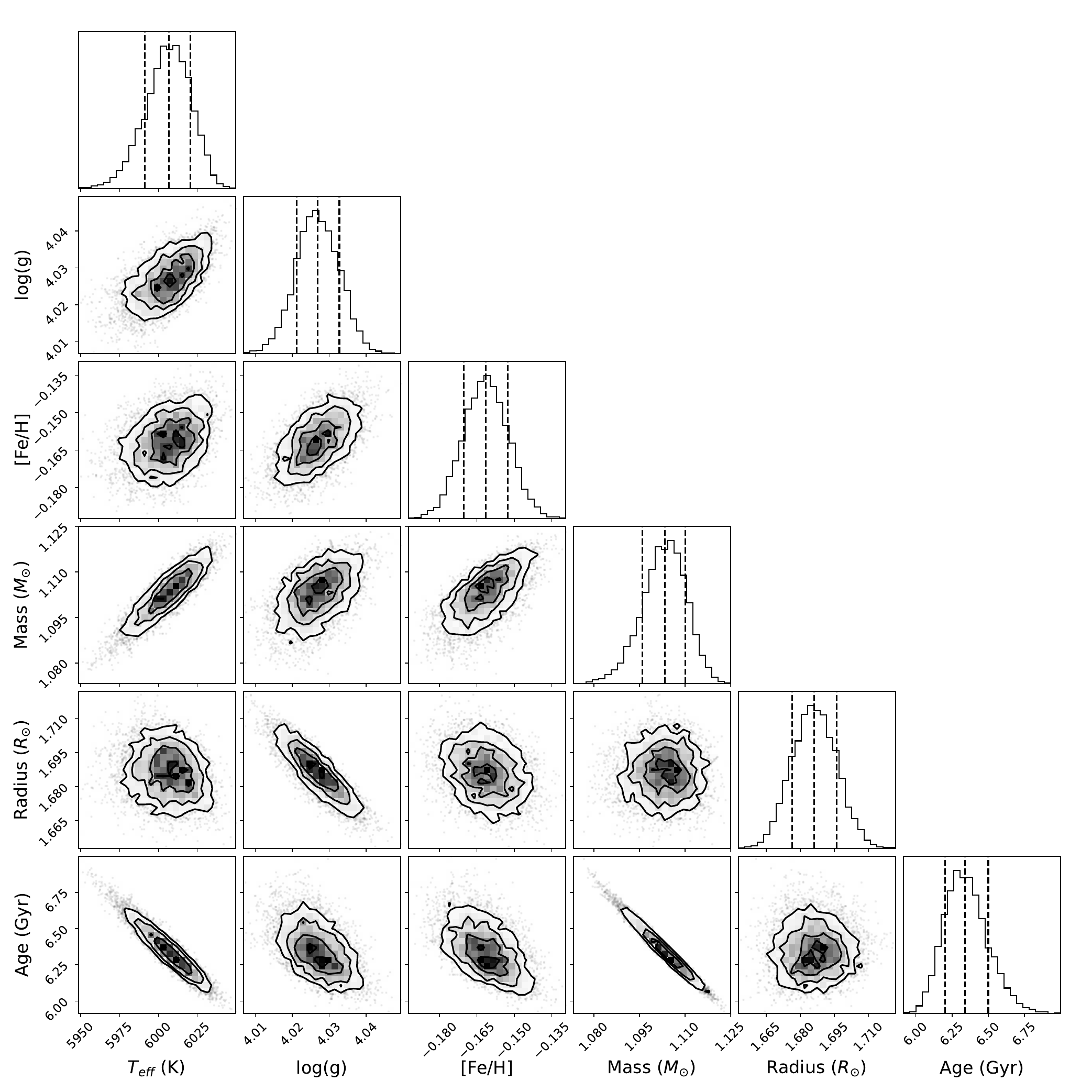}
\figsetgrpnote{Samples from the photospheric and fundamental stellar
parameter posterior resulting from our Bayesian analysis of astrometric,
photometric, and spectroscopic data.}
\figsetgrpend

\figsetgrpstart
\figsetgrpnum{2.39}
\figsetgrptitle{J.\ M.\ Brewer et al.\ Kepler-93}
\figsetplot{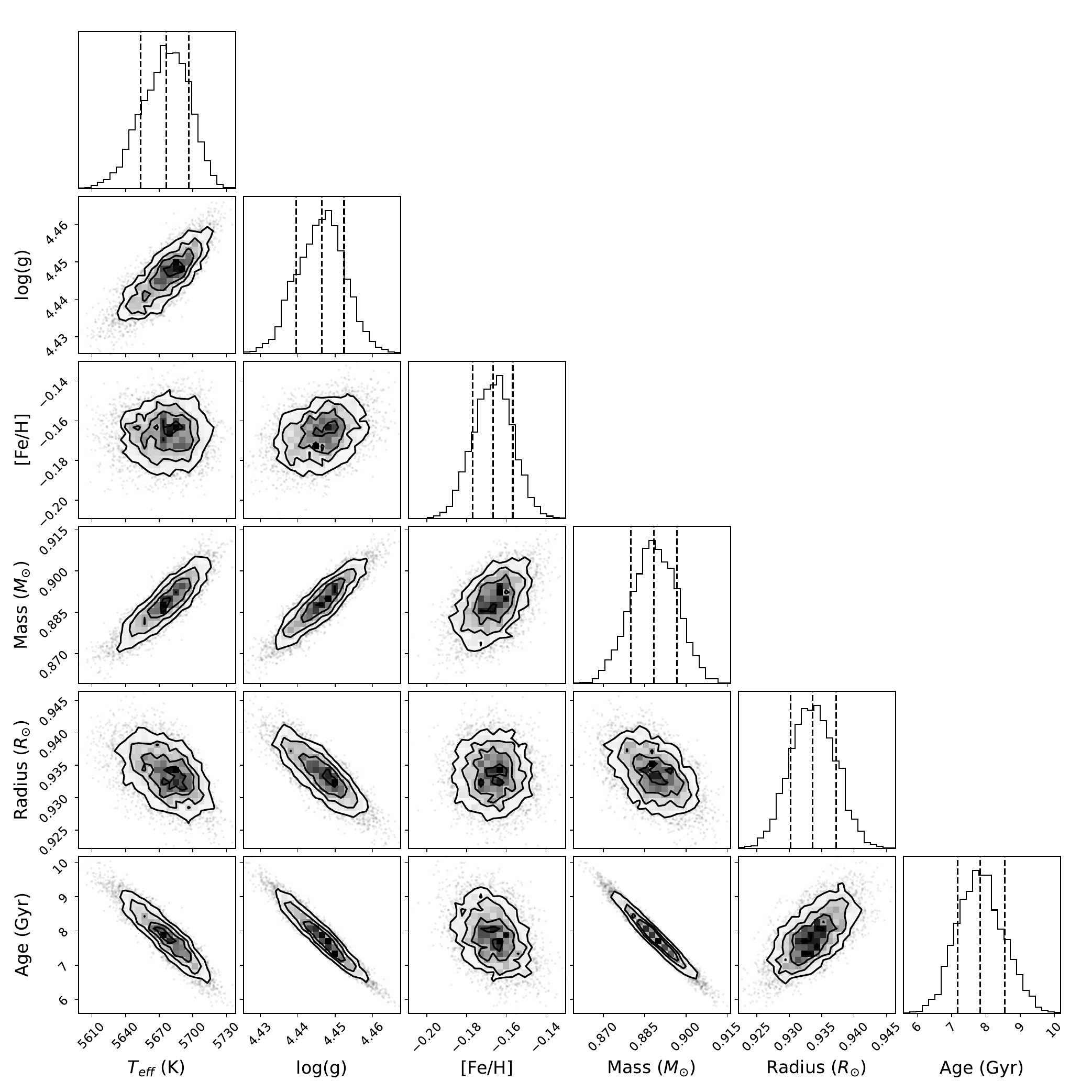}
\figsetgrpnote{Samples from the photospheric and fundamental stellar
parameter posterior resulting from our Bayesian analysis of astrometric,
photometric, and spectroscopic data.}
\figsetgrpend

\figsetgrpstart
\figsetgrpnum{2.40}
\figsetgrptitle{J.\ M.\ Brewer et al.\ Kepler-406}
\figsetplot{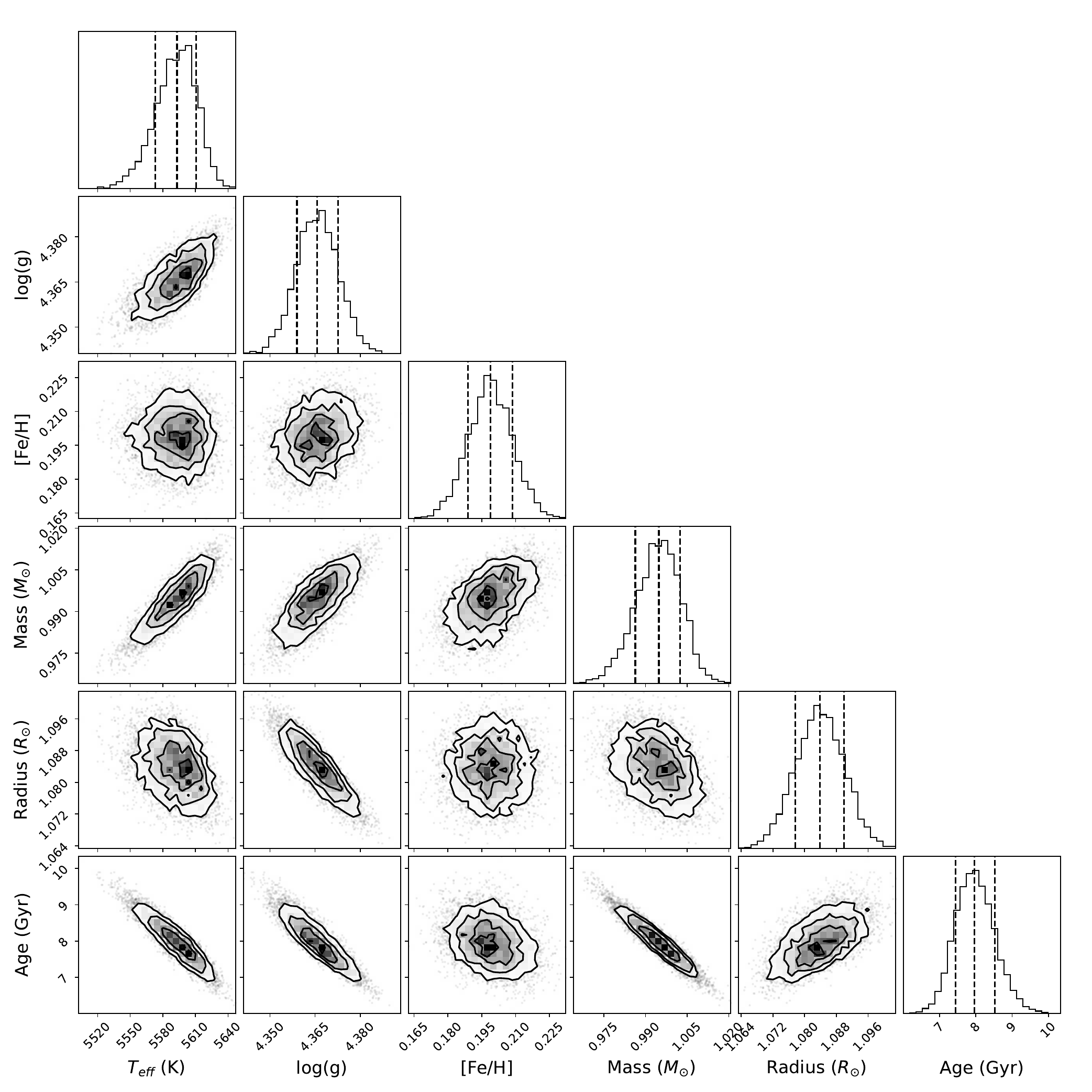}
\figsetgrpnote{Samples from the photospheric and fundamental stellar
parameter posterior resulting from our Bayesian analysis of astrometric,
photometric, and spectroscopic data.}
\figsetgrpend

\figsetgrpstart
\figsetgrpnum{2.41}
\figsetgrptitle{J.\ M.\ Brewer et al.\ Kepler-78}
\figsetplot{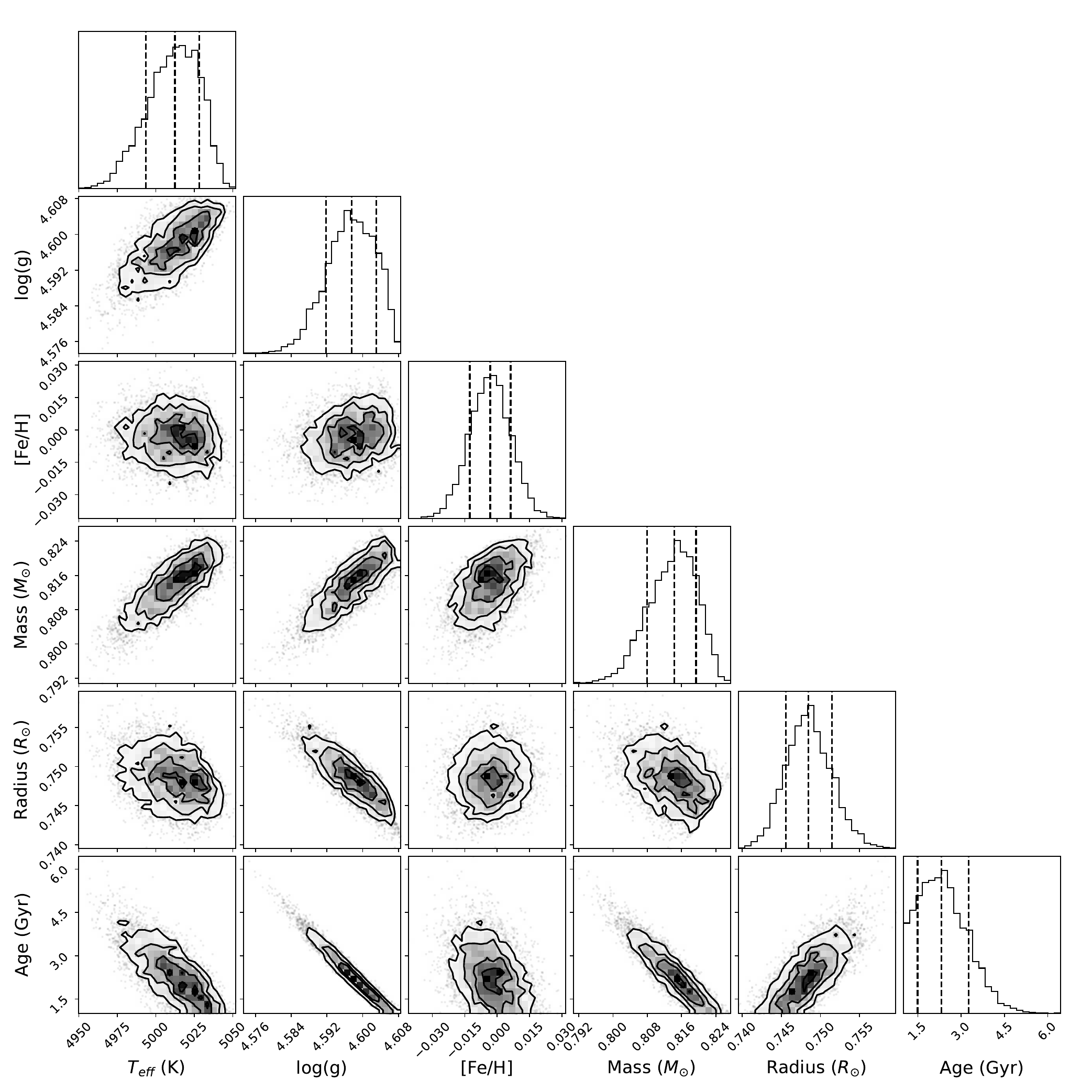}
\figsetgrpnote{Samples from the photospheric and fundamental stellar
parameter posterior resulting from our Bayesian analysis of astrometric,
photometric, and spectroscopic data.}
\figsetgrpend

\figsetgrpstart
\figsetgrpnum{2.42}
\figsetgrptitle{J.\ M.\ Brewer et al.\ Kepler-107}
\figsetplot{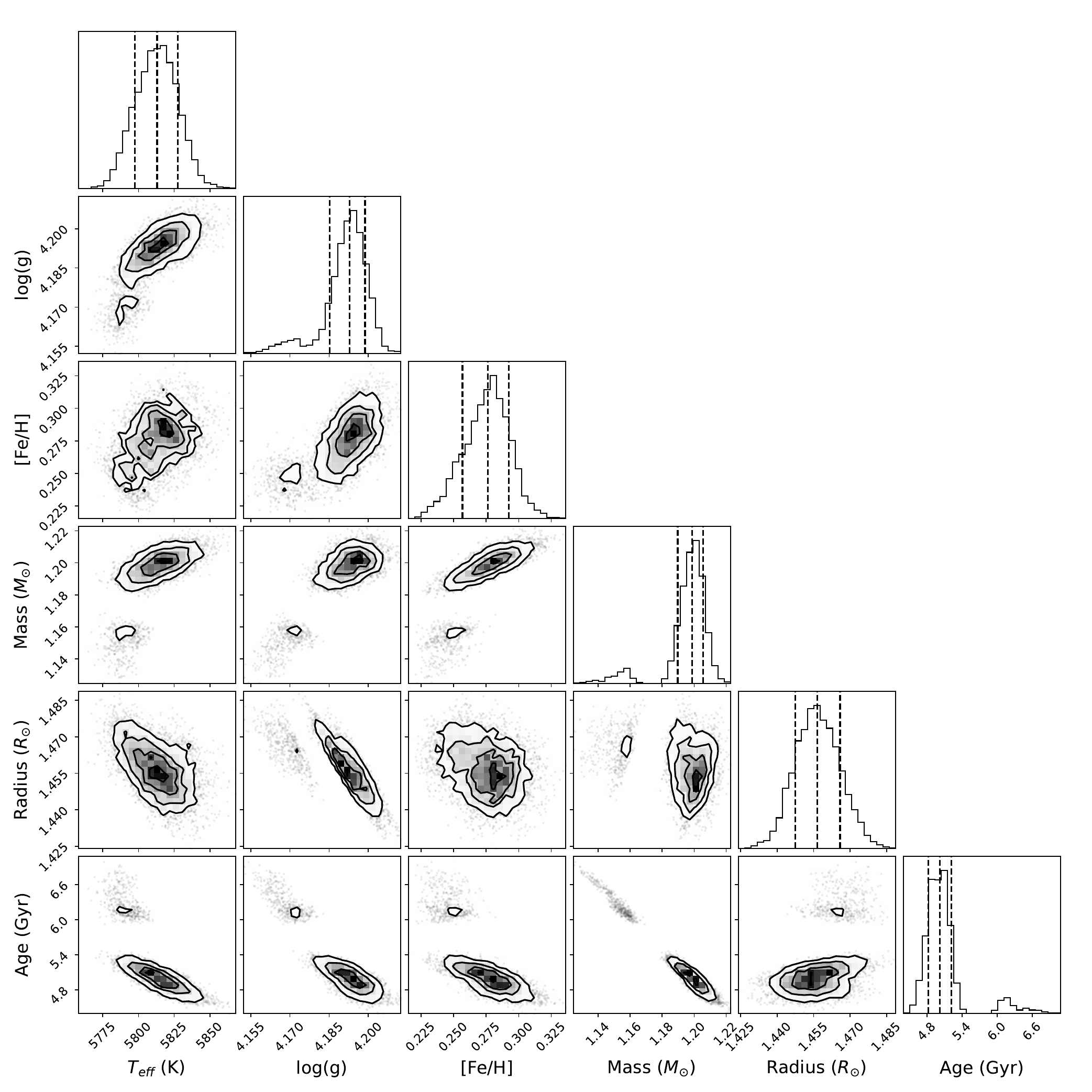}
\figsetgrpnote{Samples from the photospheric and fundamental stellar
parameter posterior resulting from our Bayesian analysis of astrometric,
photometric, and spectroscopic data.}
\figsetgrpend

\figsetgrpstart
\figsetgrpnum{2.43}
\figsetgrptitle{J.\ M.\ Brewer et al.\ Kepler-99}
\figsetplot{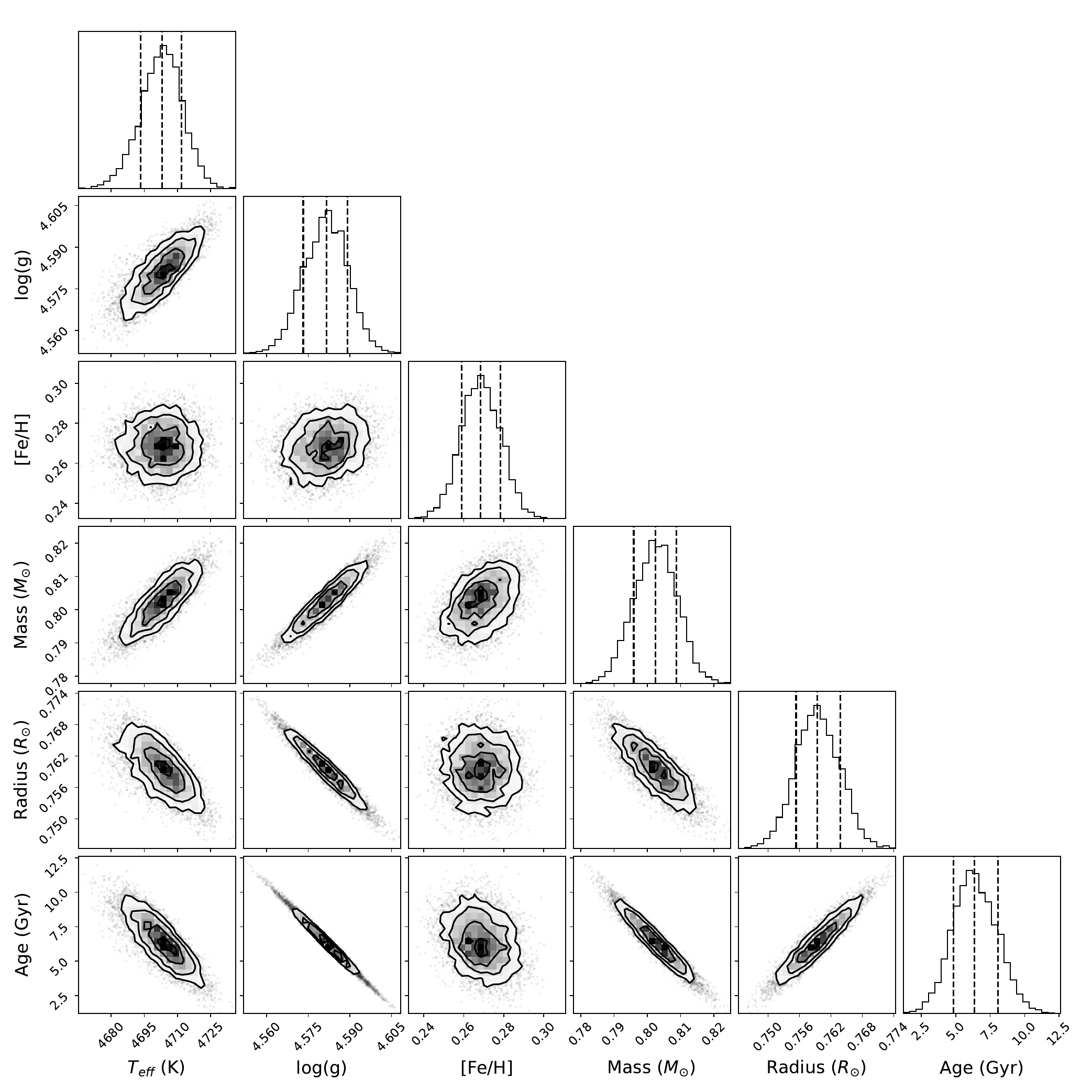}
\figsetgrpnote{Samples from the photospheric and fundamental stellar
parameter posterior resulting from our Bayesian analysis of astrometric,
photometric, and spectroscopic data.}
\figsetgrpend

\figsetgrpstart
\figsetgrpnum{2.44}
\figsetgrptitle{J.\ M.\ Brewer et al.\ K2-265}
\figsetplot{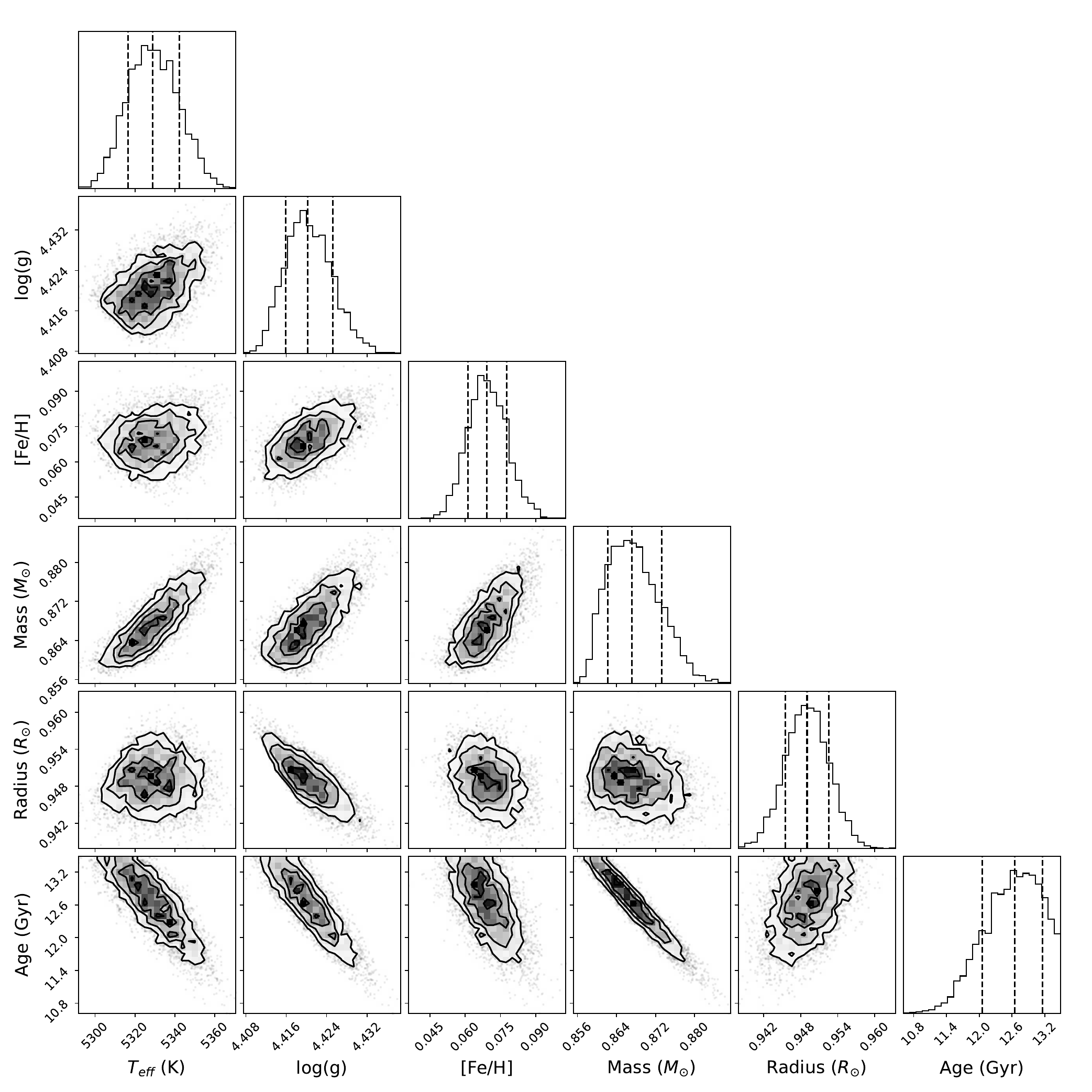}
\figsetgrpnote{Samples from the photospheric and fundamental stellar
parameter posterior resulting from our Bayesian analysis of astrometric,
photometric, and spectroscopic data.}
\figsetgrpend

To complement our own photospheric stellar parameter inferences for the
20 stars with archival spectra listed in Table \ref{tab:sample_list},
we also use the high-quality photospheric stellar parameters from SDSS
DR17 \citep{APOGEEDR17}, \citet{Brewer2016}, and \citet{Brewer2018}.
The spectra on which the SDSS DR17 stellar parameters are based were
collected with the APOGEE spectrographs \citep{Zasowski2013, Zasowski2017,
Wilson2019, Beaton2021, Santana2021} on the New Mexico State University
1-m Telescope \citep{Holtzman10}, the Sloan Foundation 2.5-m Telescope
\citep{Gunn2006}, and the 2.5-m Ir\'{e}n\'{e}e du Pont Telescope
\citep{Bowen1973}.  As part of SDSS DR17, these spectra were reduced
and analyzed with the APOGEE Stellar Parameter and Chemical Abundance
Pipeline \citep[ASPCAP;][]{AllendePrieto06, Holtzman2015, Nidever2015,
aspcap} using an $H$-band line list, MARCS model atmospheres, and
model-fitting tools optimized for the APOGEE effort \citep{Alvarez1998,
Gustafsson2008, Hubeny2011, Plez2012, Smith2013, Smith2021, Cunha2015,
Shetrone2015, Jonsson2020}.

\begin{figure*}
\centering
\plotone{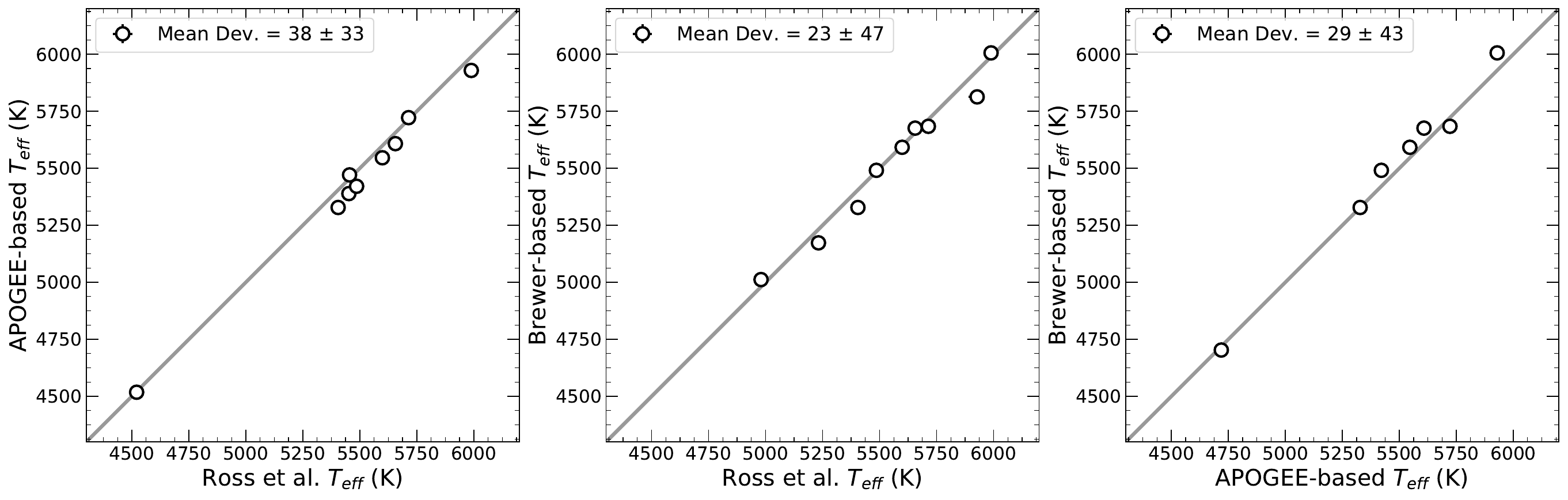}
\plotone{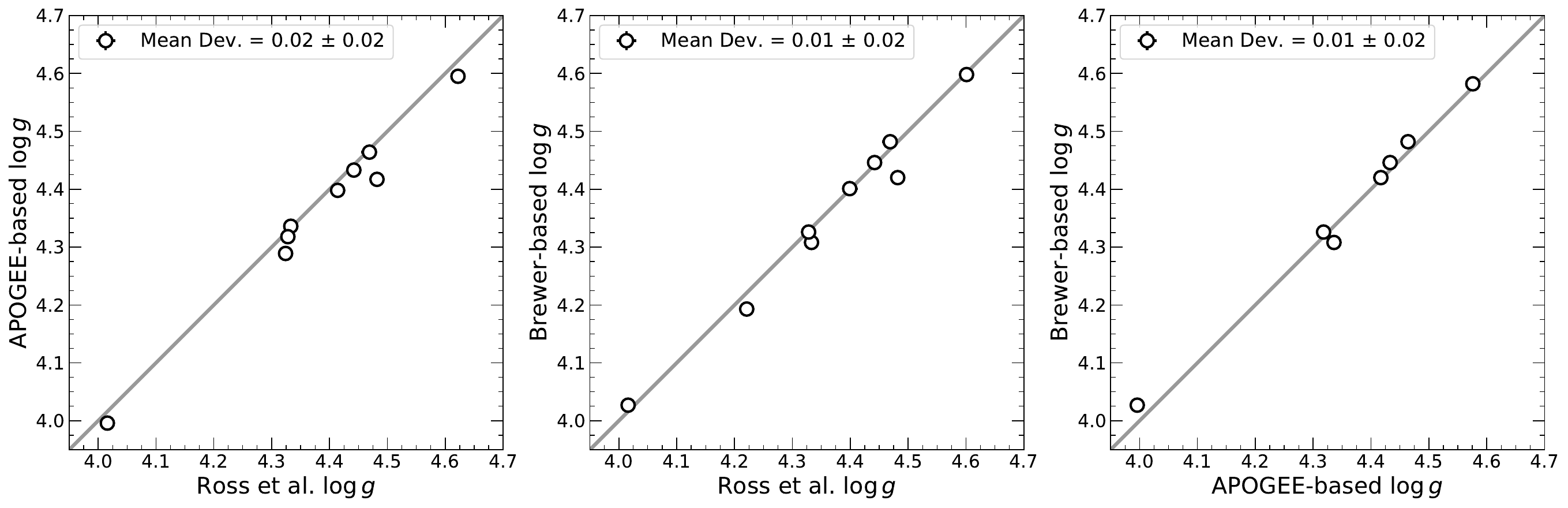}
\plotone{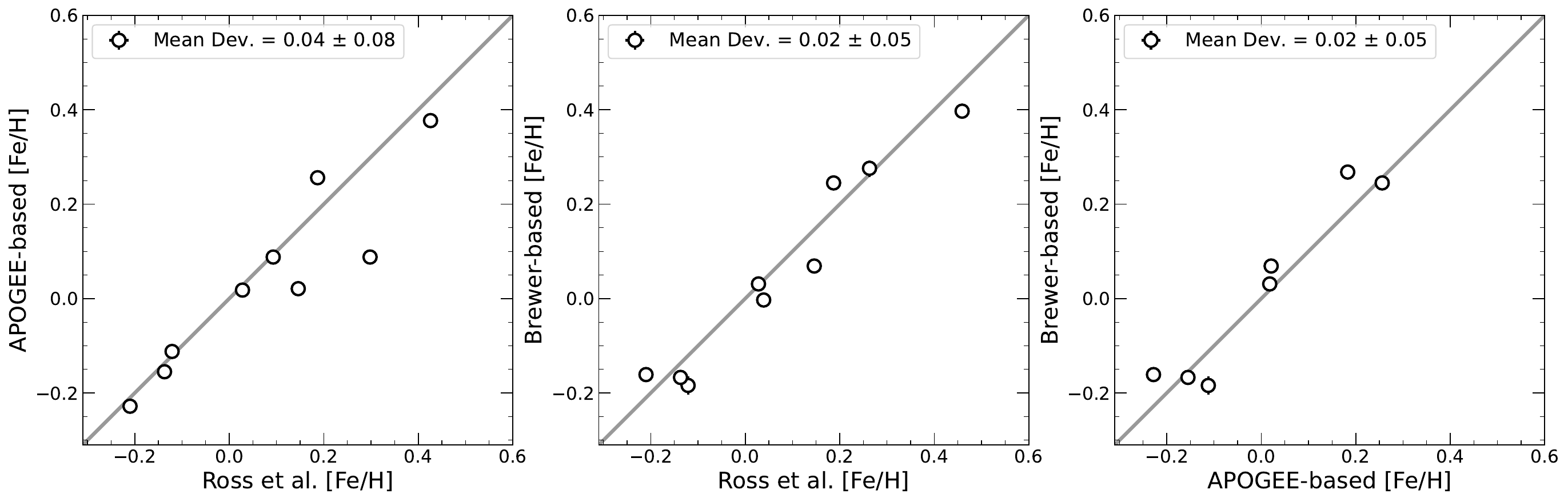}
\caption{Comparisons of inferred photospheric stellar
parameters between between our analyses, APOGEE DR17, and
\citet{Brewer2016}/\citet{Brewer2018}.  We indicate the mean deviation
about the line $y=x$ in each panel, and those values support the mutual
consistency of the photospheric stellar parameters produced by these
three independent analyses.}
\label{fig:param_compare}
\end{figure*}

\begin{figure*}
\centering
\plotone{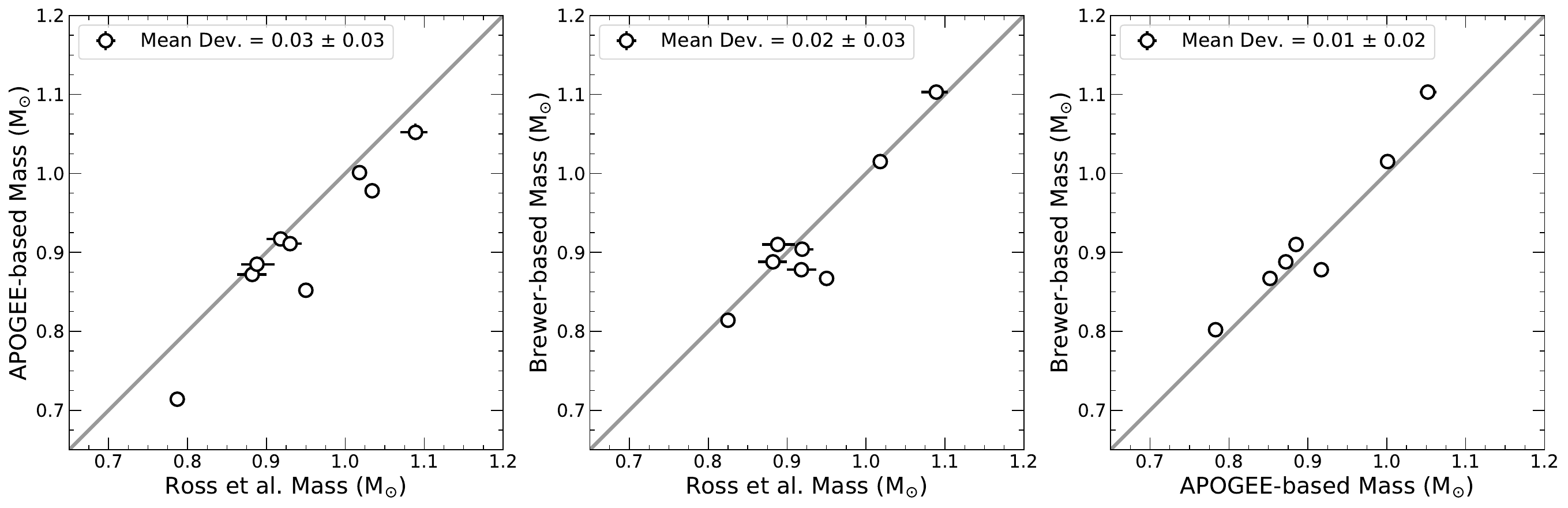}
\plotone{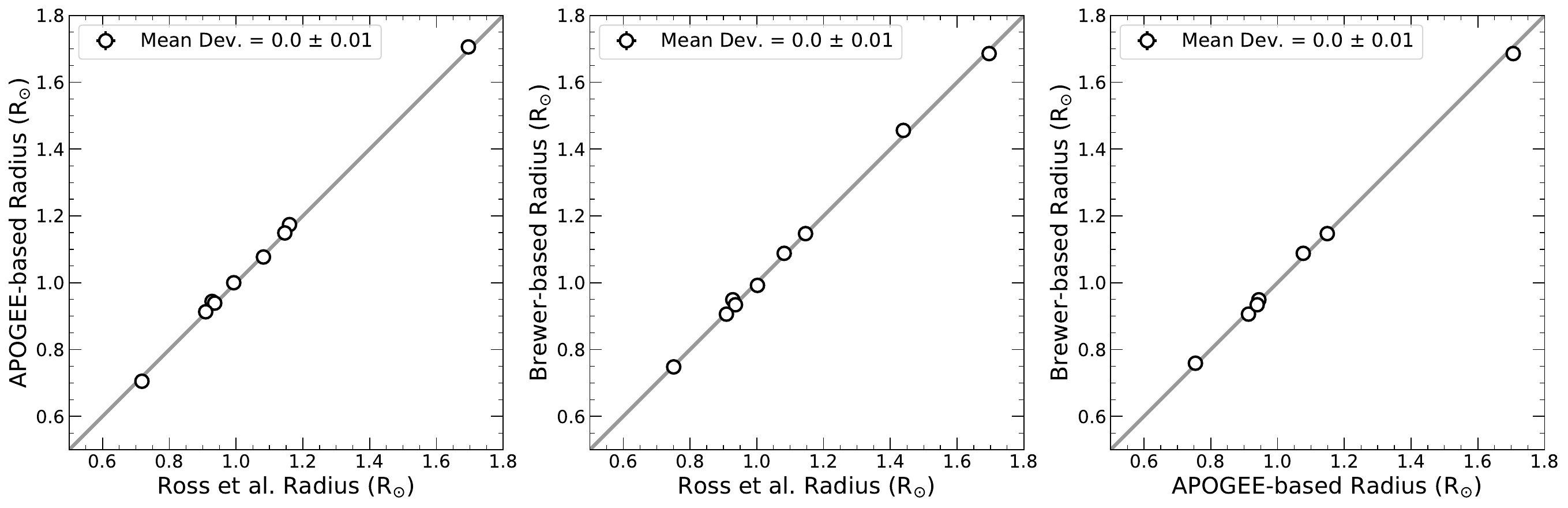}
\caption{Comparisons of inferred stellar masses and radii based on
our approach using as inputs photospheric stellar parameters from
our analyses, APOGEE DR17, and \citet{Brewer2016}/\citet{Brewer2018}.
We indicate the mean deviation about the line $y=x$ line in each panel,
and those values support the mutual consistency of the stellar masses and
radii implied by these three independent photospheric stellar parameter
analyses.}
\label{fig:star_compare}
\end{figure*}

We infer stellar masses and radii for the 12 stars with APOGEE
DR17 and the 12 stars with \citet{Brewer2016}/\citet{Brewer2018}
photospheric stellar parameters using \texttt{isochrones} to execute with
\texttt{MultiNest} a simultaneous Bayesian fit of the MIST grid to those
photospheric stellar parameters plus Gaia DR3 astrometry and the same
photometry and reddening inferences used in our analyses.  We list the
resulting photospheric stellar parameters, stellar masses, and stellar
radii in Table \ref{tab:stellar_params} and plot the intersections
between all three samples in Figures \ref{fig:param_compare} and
\ref{fig:star_compare}.  The union of our analyses with APOGEE DR17 and
\citet{Brewer2016}/\citet{Brewer2018} photospheric stellar parameters
allows us to infer homogeneous stellar mass and radii for 25 stars.
The intersection of these three samples contains six stars: Kepler-10,
Kepler-20, Kepler-36, Kepler-93, K2-38, and K2-265.  The standard
deviations of the three $T_{\text{eff}}$, $\log{g}$, and $[\text{Fe/H}]$
inferences for these six stars have median values 50 K, 0.03 dex, and
0.04 dex comparable to their random uncertainties.  The implication is
that these three sources of photospheric stellar parameters are in good
agreement and any systematic uncertainties in our analysis must be small.

Since the analyses that follow are based on our own internally
consistent stellar parameters, the random uncertainties quoted in Table
\ref{tab:stellar_params} are the correct uncertainties for the analyses
we execute in this article.  If others wish to use the photospheric
stellar parameters in Table \ref{tab:stellar_params} alongside data from
other sources, then they may want to add in quadrature $T_{\text{eff}}$,
$\log{g}$, and $[\text{Fe/H}]$ uncertainties of 50 K, 0.03 dex, and 0.04
dex to the random uncertainties given in Table \ref{tab:stellar_params}
to account for the possible impact of systematic uncertainties.

To validate our estimates of the photospheric stellar parameter systematic
uncertainties that might be present in our analyses, we compare
the $T_{\text{eff}}$ values produced by our approach with published
interferometry-based $T_{\text{eff}}$ inferences that were derived using
the Stefan-Boltzmann law for the two terrestrial exoplanet host stars
we analyzed that have those data: 55 Cnc and HD 219134.  We note that
that theoretical model atmospheres must be used in interferometry-based
$T_{\text{eff}}$ inferences to provide both (1) the inputs to bolometric
luminosity measurement via spectral-energy distribution fitting and (2)
the limb darkening models needed to interpret interferometric visibilities
as angular diameters.  The implication is that interferometry-based
$T_{\text{eff}}$ inferences, like our own $T_{\text{eff}}$ inferences,
are model dependent.

55 Cnc has been interferometrically studied by \citet{Baines2008},
\citet{vonBraun2011}, and \citet{Ligi2016}.  \citet{Baines2008} did not
independently calculate $T_{\text{eff}}$.  \citet{vonBraun2011} found
$T_{\text{eff}} = 5196 \pm 24$ K (random), while Ligi et al. (2016) found
$T_{\text{eff}} = 5165 \pm 46$ K (random).  We found $T_{\text{eff}}
= 5232 \pm 15$ K (random).  Even accounting for random uncertainties
alone, our results are consistent with \citet{vonBraun2011} and differ
from \citet{Ligi2016} by less than 1.4 $\sigma$.  Including a 50 K
$T_{\text{eff}}$ systematic would easily reconcile our result with that
of \citet{Ligi2016}.

HD 219134 has been interferometrically studied by \citet{Ligi2019}
who found $T_{\text{eff}} = 4858 \pm 50$ K (random).  We found
$T_{\text{eff}} = 4907 \pm 26$ K (random).  Again, even accounting
for random uncertainties alone our results are consistent with
\citet{Ligi2019}.  Considering both 55 Cnc and HD 219134 and accounting
for random uncertainties alone, our $T_{\text{eff}}$ inferences agree
at the 1-$\sigma$ level with two out of three independent analyses.
That is exactly what one would expect if our uncertainties have been
accurately estimated and systematics do not dominate our uncertainty
budget.  The net result of these analyses is that there is no reason
to think the systematic $T_{\text{eff}}$ uncertainties present in our
analyses are greater than 50 K.

We can also use the angular diameter measurements presented in
\citet{vonBraun2011}, \citet{Ligi2016}, and \citet{Ligi2019} to validate
our stellar radius inferences for 55 Cnc and HD 219134.  For 55 Cnc,
\citet{vonBraun2011} found $R_{\ast} = 0.943 \pm 0.010~R_{\odot}$ (random)
based on an angular diameter measured in the $H$-band.  Alternatively,
\citet{Ligi2016} found $R_{\ast} = 0.960 \pm 0.018~R_{\odot}$ (random)
for 55 Cnc based on a higher-resolution visible light data that could
be more strongly impacted by limb darkening uncertainties.  We found
$R_{\ast} = 0.99 \pm 0.01~R_{\odot}$ (random) fully consistent with the
higher-resolution measurement presented in \citet{Ligi2016}.  For HD
219134, \citet{Ligi2019} found $R_{\ast} = 0.726 \pm 0.014~R_{\odot}$
(random) based on visible light data.  We found $R_{\ast} = 0.76 \pm
0.01~R_{\odot}$ (random).  Even accounting for random uncertainties
alone, our results are consistent with the \citet{Ligi2016} $R_{\ast}$
measurement for 55 Cnc and differ from the \citet{Ligi2019} $R_{\ast}$
measurement for HD 219134 by about 1.6 $\sigma$.  The net result of these
analyses is that there is no reason to think the systematic $R_{\ast}$
uncertainties present in our analyses are greater than about 1\%.

In addition to its interferometric characterization described above,
\citet{Li2025} characterized HD 219134 with precision radial velocity
measurement-based asteroseismology.  While those asteroseismic
measurements do not directly constrain the fundamental stellar
parameters of HD 219134, they do enhance the accuracy and precision
of model-based constraints.  \citet{Li2025} found $M_{\ast} = 0.763
\pm 0.020~\text{(random)} \pm 0.014~\text{(systematic)}~M_{\odot}$,
stellar age $\tau_{\ast} = 10.2 \pm 1.5~\text{(random)}
\pm 1.0~\text{(systematic)}$ Gyr, and $R_{\ast} = 0.748 \pm
0.007~\text{(random)} \pm 0.004~\text{(systematic)}~R_{\odot}$, all of
which are consistent with our inferences unaided by asteroseismology:
$M_{\ast} = 0.80 \pm 0.02~M_{\odot}$ (random), $\tau_{\ast} = 6.0 \pm
3.0$ Gyr (random), and $M_{\ast} = 0.76 \pm 0.01~R_{\odot}$ (random).
Again, the implication is that our stellar parameter inferences are not
impacted by unquantified systematic uncertainties.

\subsection{Host Star Photospheric Elemental
Abundances}\label{sec:abundances}

The elements oxygen, magnesium, aluminum, silicon, calcium, iron,
and nickel together make up approximately 97\% of the Earth's bulk
composition \citep[e.g.,][]{Allegre1995, McDonough1995}.  The abundances
of these elements in the photospheres of the stars in our sample can be
measured.  If one assumes that the abundances of these elements in the
photospheres of exoplanet-hosting dwarf stars reflect the abundances of
these elements in the protoplanetary disks they once hosted, then in the
absence of any further processing the compositions of any terrestrial
exoplanets formed therein can be constrained by host star elemental
abundances \citep{Dorn2015, Santos2015, Brugger2017, Plotnykov2020}.
In that case, those data can then be used as constraints on models of
terrestrial exoplanet interior structures.\footnote{As pointed out by
\citet{Reggiani2024}, the abundance anomalies observed in the photospheres
of massive main-sequence stars invalidate this assumption for exoplanets
orbiting massive main-sequence stars.}  We therefore seek to infer oxygen,
magnesium, aluminum, silicon, calcium, iron, and nickel abundances using
the same spectra from which we inferred photospheric stellar parameters.

We first measure the equivalent widths of many atomic absorption lines
of these species using \texttt{iSpec} to fit Gaussian profiles to
an updated version of the \citet{Melendez2014} linelist.  We visually
verify each line mask and the resulting equivalent width measurement in
each line mask.  We exclude from our analyses absorption lines for which
the fit failed due to blending, line saturation, or some other reason.
We generally exclude from our analyses saturated lines with successfully
measured equivalent widths in excess of 120 m\AA, though for some species
with few lines we still use for abundance inferences absorption lines
with equivalent widths less than 150 m\AA.  We report our measured
equivalent widths and the atomic data we used for their interpretation
in Table \ref{tab:ews}.

We use the methodology described in detail in \citet{Reggiani2022,
Reggiani2024} to infer photospheric elemental abundances from
these equivalent width measurements.  We assume \citet{Asplund2021}
solar abundances and \citet{Castelli2003} 1D, LTE, plane-parallel,
solar-composition ATLAS9 model atmospheres.  We then use the \texttt{q2}
wrapper to the 2019 version of \texttt{MOOG} to infer equivalent
width-based photospheric elemental abundances.  We remeasure the
equivalent widths of individual lines if the implied abundances depart
significantly from the mean abundances implied by other lines of the same
species.  If a large abundance departure persists, then we exclude that
equivalent width measurement from our abundance inferences.  We report in
Table \ref{tab:abundances} our adopted photospheric elemental abundances
under the assumptions of LTE.

We also correct our abundances inferred under the assumptions of LTE for
departures from LTE by linearly interpolating published grids of these
``non-LTE corrections'' using \texttt{scipy} \citep{scipy}.  Our non-LTE
corrections for silicon, calcium, and iron are from \citet{Amarsi2017},
\citet{Amarsi2020}, and \citet{Amarsi2016}.  For oxygen, we correct for
both departures from the assumptions of LTE and 3D effects unmodeled in 1D
model atmospheres using data from \citet{Amarsi2019}.  We report in Table
\ref{tab:abundances_nlte} our adopted photospheric elemental abundances
corrected for departures from the assumptions of LTE.  The spacing of
the grid of non-LTE corrections for calcium from \citet{Amarsi2017}
is too coarse to be useful for our purposes.  While we present non-LTE
corrected calcium abundances in Table \ref{tab:abundances_nlte}, we
prefer our calcium abundances derived under the assumptions of LTE to
those corrected for non-LTE effects.

For the six stars in all three samples Kepler-10, Kepler-20, Kepler-36,
Kepler-93, K2-38, and K2-265, the standard deviations of the five
abundance ratios relatively unaffected by departures from the assumptions
of LTE [Mg/Fe], [Al/Fe], [Si/Fe], [Ca/Fe], and [Ni/Fe] have median values
0.06, 0.10,  0.04,  0.03,  and 0.04 dex.  Our [O/Fe] inferences are based
on the \ion{O}{1} triplet at 7770 \AA~and consequently significantly
effected by departures from the assumptions of LTE.  As a result, we
compare our non-LTE corrected [O/Fe] abundance ratio to the APOGEE DR17
and \citet{Brewer2016}/\citet{Brewer2018} [O/Fe] inferences that are
much less effected by departures from the assumptions of LTE.  For the
six stars in all three samples, the standard deviation of the abundance
ratio [O/Fe] has a median value 0.07.  These values are comparable to the
uncertainties in these abundance inferences, supporting the agreement
between all three analysis approaches.  The implication is that these
three sources of photospheric stellar parameters are in good agreement and
any systematic uncertainties in our analyses must be small.  For the stars
in common between our analysis and \citet{Adibekyan2021}, we compare our
[Fe/H], [Mg/H], and [Si/H] abundances without corrections for departures
from the assumptions of LTE with those reported in \citet{Adibekyan2021}.
We find that the average differences are less than 0.1 dex.

\subsection{Planetary Mass and Radius Inference}\label{sec:planets}

We leverage our accurate, precise, homogeneous, and physically self
consistent host star masses and radii to infer updated planetary masses
and radii using Doppler and transit observables from the NASA Exoplanet
Archive \citep{Akeson2013, Christiansen2025}.  We retrieve Doppler
semiamplitudes $K$, transit-inferred orbital periods $P$, transit
depths $\delta$, and inclinations $i$ for the systems listed in Table
\ref{tab:stellar_params}.  We also list in Table \ref{tab:stellar_params}
the articles that provided these Doppler and transit observables for
each system.  When multiple data sources for a system were present in
the NASA Exoplanet Archive, we select the Doppler measurement based on
the largest number of radial velocity measurements and/or the transit
observations with the highest precisions.  The latter are generally
based on Kepler, K2, or Transiting Exoplanet Survey Satellite (TESS)
light curves.  To maintain the internal consistency of these transit
observables, we require that every set of transit observables comes from
the same source under the assumption that the system eccentricity $e = 0$.

\begin{figure*}
\centering
\plotone{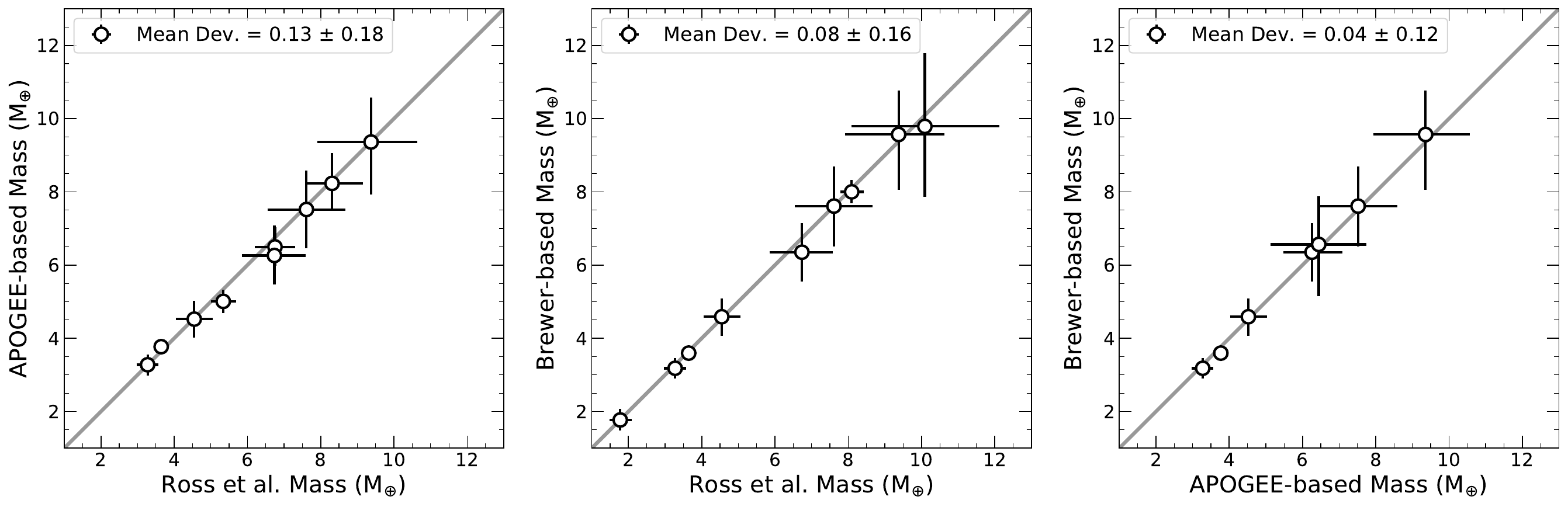}
\plotone{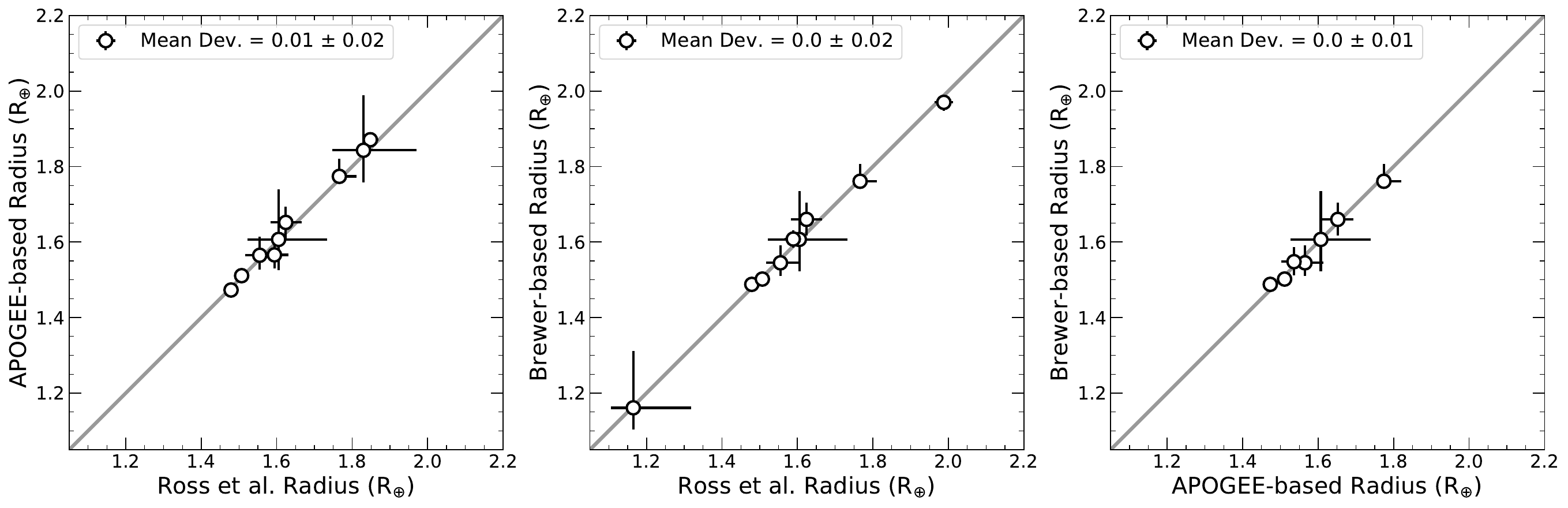}
\caption{Comparisons of updated planetary masses and radii from our
approach using Doppler and transit observables and accurate, precise,
homogeneous, and physically self consistent stellar masses and radii based
on photospheric stellar parameters from our analyses, APOGEE DR17, and
\citet{Brewer2016}/\citet{Brewer2018}.  We indicate the mean deviation
about the line $y=x$ line in each panel, and those values support the
mutual consistency of the updated planetary masses and radii implied
by Doppler/TTV and transit observables and these three independent
photospheric stellar parameter analyses.}
\label{fig:mass_radius_compare}
\end{figure*}

\begin{figure*}
\centering
\includegraphics[width=0.6\textwidth]{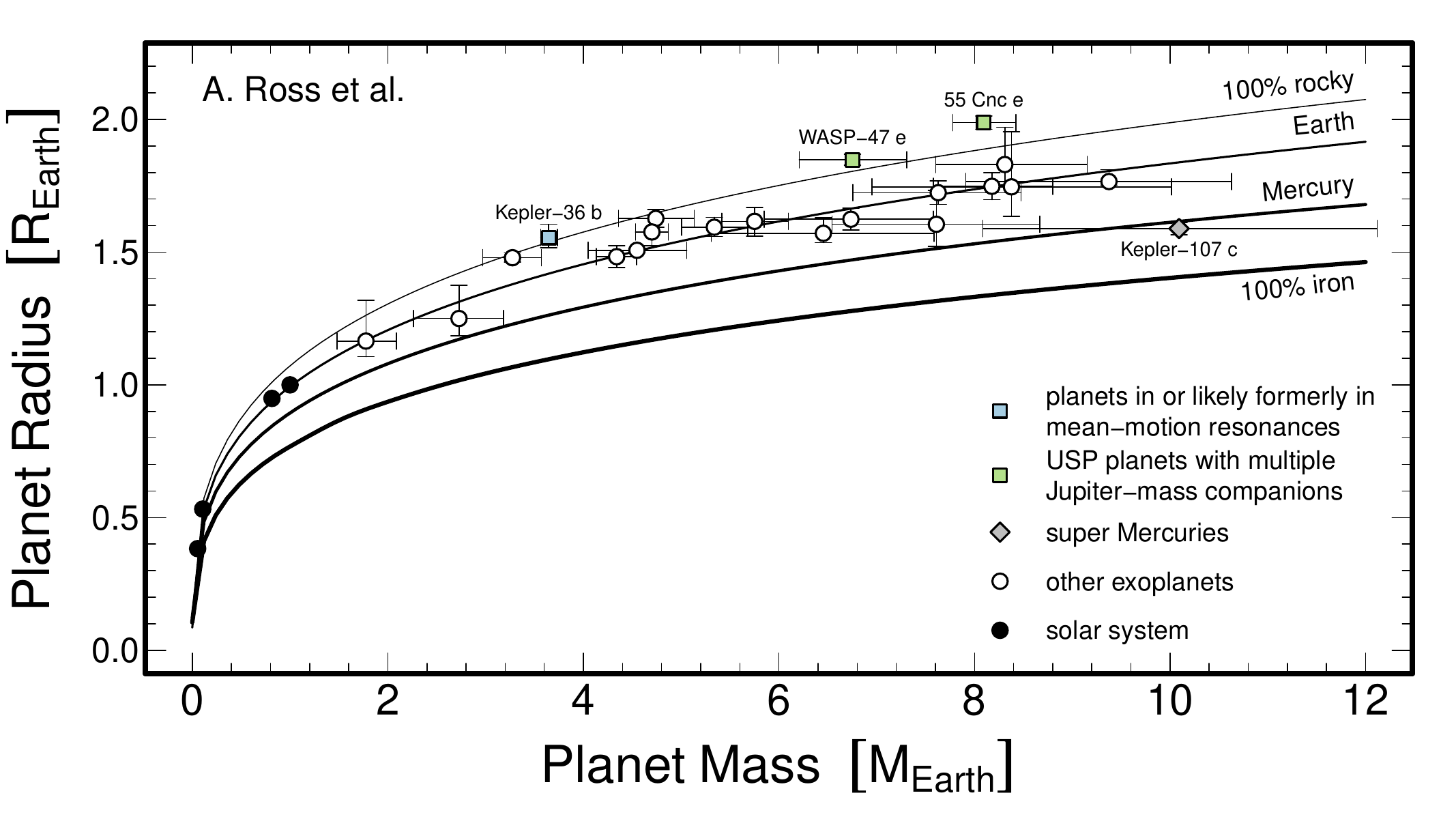}
\includegraphics[width=0.6\textwidth]{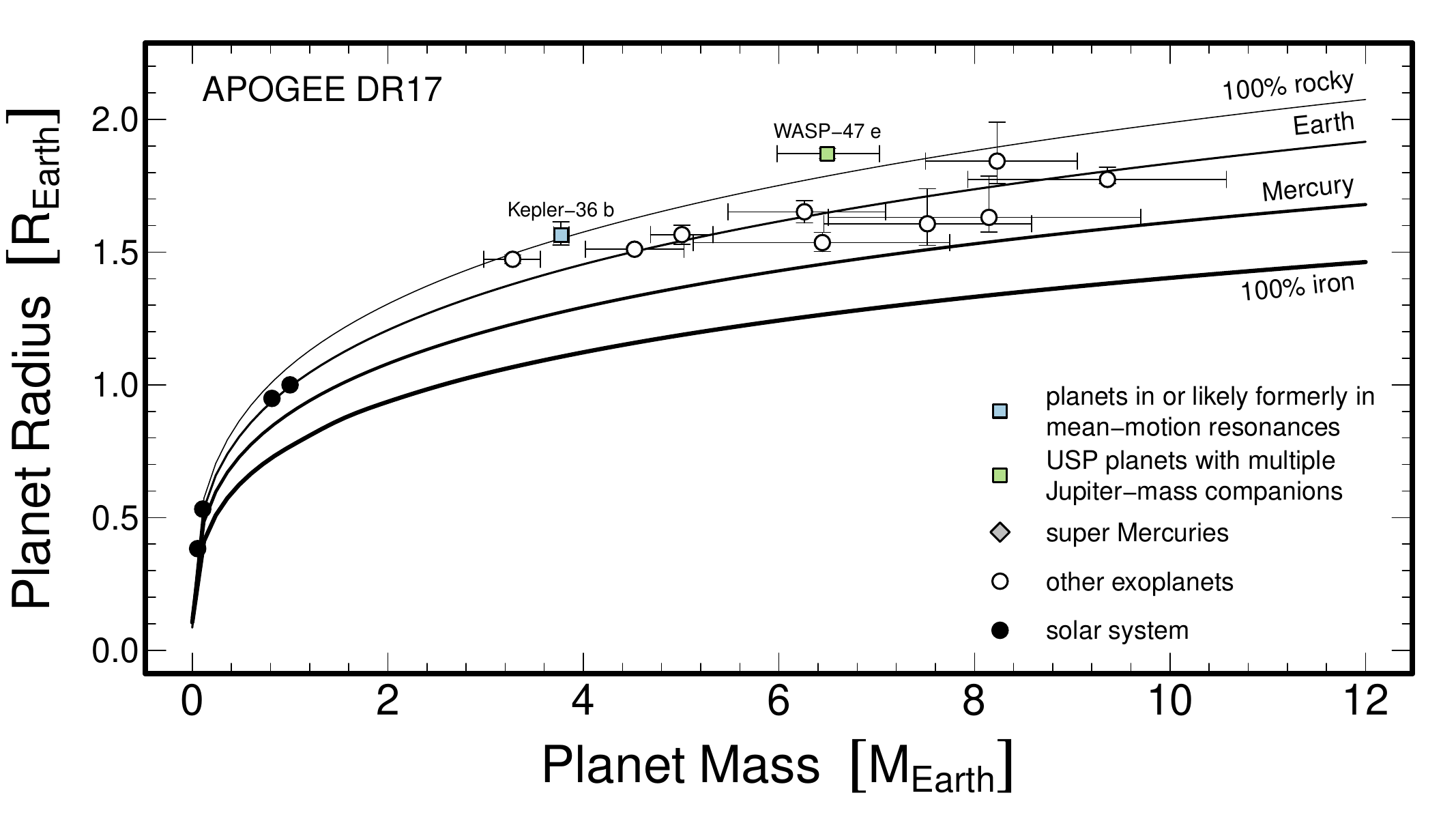}
\includegraphics[width=0.6\textwidth]{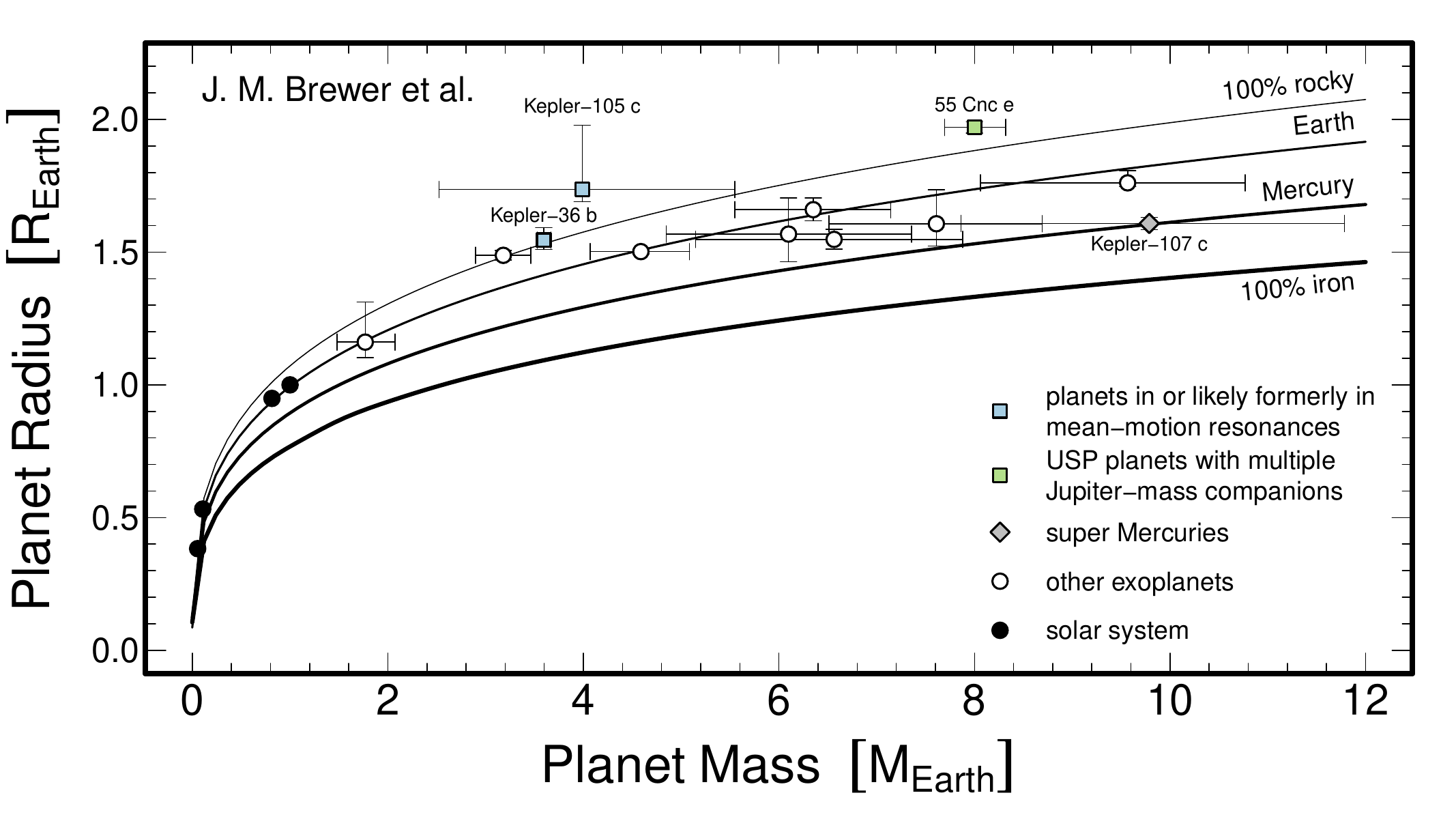}
\caption{Planet radius as a function of planet mass.  We plot the
exoplanets in our subsamples as black-boarded shapes and solar system
terrestrial planets as filled black circles.  Four exoplanets in our
sample have lower densities than predicted by \citet{Plotnykov2020}
for 100\% rock compositions: Kepler-36 b, Kepler-105 c, 55 Cnc e, and
WASP-47 e.  We indicate the exoplanets Kepler-36 b and Kepler-105 c
as black-boarded light blue squares.  Kepler-36 b was likely formerly
in a mean-motion resonance with Kepler-36 c while Kepler-105 c is
in a mean-motion resonance with Kepler-105 b.  They are the only two
terrestrial exoplanets in our sample that are or were likely formerly
mean-motion resonant.  We indicate 55 Cnc e and WASP-47 e as black-boarded
light green squares.  Both 55 Cnc e and WASP-47 e are USP exoplanets with
multiple giant exoplanet companions orbiting the two most metal-rich
stars in our sample.  We indicate the only super-Mercury in our sample
Kepler-107 c as a black-boarded gray diamond.}
\label{fig:massvradius}
\end{figure*}

To calculate updated planetary masses, radii, and densities, we use
as inputs both our own host star mass and radius inferences as well
as literature Doppler (e.g., $K$) and transit (e.g., $\delta$, $P$,
and $i$) observables.  We then execute a Monte Carlo simulation in
which we sample stellar masses and radii from the posteriors of our
analyses and planetary Doppler semiamplitudes, transit depths, orbital
periods, and orbital inclinations from normal distributions defined by
the literature mean values and 1-$\sigma$ uncertainties.  We correctly
account for asymmetric uncertainty distributions from literature sources
as appropriate.  We calculate planetary radii $R_{\text{p}}$, masses
$M_{\text{p}}$, and densities $\rho$
\begin{eqnarray}
R_{\text{p}} & = & \sqrt{\delta} R_{\ast}, \\
\frac{M_{\text{p}}}{M_{\oplus}} & = & \left(\frac{K}{0.08946~\text{m
s$^{-1}$}}\right) \left(\frac{P}{365.24~\text{d}}\right)^{1/3}
\left(\frac{1}{\sin{i}}\right) \nonumber \\
             &  & \left(\frac{M_{\ast}}{M_{\odot}}\right)^{2/3}, \\
M_{\text{p}} & = & \frac{M_{\text{p}}}{M_{\ast}} M_{\ast}, \\
\rho & = & \frac{M_{\text{p}}}{R_{\text{p}}^{3}},
\end{eqnarray}
We use Equation (2) for exoplanets with Doppler-based mass inferences and
Equation (3) for transit-timing variation (TTV)-based mass inferences.
Mass inferences in the literature for Kepler-36 b, Kepler-105 c, and
KOI-1599.01 are based on TTV observations and not Doppler measurements.
For Kepler-36 b, Kepler-105 c, and KOI-1599.01 we use the transit depths
reported in \citet{Thompson2018} and the TTV-based exoplanet-to-star
mass ratios reported in \citet{Vissapragada2020}, \citet{Hadden2017},
and \citet{Panichi2019} to calculate planetary masses and radii.
We report in Table \ref{tab:stellar_params} the 16th, 50th, and 84th
quantiles of our Monte Carlo-simulation inferred planetary masses,
radii, and densities.  We find that the radius we infer for KOI-1599.01
$R_{\text{p}} = 2.45 \pm 0.07~R_{\oplus}$ is too large to be explained
by a terrestrial composition and exclude it from consideration as a
terrestrial exoplanet.  We list these updated planetary masses and
radii in Table \ref{tab:stellar_params} and plot the intersections
between all three samples in Figure \ref{fig:mass_radius_compare}.
We also plot mass--radius diagrams for all three subsamples in Figure
\ref{fig:massvradius}.

\subsection{Terrestrial Planet Core-mass Fractions}\label{sec:cmf}

We use host star photospheric elemental abundances and updated planetary
masses and radii to constrain the core-mass fractions of the terrestrial
exoplanets in our sample using \texttt{SuperEarth} \citep{Plotnykov2020}.
While the internal structures of terrestrial exoplanets cannot be directly
measured, as described in Section \ref{sec:intro} forward modeling is
commonly used to constrain interior structures based on planetary masses
and radii.  These models assume that terrestrial exoplanet interiors
have Earth-like structures (e.g., Mg-Si-O mantles and Fe cores) but with
proportions of the elements that can differ from the Earth.

We infer core-mass fraction constraints based on three different inputs:
(1) planetary masses and radii, (2) host star photospheric iron to silica
refractory ratios $\text{Fe/Si} = 10^{\text{A(Fe)-A(Si)}}$, and (3) host
star photospheric iron to magnesium refractory ratios $\text{Fe/Mg} =
10^{\text{A(Fe)-A(Mg)}}$.  The first method is based on the physical
interior model described in \citet{Valencia2006, Valencia2007}
while the latter two methods use chemical interior models described
in \citet{Plotnykov2020}.  For all three methods, we assume a two-layer
structure consisting of an iron-alloy core surrounded by a silicate mantle
with the mantle's Mg/Si ratio fixed to that of Earth.  For the chemical
interior models, we assume planetary Fe/Mg or Fe/Si ratios based on the
abundances of those elements in the photospheres of their host stars.
These parameters Fe/Si and Fe/Mg then determine the amount of silicon
in a terrestrial exoplanet's core and the amount of iron in its mantle.
The net result is that a planet's core-mass fraction becomes only a
function of the Fe/Mg or Fe/Si ratio in the photosphere of its host star.
As we acknowledge in Section \ref{sec:intro}, the assumption of these
chemical interior models that the mean abundances of a terrestrial
exoplanet match those observed in the photosphere of its host star is
debated \citep[e.g.,][]{Thiabaud2015, Dorn2017, Plotnykov2020}.

We execute these calculations using four different sets of input data:
\begin{enumerate}
\item
the archival spectra-based host star photospheric elemental abundances
we inferred in Section \ref{sec:abundances} and the corresponding updated
planetary masses and radii we inferred in Section \ref{sec:planets};
\item
APOGEE DR17 host star photospheric elemental abundances and the
corresponding updated planetary masses and radii we inferred in Section
\ref{sec:planets};
\item
\citet{Brewer2016}/\citet{Brewer2018} host star photospheric elemental
abundances and the corresponding updated planetary masses and radii we
inferred in Section \ref{sec:planets}; and
\item
\citet{Adibekyan2021} host star photospheric elemental abundances plus
planetary masses and radii.
\end{enumerate}
If available, we use host star photospheric elemental abundances corrected
for departures from the assumptions of LTE (except for calcium as noted
in Section \ref{sec:abundances}).  We use Monte Carlo simulations to
infer the uncertainties on our core-mass fraction inferences.  We first
feed the ($M_{\text{p}}$,$R_{\text{p}}$) ordered pairs calculated in
Section \ref{sec:planets} to \texttt{SuperEarth} to calculate exoplanet
mass and radius-based core-mass fraction inferences.  We note that since
we calculate these ($M_{\text{p}}$,$R_{\text{p}}$) ordered pairs self
consistently with Equations (1)-(3) using ($M_{\ast}$,$R_{\ast}$)
ordered pairs from our stellar parameter posteriors as derived
in Section \ref{sec:stellarparams}, our exoplanet mass and
radius-based core-mass fraction inferences respect the covariances
between ($M_{\ast}$,$R_{\ast}$) imparted by Equations (1)-(3) on
($M_{\text{p}}$,$R_{\text{p}}$).  We next assume the posteriors for our
inferred stellar $\text{Fe/Si}$ and $\text{Fe/Mg}$ ratios as calculated in
Section \ref{sec:abundances} have Gaussian distributions with the means
and standard deviations as reported in Table \ref{tab:abundances_nlte}.
We sample 10,000 realizations from those distributions and feed
the resulting samples to \texttt{SuperEarth} to calculate stellar
$\text{Fe/Si}$ and $\text{Fe/Mg}$-based core-mass fraction inferences.
We report the 16th, 50th, and 84th quantiles of these core-mass fraction
distributions in Table \ref{tab:cmfs}.

The intervals defined by the 16th/84th interquantile ranges for our
$\text{Fe/Si}$- and $\text{Fe/Mg}$-based core-mass fraction inferences
overlap for about 64\% of our sample as expected if our uncertainties
are correctly inferred.  They are generally restricted to the range
between about 0.1 and 0.2.  In contrast, our mass--radius-based
core-mass fraction inferences have a broader range from about -0.2
(i.e., no core) to 0.7 (i.e., a Mercury-like core-mass fraction).
The mass--radius- and Fe/(Mg,Si)-based core-mass fraction inference
16th/84th interquantile ranges overlap for about 38\%/47\% our sample,
generally those terrestrial exoplanets for which the mass--radius-based
methodology indicates intermediate core-mass fractions.  This lack of
consistent agreement between the mass--radius- and abundance ratio-based
core-mass fraction inference approaches indicates that at least one of
our modeling approaches is imperfect or has unsatisfied assumptions.
A companion article Plotnykov et al. (2025, submitted) presents a more
detailed comparison of these differing modeling approaches for the
purposes of terrestrial exoplanet internal structure inferences.

\section{Discussion}\label{sec:discussion}

\subsection{Mass--radius-based Terrestrial Exoplanet Core-mass
Fraction Inferences Are Robust to Photospheric Stellar Parameter
Source}\label{sec:disc_cmfs}

\citet{Adibekyan2021} found a statistically significant positive linear
relationship between mass--radius-based terrestrial exoplanet core-mass
fractions and the iron contents of their host stars as well as evidence
for distinct populations of terrestrial exoplanets with Earth-like and
Mercury-like core-mass fractions.  That study was affected by a subtle
inconsistency though.  While it used homogeneously inferred host star iron
contents implied by photospheric stellar parameters, those photospheric
stellar parameters differed from the photospheric stellar parameters
used to infer the host star masses and radii needed to turn Doppler and
transit observables into planetary masses and radii.  In other words,
their input data were not internally consistent.  That systematic could
bias or increase the apparent dispersion in the core-mass fraction
distribution they reported.  On the other hand, we calculate accurate,
precise, homogeneous, and physically self consistent stellar masses and
radii using three independent sources of photospheric stellar parameters
that we then use to infer planetary masses and radii from Doppler and
transit observables.  Our data are therefore internally consistent and
free of systematics that could bias or increase the apparent dispersion
in the core-mass fraction distribution.  We are also able to evaluate
whether photospheric stellar parameter inference approach influences
the relationship between host star properties and terrestrial exoplanet
interior structures.

To investigate the statistical significance of correlations or linear
relationships between stellar metallicity and terrestrial exoplanet
core-mass fraction in this subsection, we use Monte Carlo simulations.
Since we have point estimates and uncertainties for both stellar
metallicities and terrestrial exoplanet core-mass fractions, we can use
Monte Carlo simulations to quantify the uncertainties on Kendall's $\tau$
coefficient as well as both the slopes and intercepts of our linear models
without relying on the assumptions that must be made in classical linear
modeling.  On each iteration of our Monte Carlo simulations, we sample
both stellar metallicity and terrestrial exoplanet core-mass fraction
from the distributions defining those quantities.  We then calculate
Kendall's $\tau$ coefficient, fit a linear model to those data using the
least-squares approach, and save the resulting $\tau$ coefficient, slope,
and intercept.  We then repeat this procedure 10,000 times.  We ultimately
report the resulting median, 16th, and 84th quantile values as the point
estimates and uncertainties for Kendall's $\tau$ coefficients as well
as linear model slopes and intercepts.  We plot the results of these
calculations in Figures \ref{fig:zfe_vs_mrcmf}-\ref{fig:zmg_vs_mrcmf}
in this section and in Figure \ref{fig:xfe_vs_femgsicmf} in the Appendix.

In contrast to \citet{Adibekyan2021}, our results do not consistently
support a statistically significant correlation or positive linear
relationship between the mass--radius-based core-mass fractions of
terrestrial exoplanets and the iron metallicities $Z_{\text{Fe},\ast}
= Z_{\text{Fe},\odot} 10^{[\text{Fe/H}]}$ of their host stars (Figure
\ref{fig:zfe_vs_mrcmf}).  Our Monte Carlo simulations provide Kendall's
$\tau$ coefficient and slope distributions that include zero within
the intervals bounded by the 2nd and 98th interquantile ranges
of those distributions.  In other words, we find no correlations
or linear relationships between stellar metallicity and core-mass
fraction significant at levels corresponding to the 2-$\sigma$ range
of a Gaussian for parameters based on our own work in this article,
\citet{Adibekyan2021}, or APOGEE DR17.  On the other hand, we do find a
linear relationship (but not a correlation) between stellar metallicity
and mass--radius-based core-mass fraction with slope distribution that
excludes zero from its 2nd to 98th interquantile ranges for parameters
based on data from \citet{Brewer2016}/\citet{Brewer2018}.  In other
words, we find a linear relationship significant at a level corresponding
to more than the 2-$\sigma$ range of a Gaussian for parameters based
on data from \citet{Brewer2016}/\citet{Brewer2018}.  We conclude that
there may be a positive linear relationship between mass--radius-based
core-mass fractions of terrestrial exoplanets and the iron metallicities
of their host stars, but the support for that relationship in our sample
is not strong.

\begin{figure*}
\centering
\includegraphics[width=\textwidth]{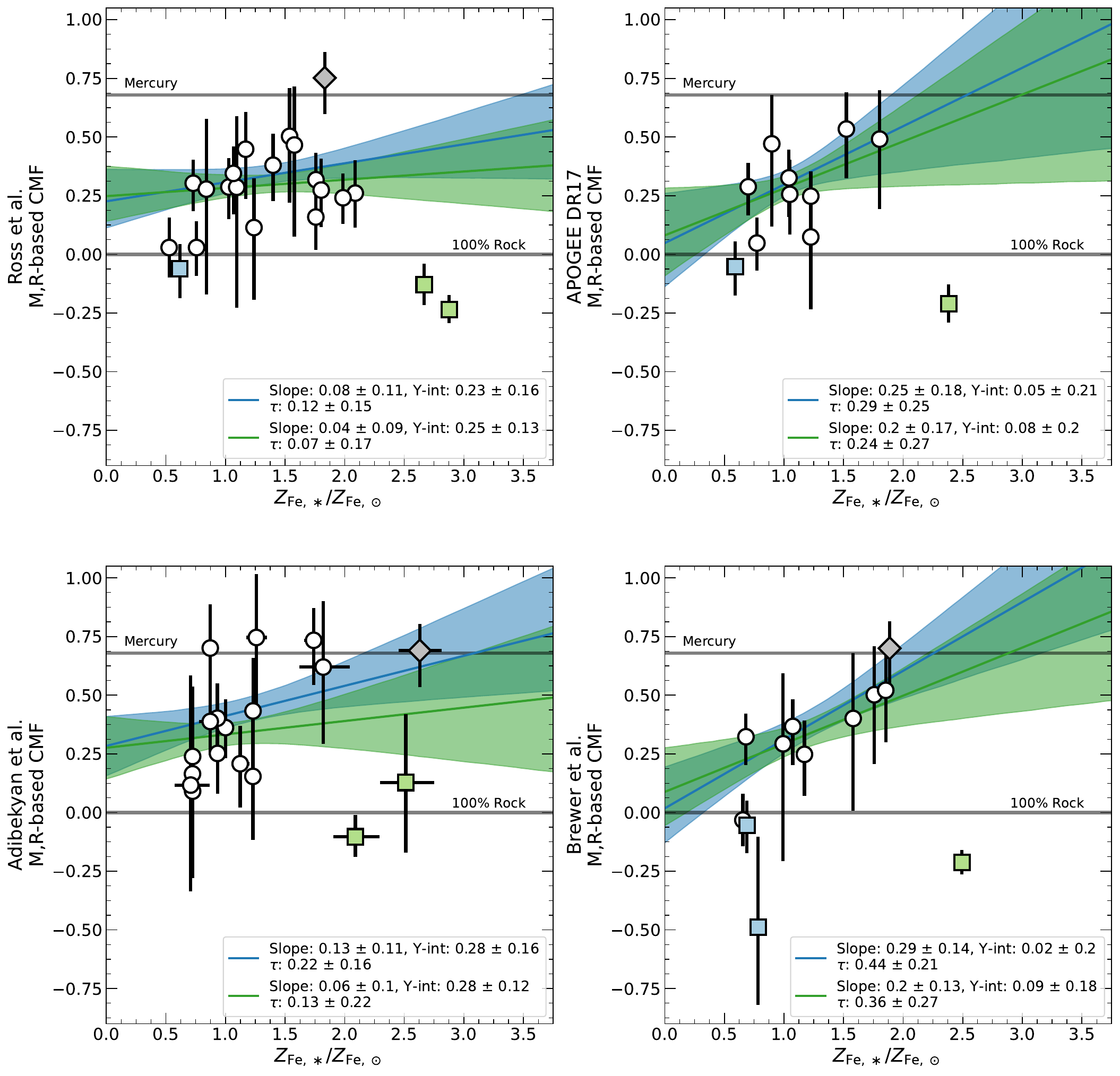}
\caption{Mass--radius-based core-mass fraction (CMF) as a function
of iron metallicity $Z_{\text{Fe},\ast}/Z_{\text{Fe},\odot}$ for
four different photospheric stellar parameter sources.  We plot the
low-density exoplanets Kepler-36 b and Kepler-105 c as black-boarded
blue squares and 55 Cnc e and WASP-47 e as black-boarded green squares.
We plot the super-Mercury exoplanet Kepler-107 c as a gray diamond.
We plot all other exoplanets as black-boarded open circles.  To evaluate
the impact of stellar iron metallicity and therefore iron mass fraction
on terrestrial exoplanet CMF, we use the Monte Carlo procedure described
in the text to calculate Kendall's $\tau$ coefficients and fit linear
models to these data after excluding exoplanets inconsistent with
pure-rock compositions (i.e., negative CMFs).  We plot the linear
models that result from our Monte Carlo simulations as blue or
green lines and indicate their 16th to 84th quantile uncertainty
regions with blue or green polygons.  We also report in the legend
of each panel the numerical values of Kendall's $\tau$ coefficients,
slopes, and intercepts as well as their uncertainties.  For the
relationship between $Z_{\text{Fe},\ast}/Z_{\text{Fe},\odot}$ and
mass--radius-based CMF depicted by the blue lines, we find positive
correlations and slopes for which zero is not excluded from the 2nd
to 98th interquantile ranges of the Monte Carlo inferred parameter
distributions (corresponding to the 2-$\sigma$ range of a Gaussian).
If we exclude exoplanets with Mercury-like CMFs, then the significances
of the correlations and slopes of the relationships depicted by the
green lines are diminished.  We conclude that the relationship between
$Z_{\text{Fe},\ast}/Z_{\text{Fe},\odot}$ and mass--radius-based CMF is
positive but not especially statistically significant given our sample.}
\label{fig:zfe_vs_mrcmf}
\end{figure*}

We also investigate the relationships between the core-mass
fractions of terrestrial exoplanets and the magnesium metallicities
$Z_{\text{Mg},\ast} = Z_{\text{Mg},\odot} 10^{[\text{Mg/H}]}$ of
their host stars (Figure \ref{fig:zmg_vs_mrcmf}).  In this case,
our Monte Carlo simulations provide Kendall's $\tau$ coefficient
and slope distributions that include zero within the 2nd and 98th
interquantile ranges of those distributions.  In other words, we
find no correlations or linear relationships significant at levels
corresponding to the 2-$\sigma$ range of a Gaussian for parameters based
on our own work in this article, \citet{Adibekyan2021}, APOGEE DR17,
or \citet{Brewer2016}/\citet{Brewer2018}.  We describe the results of
our exploration of the relationships between Fe/Mg- and Fe/Si-based
core-mass fractions of terrestrial exoplanets and the iron or magnesium
metallicities of their host stars in the Appendix. 

A more in-depth discussion and analysis on the correlations between stellar composition and planetary core-mass fractions is done in a companion paper (Plotnykov, et al., submitted).

\begin{figure*}
\centering
\includegraphics[width=\textwidth]{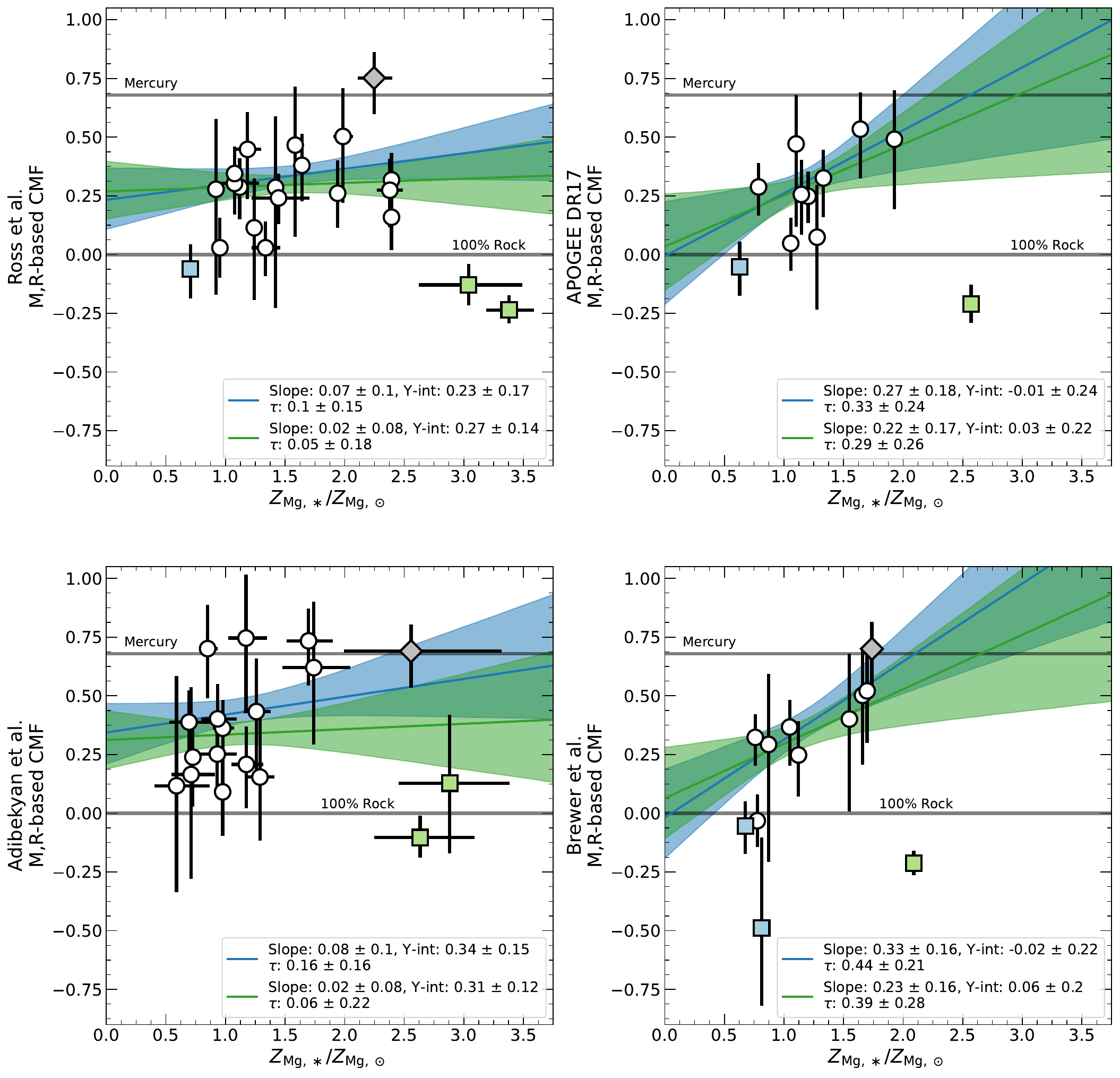}
\caption{Mass--radius-based CMF as a function of magnesium metallicity
$Z_{\text{Mg},\ast}/Z_{\text{Mg},\odot}$ for four different
photospheric stellar parameter sources.  We plot the low-density
exoplanets Kepler-36 b and Kepler-105 c as blue squares and 55 Cnc e
and WASP-47 e as green squares.  We plot the super-Mercury exoplanet
Kepler-107 c as a gray diamond.  We plot all other exoplanets as
open circles.  To evaluate the impact of stellar magnesium metallicity
and therefore magnesium mass fraction on terrestrial exoplanet CMF,
we use the Monte Carlo procedure described in the text to calculate
Kendall's $\tau$ coefficients and fit linear models to these data after
excluding exoplanets inconsistent with pure-rock compositions (i.e.,
negative CMFs).  We plot the linear models that result from our Monte
Carlo simulations as blue or green lines and indicate their 16th to 84th
quantile uncertainty regions with blue or green polygons.  We also report
in the legend of each panel the numerical values of Kendall's $\tau$
coefficients, slopes, and intercepts as well as their uncertainties.
For the relationship between $Z_{\text{Mg},\ast}/Z_{\text{Mg},\odot}$
and mass--radius-based CMF depicted by the blue lines, we find positive
correlations and slopes for which zero is not excluded from the 2nd
to 98th interquantile range of the Monte Carlo inferred parameter
distributions (corresponding to the 2-$\sigma$ range of a Gaussian).
If we exclude exoplanets with Mercury-like CMFs, then the significances
of the correlations and slopes of the relationships depicted by the
green lines are diminished.  We conclude that the relationship between
$Z_{\text{Mg},\ast}/Z_{\text{Mg},\odot}$ and mass--radius-based CMF is
positive but not especially statistically significant given our sample.}
\label{fig:zmg_vs_mrcmf}
\end{figure*}

\subsection{Terrestrial Exoplanets In or Likely Formerly In
Mean-motion Resonances Have Significant Amounts of Water in their
Interiors}\label{sec:disc_mmr}

Table \ref{tab:cmfs} shows that the exoplanets KOI-1599.01,
Kepler-36 b, Kepler-105 c, 55 Cnc e, and WASP-47 e are the
only exoplanets in our sample that have unanimously negative
mass--radius-based core-mass fraction inferences grounded on APOGEE
DR17, \citet{Brewer2016}/\citet{Brewer2018}, and our own photospheric
stellar parameters.  In this context, a negative core-mass fraction
(CMF) indicates that an  exoplanet has a larger radius than expected
for its mass even for a pure rock composition.  Their low densities
are also apparent in Figure \ref{fig:massvradius}.  These outliers
aside, the mean mass--radius-based core-mass fraction of our sample
$\overline{\text{CMF}} = 0.30 \pm 0.04$ is consistent with the Earth's
core-mass fraction $\text{CMF}_{\oplus} \approx 0.33$.

Kepler-36 b, Kepler-105 c, and KOI-1599.01 are all in or likely formerly
in mean-motion resonances.  Kepler-105 c and KOI-1599.01 are both in
mean-motion resonances \citep{MacDonald2023,Panichi2019}.  While Kepler-36
b is not currently in a mean-motion resonance \citep{Carter2012}, the
planets Kepler-36 c/Kepler-36 b have an orbital period ratio larger than
but within less than 0.6\% of 7:6 indicative of exoplanets likely to
be resonant \citep{Goldberg2023}.  Systems like Kepler-36 c/Kepler-36
b that are both plausibly first-order mean-motion resonant and close
enough to their host star for tidal dissipation to play a role in
their dynamical evolution likely formed in resonance only to diffuse
out of that resonance over time \citep{Hamer2024}.  These are the
only three exoplanets in our sample that are are currently in or were
likely in mean-motion resonances in the recent past.  As we argued in
Section \ref{sec:planets}, if KOI-1599.01 with $R_{\text{p}} = 2.45 \pm
0.7~R_{\oplus} \gtrsim 1.6~R_{\oplus}$ was composed entirely of rock
then it would have $M_{\text{p}} \approx 25~M_{\oplus}$ well in excess
of its inferred mass $M_{\text{p}} = 4.11_{-0.27}^{+0.26}~M_{\oplus}$.
It almost certainly has a significant volatile-rich atmosphere and
is likely not terrestrial \citep[e.g.,][]{Rogers2015}.  Consequently,
Kepler-36 b and Kepler-105 c are the only terrestrial exoplanets in our
sample that are currently or formerly in mean-motion resonances.

\begin{figure*}
\centering
\plotone{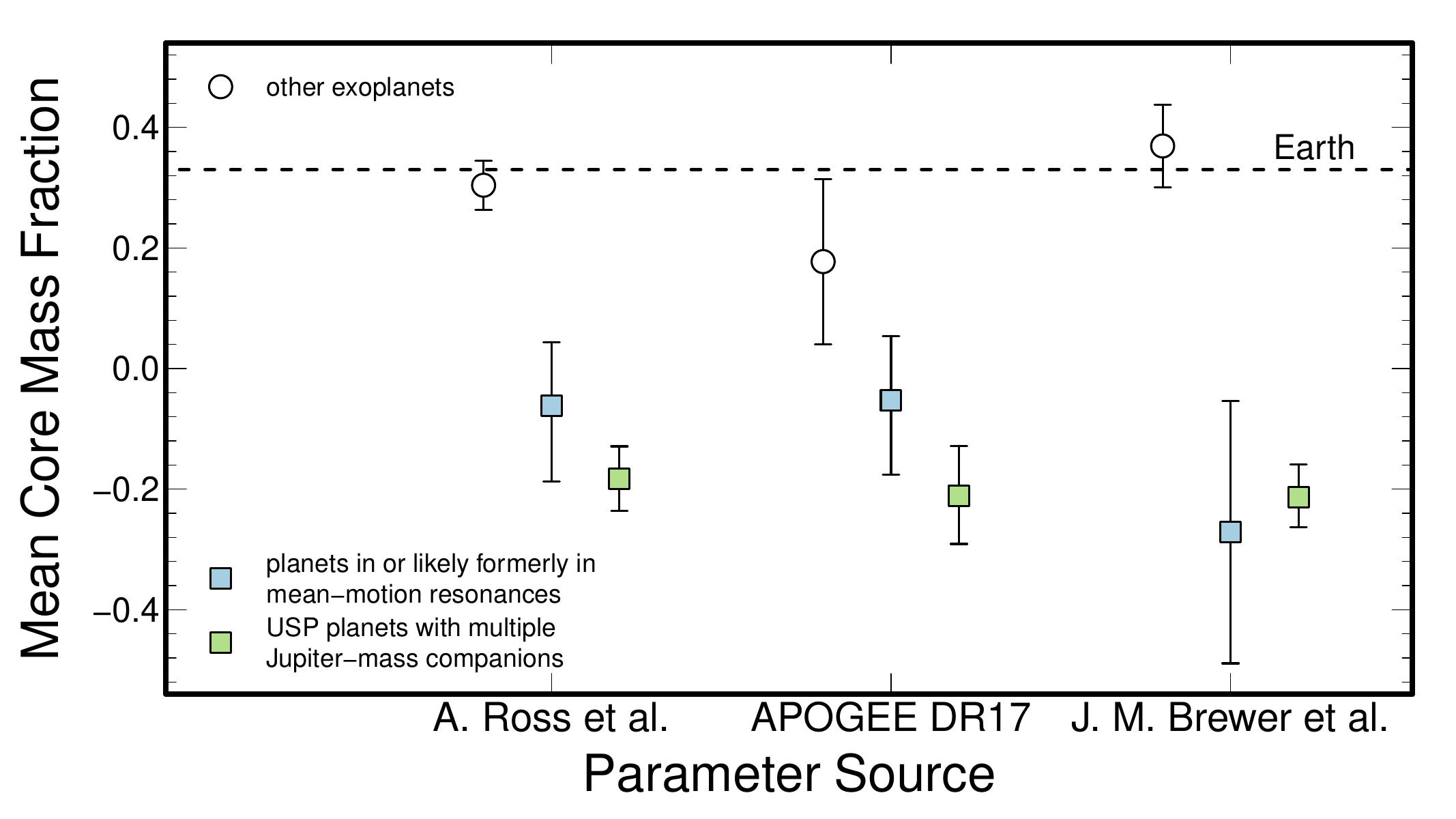}
\caption{Mean core-mass fractions for all three subsamples in our
overall sample.  In all three subsamples, the mean core-mass fraction
of the population of terrestrial exoplanets in or likely formerly in
mean-motion resonances is about $3\sigma$ lower than the mean core-mass
fraction of the other exoplanets in our sample.  Likewise, the population
of USP exoplanets orbiting metal-rich stars with multiple giant exoplanet
companions has a mean core-mass fraction about $6\sigma$ lower than
the mean core-mass fraction of the other exoplanets in our sample.
It is inconsistent with a 100\% rock compositions.  These observations
confirm that the populations of both terrestrial exoplanets that are
in or likely formerly in mean-motion resonances and USP exoplanets
orbiting metal-rich stars with multiple giant exoplanet companions are
qualitatively different than the rest of the exoplanets in our sample.
Resonant configurations suggest migration inwards from more distant
formation locations, so we attribute the low core-mass fractions of
terrestrial exoplanets in or formerly in mean-motion resonances to the
incorporation and retention of water in their interiors during their
formation.  We propose that the population of USP exoplanets orbiting
very metal-rich stars with multiple giant exoplanet companions are the
remnant cores of tidally disrupted mini-Neptunes and associate their
low densities with the presence of significant hydrogen, helium, water,
and/or other volatiles in their interiors.}
\label{fig:cmfcomp}
\end{figure*}

The low densities of Kepler-36 b and Kepler-105 c cannot easily be
explained by the presence of hydrogen/helium (H/He) or high mean molecular
weight atmospheres.  According to Figure 10 of \citet{Fulton2017},
the instellations $F_{\text{p}} = (111, 205)~F_{\oplus}$ experienced
by Kepler-36 b and Kepler-105 c combined with their radii make it
extraordinarily unlikely that they could maintain H/He atmospheres.
Conservatively assuming Mercury-like Bond albedoes $A_B = 0.068$, our
parameters imply that the equilibrium temperatures $T_{\text{eq}}$ of
Kepler-36 b and Kepler-105 c should be $T_{\text{eq}} \approx 900$ K and
$T_{\text{eq}} \approx 1000$ K respectively.  Given their equilibrium
temperatures, the density scale heights of steam atmospheres are at
least an order-of-magnitude smaller than our uncertainties in their
inferred radii.  The scale heights of atmospheres composed of molecules
more massive than H$_{2}$O would be even smaller.  Accordingly, the low
densities of Kepler-36 b and Kepler-105 c are best explained by some
difference between their interiors and the interiors of the rest of the
terrestrial exoplanets in our sample.  Indeed, the mean core-mass fraction
of the population of terrestrial exoplanets in or likely formerly in
mean-motion resonances is about $3\sigma$ lower than the mean core-mass
fraction of the other exoplanets in our sample that as a population are
consistent with the Earth's core-mass fraction (Figure \ref{fig:cmfcomp}).

Planetary systems in mean-motion resonances almost certainly migrated from
their formation locations to their currently observed locations and then
maintained their orbital architectures without experiencing episodic giant
impacts \citep[e.g.,][]{Terquem2007,Raymond2008}.  Terrestrial exoplanets
in mean-motion resonances close to their host stars could even have
formed beyond the water-ice lines of their parent protoplanetary disks,
Type I migrated to their observed locations, and then avoided desiccating
giant impacts.  As a result, such exoplanets could have incorporated
significant amounts of water into their interiors during formation and
retained that water for billions of years \citep[e.g.,][]{Emsenhuber2021a,
Emsenhuber2021b, Emsenhuber2023, Schlecker2021}.

We propose that Kepler-36 b and Kepler-105 c are the first examples of
``icy cores'', a class of super-Earth mass exoplanets composed of rock
and water formed close to or beyond the water-ice lines of their parent
protoplanetary disks predicted by \citet{Schlecker2021}.  We calculate
the water mass fractions necessary to explain the masses and radii of
Kepler-36 b and Kepler-105 c in four ways.  If we assume that their water
is in a condensed phase and interpret the masses and radii of Kepler-36
b and Kepler-105 c as a consequence of rock/water compositions, then
we can interpolate the grid of two-layer rock--water models presented
in \citet{Zeng2016}.  In that case, we find that Kepler-36 b is best
explained by a 98\% rock/2\% water composition and Kepler-105 c is
best explained by a 75\% rock/25\% water composition.  Instead, if we
interpret the masses and radii of Kepler-36 b and Kepler-105 c using a
three-layer core--mantle--water model generated assuming an Earth-like
mineralogy using \texttt{ExoPlex} \citep{Unterborn2023}, then we find
water mass fractions of 5\%/11\% for Kepler-36 b/Kepler-105 c.  If we
account for the fact that at their equilibrium temperatures most of the
water content of Kepler-36 b and Kepler-105 c will be in the vapor phase,
then by interpolating the two-layer rock--water vapor 1 millibar surface
pressure models presented in \citet{Zeng2019} we find that Kepler-36
b is best explained by a 97\% rock/3\% water vapor composition and
Kepler-105 c is best explained by a 78\% rock/22\% water composition.
Alternatively, if we use the \texttt{CEPAM} code \citep{Guillot1995}
assuming an Earth-like core-mass fraction and the \citet{French2009}
water equation of state, then we find water mass fractions of 1\%/8\%
for Kepler-36 b/Kepler-105 c.  Regardless of these model assumptions, our
conclusion is the same: Kepler-36 b and Kepler-105 c are both water rich.
We predict that future studies of the internal structures of super-Earth
mass terrestrial exoplanets in mean-motion resonances will identify more
icy cores like Kepler-36 b and Kepler-105 c.

\subsection{USP Planets Orbiting Metal-rich Stars with
Multiple Giant Planet Companions are the Cores of Disrupted
Mini-Neptunes}\label{sec:disc_usp}

As we highlighted in Section \ref{sec:disc_mmr}, Table \ref{tab:cmfs}
shows that the exoplanets 55 Cnc e and WASP-47 e have unanimously negative
mass--radius-based core-mass fraction inferences grounded on APOGEE DR17,
\citet{Brewer2016}/\citet{Brewer2018}, and our own photospheric stellar
parameters that are individually inconsistent with pure rock compositions.
Their low densities are also apparent in Figure \ref{fig:massvradius}.
There is no reason to conclude that the other USP exoplanets in our sample
Kepler-10 b, Kepler-78 b, K2-106 b, K2-131 b, K2-141 b, K2-229 b, or HD
80653 b are inconsistent with pure rock compositions.  The low densities
of 55 Cnc e and WASP-47 e inconsistent with pure rock composition cannot
be attributed to their USP status alone.

The stars 55 Cnc and WASP-47 are the two most metal-rich stars in
our sample and among the most metal-rich exoplanet host stars known.
Both host similar exoplanet systems with (1) a massive USP planet,
(2) multiple longer-period giant planets, and (3) an innermost giant
planet orbiting close to the massive USP planet.  Metal-rich stars like
these that once hosted metal-rich protoplanetary disks are the most
likely locations for in situ mini-Neptune and giant planet formation
\citep[e.g.,][]{Bodenheimer2000, Chiang2013, Schlaufman2014}.  Treating
the two exoplanets 55 Cnc e and WASP-47 e as a population, that population
of USP exoplanets orbiting metal-rich stars with multiple giant exoplanet
companions has a mean mass--radius-based core-mass fraction 6$\sigma$
below the rest of the terrestrial exoplanet population in our sample
(Figure \ref{fig:cmfcomp}).  While it is possible the MIST isochrones
grid we use to infer the radii of the host stars 55 Cnc and WASP-47
systematically overestimates the radii of very metal-rich stars, the
radius we infer for 55 Cnc $R_{\ast} = 0.99_{-0.01}^{+0.01}~R_{\odot}$
is consistent with the interferometrically derived but limb-darkening
dependent value $R_{\ast} = 0.960_{-0.018}^{+0.018}~R_{\odot}$
\citep[e.g.,][]{vonBraun2011, Ligi2016}.

Theoretical models have established that the low densities of 55
Cnc e and WASP-47 e cannot be explained by the presence of H/He
\citep[e.g.,][]{Bourrier2018}.  Likewise, the empirical constraints from
\citet{Fulton2017} have shown that the instellations experienced by 55
Cnc e and WASP-47 e make it impossible that they could sustain long-lived
H/He atmospheres.  Given their equilibrium temperatures, the density
scale heights of steam atmospheres are a factor of three smaller than
our uncertainties in their inferred radii.  Nondetections of escaping
H/He or water/other species in high-resolution cross-correlation
analyses also disfavor H/He atmospheres, steam atmospheres, or
atmospheres comprised of many other high mean-molecular weight species
\citep[e.g,][]{Ehrenreich2012, Zhang2021, Esteves2017, Jindal2020,
Tabernero2020, Deibert2021, Keles2022, Rasmussen2023}.  The density
scale heights of atmospheres composed of molecules like CO or CO$_{2}$
proposed by \citet{Angelo2017}, \citet{Hu2024}, and \citet{Patel2024}
on 55 Cnc e and WASP-47 e would be least four times smaller than our
uncertainties in their inferred radii.

The root cause(s) of the low densities of 55 Cnc e and WASP-47 e
are a topic of current debate \citep[e.g.,][]{Demory2016, Crida2018,
Bourrier2018, Dorn2019, Luo2024, Peng2024}.  Most suggestions for the
masses and radii of 55 Cnc e and WASP-47 e invoke atmospheres that have
yet to be definitively observed despite more than a decade of intense
observational scrutiny.  Alternatively, \citet{Dorn2019} proposed that
the interiors of 55 Cnc e and WASP-47 e differ from those of other
terrestrial exoplanets because they formed in the hottest parts of
their parent protoplanetary disks.  They are consequently enriched in
species like calcium and aluminum with high condensation temperatures
and depleted in iron.  If in situ formation in the hottest parts of
parent protoplanetary disks produces low-density exoplanets, then the
\citet{Dorn2019} scenario would predict that the other USP exoplanets
in our sample Kepler-10 b, Kepler-78 b, K2-106 b, K2-131 b, K2-141 b,
K2-229 b, and HD 80653 b should have similarly low densities.  This is
not the case though.

We argue that any explanation for the low densities of 55 Cnc e and
WASP-47 e must account for these planets not just as isolated objects but
also in the context of the exoplanet systems in which these exoplanets
are found.  In particular, any explanation for the low densities of 55 Cnc
e and WASP-47 e must explain (1) the reason why their internal structures
differ from the Earth-like compositions of the rest of the USP planet
population and (2) the reason why these two low-density USP exoplanets
both orbit very metal-rich stars with nearby giant planet companions.

We suggest a possible scenario that both explains the low densities of 55
Cnc e and WASP-47 e and accounts for the context of the exoplanet systems
in which they are found: both 55 Cnc e and WASP-47 e are the remnant cores
of relatively recently disrupted mini-Neptunes.  If the rocky planets 55
Cnc e and WASP-47 e spent billions of years as the cores of mini-Neptunes,
then there would have been plenty of time for interactions between the
H/He atmospheres and cores of those mini-Neptunes to push hydrogen,
helium, water, and/or other volatiles deep into the interiors of those
mini-Neptune cores.  In this scenario, the low densities of 55 Cnc e
and WASP-47 e are a consequence of significant amounts of hydrogen,
helium, water, and/or other volatiles that were incorporated into the
cores of these relatively recently disrupted mini-Neptunes via chemical
reactions, convection, or diffusion at the permeable interfaces between
H/He atmospheres and molten cores.  These interactions led to the
deposition of significant amounts of hydrogen, helium, water, and/or
other volatiles into the interiors of the now exposed cores referred to
as the USP exoplanets 55 Cnc e and WASP-47 e.

This explanation for the low densities of 55 Cnc e and WASP-47 e proposed
above is supported by detailed studies of the interactions between
the primordial atmospheres of mini-Neptunes and their rocky cores.
\citet{Rogers2024} showed that hydrogen can be efficiently sequestered
in the rocky core of a mini-Neptune before atmospheric escape, thereby
lowering its bulk density.  The possibility of hydrogen sequestration
needs to be further investigated and verified with experimental values
for hydrogen partition coefficients at high pressure and temperature.
Alternatively, \citet{Dorn2021} and \citet{Luo2024} argued that
sequestration of water in the interiors of terrestrial planets in
general and in 55 Cnc e specifically can lower their bulk densities.
If this water is trapped in the core as suggested by \citet{Luo2024},
then it would not be outgassed into the atmospheres of such planets.
The implication is that low-density planets like 55 Cnc e or WASP-47
e could retain large amounts of water even if it is not present in
observable quantities at their surfaces.

The disruption of the mini-Neptunes now referred to as 55 Cnc e and
WASP-47 e would be a straightforward consequence of the USP formation
scenario outlined in \citet{Schlaufman2010}, \citet{Hamer2020}, and
\citet{Schmidt2024}.  In that scenario, USP exoplanets tidally migrate
over billions of years from the locations of the inner edges of their
parent protoplanetary disks to their observed locations via cycles of
secular eccentricity excitation and inside-planet tidal dissipation.
Since 55 Cnc e and WASP-47 e are both more massive than the typical USP
exoplanet, they would require larger perturbations from their companion
planets to excite their eccentricities to the unobservably small but
non-zero values required for this migration mechanism.  As described
above, both 55 Cnc e and WASP-47 e have giant planet companions less than
0.1 AU away from their observed locations.  These nearby giant planet
companions can easily provide the eccentricity excitation necessary for
this tidal migration mechanism, and such short-period giant planets are
almost always found orbiting very metal-rich stars.

Unlike the higher-density USP exoplanets in our sample that we suggest
lost their primordial H/He envelopes long before this tidal migration
process turned them from proto-USP planets into USP planets, 55 Cnc e
and WASP-47 e were once mini-Neptunes massive enough to retain their
primordial atmospheres during tidal migration (perhaps scaled-down
versions of HAT-P-11 b or Kepler-4 b).  The tidal migration of 55
Cnc e and WASP-47 e accelerated as they approached their host stars
and a combination of intense instellation, tidal heating, and Hill
sphere shrinkage caused the escape of their primordial atmospheres.
They were left as the low-density cores we observe today.  These final
disruption events could be quite recent on astronomical timescales, as
\citet{Schmidt2024} showed that the typical ages of USP exoplanets are in
the range 5 Gyr $\lesssim \tau_{\ast} \lesssim$ 6 Gyr.  In our proposed
scenario, transient volatile outgassing from the now-exposed cores could
be responsible for the redistribution of heat across the surface of 55
Cnc e observed by the James Webb Space Telescope \citep[e.g.,][]{Hu2024,
Patel2024}.

An explanation for the masses and radii of 55 Cnc e and WASP-47 e
complementary to the scenario described in the rest of this subsection
could be its potential status as a ``puffy Venus'' like exoplanet with
a thick carbon-dominated atmosphere in equilibrium with a global magma
ocean \citep[e.g.,][]{Peng2024}.  The scenario described above suggests
that 55 Cnc e and WASP-47 e existed as mini-Neptunes with large H/He
atmospheres for several billion years.  If instead 55 Cnc e and WASP-47
e lost their H/He atmospheres very quickly, then they could be ``puffy
Venus'' type planets.  That is possible, especially if at that time 55
Cnc e/WASP-47 e were highly reduced such that hydrogen would not have
been easily able to interact with a global magma ocean to make water.
If the planets were oxidized, however, then without very rapid loss the
interactions between hydrogen and a global magma ocean would have produced
a lot of water which would then also need to be lost.  We acknowledge that
without detailed knowledge of the all three of (1) the possible escape
of a high mean-molecular weight atmosphere, (2) the time dependence of
that loss, and (3) the details of the magma ocean cooling process this
more detailed ``puffy Venus'' scenario is at this point speculative.

\subsection{Mercury-like Core-mass Fractions Are
Uncommon}\label{sec:disc_merc}

\begin{figure*}
\centering
\plotone{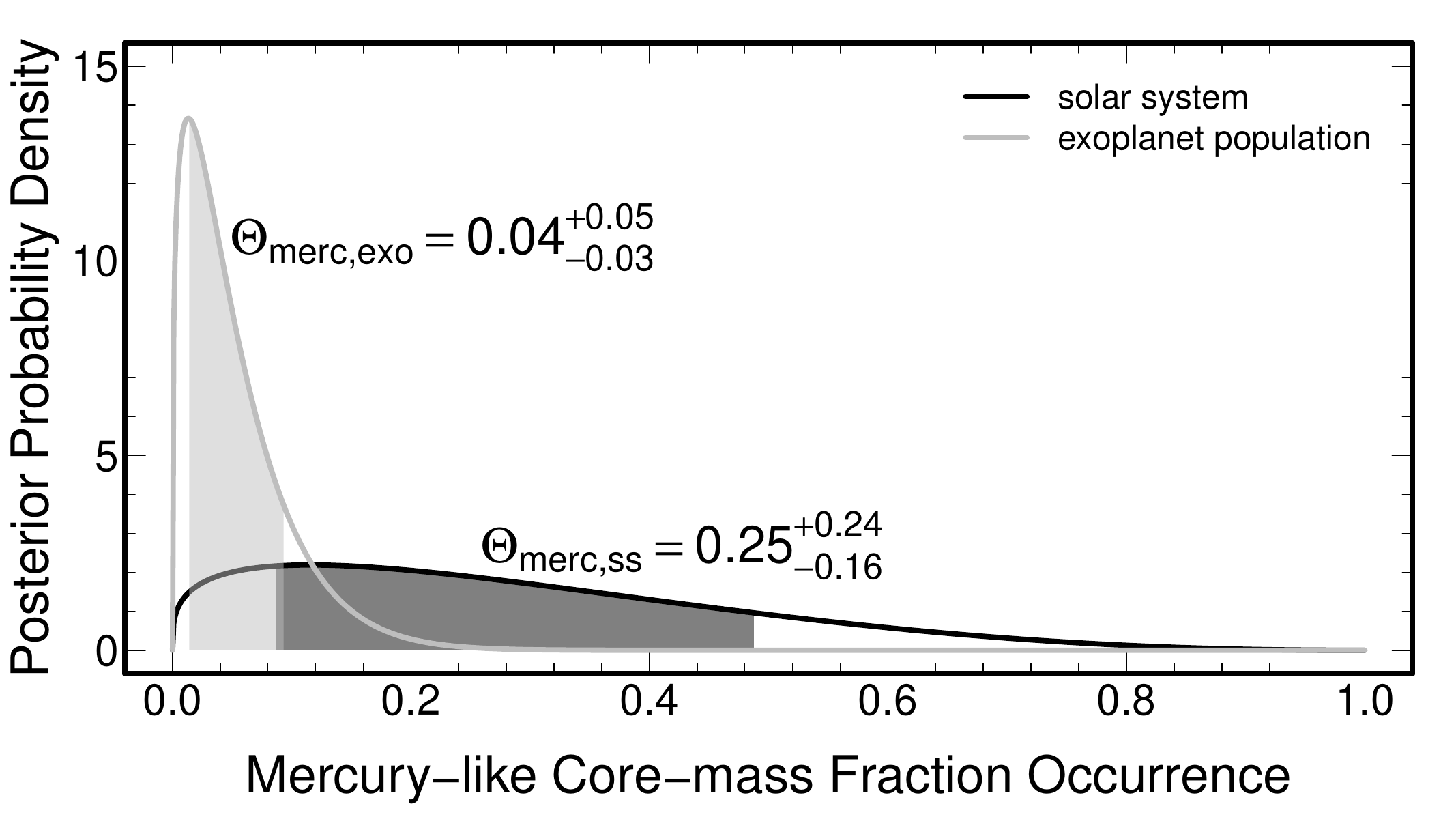}
\caption{Comparison of posterior probability density functions
for the occurrence of Mercury-like planets $\Theta_{\text{merc}}$
in the solar system and in our sample of terrestrial exoplanets.
Our inference of the occurrence of Mercury-like planets in the
solar system $\Theta_{\text{merc,ss}} = 0.25_{-0.16}^{+0.24}$ is
consistent with the occurrence of super-Mercuries in the terrestrial
exoplanet population $\Theta_{\text{merc,exo}} = 0.04_{-0.03}^{+0.05}$.
The probability that our super-Mercury occurrence inference is consistent
with the value $\Theta_{\text{merc,exo}} \approx 0.2$ put forward by
\citet{Adibekyan2021} and \citet{Unterborn2023} is about 1\% (equivalent
to 2.3$\sigma$ assuming Gaussian statistics).}
\label{fig:mercury_cmf_occurrence}
\end{figure*}

\citet{Adibekyan2021} proposed that terrestrial exoplanets with Earth-like
and Mercury-like core-mass fractions are two distinct components of
the terrestrial exoplanet population.  Likewise, \citet{Unterborn2023}
suggested that approximately 20\% of the terrestrial exoplanet population
can be thought of as super-Mercuries.  Defining terrestrial exoplanets
with Mercury-like core-mass fractions as those with mass--radius-based
CMF $> 0.7$ \citep[e.g.,][]{Margot2018}, Kepler-107 c is the only
unanimously Mercury-like exoplanet in our sample grounded on APOGEE DR17,
\citet{Brewer2016}/\citet{Brewer2018}, and our own photospheric stellar
parameters.  With the exclusion of KOI-1599.01, there are 24 terrestrial
exoplanets in our sample only one of which is Mercury-like.  A single
object in a sample of 24 cannot be considered a distinct component of
a larger population.  Using the Bayesian occurrence formalism presented
in \citet{Schlaufman2014}, we find occurrences of Mercury-like planets
in the solar system $\Theta_{\text{merc,ss}} = 0.25_{-0.16}^{+0.24}$
and in the terrestrial exoplanet population $\Theta_{\text{merc,exo}}
= 0.04_{-0.03}^{+0.05}$ (Figure \ref{fig:mercury_cmf_occurrence}).
While our inferences for the occurrence of Mercury-like planets in the
solar system and the terrestrial exoplanet population are consistent,
our latter inference is inconsistent at the $p = 0.01$ level (about
2.3$\sigma$ assuming Gaussian statistics) with a distinct population of
super-Mercuries comprising 20\% of the terrestrial exoplanet population.
A similar conclusion was recently reached by \citet{Brinkman2024}.

\subsection{No Relationship Between Host Star Age and Super-Earth Density
or Core-mass Fraction}\label{sec:age_cmf}

\citet{Weeks2025} identified an inverse correlation between host star
age and rocky exoplanet composition for super-Earth host stars in
the age range 2 Gyr $\lesssim \tau_{\ast} \lesssim$ 14 Gyr.  To infer
the stellar parameters used in their analyses, they used the BAyesian
STellar Algorithm code \citep[\texttt{BASTA},][]{Aguirre2022} to fit
a custom grid of stellar evolution models generated with the Garching
Stellar Evolution Code \citep[\texttt{GARSTEC},][]{Weiss2008} to both
Gaia DR3 parallaxes and photospheric stellar parameters based on Gaia
Radial Velocity Spectrometer (RVS) spectra \citep{Recio-Blanco2023}.
In parallel to the approach we adopted in Section \ref{sec:planets},
they then used those homogeneously derived stellar parameters to infer
updated planetary masses and radii based on observed transit depths,
orbital periods, Doppler semiamplitudes, and orbital eccentricities.
They next calculated densities using their inferred planetary masses and
radii as well as iron-mass fractions by interpolating the \citet{Zeng2019}
grid of theoretical mass--radius relations.  Using those homogeneously
inferred host star ages, exoplanet densities, and iron-mass fractions,
\citet{Weeks2025} ultimately found inverse correlations between host star
age and both the logarithm of exoplanet density and iron-mass fraction
significant at about the 3-$\sigma$ level.

\begin{figure*}
\centering
\plotone{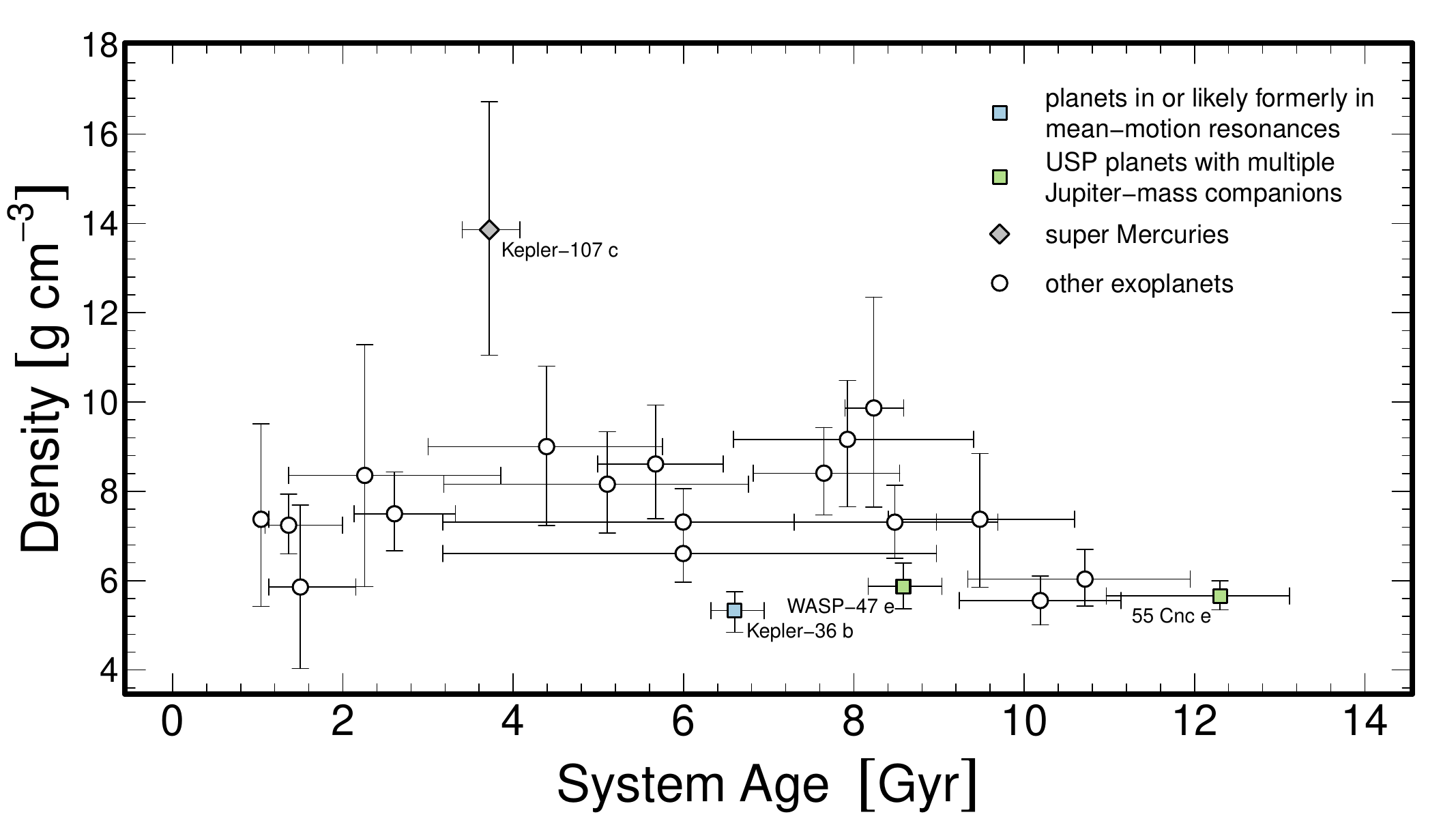}
\plotone{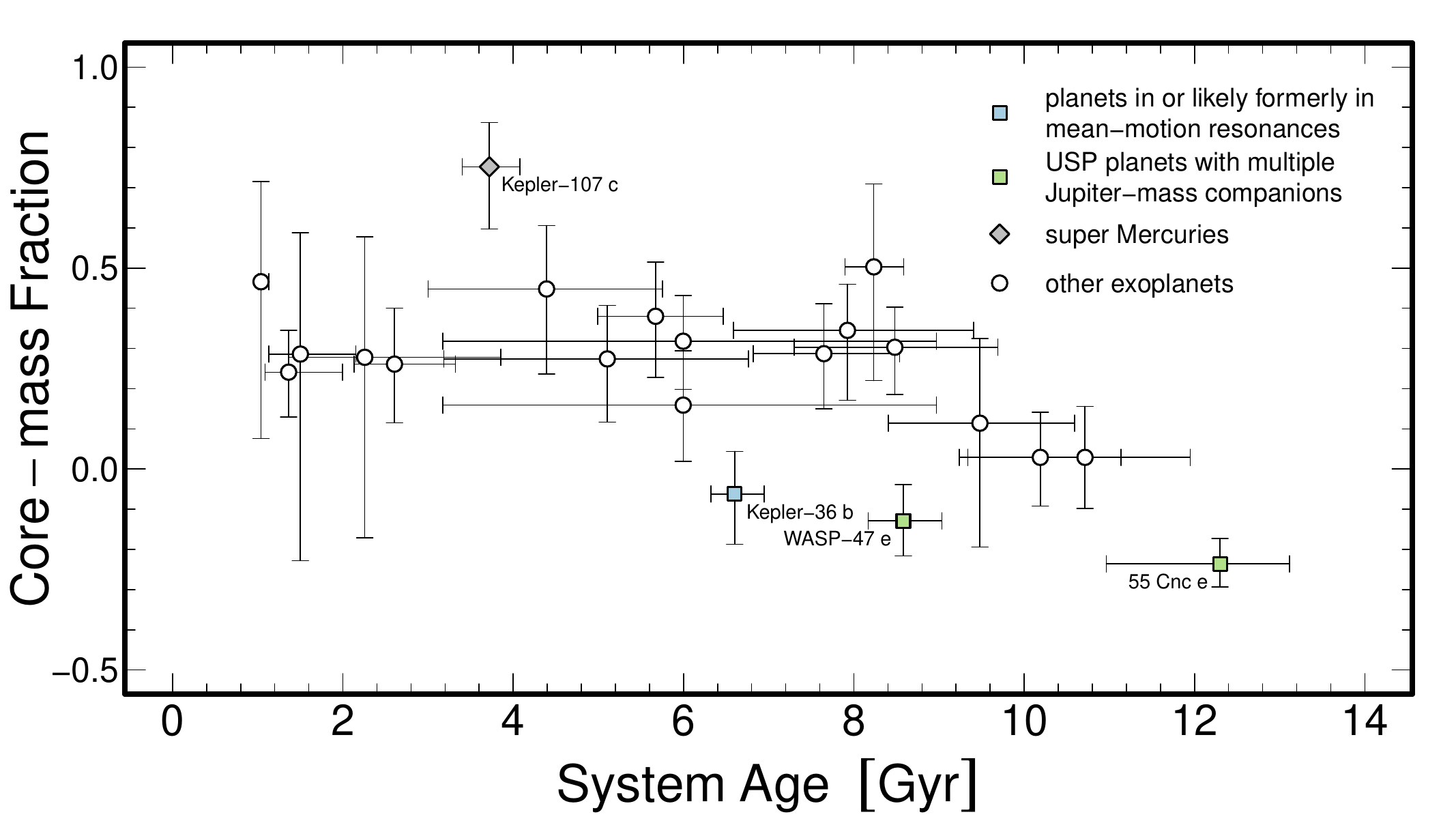}
\caption{Top: relationship between host star age and exoplanet density.
Unlike \citet{Weeks2025}, we find no statistically significant inverse
correlation between host star age and either exoplanet density or its
logarithm.  Bottom: relationship between host star age and exoplanet
core-mass fraction.  Like \citet{Weeks2025}, we find a statistically
significant inverse correlation between host star age and planetary
core-mass fraction.  Unlike \citet{Weeks2025}, we attribute this inverse
correlation to an observational bias: exoplanets in the super-Earth
radius range with low core-mass fractions will only yield Doppler-based
mass inferences if they orbit quiescent and therefore very old stars.}
\label{fig:age_cmf_density}
\end{figure*}

To follow up the \citet{Weeks2025} conclusion, we searched for possible
inverse correlations between our own system ages and planetary densities
in Table \ref{tab:stellar_params} and core-mass fractions in Table
\ref{tab:cmfs}.  We plot those data in Figure \ref{fig:age_cmf_density}.
We find no statistically significant correlations between system age and
either density or the logarithm of density.  We do identify a relationship
between system age and core-mass fraction that is significant according
to Kendall's $\tau$ at the $p = 0.03$ level (about the 2-$\sigma$
level assuming Gaussian statistics).  That relationship is influenced
by the low densities and ancient ages of 55 Cnc e and WASP-47 e though,
and as we argued in Section \ref{sec:disc_usp} those two exoplanets are
unlike the rest of our sample in that they were likely once the cores
of now-disrupted mini Neptunes.  If we put those two exoplanets aside,
the significance of the correlation between system age and core-mass
according to Kendall's $\tau$ drops to the $p = 0.18$ level (slightly more
than the 1-$\sigma$ level assuming Gaussian statistics).  The statistical
significance of the relationship can be attributed solely to the existence
of the two low core-mass fraction exoplanets HD 136352 b and Kepler-20 b
orbiting host stars with posterior median ages in excess of 10 Gyr.  It is
not present in the age range 1 Gyr $\lesssim \tau_{\ast} \lesssim 10$ Gyr.

\citet{Weeks2025} considered and rejected several possible observational
biases that could lead to a statistically significant inverse relationship
between host star age and the logarithm of exoplanet density or iron-mass
fraction.  In particular, they rejected the possibility that relatively
high radial velocity jitters known to be associated with relatively
young stars would render Doppler-based mass inferences impossible for
low-density planets that at constant radius produce smaller Doppler
semiamplitudes than denser planets.

\begin{figure*}
\centering
\plotone{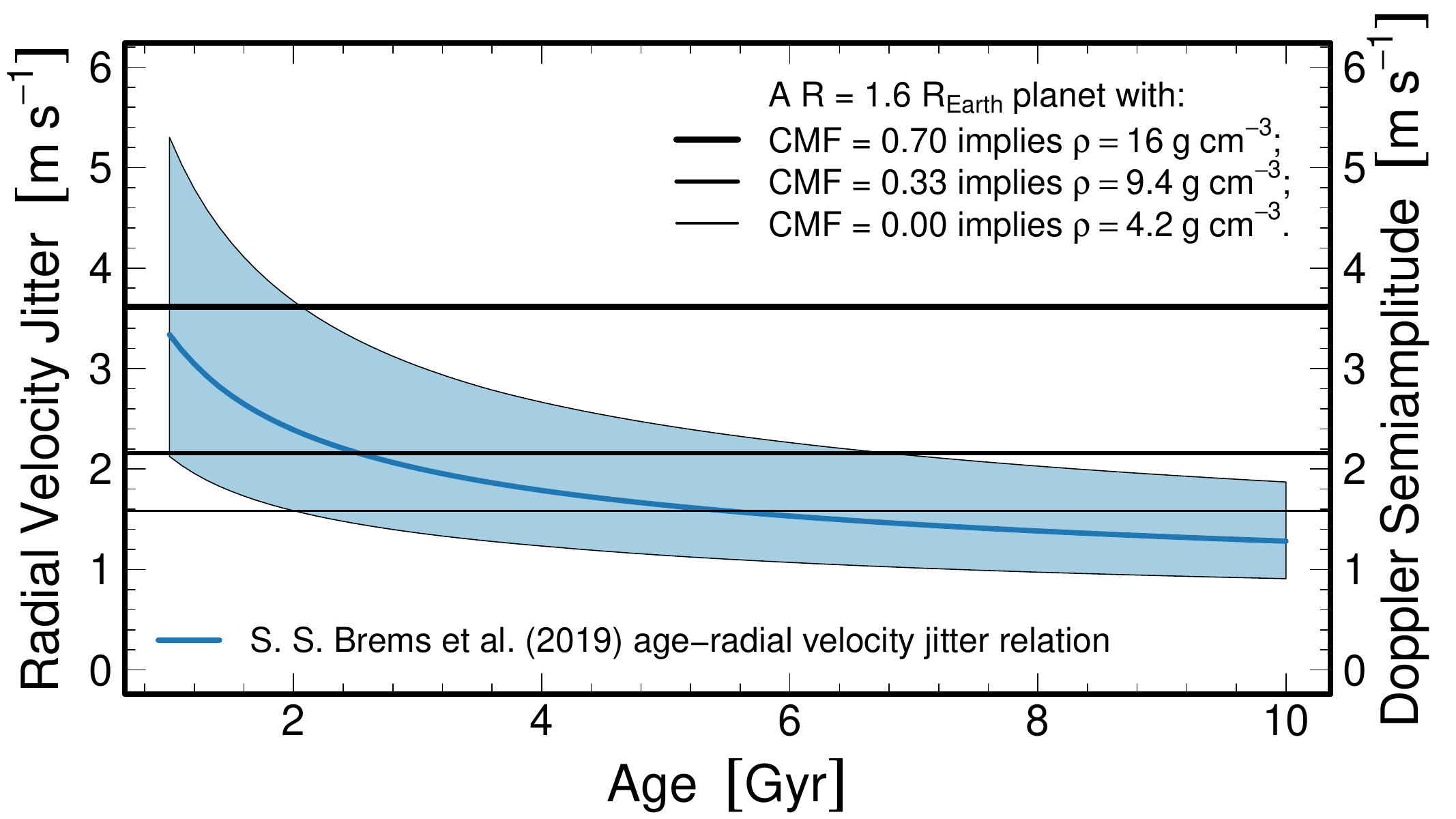}
\caption{Expected Doppler semiamplitude imparted on a solar-mass host
star by a super-Earth with orbital period $P = 2$ days and $R_{\text{p}}
= 1.6~R_{\oplus}$ as a function of core-mass fraction superimposed
on the \citet{Brems2019} relationship between stellar age and radial
velocity jitter.  The solid blue line is the maximum-likelihood
\citet{Brems2019} relationship while the light blue polygon is the
16th to 84th interquantile range; the horizontal black lines indicate
the Doppler semiamplitudes imparted by the hypothesized super-Earths.
We find that while high core-mass fraction super-Earths can be detected
in the presence of radial velocity jitter at all ages, low core-mass
fraction super-Earths can only be detected orbiting host stars with
ages in excess of 5 Gyr.  This observational bias against the detection
of low core-mass fraction terrestrial exoplanets orbiting young stars
can easily explain the statistically significant inverse correlation
identified by \citet{Weeks2025} between host star age and either the
logarithm of planet density or iron-mass fraction.}
\label{fig:age_jitter}
\end{figure*}

To reconsider this possibility that observational biases can explain
an inverse relationship between host star age and exoplanet core-mass
fraction, we use the model proposed in \citet{Brems2019} connecting
radial velocity jitter and stellar age in the interval 1 Myr $\lesssim
\tau_{\ast} \lesssim$ 10 Gyr to evaluate the effect of host star age on
the detectability of both high and low core-mass fraction super-Earths.
As function of core-mass fraction, we calculate the Doppler semiamplitude
imparted on a $M_{\ast} = 1~M_{\odot}$ host star by the orbit of a
super-Earth with $i = 90^{\circ}$, the median orbital period of the
systems in our sample $P = 2$ days, and the median planetary radius
of our sample $R_{\text{p}} = 1.6~R_{\oplus}$.  We plot the results
of these calculations in Figure \ref{fig:age_jitter}.  We find that
super-Mercuries with CMF = 0.70 are detectable at all host star ages
$\tau_{\ast} \gtrsim 1$ Gyr and that exoplanets resembling Earth with
CMF = 0.33 are detectable at all host star ages $\tau_{\ast} \gtrsim
2$ Gyr.  In contrast, we find that low-density exoplanets with CMF =
0.00 are only detectable orbiting host stars with $\tau_{\ast} \gtrsim
5$ Gyr.  This observational bias against the detection of low core-mass
fraction exoplanets orbiting stars with $\tau_{\ast} \lesssim 5$ Gyr
is an explanation for both the \citet{Weeks2025} inverse correlation
between host star age and iron-mass fraction and our low-significance
inverse correlation between system age and core-mass fraction.

\section{Conclusion}\label{sec:conclusion}

We infer accurate, precise, homogeneous, and physically self
consistent stellar masses and radii for 24 dwarf host stars of
terrestrial exoplanets.  We use these stellar masses and radii plus
Doppler/TTV and transit observables to infer updated terrestrial
exoplanet masses, radii, and core-mass fractions.  We find that both
the correlation and linear relationship between stellar metallicity
$Z_{\text{Fe},\ast}/Z_{\text{Fe},\odot}$ and mass--radius-based
core-mass fraction are positive, but the statistical significance of that
relationship depends on stellar parameter source and does not consistently
excludes a Kendall's $\tau$ coefficient or slope of zero from the 2nd
to 98th interquantile ranges of those statistics given our sample.
The connection between Fe/Mg and Fe/Si abundance ratio-based core-mass
fraction and host star metallicity is less clear with the strength and
even the direction of the relation dependent on the source of photospheric
stellar parameters.  The relationships between stellar metallicity and
the abundance ratios Fe/Mg and Fe/Si and terrestrial exoplanet core-mass
fractions therefore merits further study.

We conclude that the terrestrial exoplanets Kepler-36 b, Kepler-105
c, 55 Cnc e and WASP-47 e all are low core-mass fraction outliers.
These outliers aside, the mean mass--radius-based core-mass fraction of
our sample $\overline{\text{CMF}} = 0.30 \pm 0.04$ is consistent with the
Earth's core-mass fraction $\text{CMF}_{\oplus} \approx 0.33$.  We find
that the population of terrestrial exoplanets in or likely formerly in
mean-motion resonances represented by Kepler-36 b and Kepler-105 c has a
mean mass--radius-based core-mass fraction 3$\sigma$ lower than the rest
of terrestrial exoplanets in our sample.  The resonant configurations
of Kepler-36 b and Kepler-105 c indicate that they migrated inward from
initially more distant formation locations to their currently observed
locations, and we attribute their low densities to the incorporation and
retention of significant amounts of water in their interiors.  We propose
that they are the first examples of a class of terrestrial exoplanets
called icy cores predicted by \citet{Schlecker2021} to have formed close
to or beyond the water-ice lines of their parent protoplanetary disks.

We confirm that the ultra-short-period exoplanets 55 Cnc e and WASP-47
e have densities inconsistent with pure rock compositions.  None of
the other USP exoplanets in our sample of terrestrial exoplanets have
similarly low densities.  Their host stars 55 Cnc and WASP-47 are the
two most metal-rich host stars in our sample, and both systems include
giant planet companions close to their USP planets.  Like the rest of
the USP exoplanet population, we suggest that 55 Cnc e and WASP-47 e
tidally migrated from the inner edges of their parent protoplanetary
disks to their observed locations at $P < 1$ day over billions of years
as a consequence of eccentricity excitation and inside-planet tidal
dissipation \citep[e.g.,][]{Schlaufman2010, Hamer2020, Schmidt2024}.
Unlike the rest of the USP exoplanets in our sample that lost their
primordial atmospheres long ago and have been super-Earths for the
vast majorities of their existences, the more massive 55 Cnc e and
WASP-47 e were until relatively recently mini-Neptunes.  As a result,
when the tidal migration of 55 Cnc e and WASP-47 e accelerated as they
approached their currently observed locations a combination of intense
instellation, tidal heating, and Hill sphere shrinkage caused the escape
of their primordial atmospheres.  For that reason, unlike the rest of
the USP exoplanet population, we propose that 55 Cnc e and WASP-47 e
are the remnant cores of now disrupted mini-Neptunes that had retained
their primordial atmospheres for billions of years.  We argue that their
low densities can be explained by the presence of significant amounts of
hydrogen, helium, water, and/or other volatiles that were incorporated
into their interiors via chemical reactions, convection, or diffusion at
the permeable interfaces between the primordial atmospheres and molten
cores of mini-Neptunes over billions of years.  This idea reconciles
the low densities of 55 Cnc e and WASP-47 e with the nondetections of
atmospheres on those exoplanets.

We identify only one terrestrial exoplanet with a Mercury-like
core-mass fraction Kepler-107 c, thereby calling into question the
claim by \citet{Adibekyan2021} that super-Mercuries are a distinct
component of the terrestrial exoplanet population.  We infer an
occurrence of terrestrial exoplanets with Mercury-like core-mass
fractions $\Theta_{\text{merc,exo}} = 0.04_{-0.03}^{+0.05}$ that is
consistent with the occurrence of such planets in the solar system
$\Theta_{\text{merc,ss}} = 0.25_{-0.16}^{+0.24}$ but inconsistent
with the value $\Theta_{\text{merc,exo}} \approx 0.2$ favored by
\citet{Adibekyan2021} and \citet{Unterborn2023} at the $p = 0.01$ level
(about 2.3$\sigma$ assuming Gaussian statistics).

Like \citet{Weeks2025}, we find a correlation between system age
and terrestrial exoplanet core-mass fraction significant at the $p =
0.03$ level (about $2\sigma$ assuming Gaussian statistics).  Unlike
\citet{Weeks2025}, we attribute this correlation to a systematic bias
against precise Doppler-based mass inferences for low core-mass fraction
Earth-radius exoplanets orbiting solar-type stars.  This bias is a
consequence of the inverse relationship between radial velocity jitter
and stellar age that enables at all ages mass inferences for dense,
high core-mass fraction terrestrial exoplanets that impart large Doppler
semiamplitudes on their host stars.  On the other hand, low core-mass
fraction terrestrial exoplanets impart small Doppler semiamplitudes
on their host stars that can only be robustly detected in older stars
with low radial velocity jitter.  Indeed, the correlation between system
age and terrestrial exoplanet core-mass fraction disappears if we only
consider systems with ages between 1 and 10 Gyr or remove 55 Cnc e and
WASP-47 e from our sample as a consequence of their formation through a
process qualitatively different from the processes responsible for the
formation of the rest of the terrestrial exoplanets in our sample.

\begin{acknowledgments}

We thank Guangwei Fu, Bo Peng, David Sing, and Daniel Thorngren for
helpful conversations.  This material is based upon work supported
by the National Aeronautics and Space Administration under Grant
No. 80NSSC23K0266 issued through the Exoplanets Research Program
(XRP).  Based on data obtained from the ESO Science Archive
Facility with DOI(s): \url{https://doi.org/10.18727/archive/21},
\url{https://doi.org/10.18727/archive/24}, and
\url{https://doi.org/10.18727/archive/50}.  This research has made
use of the Keck Observatory Archive (KOA), which is operated by the
W. M. Keck Observatory and the NASA Exoplanet Science Institute (NExScI),
under contract with the National Aeronautics and Space Administration.
This research used the facilities of the Italian Center for Astronomical
Archive (IA2) operated by INAF at the Astronomical Observatory of Trieste.
This research used the facilities of the Canadian Astronomy Data Centre
operated by the National Research Council of Canada with the support of
the Canadian Space Agency.  Based on observations obtained with the Apache
Point Observatory 3.5-meter telescope, which is owned and operated by
the Astrophysical Research Consortium.  Funding for the SDSS and SDSS-II
has been provided by the Alfred P. Sloan Foundation, the Participating
Institutions, the National Science Foundation, the U.S. Department of
Energy, the National Aeronautics and Space Administration, the Japanese
Monbukagakusho, the Max Planck Society, and the Higher Education Funding
Council for England.  The SDSS Web Site is \url{http://www.sdss.org/}.
The SDSS is managed by the Astrophysical Research Consortium for the
Participating Institutions.  The Participating Institutions are the
American Museum of Natural History, Astrophysical Institute Potsdam,
University of Basel, University of Cambridge, Case Western Reserve
University, University of Chicago, Drexel University, Fermilab, the
Institute for Advanced Study, the Japan Participation Group, Johns
Hopkins University, the Joint Institute for Nuclear Astrophysics, the
Kavli Institute for Particle Astrophysics and Cosmology, the Korean
Scientist Group, the Chinese Academy of Sciences (LAMOST), Los Alamos
National Laboratory, the Max-Planck-Institute for Astronomy (MPIA), the
Max-Planck-Institute for Astrophysics (MPA), New Mexico State University,
Ohio State University, University of Pittsburgh, University of Portsmouth,
Princeton University, the United States Naval Observatory, and the
University of Washington.  Funding for SDSS-III has been provided by the
Alfred P. Sloan Foundation, the Participating Institutions, the National
Science Foundation, and the U.S. Department of Energy Office of Science.
The SDSS-III web site is \url{http://www.sdss3.org/}.  SDSS-III is
managed by the Astrophysical Research Consortium for the Participating
Institutions of the SDSS-III Collaboration including the University
of Arizona, the Brazilian Participation Group, Brookhaven National
Laboratory, Carnegie Mellon University, University of Florida, the French
Participation Group, the German Participation Group, Harvard University,
the Instituto de Astrofisica de Canarias, the Michigan State/Notre
Dame/JINA Participation Group, Johns Hopkins University, Lawrence
Berkeley National Laboratory, Max Planck Institute for Astrophysics,
Max Planck Institute for Extraterrestrial Physics, New Mexico State
University, New York University, Ohio State University, Pennsylvania
State University, University of Portsmouth, Princeton University, the
Spanish Participation Group, University of Tokyo, University of Utah,
Vanderbilt University, University of Virginia, University of Washington,
and Yale University.  Funding for the Sloan Digital Sky Survey IV has
been provided by the Alfred P. Sloan Foundation, the U.S. Department
of Energy Office of Science, and the Participating Institutions.
SDSS-IV acknowledges support and resources from the Center for High
Performance Computing  at the University of Utah.  The SDSS website is
\url{www.sdss4.org}.  SDSS-IV is managed by the Astrophysical Research
Consortium for the Participating Institutions of the SDSS Collaboration
including the Brazilian Participation Group, the Carnegie Institution
for Science, Carnegie Mellon University, Center for Astrophysics |
Harvard \& Smithsonian, the Chilean Participation Group, the French
Participation Group, Instituto de Astrof\'isica de Canarias, The Johns
Hopkins University, Kavli Institute for the Physics and Mathematics of
the Universe (IPMU) / University of Tokyo, the Korean Participation Group,
Lawrence Berkeley National Laboratory, Leibniz Institut f\"ur Astrophysik
Potsdam (AIP), Max-Planck-Institut f\"ur Astronomie (MPIA Heidelberg),
Max-Planck-Institut f\"ur Astrophysik (MPA Garching), Max-Planck-Institut
f\"ur Extraterrestrische Physik (MPE), National Astronomical Observatories
of China, New Mexico State University, New York University, University
of Notre Dame, Observat\'ario Nacional / MCTI, The Ohio State University,
Pennsylvania State University, Shanghai Astronomical Observatory, United
Kingdom Participation Group, Universidad Nacional Aut\'onoma de M\'exico,
University of Arizona, University of Colorado Boulder, University of
Oxford, University of Portsmouth, University of Utah, University of
Virginia, University of Washington, University of Wisconsin, Vanderbilt
University, and Yale University.  The national facility capability for
SkyMapper has been funded through ARC LIEF grant LE130100104 from the
Australian Research Council, awarded to the University of Sydney, the
Australian National University, Swinburne University of Technology,
the University of Queensland, the University of Western Australia,
the University of Melbourne, Curtin University of Technology, Monash
University and the Australian Astronomical Observatory.  SkyMapper is
owned and operated by The Australian National University's Research
School of Astronomy and Astrophysics.  The survey data were processed
and provided by the SkyMapper Team at ANU.  The SkyMapper node of the
All-Sky Virtual Observatory (ASVO) is hosted at the National Computational
Infrastructure (NCI).  Development and support of the SkyMapper node of
the ASVO has been funded in part by Astronomy Australia Limited (AAL) and
the Australian Government through the Commonwealth's Education Investment
Fund (EIF) and National Collaborative Research Infrastructure Strategy
(NCRIS), particularly the National eResearch Collaboration Tools and
Resources (NeCTAR) and the Australian National Data Service Projects
(ANDS).  This work has made use of data from the European Space
Agency (ESA) mission Gaia (\url{https://www.cosmos.esa.int/gaia}),
processed by the Gaia Data Processing and Analysis Consortium (DPAC,
\url{https://www.cosmos.esa.int/web/gaia/dpac/consortium}).  Funding for
the DPAC has been provided by national institutions, in particular
the institutions participating in the Gaia Multilateral Agreement.
This publication makes use of data products from the Two Micron All
Sky Survey, which is a joint project of the University of Massachusetts
and the Infrared Processing and Analysis Center/California Institute of
Technology, funded by the National Aeronautics and Space Administration
and the National Science Foundation.  This publication makes use of
data products from the Wide-field Infrared Survey Explorer, which is a
joint project of the University of California, Los Angeles, and the Jet
Propulsion Laboratory/California Institute of Technology, funded by the
National Aeronautics and Space Administration.  This research has made
use of the NASA Exoplanet Archive, which is operated by the California
Institute of Technology, under contract with the National Aeronautics
and Space Administration under the Exoplanet Exploration Program.  This
research has made use of the SIMBAD database, operated at CDS, Strasbourg,
France \citep{Wenger2000}.  This research has made use of the VizieR
catalog access tool, CDS, Strasbourg, France.  The original description of
the VizieR service was published in \citet{Ochsenbein2000}.  This research
has made use of NASA's Astrophysics Data System Bibliographic Services.

\end{acknowledgments}

\facilities{ARC (ARCES), CDS, CFHT (ESPaDOnS), CTIO:2MASS, Du Pont
(APOGEE), FLWO:2MASS, Exoplanet Archive, FLWO:2MASS, Gaia, GALEX, IRSA,
Keck:I (HIRES), Max Planck:2.2m (FEROS), NEOWISE, Skymapper, Sloan
(APOGEE), TNG (HARPS-N), VLT (ESPRESSO), VLT:Kueyen (UVES), WISE}

\software{\texttt{astropy} \citep{astropy2013, astropy2018, astropy2022},
          \texttt{isochrones} \citep{isochrones},
          \texttt{iSpec} \citep{iSpec1, iSpec2}, 
          \texttt{Matplotlib} \citep{matplotlib},
          \texttt{MOOG} \citep{Sneden1973},
          \texttt{numpy} \citep{Harris2020},
          \texttt{pandas} \citep{pandas1, pandas2},
          \texttt{q2} \citep{q2},
          \texttt{R} \citep{r25},
          \texttt{scipy} \citep{scipy},
          \texttt{SuperEarth} \citep{Plotnykov2020}
          }

\begin{figure*}
\centering
\plottwo{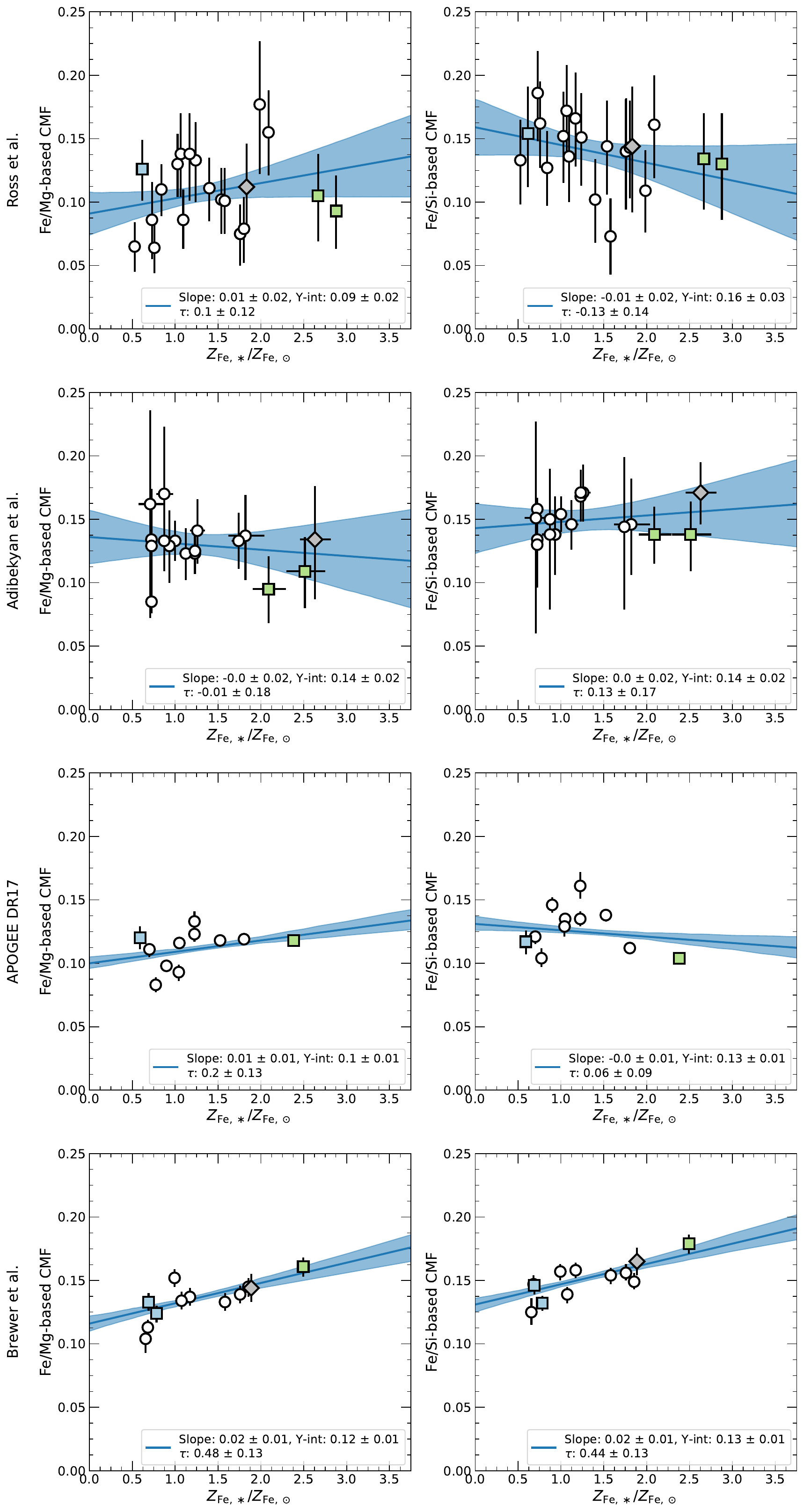}{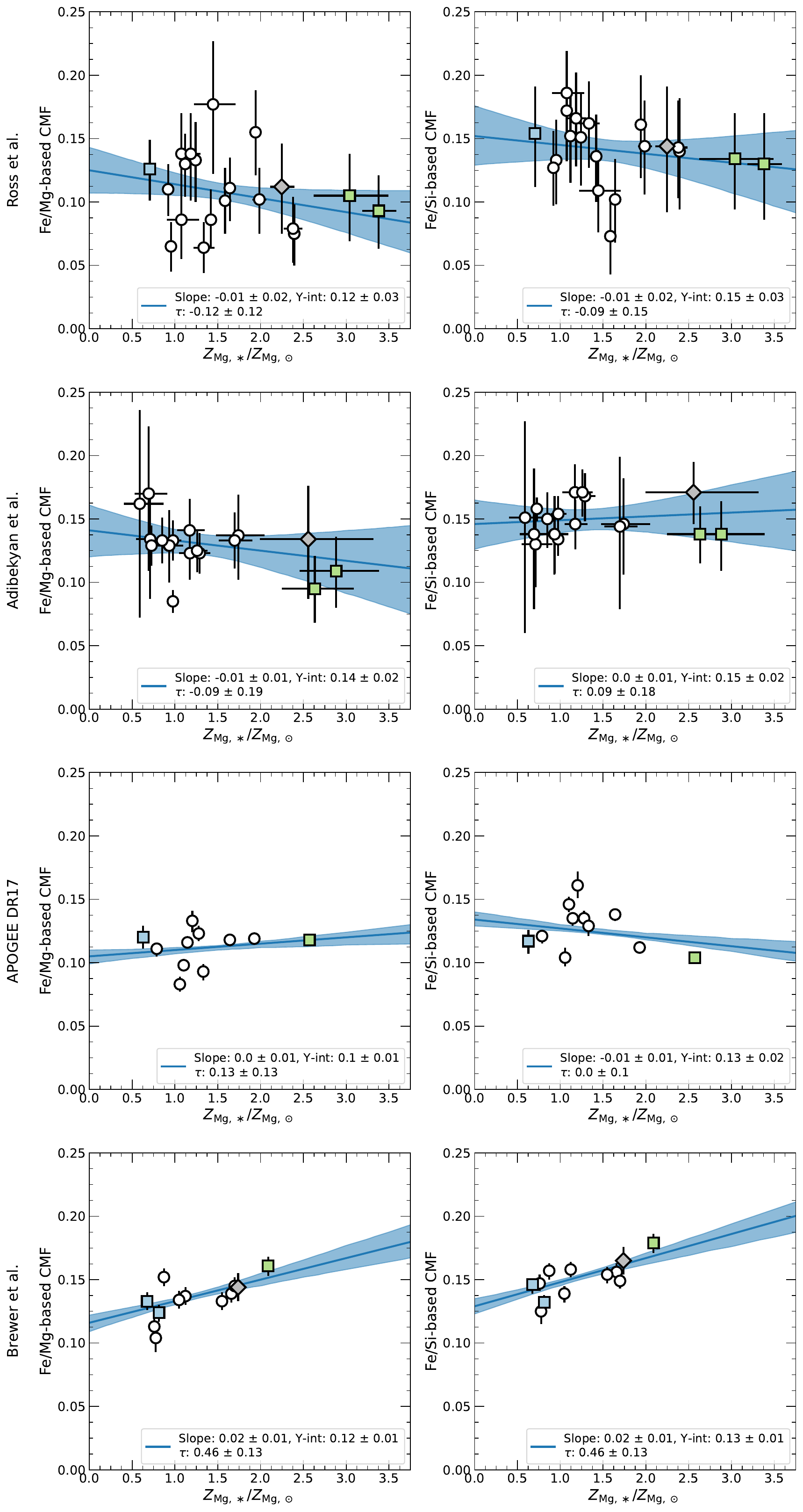}
\caption{Fe/Mg- and Fe/Si-based CMF as a function of iron metallicity
$Z_{\text{Fe},\ast}/Z_{\text{Fe},\odot}$ and for four different
photospheric stellar parameter sources.  We plot the low-density
exoplanets Kepler-36 b and Kepler-105 c as black-boarded blue squares
and 55 Cnc e and WASP-47 e as black-boarded green squares.  We plot
the super-Mercury exoplanet Kepler-107 c as a gray diamond.  We plot
all other exoplanets as black-boarded open circles.  To evaluate the
impact of stellar iron metallicity and therefore iron mass fraction on
terrestrial exoplanet CMF, we use the Monte Carlo procedure described
in the text to calculate Kendall's $\tau$ coefficients and fit linear
models to these data after excluding exoplanets inconsistent with
pure-rock compositions (i.e., negative CMFs).  We plot the linear
models that result from our Monte Carlo simulations as blue or green
lines and indicate their 16th to 84th quantile uncertainty regions with
blue or green polygons.  We also report in the legend of each panel
the numerical values of Kendall's $\tau$ coefficients, slopes, and
intercepts as well as their uncertainties.  Left: for the relationship
between $Z_{\text{Fe},\ast}/Z_{\text{Fe},\odot}$ and Fe/Mg-based
CMF, we find both positive and negative correlations and slopes for
which zero is excluded from the 0.1 to 99.9 interquantile ranges
of the Monte Carlo inferred parameter distributions (corresponding
to the 3-$\sigma$ range of a Gaussian) only for abundances based on
\citet{Brewer2016}/\citet{Brewer2018}.  Middle left: for the relationship
between $Z_{\text{Fe},\ast}/Z_{\text{Fe},\odot}$ and Fe/Si-based CMF,
we find both positive and negative correlations and slopes for which
zero is only excluded from the 0.1 to 99.9 interquantile ranges of
the Monte Carlo inferred parameter distributions only for abundances
based on \citet{Brewer2016}/\citet{Brewer2018}.  Middle right: for
the relationship between $Z_{\text{Mg},\ast}/Z_{\text{Mg},\odot}$ and
Fe/Mg-based CMF, we find both positive and negative correlations and
slopes for which zero is excluded from the 0.1 to 99.9 interquantile
ranges of the Monte Carlo inferred parameter distributions (corresponding
to the 3-$\sigma$ range of a Gaussian) only abundances based on
\citet{Brewer2016}/\citet{Brewer2018}.  Right: for the relationship
between $Z_{\text{Mg},\ast}/Z_{\text{Mg},\odot}$ and Fe/Si-based
CMF, we find both positive and negative correlations and slopes for
which zero is excluded from the 0.1 to 99.9 interquantile range of
the Monte Carlo inferred parameter distributions only abundances
based on \citet{Brewer2016}/\citet{Brewer2018}.  We conclude that
the relationships between $Z_{\text{Fe},\ast}/Z_{\text{Fe},\odot}$,
$Z_{\text{Mg},\ast}/Z_{\text{Mg},\odot}$, and Fe/(Mg,Si)-based CMF are
not straightforward and merit further study.}
\label{fig:xfe_vs_femgsicmf}
\end{figure*}

\appendix
The relationships between Fe/Mg- and Fe/Si-based core-mass fractions of
terrestrial exoplanets and the iron or magnesium metallicities of their
host stars are more complicated than the corresponding relationships
between mass--radius-based core-mass fractions and metallicities (Figure
\ref{fig:xfe_vs_femgsicmf}).  For Fe/(Mg,Si)-based core-mass fractions,
our Monte Carlo simulations provide Kendall's $\tau$ coefficient
and slope distributions that include zero within the 2nd to 98th
interquantile ranges of those distributions for parameters based on
our own work in this article or \citet{Adibekyan2021}.  In other words,
we find no correlations or linear relationships significant at levels
corresponding to the 2-$\sigma$ range of a Gaussian for parameters
based on our own work in this article or \citet{Adibekyan2021}.
The same is true for the Kendall's $\tau$ coefficient distribution for
parameters based on APOGEE DR17.  On the other hand, for parameters
based on data from \citet{Brewer2016}/\citet{Brewer2018}, we do find
correlations and linear relationships between both stellar iron and
magnesium metallicities and Fe/(Mg,Si)-based core-mass fraction with
Kendall's $\tau$ coefficient and slope distributions that exclude
zero from their 0.1 to 99.9 interquantile ranges.  In other words,
we find a correlation and linear relationship significant at a level
corresponding to about the 3-$\sigma$ level range of a Gaussian.
Likewise, we find that for iron metallicity and Fe/Mg-based core-mass
fractions, our Monte Carlo simulations provide slope distributions
that exclude zero from the 2nd to 98th interquantile ranges of those
distributions for parameters based based on APOGEE DR17.  We conclude
that the relationships between $Z_{\text{Fe},\ast}/Z_{\text{Fe},\odot}$,
$Z_{\text{Mg},\ast}/Z_{\text{Mg},\odot}$, and Fe/(Mg,Si)-based CMF are
not straightforward and merit further study.

\begin{longrotatetable}
\begin{deluxetable*}{ccccccccccc}
\tablecaption{Photometry\label{tab:photometry}}
\tablewidth{0pt}
\tabletypesize{\tiny}
\tablehead{
\colhead{} & \colhead{} & \colhead{} & \colhead{} & \colhead{} & \colhead{} & \colhead{} & \colhead{} & \colhead{} & \colhead{} & \colhead{}
}
\startdata
Property	& K2-216	& K2-106	& HD 15337	& K2-291	& 55 Cnc	& HD 80653	& K2-131	& K2-229	& HD 136352	& Units\\
\hline
Gaia DR3 Parallax	& $8.688\pm0.023$	& $4.085\pm0.018$	& $22.292\pm0.016$	& $11.171\pm0.016$	& $79.448\pm0.043$	& $9.26\pm0.022$	& $6.479\pm0.021$	& $9.762\pm0.021$	& $67.847\pm0.06$	& mas\\
GALEX $FUV$	& $\cdots$	& $\cdots$	& $\cdots$	& $\cdots$	& $\cdots$	& $\cdots$	& $\cdots$	& $\cdots$	& $\cdots$	& AB mag\\
GALEX $NUV$	& $20.799\pm0.14$	& $18.388\pm0.033$	& $16.099\pm0.013$	& $\cdots$	& $\cdots$	& $15.045\pm0.008$	& $18.952\pm0.08$	& $17.699\pm0.03$	& $\cdots$	& AB mag\\
SkyMapper $u$	& $\cdots$	& $\cdots$	& $\cdots$	& $\cdots$	& $\cdots$	& $\cdots$	& $14.395\pm0.012$	& $\cdots$	& $\cdots$	& AB mag\\
SkyMapper $v$	& $\cdots$	& $\cdots$	& $\cdots$	& $\cdots$	& $\cdots$	& $\cdots$	& $\cdots$	& $\cdots$	& $\cdots$	& AB mag\\
SkyMapper $g$	& $\cdots$	& $\cdots$	& $\cdots$	& $\cdots$	& $\cdots$	& $\cdots$	& $\cdots$	& $\cdots$	& $\cdots$	& AB mag\\
SkyMapper $r$	& $\cdots$	& $\cdots$	& $\cdots$	& $\cdots$	& $\cdots$	& $\cdots$	& $\cdots$	& $\cdots$	& $\cdots$	& AB mag\\
SkyMapper $i$	& $\cdots$	& $\cdots$	& $\cdots$	& $\cdots$	& $\cdots$	& $\cdots$	& $\cdots$	& $\cdots$	& $\cdots$	& AB mag\\
SkyMapper $z$	& $\cdots$	& $\cdots$	& $\cdots$	& $\cdots$	& $\cdots$	& $\cdots$	& $\cdots$	& $\cdots$	& $\cdots$	& AB mag\\
SDSS $u$	& $\cdots$	& $14.038\pm0.028$	& $\cdots$	& $\cdots$	& $\cdots$	& $\cdots$	& $\cdots$	& $\cdots$	& $\cdots$	& AB mag\\
SDSS $g$	& $\cdots$	& $\cdots$	& $\cdots$	& $\cdots$	& $\cdots$	& $\cdots$	& $\cdots$	& $\cdots$	& $\cdots$	& AB mag\\
SDSS $r$	& $\cdots$	& $\cdots$	& $\cdots$	& $\cdots$	& $\cdots$	& $\cdots$	& $\cdots$	& $\cdots$	& $\cdots$	& AB mag\\
SDSS $i$	& $\cdots$	& $\cdots$	& $\cdots$	& $\cdots$	& $\cdots$	& $\cdots$	& $\cdots$	& $\cdots$	& $\cdots$	& AB mag\\
SDSS $z$	& $\cdots$	& $11.648\pm0.017$	& $\cdots$	& $\cdots$	& $\cdots$	& $\cdots$	& $\cdots$	& $\cdots$	& $\cdots$	& AB mag\\
Tycho-2 $B_T$	& $\cdots$	& $\cdots$	& $10.17\pm0.027$	& $\cdots$	& $7.037\pm0.015$	& $10.269\pm0.031$	& $\cdots$	& $\cdots$	& $\cdots$	& Vega mag\\
Tycho-2 $V_T$	& $\cdots$	& $\cdots$	& $9.184\pm0.018$	& $\cdots$	& $6.036\pm0.009$	& $9.528\pm0.023$	& $\cdots$	& $\cdots$	& $\cdots$	& Vega mag\\
Gaia DR2 $G$	& $12.058\pm0.002$	& $11.954\pm0.002$	& $8.856\pm0.002$	& $9.866\pm0.002$	& $5.714\pm0.002$	& $9.307\pm0.002$	& $11.9\pm0.002$	& $10.786\pm0.002$	& $5.462\pm0.002$	& Vega mag\\
2MASS $J$	& $10.394\pm0.019$	& $10.77\pm0.02$	& $7.553\pm0.009$	& $8.765\pm0.027$	& $4.768\pm0.244$	& $8.315\pm0.015$	& $10.57\pm0.023$	& $9.518\pm0.018$	& $4.308\pm0.226$	& Vega mag\\
2MASS $H$	& $9.856\pm0.031$	& $10.454\pm0.024$	& $7.215\pm0.031$	& $8.407\pm0.009$	& $4.265\pm0.234$	& $8.079\pm0.025$	& $10.122\pm0.02$	& $9.126\pm0.02$	& $3.898\pm0.192$	& Vega mag\\
2MASS $K_{\text{s}}$	& $9.721\pm0.016$	& $10.344\pm0.019$	& $7.044\pm0.011$	& $8.353\pm0.018$	& $4.015\pm0.033$	& $8.018\pm0.015$	& $10.051\pm0.024$	& $9.05\pm0.021$	& $4.159\pm0.033$	& Vega mag\\
WISE $W1$	& $9.65\pm0.021$	& $10.299\pm0.023$	& $6.976\pm0.033$	& $8.291\pm0.024$	& $4.077\pm0.097$	& $7.959\pm0.023$	& $9.95\pm0.024$	& $8.969\pm0.024$	& $\cdots$	& Vega mag\\
WISE $W2$	& $9.744\pm0.019$	& $10.355\pm0.021$	& $7.038\pm0.02$	& $8.346\pm0.021$	& $3.481\pm0.059$	& $8.012\pm0.021$	& $9.997\pm0.021$	& $9.036\pm0.021$	& $\cdots$	& Vega mag\\
WISE $W3$	& $\cdots$	& $\cdots$	& $\cdots$	& $\cdots$	& $\cdots$	& $\cdots$	& $\cdots$	& $\cdots$	& $\cdots$	& Vega mag\\
WISE $W4$	& $\cdots$	& $\cdots$	& $\cdots$	& $\cdots$	& $\cdots$	& $\cdots$	& $\cdots$	& $\cdots$	& $\cdots$	& Vega mag\\
\hline
\hline
Property	& K2-38	& Kepler-10	& Kepler-20	& Kepler-105	& Kepler-36	& Kepler-93	& Kepler-406	& Kepler-78	& Kepler-107	& Units\\
\hline
Gaia DR3 Parallax	& $5.247\pm0.023$	& $5.37\pm0.01$	& $3.494\pm0.009$	& $2.149\pm0.012$	& $1.853\pm0.009$	& $10.415\pm0.01$	& $2.743\pm0.01$	& $8.008\pm0.01$	& $1.926\pm0.009$	& mas\\
GALEX $FUV$	& $\cdots$	& $\cdots$	& $\cdots$	& $\cdots$	& $\cdots$	& $\cdots$	& $\cdots$	& $\cdots$	& $\cdots$	& AB mag\\
GALEX $NUV$	& $\cdots$	& $16.604\pm0.009$	& $\cdots$	& $17.664\pm0.031$	& $\cdots$	& $\cdots$	& $\cdots$	& $\cdots$	& $18.33\pm0.046$	& AB mag\\
SkyMapper $u$	& $13.433\pm0.004$	& $\cdots$	& $\cdots$	& $\cdots$	& $\cdots$	& $\cdots$	& $\cdots$	& $\cdots$	& $\cdots$	& AB mag\\
SkyMapper $v$	& $13.076\pm0.006$	& $\cdots$	& $\cdots$	& $\cdots$	& $\cdots$	& $\cdots$	& $\cdots$	& $\cdots$	& $\cdots$	& AB mag\\
SkyMapper $g$	& $11.565\pm0.006$	& $\cdots$	& $\cdots$	& $\cdots$	& $\cdots$	& $\cdots$	& $\cdots$	& $\cdots$	& $\cdots$	& AB mag\\
SkyMapper $r$	& $11.175\pm0.005$	& $\cdots$	& $\cdots$	& $\cdots$	& $\cdots$	& $\cdots$	& $\cdots$	& $\cdots$	& $\cdots$	& AB mag\\
SkyMapper $i$	& $\cdots$	& $\cdots$	& $\cdots$	& $\cdots$	& $\cdots$	& $\cdots$	& $\cdots$	& $\cdots$	& $\cdots$	& AB mag\\
SkyMapper $z$	& $10.874\pm0.008$	& $\cdots$	& $\cdots$	& $\cdots$	& $\cdots$	& $\cdots$	& $\cdots$	& $\cdots$	& $\cdots$	& AB mag\\
SDSS $u$	& $\cdots$	& $\cdots$	& $\cdots$	& $\cdots$	& $\cdots$	& $\cdots$	& $\cdots$	& $\cdots$	& $\cdots$	& AB mag\\
SDSS $g$	& $\cdots$	& $11.388\pm0.02$	& $12.997\pm0.02$	& $13.111\pm0.02$	& $12.629\pm0.02$	& $10.38\pm0.02$	& $13.005\pm0.02$	& $12.182\pm0.02$	& $12.904\pm0.02$	& AB mag\\
SDSS $r$	& $\cdots$	& $10.92\pm0.02$	& $12.423\pm0.02$	& $12.732\pm0.02$	& $11.95\pm0.02$	& $9.862\pm0.02$	& $12.461\pm0.02$	& $11.46\pm0.02$	& $12.427\pm0.02$	& AB mag\\
SDSS $i$	& $\cdots$	& $10.778\pm0.02$	& $12.284\pm0.02$	& $12.654\pm0.02$	& $\cdots$	& $9.739\pm0.02$	& $12.312\pm0.02$	& $11.281\pm0.02$	& $12.309\pm0.02$	& AB mag\\
SDSS $z$	& $\cdots$	& $10.729\pm0.02$	& $12.209\pm0.02$	& $12.648\pm0.02$	& $\cdots$	& $9.705\pm0.02$	& $12.26\pm0.02$	& $11.129\pm0.02$	& $12.274\pm0.02$	& AB mag\\
Tycho-2 $B_T$	& $\cdots$	& $\cdots$	& $\cdots$	& $\cdots$	& $\cdots$	& $\cdots$	& $\cdots$	& $\cdots$	& $\cdots$	& Vega mag\\
Tycho-2 $V_T$	& $\cdots$	& $\cdots$	& $\cdots$	& $\cdots$	& $\cdots$	& $\cdots$	& $\cdots$	& $\cdots$	& $\cdots$	& Vega mag\\
Gaia DR2 $G$	& $11.184\pm0.002$	& $10.919\pm0.002$	& $12.453\pm0.002$	& $12.802\pm0.002$	& $12.077\pm0.002$	& $9.862\pm0.002$	& $12.49\pm0.002$	& $11.529\pm0.002$	& $12.463\pm0.002$	& Vega mag\\
2MASS $J$	& $9.911\pm0.021$	& $9.889\pm0.018$	& $11.252\pm0.017$	& $11.811\pm0.017$	& $11.124\pm0.02$	& $8.771\pm0.009$	& $11.34\pm0.017$	& $10.184\pm0.021$	& $11.392\pm0.018$	& Vega mag\\
2MASS $H$	& $9.6\pm0.024$	& $9.563\pm0.024$	& $10.91\pm0.017$	& $11.555\pm0.018$	& $10.85\pm0.028$	& $8.446\pm0.009$	& $11.035\pm0.015$	& $9.675\pm0.015$	& $11.114\pm0.015$	& Vega mag\\
2MASS $K_{\text{s}}$	& $9.47\pm0.019$	& $9.496\pm0.02$	& $10.871\pm0.01$	& $11.503\pm0.01$	& $10.788\pm0.02$	& $8.37\pm0.017$	& $10.97\pm0.015$	& $9.586\pm0.012$	& $11.06\pm0.013$	& Vega mag\\
WISE $W1$	& $9.42\pm0.023$	& $9.44\pm0.023$	& $10.799\pm0.022$	& $\cdots$	& $10.753\pm0.023$	& $8.338\pm0.023$	& $10.934\pm0.023$	& $9.529\pm0.022$	& $11.028\pm0.022$	& Vega mag\\
WISE $W2$	& $9.466\pm0.021$	& $9.491\pm0.02$	& $10.85\pm0.02$	& $\cdots$	& $10.78\pm0.02$	& $8.37\pm0.021$	& $10.984\pm0.021$	& $9.584\pm0.019$	& $11.09\pm0.02$	& Vega mag\\
WISE $W3$	& $\cdots$	& $\cdots$	& $\cdots$	& $\cdots$	& $\cdots$	& $\cdots$	& $\cdots$	& $\cdots$	& $\cdots$	& Vega mag\\
WISE $W4$	& $\cdots$	& $\cdots$	& $\cdots$	& $\cdots$	& $\cdots$	& $\cdots$	& $\cdots$	& $\cdots$	& $\cdots$	& Vega mag\\
\hline
\hline
Property	& Kepler-99	& KOI-1599	& WASP-47	& HD 213885	& K2-265	& HD 219134	& K2-141	& 	& 	& Units\\
\hline
Gaia DR3 Parallax	& $4.801\pm0.014$	& $0.879\pm0.018$	& $3.701\pm0.02$	& $20.932\pm0.022$	& $7.155\pm0.016$	& $152.864\pm0.049$	& $16.13\pm0.018$	& $\cdots$	& $\cdots$	& mas\\
GALEX $FUV$	& $\cdots$	& $\cdots$	& $\cdots$	& $19.897\pm0.166$	& $\cdots$	& $\cdots$	& $\cdots$	& $\cdots$	& $\cdots$	& AB mag\\
GALEX $NUV$	& $\cdots$	& $\cdots$	& $18.635\pm0.037$	& $\cdots$	& $\cdots$	& $\cdots$	& $\cdots$	& $\cdots$	& $\cdots$	& AB mag\\
SkyMapper $u$	& $\cdots$	& $\cdots$	& $14.122\pm0.005$	& $\cdots$	& $\cdots$	& $\cdots$	& $14.079\pm0.012$	& $\cdots$	& $\cdots$	& AB mag\\
SkyMapper $v$	& $\cdots$	& $\cdots$	& $13.723\pm0.004$	& $\cdots$	& $12.806\pm0.006$	& $\cdots$	& $13.74\pm0.004$	& $\cdots$	& $\cdots$	& AB mag\\
SkyMapper $g$	& $\cdots$	& $\cdots$	& $12.134\pm0.008$	& $\cdots$	& $11.274\pm0.005$	& $\cdots$	& $11.356\pm0.007$	& $\cdots$	& $\cdots$	& AB mag\\
SkyMapper $r$	& $\cdots$	& $\cdots$	& $11.754\pm0.008$	& $\cdots$	& $10.906\pm0.005$	& $\cdots$	& $10.708\pm0.005$	& $\cdots$	& $\cdots$	& AB mag\\
SkyMapper $i$	& $\cdots$	& $\cdots$	& $11.571\pm0.007$	& $\cdots$	& $10.701\pm0.005$	& $\cdots$	& $10.301\pm0.007$	& $\cdots$	& $\cdots$	& AB mag\\
SkyMapper $z$	& $\cdots$	& $\cdots$	& $\cdots$	& $\cdots$	& $10.656\pm0.008$	& $\cdots$	& $\cdots$	& $\cdots$	& $\cdots$	& AB mag\\
SDSS $u$	& $\cdots$	& $\cdots$	& $\cdots$	& $\cdots$	& $\cdots$	& $\cdots$	& $\cdots$	& $\cdots$	& $\cdots$	& AB mag\\
SDSS $g$	& $13.819\pm0.02$	& $15.353\pm0.02$	& $\cdots$	& $\cdots$	& $\cdots$	& $\cdots$	& $\cdots$	& $\cdots$	& $\cdots$	& AB mag\\
SDSS $r$	& $12.898\pm0.02$	& $14.724\pm0.02$	& $\cdots$	& $\cdots$	& $\cdots$	& $\cdots$	& $\cdots$	& $\cdots$	& $\cdots$	& AB mag\\
SDSS $i$	& $12.606\pm0.02$	& $14.565\pm0.02$	& $\cdots$	& $\cdots$	& $\cdots$	& $\cdots$	& $\cdots$	& $\cdots$	& $\cdots$	& AB mag\\
SDSS $z$	& $12.441\pm0.02$	& $14.455\pm0.02$	& $\cdots$	& $\cdots$	& $\cdots$	& $\cdots$	& $\cdots$	& $\cdots$	& $\cdots$	& AB mag\\
Tycho-2 $B_T$	& $\cdots$	& $\cdots$	& $\cdots$	& $\cdots$	& $\cdots$	& $\cdots$	& $\cdots$	& $\cdots$	& $\cdots$	& Vega mag\\
Tycho-2 $V_T$	& $\cdots$	& $\cdots$	& $\cdots$	& $\cdots$	& $\cdots$	& $\cdots$	& $\cdots$	& $\cdots$	& $\cdots$	& Vega mag\\
Gaia DR2 $G$	& $12.962\pm0.002$	& $14.854\pm0.002$	& $11.789\pm0.002$	& $7.802\pm0.002$	& $10.928\pm0.002$	& $5.208\pm0.002$	& $10.722\pm0.002$	& $\cdots$	& $\cdots$	& Vega mag\\
2MASS $J$	& $11.368\pm0.016$	& $\cdots$	& $10.613\pm0.018$	& $6.806\pm0.015$	& $9.726\pm0.023$	& $3.981\pm0.258$	& $9.086\pm0.017$	& $\cdots$	& $\cdots$	& Vega mag\\
2MASS $H$	& $10.852\pm0.021$	& $13.266\pm0.034$	& $10.31\pm0.02$	& $6.501\pm0.031$	& $9.312\pm0.02$	& $3.469\pm0.226$	& $8.524\pm0.051$	& $\cdots$	& $\cdots$	& Vega mag\\
2MASS $K_{\text{s}}$	& $10.757\pm0.016$	& $13.238\pm0.029$	& $10.192\pm0.024$	& $6.419\pm0.019$	& $9.259\pm0.026$	& $3.261\pm0.304$	& $8.401\pm0.021$	& $\cdots$	& $\cdots$	& Vega mag\\
WISE $W1$	& $10.684\pm0.023$	& $12.969\pm0.024$	& $10.165\pm0.023$	& $\cdots$	& $9.178\pm0.022$	& $3.346\pm0.098$	& $8.311\pm0.024$	& $\cdots$	& $\cdots$	& Vega mag\\
WISE $W2$	& $10.772\pm0.02$	& $13.019\pm0.026$	& $10.236\pm0.019$	& $\cdots$	& $9.213\pm0.02$	& $2.584\pm0.054$	& $8.391\pm0.02$	& $\cdots$	& $\cdots$	& Vega mag\\
WISE $W3$	& $\cdots$	& $\cdots$	& $\cdots$	& $\cdots$	& $\cdots$	& $3.137\pm0.011$	& $\cdots$	& $\cdots$	& $\cdots$	& Vega mag\\
WISE $W4$	& $\cdots$	& $\cdots$	& $\cdots$	& $\cdots$	& $\cdots$	& $3.097\pm0.018$	& $\cdots$	& $\cdots$	& $\cdots$	& Vega mag
\enddata
\tablecomments{This table is also published in a machine-readable format.}
\end{deluxetable*}
\end{longrotatetable}

\begin{longrotatetable}
\begin{deluxetable*}{c|ccccc|ccccc|ccccc|ccccc}
\tablecaption{Elemental Abundances\label{tab:abundances}}
\tablewidth{0pt}
\tablehead{
\colhead{Species} &
\colhead{A(X)} & \colhead{[X/H]} & \colhead{[X/Fe]} & \colhead{$\sigma_{\text{[X/Fe]}}$} & \colhead{n$_{\text{X}}$} & \colhead{A(X)} & \colhead{[X/H]} & \colhead{[X/Fe]} & \colhead{$\sigma_{\text{[X/Fe]}}$} & \colhead{n$_{\text{X}}$} & \colhead{A(X)} & \colhead{[X/H]} & \colhead{[X/Fe]} & \colhead{$\sigma_{\text{[X/Fe]}}$} & \colhead{n$_{\text{X}}$} & \colhead{A(X)} & \colhead{[X/H]} & \colhead{[X/Fe]} & \colhead{$\sigma_{\text{[X/Fe]}}$} & \colhead{n$_{\text{X}}$}
}
\startdata
 	& \multicolumn{5}{c}{K2-106} & \multicolumn{5}{c}{HD 15337} & \multicolumn{5}{c}{K2-291} & \multicolumn{5}{c}{55 Cnc}\\
\hline
\ion{O}{1} 	& 9.017 	& 0.327 	& 0.234 	& 0.001 	& 3 	& 9.061 	& 0.371 	& 0.115 	& 0.033 	& 3 	& 9.152 	& 0.462 	& 0.394 	& 0.085 	& 3 	& 9.271 	& 0.581 	& 0.122 	& 0.043 	& 3\\
\ion{Mg}{1} 	& 7.644 	& 0.094 	& 0.001 	& 0.014 	& 3 	& 7.926 	& 0.376 	& 0.12 	& 0.02 	& 3 	& 7.624 	& 0.074 	& 0.006 	& 0.041 	& 2 	& 8.079 	& 0.529 	& 0.07 	& 0.026 	& 2\\
\ion{Al}{1} 	& 6.479 	& 0.05 	& -0.043 	& 0.014 	& 4 	& 6.624 	& 0.194 	& -0.062 	& 0.017 	& 3 	& 6.434 	& 0.004 	& -0.064 	& 0.034 	& 2 	& 6.886 	& 0.456 	& -0.003 	& 0.022 	& 4\\
\ion{Si}{1} 	& 7.637 	& 0.127 	& 0.034 	& 0.013 	& 12 	& 7.805 	& 0.295 	& 0.039 	& 0.016 	& 14 	& 7.59 	& 0.08 	& 0.012 	& 0.01 	& 13 	& 8.084 	& 0.574 	& 0.115 	& 0.027 	& 10\\
\ion{Ca}{1} 	& 6.369 	& 0.069 	& -0.024 	& 0.014 	& 11 	& 6.477 	& 0.177 	& -0.079 	& 0.018 	& 6 	& 6.375 	& 0.075 	& 0.007 	& 0.011 	& 10 	& 6.64 	& 0.34 	& -0.119 	& 0.016 	& 8\\
\ion{Fe}{1} 	& 7.534 	& 0.074 	& -0.019 	& 0.006 	& 86 	& 7.698 	& 0.238 	& -0.018 	& 0.008 	& 84 	& 7.516 	& 0.056 	& -0.012 	& 0.006 	& 90 	& 7.901 	& 0.441 	& -0.018 	& 0.01 	& 31\\
\ion{Fe}{2} 	& 7.686 	& 0.226 	& 0.133 	& 0.02 	& 13 	& 7.873 	& 0.413 	& 0.157 	& 0.028 	& 9 	& 7.595 	& 0.135 	& 0.067 	& 0.016 	& 13 	& 8.054 	& 0.594 	& 0.135 	& 0.024 	& 4\\
\ion{Ni}{1} 	& 6.365 	& 0.165 	& 0.072 	& 0.012 	& 20 	& 6.583 	& 0.383 	& 0.127 	& 0.01 	& 16 	& 6.311 	& 0.111 	& 0.043 	& 0.015 	& 18 	& 6.803 	& 0.603 	& 0.144 	& 0.024 	& 15\\
Fe/Mg 	& 0.776 	& $\cdots$ 	& $\cdots$ 	& 0.027 	& $\cdots$ 	& 0.592 	& $\cdots$ 	& $\cdots$ 	& 0.029 	& $\cdots$ 	& 0.78 	& $\cdots$ 	& $\cdots$ 	& 0.074 	& $\cdots$ 	& 0.664 	& $\cdots$ 	& $\cdots$ 	& 0.043 	& $\cdots$\\
Fe/Si 	& 0.789 	& $\cdots$ 	& $\cdots$ 	& 0.026 	& $\cdots$ 	& 0.782 	& $\cdots$ 	& $\cdots$ 	& 0.032 	& $\cdots$ 	& 0.843 	& $\cdots$ 	& $\cdots$ 	& 0.023 	& $\cdots$ 	& 0.656 	& $\cdots$ 	& $\cdots$ 	& 0.044 	& $\cdots$\\
\hline
\hline
 	& \multicolumn{5}{c}{HD 80653} & \multicolumn{5}{c}{K2-131} & \multicolumn{5}{c}{K2-229} & \multicolumn{5}{c}{HD 136352}\\
\hline
\ion{O}{1} 	& 9.37 	& 0.68 	& 0.36 	& 0.04 	& 3 	& 9.498 	& 0.808 	& 0.884 	& 0.026 	& 3 	& 9.694 	& 1.004 	& 0.806 	& 0.01 	& 3 	& 8.832 	& 0.142 	& 0.42 	& 0.012 	& 3\\
\ion{Mg}{1} 	& 7.838 	& 0.288 	& -0.032 	& 0.0 	& 1 	& 7.514 	& -0.036 	& 0.04 	& 0.017 	& 2 	& 7.75 	& 0.2 	& 0.002 	& 0.018 	& 2 	& 7.529 	& -0.021 	& 0.257 	& 0.022 	& 5\\
\ion{Al}{1} 	& 6.703 	& 0.273 	& -0.047 	& 0.037 	& 2 	& 6.24 	& -0.189 	& -0.113 	& 0.002 	& 2 	& 6.468 	& 0.038 	& -0.16 	& 0.038 	& 2 	& 6.304 	& -0.126 	& 0.152 	& 0.015 	& 4\\
\ion{Si}{1} 	& 7.863 	& 0.353 	& 0.033 	& 0.018 	& 9 	& 7.499 	& -0.011 	& 0.065 	& 0.017 	& 11 	& 7.938 	& 0.428 	& 0.23 	& 0.018 	& 13 	& 7.361 	& -0.149 	& 0.129 	& 0.015 	& 13\\
\ion{Ca}{1} 	& 6.611 	& 0.311 	& -0.009 	& 0.012 	& 8 	& 6.145 	& -0.155 	& -0.079 	& 0.026 	& 7 	& 6.261 	& -0.039 	& -0.237 	& 0.015 	& 10 	& 6.162 	& -0.138 	& 0.14 	& 0.01 	& 10\\
\ion{Fe}{1} 	& 7.767 	& 0.307 	& -0.013 	& 0.008 	& 66 	& 7.365 	& -0.095 	& -0.019 	& 0.008 	& 74 	& 7.579 	& 0.119 	& -0.079 	& 0.009 	& 87 	& 7.18 	& -0.28 	& -0.002 	& 0.007 	& 81\\
\ion{Fe}{2} 	& 7.887 	& 0.427 	& 0.107 	& 0.018 	& 8 	& 7.518 	& 0.058 	& 0.134 	& 0.027 	& 10 	& 8.235 	& 0.775 	& 0.577 	& 0.026 	& 12 	& 7.181 	& -0.279 	& -0.001 	& 0.01 	& 12\\
\ion{Ni}{1} 	& 6.638 	& 0.438 	& 0.118 	& 0.015 	& 15 	& 6.153 	& -0.047 	& 0.029 	& 0.016 	& 14 	& 6.579 	& 0.379 	& 0.181 	& 0.017 	& 20 	& 6.001 	& -0.199 	& 0.079 	& 0.009 	& 17\\
Fe/Mg 	& 0.849 	& $\cdots$ 	& $\cdots$ 	& 0.016 	& $\cdots$ 	& 0.71 	& $\cdots$ 	& $\cdots$ 	& 0.031 	& $\cdots$ 	& 0.675 	& $\cdots$ 	& $\cdots$ 	& 0.031 	& $\cdots$ 	& 0.448 	& $\cdots$ 	& $\cdots$ 	& 0.024 	& $\cdots$\\
Fe/Si 	& 0.802 	& $\cdots$ 	& $\cdots$ 	& 0.036 	& $\cdots$ 	& 0.735 	& $\cdots$ 	& $\cdots$ 	& 0.032 	& $\cdots$ 	& 0.438 	& $\cdots$ 	& $\cdots$ 	& 0.02 	& $\cdots$ 	& 0.659 	& $\cdots$ 	& $\cdots$ 	& 0.025 	& $\cdots$\\
\hline
\hline
 	& \multicolumn{5}{c}{K2-38} & \multicolumn{5}{c}{Kepler-10} & \multicolumn{5}{c}{Kepler-20} & \multicolumn{5}{c}{Kepler-36}\\
\hline
\ion{O}{1} 	& 9.292 	& 0.602 	& 0.415 	& 0.012 	& 3 	& 8.932 	& 0.242 	& 0.363 	& 0.036 	& 3 	& $\cdots$ 	& $\cdots$ 	& $\cdots$ 	& $\cdots$ 	& $\cdots$ 	& $\cdots$ 	& $\cdots$ 	& $\cdots$ 	& $\cdots$ 	& $\cdots$\\
\ion{Mg}{1} 	& 7.848 	& 0.298 	& 0.111 	& 0.018 	& 2 	& 7.675 	& 0.125 	& 0.246 	& 0.04 	& 4 	& 7.582 	& 0.032 	& 0.004 	& 0.02 	& 3 	& 7.399 	& -0.151 	& 0.059 	& 0.01 	& 2\\
\ion{Al}{1} 	& 6.662 	& 0.232 	& 0.045 	& 0.014 	& 4 	& 6.404 	& -0.026 	& 0.095 	& 0.012 	& 6 	& 6.421 	& -0.009 	& -0.037 	& 0.012 	& 2 	& $\cdots$ 	& $\cdots$ 	& $\cdots$ 	& $\cdots$ 	& $\cdots$\\
\ion{Si}{1} 	& 7.838 	& 0.328 	& 0.141 	& 0.015 	& 13 	& 7.457 	& -0.053 	& 0.068 	& 0.014 	& 12 	& 7.582 	& 0.072 	& 0.044 	& 0.012 	& 12 	& 7.371 	& -0.139 	& 0.071 	& 0.024 	& 9\\
\ion{Ca}{1} 	& 6.477 	& 0.177 	& -0.01 	& 0.013 	& 11 	& 6.255 	& -0.045 	& 0.076 	& 0.026 	& 9 	& 6.313 	& 0.013 	& -0.015 	& 0.014 	& 11 	& 6.188 	& -0.112 	& 0.098 	& 0.021 	& 9\\
\ion{Fe}{1} 	& 7.645 	& 0.185 	& -0.002 	& 0.006 	& 87 	& 7.331 	& -0.129 	& -0.008 	& 0.018 	& 78 	& 7.478 	& 0.018 	& -0.01 	& 0.008 	& 84 	& 7.23 	& -0.23 	& -0.02 	& 0.009 	& 49\\
\ion{Fe}{2} 	& 7.86 	& 0.4 	& 0.213 	& 0.013 	& 13 	& 7.388 	& -0.072 	& 0.049 	& 0.022 	& 12 	& 7.553 	& 0.093 	& 0.065 	& 0.015 	& 12 	& 7.344 	& -0.116 	& 0.094 	& 0.025 	& 9\\
\ion{Ni}{1} 	& 6.569 	& 0.369 	& 0.182 	& 0.012 	& 20 	& 6.14 	& -0.06 	& 0.061 	& 0.018 	& 14 	& 6.3 	& 0.1 	& 0.072 	& 0.019 	& 18 	& 6.002 	& -0.198 	& 0.012 	& 0.037 	& 10\\
Fe/Mg 	& 0.627 	& $\cdots$ 	& $\cdots$ 	& 0.027 	& $\cdots$ 	& 0.453 	& $\cdots$ 	& $\cdots$ 	& 0.046 	& $\cdots$ 	& 0.787 	& $\cdots$ 	& $\cdots$ 	& 0.039 	& $\cdots$ 	& 0.678 	& $\cdots$ 	& $\cdots$ 	& 0.021 	& $\cdots$\\
Fe/Si 	& 0.641 	& $\cdots$ 	& $\cdots$ 	& 0.024 	& $\cdots$ 	& 0.748 	& $\cdots$ 	& $\cdots$ 	& 0.039 	& $\cdots$ 	& 0.787 	& $\cdots$ 	& $\cdots$ 	& 0.026 	& $\cdots$ 	& 0.723 	& $\cdots$ 	& $\cdots$ 	& 0.043 	& $\cdots$\\
\hline
\hline
 	& \multicolumn{5}{c}{Kepler-93} & \multicolumn{5}{c}{Kepler-78} & \multicolumn{5}{c}{Kepler-107} & \multicolumn{5}{c}{WASP-47}\\
\hline
\ion{O}{1} 	& 8.822 	& 0.132 	& 0.269 	& 0.013 	& 2 	& 9.062 	& 0.372 	& 0.333 	& 0.027 	& 2 	& 9.105 	& 0.415 	& 0.152 	& 0.005 	& 3 	& 9.343 	& 0.653 	& 0.227 	& 0.028 	& 3\\
\ion{Mg}{1} 	& 7.582 	& 0.032 	& 0.169 	& 0.076 	& 3 	& 7.702 	& 0.152 	& 0.113 	& 0.008 	& 2 	& 7.902 	& 0.352 	& 0.089 	& 0.028 	& 2 	& 8.032 	& 0.482 	& 0.056 	& 0.062 	& 2\\
\ion{Al}{1} 	& 6.303 	& -0.127 	& 0.01 	& 0.034 	& 2 	& 6.297 	& -0.133 	& -0.172 	& 0.013 	& 6 	& 6.758 	& 0.329 	& 0.066 	& 0.038 	& 2 	& 6.871 	& 0.441 	& 0.015 	& 0.022 	& 3\\
\ion{Si}{1} 	& 7.407 	& -0.103 	& 0.034 	& 0.013 	& 13 	& 7.621 	& 0.111 	& 0.072 	& 0.019 	& 10 	& 7.866 	& 0.356 	& 0.093 	& 0.019 	& 13 	& 8.053 	& 0.543 	& 0.117 	& 0.017 	& 13\\
\ion{Ca}{1} 	& 6.181 	& -0.119 	& 0.018 	& 0.011 	& 8 	& 6.271 	& -0.029 	& -0.068 	& 0.013 	& 5 	& 6.554 	& 0.254 	& -0.009 	& 0.025 	& 8 	& 6.643 	& 0.343 	& -0.083 	& 0.017 	& 8\\
\ion{Fe}{1} 	& 7.31 	& -0.15 	& -0.013 	& 0.006 	& 86 	& 7.5 	& 0.04 	& 0.001 	& 0.009 	& 64 	& 7.718 	& 0.258 	& -0.005 	& 0.01 	& 78 	& 7.859 	& 0.399 	& -0.027 	& 0.007 	& 72\\
\ion{Fe}{2} 	& 7.394 	& -0.066 	& 0.071 	& 0.012 	& 13 	& 7.805 	& 0.345 	& 0.306 	& 0.042 	& 6 	& 7.766 	& 0.306 	& 0.043 	& 0.031 	& 8 	& 8.068 	& 0.608 	& 0.182 	& 0.019 	& 9\\
\ion{Ni}{1} 	& 6.111 	& -0.089 	& 0.048 	& 0.009 	& 16 	& 6.318 	& 0.118 	& 0.079 	& 0.021 	& 15 	& 6.615 	& 0.415 	& 0.152 	& 0.022 	& 19 	& 6.779 	& 0.579 	& 0.153 	& 0.014 	& 19\\
Fe/Mg 	& 0.535 	& $\cdots$ 	& $\cdots$ 	& 0.094 	& $\cdots$ 	& 0.628 	& $\cdots$ 	& $\cdots$ 	& 0.017 	& $\cdots$ 	& 0.655 	& $\cdots$ 	& $\cdots$ 	& 0.045 	& $\cdots$ 	& 0.671 	& $\cdots$ 	& $\cdots$ 	& 0.096 	& $\cdots$\\
Fe/Si 	& 0.8 	& $\cdots$ 	& $\cdots$ 	& 0.026 	& $\cdots$ 	& 0.757 	& $\cdots$ 	& $\cdots$ 	& 0.037 	& $\cdots$ 	& 0.711 	& $\cdots$ 	& $\cdots$ 	& 0.035 	& $\cdots$ 	& 0.64 	& $\cdots$ 	& $\cdots$ 	& 0.027 	& $\cdots$\\
\hline
\hline
 	& \multicolumn{5}{c}{HD 213885} & \multicolumn{5}{c}{K2-265} & \multicolumn{5}{c}{HD 219134} & \multicolumn{5}{c}{K2-141}\\
\hline
\ion{O}{1} 	& 9.031 	& 0.341 	& 0.329 	& 0.02 	& 3 	& 9.54 	& 0.85 	& 0.704 	& 0.007 	& 3 	& 9.089 	& 0.399 	& 0.154 	& 0.066 	& 3 	& $\cdots$ 	& $\cdots$ 	& $\cdots$ 	& $\cdots$ 	& $\cdots$\\
\ion{Mg}{1} 	& 7.599 	& 0.049 	& 0.037 	& 0.009 	& 2 	& 7.765 	& 0.215 	& 0.069 	& 0.0 	& 1 	& 7.929 	& 0.379 	& 0.134 	& 0.0 	& 1 	& 7.71 	& 0.16 	& -0.138 	& 0.073 	& 2\\
\ion{Al}{1} 	& 6.449 	& 0.019 	& 0.007 	& 0.02 	& 4 	& 6.601 	& 0.171 	& 0.025 	& 0.016 	& 4 	& 6.643 	& 0.213 	& -0.032 	& 0.026 	& 4 	& 6.496 	& 0.066 	& -0.232 	& 0.021 	& 2\\
\ion{Si}{1} 	& 7.578 	& 0.068 	& 0.056 	& 0.019 	& 10 	& 7.887 	& 0.377 	& 0.231 	& 0.016 	& 14 	& 7.802 	& 0.292 	& 0.047 	& 0.022 	& 8 	& 7.949 	& 0.439 	& 0.141 	& 0.017 	& 6\\
\ion{Ca}{1} 	& 6.35 	& 0.05 	& 0.038 	& 0.028 	& 10 	& 6.361 	& 0.061 	& -0.085 	& 0.015 	& 10 	& 6.509 	& 0.209 	& -0.036 	& 0.034 	& 4 	& 6.498 	& 0.198 	& -0.1 	& 0.044 	& 2\\
\ion{Fe}{1} 	& 7.464 	& 0.004 	& -0.008 	& 0.007 	& 85 	& 7.605 	& 0.145 	& -0.001 	& 0.007 	& 85 	& 7.696 	& 0.236 	& -0.009 	& 0.01 	& 70 	& 7.738 	& 0.278 	& -0.02 	& 0.014 	& 49\\
\ion{Fe}{2} 	& 7.528 	& 0.068 	& 0.056 	& 0.015 	& 12 	& 8.07 	& 0.61 	& 0.464 	& 0.019 	& 12 	& 7.823 	& 0.363 	& 0.118 	& 0.036 	& 6 	& 8.104 	& 0.644 	& 0.346 	& 0.042 	& 3\\
\ion{Ni}{1} 	& 6.288 	& 0.088 	& 0.076 	& 0.021 	& 16 	& 6.578 	& 0.378 	& 0.232 	& 0.014 	& 19 	& 6.538 	& 0.338 	& 0.093 	& 0.011 	& 16 	& 6.569 	& 0.369 	& 0.071 	& 0.028 	& 11\\
Fe/Mg 	& 0.733 	& $\cdots$ 	& $\cdots$ 	& 0.019 	& $\cdots$ 	& 0.692 	& $\cdots$ 	& $\cdots$ 	& 0.011 	& $\cdots$ 	& 0.585 	& $\cdots$ 	& $\cdots$ 	& 0.013 	& $\cdots$ 	& 1.067 	& $\cdots$ 	& $\cdots$ 	& 0.183 	& $\cdots$\\
Fe/Si 	& 0.769 	& $\cdots$ 	& $\cdots$ 	& 0.036 	& $\cdots$ 	& 0.522 	& $\cdots$ 	& $\cdots$ 	& 0.021 	& $\cdots$ 	& 0.783 	& $\cdots$ 	& $\cdots$ 	& 0.044 	& $\cdots$ 	& 0.615 	& $\cdots$ 	& $\cdots$ 	& 0.031 	& $\cdots$
\enddata
\tablecomments{This table is also published in a machine-readable format.}
\end{deluxetable*}
\end{longrotatetable}

\begin{longrotatetable}
\begin{deluxetable*}{c|ccccc|ccccc|ccccc|ccccc}
\tablecaption{Elemental Abundances Corrected for non-LTE Effects\label{tab:abundances_nlte}}
\tablewidth{0pt}
\tablehead{
\colhead{Species} &
\colhead{A(X)} & \colhead{[X/H]} & \colhead{[X/Fe]} & \colhead{$\sigma_{\text{[X/H]}}$} & \colhead{n$_{\text{X}}$} & \colhead{A(X)} & \colhead{[X/H]} & \colhead{[X/Fe]} & \colhead{$\sigma_{\text{[X/H]}}$} & \colhead{n$_{\text{X}}$} & \colhead{A(X)} & \colhead{[X/H]} & \colhead{[X/Fe]} & \colhead{$\sigma_{\text{[X/H]}}$} & \colhead{n$_{\text{X}}$} & \colhead{A(X)} & \colhead{[X/H]} & \colhead{[X/Fe]} & \colhead{$\sigma_{\text{[X/H]}}$} & \colhead{n$_{\text{X}}$}
}
\startdata
 	& \multicolumn{5}{c}{K2-106} & \multicolumn{5}{c}{HD 15337} & \multicolumn{5}{c}{K2-291} & \multicolumn{5}{c}{55 Cnc}\\
\hline
\ion{O}{1} 	& 8.869 	& 0.179 	& 0.058 	& 0.012 	& 3 	& 8.987 	& 0.297 	& 0.054 	& 0.049 	& 3 	& 9.009 	& 0.319 	& 0.217 	& 0.128 	& 3 	& 9.171 	& 0.481 	& 0.04 	& 0.063 	& 3\\
\ion{Si}{1} 	& 7.611 	& 0.1 	& -0.02 	& 0.042 	& 2 	& 7.752 	& 0.242 	& 0.0 	& 0.045 	& 4 	& 7.566 	& 0.056 	& -0.046 	& 0.042 	& 17 	& 7.986 	& 0.476 	& 0.035 	& 0.069 	& 3\\
\ion{Al}{1} 	& 6.447 	& 0.017 	& -0.104 	& 0.03 	& 2 	& 6.59 	& 0.16 	& -0.082 	& 0.025 	& 2 	& 6.373 	& -0.057 	& -0.159 	& 0.0 	& 1 	& 6.869 	& 0.439 	& -0.002 	& 0.049 	& 2\\
\ion{Ca}{1} 	& 6.13 	& -0.17 	& -0.291 	& 0.052 	& 2 	& 6.345 	& 0.045 	& -0.197 	& 0.05 	& 2 	& 6.109 	& -0.191 	& -0.293 	& 0.03 	& 2 	& 6.569 	& 0.27 	& -0.171 	& 0.05 	& 2\\
\ion{Fe}{1} 	& 7.563 	& 0.103 	& -0.017 	& 0.08 	& 86 	& 7.685 	& 0.225 	& -0.018 	& 0.085 	& 84 	& 7.554 	& 0.094 	& -0.008 	& 0.076 	& 90 	& 7.885 	& 0.425 	& -0.016 	& 0.087 	& 31\\
\ion{Fe}{2} 	& 7.696 	& 0.236 	& 0.115 	& 0.055 	& 13 	& 7.868 	& 0.408 	& 0.166 	& 0.086 	& 9 	& 7.614 	& 0.154 	& 0.053 	& 0.048 	& 13 	& 8.026 	& 0.566 	& 0.125 	& 0.042 	& 4\\
Fe/Mg 	& 0.83 	& $\cdots$ 	& $\cdots$ 	& 0.155 	& $\cdots$ 	& 0.574 	& $\cdots$ 	& $\cdots$ 	& 0.115 	& $\cdots$ 	& 0.851 	& $\cdots$ 	& $\cdots$ 	& 0.169 	& $\cdots$ 	& 0.64 	& $\cdots$ 	& $\cdots$ 	& 0.134 	& $\cdots$\\
Fe/Si 	& 0.895 	& $\cdots$ 	& $\cdots$ 	& 0.186 	& $\cdots$ 	& 0.857 	& $\cdots$ 	& $\cdots$ 	& 0.19 	& $\cdots$ 	& 0.973 	& $\cdots$ 	& $\cdots$ 	& 0.194 	& $\cdots$ 	& 0.793 	& $\cdots$ 	& $\cdots$ 	& 0.203 	& $\cdots$\\
\hline
\hline
 	& \multicolumn{5}{c}{HD 80653} & \multicolumn{5}{c}{K2-131} & \multicolumn{5}{c}{K2-229} & \multicolumn{5}{c}{HD 136352}\\
\hline
\ion{O}{1} 	& 9.12 	& 0.43 	& 0.067 	& 0.074 	& 3 	& 9.458 	& 0.768 	& 0.837 	& 0.041 	& 3 	& 9.615 	& 0.925 	& 0.725 	& 0.021 	& 3 	& 8.65 	& -0.04 	& 0.182 	& 0.027 	& 3\\
\ion{Si}{1} 	& 7.834 	& 0.324 	& -0.039 	& 0.054 	& 12 	& 7.479 	& -0.031 	& 0.038 	& 0.057 	& 14 	& 7.85 	& 0.34 	& 0.14 	& 0.069 	& 4 	& 7.334 	& -0.176 	& 0.046 	& 0.056 	& 17\\
\ion{Al}{1} 	& 6.637 	& 0.207 	& -0.156 	& 0.0 	& 1 	& 6.213 	& -0.217 	& -0.148 	& 0.0 	& 1 	& 6.404 	& -0.026 	& -0.226 	& 0.0 	& 1 	& 6.277 	& -0.152 	& 0.069 	& 0.027 	& 2\\
\ion{Ca}{1} 	& 6.437 	& 0.137 	& -0.226 	& 0.0 	& 1 	& 5.942 	& -0.358 	& -0.289 	& 0.002 	& 2 	& 6.056 	& -0.243 	& -0.444 	& 0.076 	& 2 	& 5.691 	& -0.608 	& -0.387 	& 0.047 	& 2\\
\ion{Fe}{1} 	& 7.812 	& 0.352 	& -0.011 	& 0.081 	& 66 	& 7.371 	& -0.089 	& -0.02 	& 0.056 	& 74 	& 7.581 	& 0.121 	& -0.079 	& 0.076 	& 87 	& 7.242 	& -0.218 	& 0.004 	& 0.068 	& 81\\
\ion{Fe}{2} 	& 7.915 	& 0.455 	& 0.092 	& 0.051 	& 8 	& 7.539 	& 0.079 	& 0.148 	& 0.064 	& 10 	& 8.231 	& 0.771 	& 0.571 	& 0.07 	& 12 	& 7.211 	& -0.249 	& -0.027 	& 0.034 	& 12\\
Fe/Mg 	& 0.942 	& $\cdots$ 	& $\cdots$ 	& 0.176 	& $\cdots$ 	& 0.719 	& $\cdots$ 	& $\cdots$ 	& 0.097 	& $\cdots$ 	& 0.678 	& $\cdots$ 	& $\cdots$ 	& 0.122 	& $\cdots$ 	& 0.516 	& $\cdots$ 	& $\cdots$ 	& 0.085 	& $\cdots$\\
Fe/Si 	& 0.951 	& $\cdots$ 	& $\cdots$ 	& 0.213 	& $\cdots$ 	& 0.78 	& $\cdots$ 	& $\cdots$ 	& 0.143 	& $\cdots$ 	& 0.538 	& $\cdots$ 	& $\cdots$ 	& 0.127 	& $\cdots$ 	& 0.809 	& $\cdots$ 	& $\cdots$ 	& 0.164 	& $\cdots$\\
\hline
\hline
 	& \multicolumn{5}{c}{K2-38} & \multicolumn{5}{c}{Kepler-10} & \multicolumn{5}{c}{Kepler-20} & \multicolumn{5}{c}{Kepler-36}\\
\hline
\ion{O}{1} 	& 9.094 	& 0.404 	& 0.157 	& 0.025 	& 3 	& 8.742 	& 0.052 	& 0.124 	& 0.011 	& 3 	& $\cdots$ 	& $\cdots$ 	& $\cdots$ 	& $\cdots$ 	& $\cdots$ 	& $\cdots$ 	& $\cdots$ 	& $\cdots$ 	& $\cdots$ 	& $\cdots$\\
\ion{Si}{1} 	& 7.744 	& 0.234 	& -0.013 	& 0.055 	& 3 	& 7.403 	& -0.107 	& -0.036 	& 0.054 	& 4 	& 7.511 	& 0.001 	& -0.06 	& 0.032 	& 3 	& 7.339 	& -0.171 	& -0.022 	& 0.074 	& 10\\
\ion{Al}{1} 	& 6.619 	& 0.189 	& -0.058 	& 0.027 	& 2 	& 6.38 	& -0.05 	& 0.022 	& 0.024 	& 2 	& 6.381 	& -0.049 	& -0.111 	& 0.0 	& 1 	& $\cdots$ 	& $\cdots$ 	& $\cdots$ 	& $\cdots$ 	& $\cdots$\\
\ion{Ca}{1} 	& 6.251 	& -0.049 	& -0.296 	& 0.077 	& 2 	& 5.886 	& -0.414 	& -0.343 	& 0.129 	& 2 	& 6.012 	& -0.288 	& -0.35 	& 0.006 	& 2 	& 5.7 	& -0.6 	& -0.451 	& 0.0 	& 1\\
\ion{Fe}{1} 	& 7.681 	& 0.221 	& -0.026 	& 0.075 	& 87 	& 7.383 	& -0.077 	& -0.005 	& 0.06 	& 78 	& 7.514 	& 0.054 	& -0.007 	& 0.081 	& 84 	& 7.299 	& -0.161 	& -0.012 	& 0.064 	& 49\\
\ion{Fe}{2} 	& 7.879 	& 0.419 	& 0.172 	& 0.042 	& 13 	& 7.421 	& -0.039 	& 0.032 	& 0.047 	& 12 	& 7.573 	& 0.113 	& 0.051 	& 0.037 	& 12 	& 7.375 	& -0.085 	& 0.064 	& 0.077 	& 9\\
Fe/Mg 	& 0.681 	& $\cdots$ 	& $\cdots$ 	& 0.121 	& $\cdots$ 	& 0.511 	& $\cdots$ 	& $\cdots$ 	& 0.085 	& $\cdots$ 	& 0.855 	& $\cdots$ 	& $\cdots$ 	& 0.164 	& $\cdots$ 	& 0.794 	& $\cdots$ 	& $\cdots$ 	& 0.118 	& $\cdots$\\
Fe/Si 	& 0.865 	& $\cdots$ 	& $\cdots$ 	& 0.185 	& $\cdots$ 	& 0.955 	& $\cdots$ 	& $\cdots$ 	& 0.178 	& $\cdots$ 	& 1.007 	& $\cdots$ 	& $\cdots$ 	& 0.202 	& $\cdots$ 	& 0.912 	& $\cdots$ 	& $\cdots$ 	& 0.205 	& $\cdots$\\
\hline
\hline
 	& \multicolumn{5}{c}{Kepler-93} & \multicolumn{5}{c}{Kepler-78} & \multicolumn{5}{c}{Kepler-107} & \multicolumn{5}{c}{WASP-47}\\
\hline
\ion{O}{1} 	& 8.663 	& -0.027 	& 0.058 	& 0.01 	& 2 	& 9.0 	& 0.31 	& 0.256 	& 0.029 	& 2 	& 8.819 	& 0.129 	& -0.187 	& 0.026 	& 3 	& 9.187 	& 0.497 	& 0.064 	& 0.044 	& 3\\
\ion{Si}{1} 	& 7.333 	& -0.177 	& -0.092 	& 0.044 	& 4 	& 7.571 	& 0.061 	& 0.007 	& 0.053 	& 3 	& 7.826 	& 0.316 	& 0.0 	& 0.078 	& 15 	& 7.964 	& 0.454 	& 0.02 	& 0.056 	& 3\\
\ion{Al}{1} 	& 6.242 	& -0.188 	& -0.102 	& 0.0 	& 1 	& 6.274 	& -0.156 	& -0.21 	& 0.042 	& 2 	& 6.687 	& 0.257 	& -0.05 	& 0.0 	& 1 	& 6.832 	& 0.401 	& -0.032 	& 0.034 	& 2\\
\ion{Ca}{1} 	& 5.713 	& -0.587 	& -0.502 	& 0.042 	& 2 	& 6.127 	& -0.173 	& -0.227 	& 0.0 	& 1 	& 6.171 	& -0.129 	& -0.445 	& 0.0 	& 1 	& 6.565 	& 0.265 	& -0.169 	& 0.038 	& 2\\
\ion{Fe}{1} 	& 7.366 	& -0.094 	& -0.008 	& 0.064 	& 86 	& 7.487 	& 0.027 	& -0.027 	& 0.072 	& 64 	& 7.763 	& 0.303 	& -0.004 	& 0.099 	& 78 	& 7.873 	& 0.413 	& -0.021 	& 0.08 	& 72\\
\ion{Fe}{2} 	& 7.426 	& -0.034 	& 0.052 	& 0.042 	& 13 	& 7.797 	& 0.337 	& 0.283 	& 0.104 	& 6 	& 7.808 	& 0.348 	& 0.04 	& 0.08 	& 8 	& 8.063 	& 0.603 	& 0.169 	& 0.06 	& 9\\
Fe/Mg 	& 0.608 	& $\cdots$ 	& $\cdots$ 	& 0.139 	& $\cdots$ 	& 0.61 	& $\cdots$ 	& $\cdots$ 	& 0.102 	& $\cdots$ 	& 0.741 	& $\cdots$ 	& $\cdots$ 	& 0.192 	& $\cdots$ 	& 0.693 	& $\cdots$ 	& $\cdots$ 	& 0.162 	& $\cdots$\\
Fe/Si 	& 1.079 	& $\cdots$ 	& $\cdots$ 	& 0.193 	& $\cdots$ 	& 0.824 	& $\cdots$ 	& $\cdots$ 	& 0.17 	& $\cdots$ 	& 0.883 	& $\cdots$ 	& $\cdots$ 	& 0.273 	& $\cdots$ 	& 0.811 	& $\cdots$ 	& $\cdots$ 	& 0.182 	& $\cdots$\\
\hline
\hline
 	& \multicolumn{5}{c}{HD 213885} & \multicolumn{5}{c}{K2-265} & \multicolumn{5}{c}{HD 219134} & \multicolumn{5}{c}{K2-141}\\
\hline
\ion{O}{1} 	& 8.802 	& 0.112 	& 0.058 	& 0.033 	& 3 	& 9.403 	& 0.713 	& 0.493 	& 0.019 	& 3 	& 9.039 	& 0.349 	& 0.115 	& 0.094 	& 3 	& $\cdots$ 	& $\cdots$ 	& $\cdots$ 	& $\cdots$ 	& $\cdots$\\
\ion{Si}{1} 	& 7.552 	& 0.042 	& -0.013 	& 0.057 	& 13 	& 7.801 	& 0.292 	& 0.072 	& 0.063 	& 4 	& 7.749 	& 0.24 	& 0.005 	& 0.078 	& 2 	& 7.892 	& 0.382 	& 0.084 	& 0.0 	& 1\\
\ion{Al}{1} 	& 6.399 	& -0.03 	& -0.085 	& 0.035 	& 2 	& 6.562 	& 0.132 	& -0.088 	& 0.034 	& 2 	& 6.619 	& 0.189 	& -0.046 	& 0.006 	& 4 	& 6.451 	& 0.021 	& -0.277 	& 0.0 	& 1\\
\ion{Ca}{1} 	& 5.954 	& -0.346 	& -0.401 	& 0.064 	& 2 	& 6.151 	& -0.149 	& -0.369 	& 0.084 	& 2 	& 6.448 	& 0.148 	& -0.086 	& 0.0 	& 1 	& 6.385 	& 0.085 	& -0.213 	& 0.0 	& 1\\
\ion{Fe}{1} 	& 7.509 	& 0.049 	& -0.006 	& 0.066 	& 85 	& 7.624 	& 0.164 	& -0.056 	& 0.072 	& 85 	& 7.674 	& 0.214 	& -0.02 	& 0.082 	& 70 	& 7.736 	& 0.276 	& -0.022 	& 0.094 	& 48\\
\ion{Fe}{2} 	& 7.554 	& 0.094 	& 0.04 	& 0.053 	& 12 	& 8.079 	& 0.619 	& 0.399 	& 0.05 	& 12 	& 7.813 	& 0.353 	& 0.118 	& 0.081 	& 12 	& 8.113 	& 0.653 	& 0.355 	& 0.04 	& 3\\
Fe/Mg 	& 0.813 	& $\cdots$ 	& $\cdots$ 	& 0.125 	& $\cdots$ 	& 0.723 	& $\cdots$ 	& $\cdots$ 	& 0.12 	& $\cdots$ 	& 0.556 	& $\cdots$ 	& $\cdots$ 	& 0.105 	& $\cdots$ 	& 1.062 	& $\cdots$ 	& $\cdots$ 	& 0.291 	& $\cdots$\\
Fe/Si 	& 0.906 	& $\cdots$ 	& $\cdots$ 	& 0.182 	& $\cdots$ 	& 0.665 	& $\cdots$ 	& $\cdots$ 	& 0.147 	& $\cdots$ 	& 0.841 	& $\cdots$ 	& $\cdots$ 	& 0.219 	& $\cdots$ 	& 0.698 	& $\cdots$ 	& $\cdots$ 	& 0.151 	& $\cdots$
\enddata
\tablecomments{This table is also published in a machine-readable format.}
\end{deluxetable*}
\end{longrotatetable}

\startlongtable
\begin{deluxetable*}{lcCCCC}
\tablecaption{Core-mass Fractions\label{tab:cmfs}}
\tabletypesize{\small}
\tablewidth{0pt}
\tablehead{
\colhead{Designation} & \colhead{Planet} & \colhead{This Study} &
\colhead{Adibekyan} & \colhead{APOGEE DR17} & \colhead{Brewer}
}
\startdata
\multicolumn{6}{l}{\textbf{Planet Mass and Radius-based CMF}} \\
K2-216	& b	& \cdots	& 0.117_{-0.452}^{+0.467}	& 0.471_{-0.353}^{+0.208}	& \cdots\\
K2-106	& b	& 0.114_{-0.308}^{+0.211}	& 0.746_{-0.313}^{+0.270}	& 0.074_{-0.308}^{+0.223}	& \cdots\\
HD 15337	& b	& 0.274_{-0.157}^{+0.133}	& \cdots	& \cdots	& \cdots\\
K2-291	& b	& 0.448_{-0.212}^{+0.158}	& 0.433_{-0.251}^{+0.225}	& \cdots	& \cdots\\
55 Cnc	& e	& -0.236_{-0.057}^{+0.063}	& -0.103_{-0.085}^{+0.092}	& \cdots	& -0.213_{-0.050}^{+0.054}\\
HD 80653	& b	& 0.261_{-0.146}^{+0.139}	& \cdots	& \cdots	& \cdots\\
K2-131	& b	& 0.278_{-0.449}^{+0.300}	& \cdots	& \cdots	& \cdots\\
K2-229	& b	& 0.466_{-0.390}^{+0.249}	& 0.701_{-0.212}^{+0.187}	& \cdots	& \cdots\\
HD 136352	& b	& 0.029_{-0.127}^{+0.127}	& \cdots	& \cdots	& \cdots\\
K2-38	& b	& 0.503_{-0.283}^{+0.206}	& 0.620_{-0.328}^{+0.280}	& 0.492_{-0.300}^{+0.208}	& 0.502_{-0.296}^{+0.206}\\
Kepler-10	& b	& 0.029_{-0.121}^{+0.112}	& 0.091_{-0.188}^{+0.165}	& 0.048_{-0.117}^{+0.108}	& -0.032_{-0.111}^{+0.112}\\
Kepler-20	& b	& 0.345_{-0.174}^{+0.115}	& 0.208_{-0.186}^{+0.161}	& 0.326_{-0.167}^{+0.120}	& 0.367_{-0.165}^{+0.116}\\
Kepler-105	& c	& \cdots	& \cdots	& \cdots	& -0.489_{-0.331}^{+0.387}\\
Kepler-36	& b	& -0.062_{-0.125}^{+0.106}	& \cdots	& -0.053_{-0.123}^{+0.107}	& -0.054_{-0.120}^{+0.105}\\
Kepler-93	& b	& 0.303_{-0.118}^{+0.100}	& 0.239_{-0.210}^{+0.160}	& 0.288_{-0.121}^{+0.101}	& 0.323_{-0.120}^{+0.098}\\
Kepler-406	& b	& \cdots	& 0.734_{-0.191}^{+0.137}	& \cdots	& 0.401_{-0.394}^{+0.279}\\
Kepler-78	& b	& 0.286_{-0.514}^{+0.302}	& 0.164_{-0.443}^{+0.373}	& \cdots	& 0.293_{-0.501}^{+0.300}\\
Kepler-107	& c	& 0.752_{-0.155}^{+0.110}	& 0.690_{-0.156}^{+0.113}	& \cdots	& 0.700_{-0.159}^{+0.115}\\
Kepler-99	& b	& \cdots	& \cdots	& 0.534_{-0.210}^{+0.156}	& 0.521_{-0.221}^{+0.158}\\
KOI-1599	& .01	& \cdots	& \cdots	& -0.963_{-0.021}^{+0.026}	& \cdots\\
WASP-47	& e	& -0.129_{-0.087}^{+0.090}	& 0.128_{-0.299}^{+0.292}	& -0.211_{-0.080}^{+0.083}	& \cdots\\
HD 213885	& b	& 0.287_{-0.137}^{+0.124}	& 0.362_{-0.130}^{+0.121}	& \cdots	& \cdots\\
K2-265	& b	& 0.380_{-0.152}^{+0.135}	& 0.154_{-0.271}^{+0.258}	& 0.256_{-0.172}^{+0.147}	& 0.248_{-0.176}^{+0.144}\\
HD 219134	& b	& 0.159_{-0.140}^{+0.135}	& 0.252_{-0.171}^{+0.166}	& \cdots	& \cdots\\
HD 219134	& c	& 0.318_{-0.120}^{+0.114}	& 0.402_{-0.152}^{+0.149}	& \cdots	& \cdots\\
K2-141	& b	& 0.241_{-0.112}^{+0.104}	& 0.388_{-0.142}^{+0.135}	& 0.248_{-0.114}^{+0.105}	& \cdots\\
\hline
\multicolumn{6}{l}{\textbf{Fe/Mg-based CMF}} \\
K2-216	& b	& \cdots	& 0.162_{-0.090}^{+0.074}	& 0.098_{-0.005}^{+0.005}	& \cdots\\
K2-106	& b	& 0.133_{-0.033}^{+0.030}	& 0.141_{-0.026}^{+0.025}	& 0.123_{-0.006}^{+0.006}	& \cdots\\
HD 15337	& b	& 0.079_{-0.027}^{+0.025}	& \cdots	& \cdots	& \cdots\\
K2-291	& b	& 0.138_{-0.037}^{+0.032}	& 0.125_{-0.017}^{+0.016}	& \cdots	& \cdots\\
55 Cnc	& e	& 0.093_{-0.030}^{+0.028}	& 0.095_{-0.027}^{+0.026}	& \cdots	& 0.161_{-0.008}^{+0.007}\\
HD 80653	& b	& 0.155_{-0.034}^{+0.033}	& \cdots	& \cdots	& \cdots\\
K2-131	& b	& 0.110_{-0.021}^{+0.020}	& \cdots	& \cdots	& \cdots\\
K2-229	& b	& 0.101_{-0.026}^{+0.026}	& 0.133_{-0.018}^{+0.018}	& \cdots	& \cdots\\
HD 136352	& b	& 0.065_{-0.020}^{+0.019}	& \cdots	& \cdots	& \cdots\\
K2-38	& b	& 0.102_{-0.027}^{+0.025}	& 0.137_{-0.035}^{+0.032}	& 0.119_{-0.005}^{+0.004}	& 0.139_{-0.007}^{+0.007}\\
Kepler-10	& b	& 0.064_{-0.020}^{+0.020}	& 0.085_{-0.009}^{+0.009}	& 0.083_{-0.006}^{+0.006}	& 0.104_{-0.011}^{+0.009}\\
Kepler-20	& b	& 0.138_{-0.034}^{+0.032}	& 0.123_{-0.021}^{+0.020}	& 0.093_{-0.007}^{+0.006}	& 0.134_{-0.007}^{+0.007}\\
Kepler-105	& c	& \cdots	& \cdots	& \cdots	& 0.124_{-0.007}^{+0.006}\\
Kepler-36	& b	& 0.126_{-0.025}^{+0.023}	& \cdots	& 0.120_{-0.009}^{+0.009}	& 0.133_{-0.007}^{+0.007}\\
Kepler-93	& b	& 0.086_{-0.031}^{+0.030}	& 0.129_{-0.016}^{+0.016}	& 0.111_{-0.006}^{+0.005}	& 0.113_{-0.007}^{+0.006}\\
Kepler-406	& b	& \cdots	& 0.133_{-0.022}^{+0.022}	& \cdots	& 0.133_{-0.007}^{+0.007}\\
Kepler-78	& b	& 0.086_{-0.023}^{+0.024}	& 0.134_{-0.047}^{+0.040}	& \cdots	& 0.152_{-0.007}^{+0.007}\\
Kepler-107	& c	& 0.112_{-0.037}^{+0.034}	& 0.134_{-0.047}^{+0.042}	& \cdots	& 0.144_{-0.011}^{+0.011}\\
Kepler-99	& b	& \cdots	& \cdots	& 0.118_{-0.005}^{+0.005}	& 0.145_{-0.008}^{+0.007}\\
KOI-1599	& .01	& \cdots	& \cdots	& 0.111_{-0.008}^{+0.008}	& \cdots\\
WASP-47	& e	& 0.105_{-0.036}^{+0.033}	& 0.109_{-0.029}^{+0.027}	& 0.118_{-0.005}^{+0.004}	& \cdots\\
HD 213885	& b	& 0.130_{-0.026}^{+0.024}	& 0.133_{-0.016}^{+0.016}	& \cdots	& \cdots\\
K2-265	& b	& 0.111_{-0.026}^{+0.024}	& 0.123_{-0.016}^{+0.016}	& 0.116_{-0.005}^{+0.004}	& 0.137_{-0.007}^{+0.007}\\
HD 219134	& b	& 0.075_{-0.025}^{+0.023}	& 0.129_{-0.029}^{+0.028}	& \cdots	& \cdots\\
HD 219134	& c	& 0.075_{-0.025}^{+0.023}	& 0.129_{-0.029}^{+0.028}	& \cdots	& \cdots\\
K2-141	& b	& 0.177_{-0.055}^{+0.050}	& 0.170_{-0.061}^{+0.053}	& 0.133_{-0.009}^{+0.008}	& \cdots\\
\hline
\multicolumn{6}{l}{\textbf{Fe/Si-based CMF}} \\
K2-216	& b	& \cdots	& 0.151_{-0.091}^{+0.076}	& 0.146_{-0.006}^{+0.006}	& \cdots\\
K2-106	& b	& 0.151_{-0.038}^{+0.035}	& 0.171_{-0.023}^{+0.022}	& 0.135_{-0.006}^{+0.006}	& \cdots\\
HD 15337	& b	& 0.143_{-0.040}^{+0.037}	& \cdots	& \cdots	& \cdots\\
K2-291	& b	& 0.166_{-0.038}^{+0.036}	& 0.171_{-0.019}^{+0.018}	& \cdots	& \cdots\\
55 Cnc	& e	& 0.130_{-0.044}^{+0.040}	& 0.138_{-0.023}^{+0.022}	& \cdots	& 0.179_{-0.008}^{+0.007}\\
HD 80653	& b	& 0.161_{-0.042}^{+0.039}	& \cdots	& \cdots	& \cdots\\
K2-131	& b	& 0.127_{-0.030}^{+0.029}	& \cdots	& \cdots	& \cdots\\
K2-229	& b	& 0.073_{-0.030}^{+0.030}	& 0.150_{-0.023}^{+0.021}	& \cdots	& \cdots\\
HD 136352	& b	& 0.133_{-0.035}^{+0.032}	& \cdots	& \cdots	& \cdots\\
K2-38	& b	& 0.144_{-0.038}^{+0.036}	& 0.146_{-0.040}^{+0.036}	& 0.112_{-0.005}^{+0.005}	& 0.156_{-0.006}^{+0.007}\\
Kepler-10	& b	& 0.162_{-0.035}^{+0.033}	& 0.134_{-0.013}^{+0.012}	& 0.104_{-0.007}^{+0.008}	& 0.125_{-0.010}^{+0.011}\\
Kepler-20	& b	& 0.172_{-0.040}^{+0.036}	& 0.146_{-0.020}^{+0.019}	& 0.129_{-0.008}^{+0.009}	& 0.139_{-0.007}^{+0.006}\\
Kepler-105	& c	& \cdots	& \cdots	& \cdots	& 0.132_{-0.006}^{+0.006}\\
Kepler-36	& b	& 0.154_{-0.042}^{+0.037}	& \cdots	& 0.117_{-0.010}^{+0.009}	& 0.146_{-0.007}^{+0.005}\\
Kepler-93	& b	& 0.186_{-0.038}^{+0.033}	& 0.158_{-0.010}^{+0.009}	& 0.121_{-0.006}^{+0.005}	& 0.147_{-0.006}^{+0.007}\\
Kepler-406	& b	& \cdots	& 0.144_{-0.065}^{+0.055}	& \cdots	& 0.154_{-0.007}^{+0.006}\\
Kepler-78	& b	& 0.136_{-0.036}^{+0.033}	& 0.130_{-0.034}^{+0.032}	& \cdots	& 0.157_{-0.007}^{+0.006}\\
Kepler-107	& c	& 0.144_{-0.052}^{+0.047}	& 0.171_{-0.025}^{+0.024}	& \cdots	& 0.165_{-0.011}^{+0.011}\\
Kepler-99	& b	& \cdots	& \cdots	& 0.138_{-0.005}^{+0.004}	& 0.149_{-0.006}^{+0.007}\\
KOI-1599	& .01	& \cdots	& \cdots	& 0.116_{-0.009}^{+0.008}	& \cdots\\
WASP-47	& e	& 0.134_{-0.040}^{+0.036}	& 0.138_{-0.029}^{+0.026}	& 0.104_{-0.004}^{+0.004}	& \cdots\\
HD 213885	& b	& 0.152_{-0.037}^{+0.035}	& 0.154_{-0.014}^{+0.014}	& \cdots	& \cdots\\
K2-265	& b	& 0.102_{-0.034}^{+0.032}	& 0.168_{-0.020}^{+0.018}	& 0.135_{-0.006}^{+0.005}	& 0.158_{-0.006}^{+0.006}\\
HD 219134	& b	& 0.140_{-0.046}^{+0.041}	& 0.138_{-0.032}^{+0.029}	& \cdots	& \cdots\\
HD 219134	& c	& 0.140_{-0.046}^{+0.042}	& 0.138_{-0.031}^{+0.030}	& \cdots	& \cdots\\
K2-141	& b	& 0.109_{-0.033}^{+0.032}	& 0.138_{-0.059}^{+0.052}	& 0.161_{-0.010}^{+0.011}	& \cdots
\enddata
\tablecomments{Some mass--radius-based core-mass fractions are negative,
indicating a terrestrial exoplanet that is inconsistent with pure rock
compositions according to \texttt{SuperEarth}.}
\end{deluxetable*}

\bibliography{article}
\bibliographystyle{aasjournalv7}

\end{document}